\documentclass[a4paper,11pt]{article}
\usepackage{jheppub}
\usepackage{appendix}
\usepackage{color}
\usepackage{caption}
\usepackage{amssymb}
\usepackage{braket}
\usepackage{amsthm}
\usepackage{mathtools}
\usepackage{physics}
\usepackage{hyperref}
\usepackage{etoolbox}
\usepackage[utf8]{inputenc} 

\renewcommand\arraystretch{1.4}
\patchcmd{\pprintMaketitle}
  {\fi\hrule}
  {\fi\ifvoid\extrainfobox\else\unvbox\extrainfobox\par\vskip10pt\fi\hrule}
  {}{}

\newsavebox\extrainfobox

\title{Model-independent bounds on new physics effects in  non-leptonic tree-level decays
	of B-mesons}

\author[a]{Alexander Lenz}
\author[b,c]{Gilberto Tetlalmatzi-Xolocotzi\footnote{Corresponding author.}}

\affiliation[a]{Institute for Particle Physics Phenomenology, Durham University,\\DH1 3LE Durham, United Kingdom}
\affiliation[b]{Theoretische Physik 1,  Naturwissenschaftlich-Technische Fakult\"{a}t, Universit\"{a}t Siegen,\\Walter-Flex-Strasse 3, D-57068 Siegen, Germany}
\affiliation[c]{Nikhef, Theory Group, Science Park 105, 1098 XG, Amsterdam, The Netherlands}

\emailAdd{alexander.lenz@durham.ac.uk}
\emailAdd{gtx@physik.uni-siegen.de}


\abstract{
	We present a considerably improved analysis of model-independent bounds on new physics effects 
	in  non-leptonic tree-level decays of B-mesons.
	Our main finding is that  contributions of about $\pm 0.1 $ 
	to the Wilson coefficient of the colour-singlet operator $Q_2$ of the effective weak Hamiltonian 
	and contributions in the range of $\pm 0.5$ (both for real and imaginary part) to $Q_1$ can currently not be excluded at the $90\%$ C.L..
	Effects of such a size can modify the direct experimental extraction of the CKM angle
	$\gamma$ by up to $10^{\circ}$ and they could lead to an enhancement of the decay rate
	difference $\Delta \Gamma_d$ of up to a factor of 5 over its SM value - a size that could explain the D0 dimuon asymmetry. 
	Future more precise measurements of the semi-leptonic asymmetries $a_{sl}^q$
	and the lifetime ratio $\tau (B_s) /  \tau (B_d)$ will allow to shrink the 
	bounds on tree-level new physics effects considerably.
	Due to significant improvements in the precision of the non-perturbative input
	we update all SM predictions for the mixing obervables in the course of this analysis, obtaining:
	$\Delta M_s = (18.77 \pm 0.86 ) \, \mbox{ps}^{-1}$, 
	$\Delta M_d = (0.543 \pm 0.029) \, \mbox{ps}^{-1}$,
	$\Delta \Gamma_s = (9.1 \pm 1.3 ) \cdot 10^{-2} \, \mbox{ps}^{-1}$, 
	$\Delta \Gamma_d = (2.6 \pm 0.4 ) \cdot 10^{-3} \, \mbox{ps}^{-1}$,
	$a_{sl}^s = (2.06 \pm 0.18) \cdot 10^{-5}$ and
	$a_{sl}^d = (-4.73 \pm 0.42) \cdot 10^{-4}$. }

\preprint{ IPPP/19/49
	
	\vspace{-0.45cm}
	
	\begin{flushright}
		Nikhef-2019-054 \\
	    SI-HEP-2019     \\
	    P3H-19-044
	\end{flushright}
}

\begin{document}
\maketitle
\flushbottom

\section{Introduction}
\label{sec:intro}
Motivations for flavour physics are manifold.
Standard model parameters, like the  elements of the Cabibbo-Kobayashi-Maskawa (CKM) matrix 
\cite{Cabibbo:1963yz,Kobayashi:1973fv}
or quark masses are determined very accurately in this field. Moreover the quark-sector is the only sector, where 
CP violating  effects have been detected so far - since 1964 in the Kaon sector \cite{Christenson:1964fg}
and since 2001 also in the B-sector \cite{Aubert:2002ic,Abe:2002px}.
Very recently CP violation has been measured for the first time in the charm sector \cite{Aaij:2019kcg}, which might
actually be an indication for physics beyond the standard model (BSM) \cite{Chala:2019fdb,Dery:2019ysp}.
Considering that CP violation is a 
necessary ingredient for creating a baryon asymmetry in the universe \cite{Sakharov:1967dj}, 
flavour physics might shed some light on this unsolved problem.
In addition flavour physics is perfectly suited for indirect new physics (NP) searches,
because there are many processes strongly suppressed in the standard 
model (SM) but not necessarily in hypothetical NP models.
And, last but not least, a comparison between experiment and theory predictions
can provide a deeper insight into the dynamics of QCD.
\\
In recent years experimental flavour physics entered a new precision era,
which was initiated by the B-factories at KEK and SLAC (see e.g. \cite{Bevan:2014iga}) and the Tevatron at
Fermilab \cite{Anikeev:2001rk,Borissov:2013fwa}. Currently this field is dominated by the results of the LHCb 
collaboration \cite{Bediaga:2012py,Koppenburg:2015pca}, but also  complemented by competing results from the 
general purpose detectors ATLAS and CMS, see e.g. \cite{Aad:2016tdj,Khachatryan:2015nza}.
\\
The corresponding dramatic increase in experimental precision, demands complementary improvements in
theory. Besides calculating higher orders in perturbative QCD or more precise lattice evaluations, this also
means revisiting some common approximations by investigating questions like:
How large are penguin contributions? 
How well does QCD-factorization \cite{Beneke:1999br,Beneke:2000ry,Beneke:2001ev, Beneke:2003zv} work?
How large can duality violation in the Heavy Quark Expansion (HQE)
(see e.g. \cite{Khoze:1983yp,Shifman:1984wx,Bigi:1991ir,Bigi:1992su,Blok:1992hw,Blok:1992he,Chay:1990da,Luke:1990eg}
for pioneering papers and \cite{Lenz:2015dra} for a recent review) be? 
How sizeable NP effects in tree-level decays can be?
Some of these questions have been studied in detail for quite some time.
There is e.g. a huge literature on penguin contributions, see 
e.g. \cite{Fleischer:2015mla,Artuso:2015swg} for reviews. Others gained interest recently, for instance duality
violations \cite{Jubb:2016mvq}. 
In principle all these questions are interwoven, but as a starting point it is reasonable to consider them
separately.
The assumption of no NP effects at tree-level in non-leptonic $b$-decays was already challenged after the 
measurement of the dimuon asymmetry by the D0-collaboration \cite{Abazov:2010hv,Abazov:2010hj,Abazov:2011yk,Abazov:2013uma},
see e.g. \cite{Bauer:2010dga}. And after the measurements of  $B \to D^{(*)} \tau \nu$ by BaBar, Belle and LHCb
\cite{Lees:2012xj,Aaij:2015yra,Huschle:2015rga,Abdesselam:2016cgx} for the case of semi-leptonic $b$-decays.
\\
Compared to numerous systematic studies of NP effects in the 
Wilson coefficients of the electromagnetic dipole and the
semi-leptonic penguin operators $Q_{7\gamma}$ and  
$Q_9$, $Q_{10}$ respectively, see e.g. 
\cite{Jager:2014rwa, Altmannshofer:2015sma, Descotes-Genon:2015uva,Ciuchini:2015qxb, Hurth:2016fbr,Alguero:2019ptt,Ciuchini:2019usw,Datta:2019zca,Aebischer:2019mlg,Kowalska:2019ley,Arbey:2019duh,Biswas:2020uaq,Bhom:2020lmk},
we are not aware of systematic studies for NP effects in the Wilson coefficients for
non-leptonic tree-level decays,
except the ones in \cite{Bobeth:2014rda,Bobeth:2014rra,Brod:2014bfa,
Jager:2017gal,
Aebischer:2018csl, Jager:2019bgk}
\footnote{
In \cite{Crivellin:2019isj} NP entering inside $Q_5$ and $Q_6$ is explored, establishing a link between the $B\rightarrow K \pi$ puzzle and 
the $\varepsilon'/\varepsilon$ ratio. This is further addressed in \cite{Calibbi:2019lvs} within the context of simplified $Z'$ models with 
$U(2)^3$ flavour symmetry.}.
\\
The aim of the current paper is to considerably extend the studies in \cite{Bobeth:2014rda,Brod:2014bfa} by
incorporating two main improvements:
\begin{enumerate}
\item A full $\chi^2$-fit is performed instead of a simple parameter scan. To implement this step we use the package MyFitter
      \cite{Wiebusch:2012en} and allow the different nuisance parameters to run independently. This will allow us to account
      properly for the corresponding statistical correlations.\\
\item Instead of simplified theoretical equations we include  full expressions for the observables under investigation.
\end{enumerate}
The recent work in \cite{Jager:2017gal,Jager:2019bgk} concentrates exclusively on the transition $b \to c \bar{c} s$, while we consider in this paper
all different hadronic decays, that occur in the SM on tree-level. Moreover in this work we consider only BSM effects to the tree-level operators $Q_1$
and $Q_2$, while \cite{Jager:2017gal,Jager:2019bgk} investigates also effects of four-quark operators that do not exist in the SM.
Whenever there is some direct overlap between the work in \cite{Jager:2017gal,Jager:2019bgk} we directly compare the results.
Any realistic BSM model that gives rise to new tree-level effects will  also give new effects at the loop-level, which are not considered
in the current model independent approach. In that respect this work can be considered as an important building block of future  model dependent studies.

The paper is organised as follows:
In Section \ref{sec:basic} we describe briefly the theoretical tools to be used: we start with  the effective Hamiltonian in 
Section \ref{sec:Heff}, then in Section \ref{sec:HQE} we introduce the Heavy Quark Expansion and in Section \ref{sec:QCDF}
we review basic concepts in QCD factorization relevant to this project.
Next in Section \ref{sec:strategy},  we outline our strategy for performing the $\chi^2$-fit.  
We discuss all our different constraints on NP effects in non-leptonic 
tree-level decays in Section \ref{sec:constraints}. The bounds on individual decay channels are organized as follows: 
$b \to c \bar{u} d$ in Section \ref{sec:bcud},
$b \to u \bar{u} d$ in Section \ref{sec:buud},
$b \to c \bar{c} s$ in Section \ref{sec:bccs},
$b \to c \bar{c} d$ in Section \ref{sec:bccd}. Additionally, in
Section \ref{sec:multiple_channels} we present observables constraining more decay channels.
Our main results are presented in Section
\ref{sec:Globalfitresults}:
fits for the allowed size of BSM effects in the tree-level Wilson coefficients  based on individual decay channels will be discussed in Sections
\ref{sec:buudfit} -
\ref{sec:bccdfit}.
In particular we focus on the channels which can enhance the decay rate difference of 
neutral $B_d^0$-mesons $\Delta \Gamma_d$ and we calculate these enhancements. 
Flavour-universal bounds on the tree level Wilson coefficients will be presented in
Section \ref{sec:CKMgamma}, with an emphasis on the consequences of tree-level NP
effects on the precision in
the direct extraction of the CKM angle $\gamma$.
In Section \ref{sec:future} we study observables that seem to be most promising in shrinking the 
space for new effects in $C_1$ and $C_2$.
Finally we conclude in Section \ref{sec:conclusion} and give additional information in the appendices.
\\
Since there has been tremendous progress (see e.g.\cite{King:2019rvk,DiLuzio:2019jyq})
in the theoretical precision of the mixing observables, we will present in this work numerical updates
of all mixing observables:
$\Delta \Gamma_q$ in Section \ref{subsec:DGs},
$\Delta M_q$ in Section \ref{sec:sin2betaM12}
and the semi-leptonic CP asymmetries
$a_{sl}^q$ and mixing phases $\phi_q$ in Section \ref{sec:multiple_channels}.

%
%
%
%
%
%
%
%
%
%
%
%
%
%
%
%
%
%
%
%
\section{Basic formalism}
\label{sec:basic}
In this section we provide an overview of the basic theoretical tools required for the description of our different flavour observables,
this includes: the effective Hamiltonian, the Heavy Quark Expansion for 
inclusive decays and mixing observables. A quick review of QCD factorization for exclusive, non-leptonic decays is also provided.
In addition we fix the notation to be used during this work.

\subsection{Effective Hamiltonian}
\label{sec:Heff}
We start by introducing the effective Hamiltonian describing a $b$-quark decay into a $p \bar{p}' q$ final state via electroweak interactions, with 
$p, p'= u, c$ and $q=s, d $: 

\begin{eqnarray}
{\cal \hat{H}}^{|\Delta B|=1}_{eff}&=&
\frac{G_F}{\sqrt{2}}\left\{
\sum_{p,p'=u,c}\lambda^{(q)}_{pp'}\sum_{i=1,2}C^{q, \, pp'}_{i}(\mu)\hat{Q}^{q, \, pp'}_{i}
\right.
\nonumber\\
&&
\left.
+ \sum_{p=u,c}\lambda^{(q)}_p \left[
\sum^{10}_{i=3} C^q_{i}(\mu)\hat{Q}^q_{i} + C^q_{7\gamma} \hat{Q}^q_{7\gamma} + C^q_{8 g} \hat{Q}^q_{8 g} \right]
\right\} + h. c. \, .
\label{eq:Hamiltonian}
\end{eqnarray}
The Fermi constant is denoted by $G_F$, additionally we have introduced  the following CKM combinations
\begin{eqnarray}
\lambda^{(q)}_p&=&V_{pb}V^*_{pq} \, ,
\nonumber\\
\lambda^{(q)}_{pp'}&=&V_{pb}V^*_{p'q} \, .
\label{eq:lambmdadef}
\end{eqnarray}
Moreover $C_i$ denote the Wilson coefficients of the following dimension six operators:

\begin{align}
\hat{Q}_{1}^{q, \, pp'}&=\Bigl(\bar{\hat{p}}_{\beta}\hat{b}_{\alpha}\Bigl)_{V-A}
                      \Bigl(\bar{ \hat{q} }_{\alpha}   \hat{p}'_{\beta}\Bigl)_{V-A}
                      \, , & 
\hat{Q}_{2}^{q, \, pp'}=& \Bigl(\bar{\hat{p}} \hat{b}\Bigl)_{V-A} \Bigl(\bar{\hat{q}} \hat{p}'\Bigl)_{V-A}\, ,
\nonumber    \\
\hat{Q}^q_{3}&=\Bigl(\bar{\hat{q}}\hat{b}\Bigl)_{V-A}            \sum_k\Bigl(\bar{\hat{k}}\hat{k}\Bigl)_{V-A}  \, ,
& 
\hat{Q}^q_{4}=&\Bigl(\bar{\hat{q}}_\alpha \hat{b}_\beta\Bigl)_{V-A}\sum_k\Bigl(\bar{\hat{k}}_\beta \hat{k}_\alpha\Bigl)_{V-A}\, ,
\nonumber\\
\hat{Q}^q_{5}&=\Bigl(\bar{\hat{q}} \hat{b}\bigl)_{V-A}           \sum_k\Bigl(\bar{\hat{k}}\hat{k}\Bigl)_{V+A} \, ,
& \hat{Q}^q_{6}=&\Bigl(\bar{\hat{q}}_\alpha \hat{b}_\beta\bigl)_{V-A}\sum_k\Bigl(\bar{\hat{k}}_\beta \hat{k}_\alpha\Bigl)_{V+A}\, ,
\nonumber\\
\hat{Q}^q_{7}&=\Bigl(\bar{\hat{q}} \hat{b}\bigl)_{V-A}\sum_k\frac{3}{2}e_k\Bigl(\bar{\hat{k}} \hat{k}\Bigl)_{V+A} \, ,
& \hat{Q}^q_{8}=&\Bigl(\bar{\hat{q}}_\alpha \hat{b}_\beta\bigl)_{V-A}\sum_k\frac{3}{2}e_k\Bigl(\bar{\hat{k}}_\beta \hat{k}_\alpha\Bigl)_{V+A}\, ,
\nonumber\\
 \hat{Q}^q_{9 }&=\Bigl(\bar{\hat{q}}       \hat{b}      \Bigl)_{V-A}\sum_{k}\frac{3}{2}e_k\Bigl(\bar{\hat{k}}       \hat{k}    \Bigl)_{V-A}\, ,
&\hat{Q}^q_{10}=&\Bigl(\bar{\hat{q}}_\alpha \hat{b}_\beta \Bigl)_{V-A}\sum_{k}\frac{3}{2}e_k\Bigl(\bar{\hat{k}}_\beta \hat{k}_{\alpha} \Bigl)_{V-A}\, ,
\nonumber\\
\hat{Q}^q_{7\gamma}=&\frac{e}{8\pi^2}m_b\bar{\hat{q}}\sigma_{\mu\nu}\Bigl(1+\gamma_5 \Bigl)\hat{F}^{\mu\nu}\hat{b}\, ,
&\hat{Q}^q_{8 g}=&\frac{g_s}{8\pi^2}m_b\bar{\hat{q}}\sigma_{\mu\nu}\Bigl(1+\gamma_5 \Bigl)\hat{G}^{\mu\nu}\hat{b}\, .
\label{eq:mainbasis}
\end{align}

Here $\alpha$ and $\beta$ are colour indices, $e_k$ is the electric charge of the quark $k$
(in the penguin operators the quark flavours are summed over $k= u,d,s,c,b$),
$e$ is the $U(1)_{\rm em}$ coupling and $g_s$ the $SU(3)_C$ one, $m_b$ is the mass of the $b$-quark 
and $F^{\mu\nu}$ and $G^{\mu\nu}$ are the electro-magnetic and chromo-magnetic field strength tensors respectively.
In this work we consider NP effects that will affect the tree-level operators 
$\hat{Q}_{1}^{q, \, pp'}$ and $\hat{Q}_{2}^{q, \, pp'}$ by modifying their corresponding Wilson coefficients.
In our notation  $ \hat{Q}_{1}^{q, \, pp'}$ is colour non diagonal and $ \hat{Q}_{2}^{q, \, pp'}$ is the colour singlet,
the QCD penguin operators correspond to $\hat{Q}^q_{3-6}$ and the electro-weak penguin 
interactions are described by $\hat{Q}^q_{7-10}$.
Different bases compared to the one in Eq.~(\ref{eq:mainbasis}) 
 are used in the literature. Our notation agrees with the one used in \cite{Buchalla:1995vs} and \cite{Beneke:1998sy}, here $C_{8g}$ is negative 
because we are considering $-i g \gamma_\mu T^a$ as the Feynman 
rule for the quark-gluon vertex.
In \cite{Beneke:2001ev} a different basis is used, where $\hat{Q}_1$ and $\hat{Q}_2$ are interchanged 
and $\hat{Q}_{7 \gamma}$ and $\hat{Q}_{8g}$ have a different sign (this is equivalent 
to the sign convention $i D^\mu = i \partial^\mu + g_s A_a^\mu T^a$ for the gauge-covariant derivative)
\footnote{A minimal basis  of dimension six operators for $\Delta B\neq 0$ processes 
has been introduced in \cite{Aebischer:2017gaw}. This extends our set of operators in Eq.~(\ref{eq:mainbasis}). For  the purposes of studying NP in tree-level non-leptonic 
operators the basis in Eq.~(\ref{eq:mainbasis}) is enough. However, future extension which include NP in other operators as well, should be done paying attention to the results presented in \cite{Aebischer:2017gaw}.}.
A nice introduction on effective Hamiltonians can be found in \cite{Buras:1998raa}, and a concise review up to NLO-QCD in \cite{Buchalla:1995vs}.
\\
\label{RGE}
The Wilson coefficients $C_i$ with $i= 1,2,...,10,7\gamma,8g$ in Eq.~(\ref{eq:Hamiltonian}) are obtained by matching the calculations 
of the effective theory and the full SM
at the scale $\mu=M_W$ and then evolving down to the scale $\mu \sim m_b$ using 
the renormalisation group equations according to
\begin{eqnarray}
\vec{C}(\mu)&=&\textit{\textbf{U}}(\mu, M_W, \alpha)\vec{C}(M_W) \, ,
\label{eq:fullevo}
\end{eqnarray}
where the  NLO evolution matrix is given by \cite{Beneke:2001ev}
\vspace{-0.2cm}
\begin{eqnarray}
\textit{\textbf{U}}(\mu, M_W, \alpha)&=&\textit{\textbf{U}}(\mu,\mu_W) +
\frac{\alpha}{4\pi} \textit{\textbf{R}}(\mu,\mu_W).
\label{eq:fullevoMat}
\end{eqnarray}
The matrix $\textbf{U}(\mu,\mu_W)$ accounts for pure QCD evolution, on the other hand $\textbf{R}(\mu,\mu_W)$ introduces
QED effects as well. We write at NLO \cite{Beneke:2001ev}
\begin{eqnarray}
\textit{\textbf{U}}(\mu, M_W, \alpha)
&=&
\Bigl[\textit{\textbf{U}}_0 + \frac{\alpha_{s}(\mu)}{4\pi}\textit{\textbf{J}\textbf{U}}_0 -\frac{\alpha_s(M_W)}{4\pi}\textit{\textbf{U}}_0 
\textit{\textbf{J}} 
\nonumber\\
&& 
+ \frac{\alpha}{4\pi}\Bigl(\frac{4\pi}{\alpha_s(\mu)}\textit{\textbf{R}}_0 + \textit{\textbf{R}}_1\Bigl) \Bigl],
\label{eq:ev_matrix}
\end{eqnarray}
where $\alpha_s(\mu)$ denotes the strong coupling at the scale $\mu$  calculated up to NLO-QCD precision and
$\alpha$ is the electro-magnetic coupling. The matrix $\textit{\textbf{U}}_0$ is the LO of the 
pure QCD evolution component $\textbf{U}(\mu,\mu_W)$. At LO the evolution matrix  $\textit{\textbf{U}}(\mu, M_W, \alpha)$ reduces to 

\begin{eqnarray}
\label{Eq:LOevo}
\textit{\textbf{U}}^{\rm LO}(\mu,\mu_W, \alpha) &=&\textit{\textbf{U}}_0 +  \frac{\alpha}{\alpha_s(\mu)} \textit{\textbf{R}}_0.
\end{eqnarray}

The NLO-QCD corrections are then introduced through $\textit{\textbf{J}}$.
The explicit expressions for $\textit{\textbf{U}}_0$ and 
$\textit{\textbf{J}}$ are given in Eqns.~(3.94)-(3.98) of \cite{Buchalla:1995vs}. The anomalous dimension matrices 
$\mathbf{\gamma}_s^{(0)}$ and ${\bf \gamma}_s^{(1)}$ required for these evaluations can be found in Eqn.~(6.25) and
Tables XIV and XV of 
\cite{Buchalla:1995vs}.
To introduce QED corrections we calculate ${\bf{R_0}}$ and  ${\bf{R_1}}$ using Eqns.~(7.24)-(7.28)
of \cite{Buchalla:1995vs}, the anomalous dimension matrices used are  $\gamma^{(0)}_{e}$  and $\gamma^{(1)}_{e}$
and are given in Tables XVI and XVII of \cite{Buchalla:1995vs}.
\\
The initial conditions for the Wilson coefficients have the following expansion at NLO
\begin{eqnarray}
\vec{C}(M_W)&=&\vec{C}^{(0)}_s(M_W) + \frac{\alpha_s(M_W)}{4\pi}\vec{C}^{(1)}_s(M_W)\nonumber\\
&&+
\frac{\alpha}{4\pi}
\left[
\vec{C}^{(0)}_e(M_W) + \frac{\alpha_s(M_W)}{4\pi}\vec{C}^{(1)}_e(M_W) + \vec{R}^{(0)}_e(M_W)
\right]
\label{eq:CMW},
\end{eqnarray}
as pointed out in \cite{Beneke:2001ev} the electroweak contributions $\vec{C}^{(0)}_e$ and $\vec{C}^{(1)}_e$ in Eq.~(\ref{eq:CMW}) can be
$x_t$ and/or $1/\sin^2\theta_W$ enhanced.
Consequently it is fair to treat the product between  $\alpha$ and $\vec{C}^{(0)}_e$
as a LO contribution and the product between $\alpha$ and $\vec{C}^{(1)}_e$ as a NLO effect.
The remainder, denoted by $\vec{R}^{(0)}_e$, is numerically smaller in comparison with
$\vec{C}^{(0)}_e$ and it is therefore treated as a NLO effect,
it contains the NLO scheme dependency.
This approach differs from the one followed by \cite{Buchalla:1995vs}, where the contribution of
$\vec{C}^{(0)}_e(M_W) + \vec{R}^{(0)}_e(M_W)$ is introduced as a NLO effect and then $\vec{C}^{(1)}_e$ is omitted.
The explicit expressions for $\vec{C}^{(0)}_s$, $\vec{C}_s^{(1)}$, $\vec{C}^{(0)}_e$, $\vec{C}^{(1)}_e$
and $\vec{R}^{(0)}_e$ of $\vec{C}(M_W)$ are given in Section VII.B of \cite{Buchalla:1995vs} and Section 3.1 of \cite{Beneke:2001ev}, 
the results presented for $\vec{C}^{(1)}_e$ in \cite{Beneke:2001ev} are based on the calculations of \cite{Buras:1999st}.
\\
It should be further stressed that when applying Eq.~(\ref{eq:fullevo})
 we consistently dropped products between NLO contributions from
$\textit{\textbf{U}}(\mu, M_W, \alpha)$ and NLO effects from $\vec{C}(M_W)$ but we have taken into account
 products between NLO contributions
from $\textit{\textbf{U}}(\mu, M_W, \alpha)$ and LO contributions from $\vec{C}(M_W)$ and vice versa.
\subsection{Heavy Quark Expansion}
\label{sec:HQE}
The effective Hamiltonian can be used to calculate inclusive decays of a heavy hadron $B_q$ into
an inclusive final state $X$ via
\begin{equation}
 \Gamma ( B_q \to X) = \frac{1}{2 m_{B_q}} \sum \limits_{X} \int_{\rm PS} (2 \pi)^4 \delta^{(4)}
 (p_{B_q} - p_X) | \langle X | {\cal \hat{H}}_{eff} | B_q \rangle |^2 
\, .
\label{total}
 \end{equation}
With the help of the optical theorem the total decay rate in Eq.~(\ref{total})  can be rewritten as
 \begin{equation}
 \Gamma(B_q \to X) = \frac{1}{2 m_{B_q}} \langle B_q |{\cal \hat{T} } | B_q \rangle
\, ,
 \end{equation}
with the transition operator
 \begin{equation}
 {\cal \hat{T}} = \mbox{Im} \; i \int d^4x 
 \hat{T} \left[ {\cal  \hat{H} }_{eff}(x) {\cal \hat{H} }_{eff} (0) \right] 
\, ,
\label{trans}
 \end{equation}
consisting of a non-local double insertion of the effective Hamiltonian.
Expanding this bi-local object in local operators gives the
 Heavy Quark Expansion
(see e.g. \cite{Khoze:1983yp,Shifman:1984wx,Bigi:1991ir,Bigi:1992su,Blok:1992hw,Blok:1992he,Chay:1990da,Luke:1990eg}
for pioneering papers and \cite{Lenz:2015dra} for a recent review).
The total decay rate $\Gamma$ of a $b$-hadron can then be expressed as products of perturbatively calculable
coefficients $\Gamma_i$ times non-perturbative matrix elements $\langle O_{D} \rangle$ of $\Delta B = 0$-operators of dimension
$D = i+3$:
\begin{eqnarray}
\Gamma & = & \Gamma_0     \langle \hat{O}_{D=3} \rangle 
+ \Gamma_2          \frac{\langle \hat{O}_{D=5} \rangle}{m_b^2} 
	+ \tilde{\Gamma}_3  \frac{\langle \tilde{\hat{O}}_{D=6} \rangle}{m_b^3} 
+ ...
\nonumber
\\
&&+ 16 \pi^2 \left[
\Gamma_3 \frac{\langle \hat{O}_{D=6} \rangle}{m_b^3} 
	+  \Gamma_4 \frac{\langle \hat{O}_{D=7} \rangle}{m_b^4} 
	+  \Gamma_5 \frac{\langle \hat{O}_{D=8} \rangle}{m_b^5} 
+ ...
\right]\, ,
\end{eqnarray}
with $\langle \hat{O}_{D} \rangle = \langle B_q | \hat{O}_{D} | B_q  \rangle/(2 M_{B_q})$.
The leading term $\Gamma_0$ describes the decay of a free $b$-quark and is free of non-perturbative uncertainties, since
$\langle \hat{O}_{D=3} \rangle = 1 + {\cal O} (\langle \hat{O}_{D=5} \rangle/ m_b^2 )$.
At order $1/m_b^2$ small corrections due to the kinetic and chromomagnetic operator are arising, at order  $1/m_b^3$ we get e.g.
the Darwin term in $\tilde{\Gamma}_3$, but also phase space enhanced terms $\Gamma_3$, stemming from weak exchange,
weak annihilation and Pauli interference.
The numerical values of the matrix elements are expected to be of the order the hadronic scale $\Lambda_{QCD}$, thus the HQE
is an expansion in the small parameter $\Lambda_{QCD}/m_b$.
Each of the terms  $\Gamma_i$ with $i=0,2,3,...$ can be expanded as 
\begin{equation}
\Gamma_i = \Gamma_i^{(0)} + \frac{\alpha_s}{4 \pi} \Gamma_i^{(1)} 
                          + \left( \frac{\alpha_s}{4 \pi} \right)^2 \Gamma_i^{(2)} +
                           ... \, .
\end{equation}
In our investigation of the  lifetimes we will use $\Gamma_0^{(0)}$ and  $\Gamma_0^{(1)}$  from
\cite{Krinner:2013cja}, which is based on \cite{Bagan:1994zd,Bagan:1995yf,Lenz:1997aa,Lenz:1998qp,Greub:2000an,Greub:2000sy},
$\Gamma_3^{(0)}$ from \cite{Jager:2017gal} based on \cite{Uraltsev:1996ta,Neubert:1996we} and
$\Gamma_3^{(1)}$ from \cite{Beneke:2002rj,Franco:2002fc}.
The matrix elements of the dimension six operators were recently determined in \cite{Kirk:2017juj}.
\\
The HQE can also be used to describe the off-diagonal element $\Gamma_{12}$ of the meson mixing matrix
\begin{equation}
  \Gamma_{12}^q =  \left[ \Gamma_{12,3}^{q,(0)} +  \frac{\alpha_s}{4 \pi} \Gamma_{12,3}^{q,(1)} + ... \right]
		  \frac{\langle \hat{Q}_{D=6} \rangle}{m_b^3}  
                +  \left[ \Gamma_{12,4}^{q,(0)} +  \frac{\alpha_s}{4 \pi} \Gamma_{12,4}^{q,(1)} + ... \right]
		  \frac{\langle \hat{Q}_{D=7} \rangle}{m_b^4}  
             + ... \, ,
\end{equation}
with $\langle \hat{Q}_{D} \rangle = \langle B_q | Q_{D} | \bar{B}_q  \rangle/(2 M_{B_q})$, where 
$\hat{Q}_D$ are $\Delta B = 2$-operators of dimension $D$.
The matrix element $\Gamma_{12}^q$ can be used together with $M_{12}^q$ to predict physical observables like mass differences,  decay rate differences
or semi-leptonic CP-asymmetries, see e.g. \cite{Artuso:2015swg}
\begin{eqnarray}
  \Delta M_q & = & 2 |M_{12}^q|  \, ,\label{eq:dMq}
\\
\Delta \Gamma_q & = & 2 |\Gamma_{12}^q| \cos \phi_{12}^q = - {\rm Re}\left( \frac{\Gamma_{12}^q}{M_{12}^q} \right) \Delta M_q \, ,\label{eq:dGammaq}
\\
a_{sl}^q & = &  \left| \frac{\Gamma_{12}^q}{M_{12}^q} \right| \sin \phi_{12}^q = {\rm Im} \left( \frac{\Gamma_{12}^q}{M_{12}^q} \right) \, ,\label{eq:aslq}
\end{eqnarray}
with the phase $\phi_{12}^q = \arg (- M_{12}^q / \Gamma_{12}^q)$.
For our numerical analysis we use
results for $\Gamma_{12,3}^{q,(0)}$, $\Gamma_{12,3}^{q,(1)}$ and $\Gamma_{12,4}^{q,(0)}$ from 
\cite{Beneke:1998sy,Beneke:2002rj,Beneke:1996gn,Dighe:2001gc,Ciuchini:2003ww,Beneke:2003az,Lenz:2006hd},
results for $M_{12}^q$ from
\cite{Inami:1980fz,Buras:1990fn}
and for
the hadronic matrix elements of dimension six the averages presented in
\cite{DiLuzio:2019jyq}
based on \cite{Grozin:2016uqy,Kirk:2017juj,King:2019lal} and \cite{Christ:2014uea,Bussone:2016iua,Hughes:2017spc,Bazavov:2017lyh}.
Recently also the first non-perturbative evaluation of dimension seven matrix elements
became available \cite{Davies:2019gnp}, which we will use for $\Gamma_{12}^q$.
\subsection{QCD Factorization}
\label{sec:QCDF}
\begin{figure}[h]
\centering
\includegraphics[height=3.5cm]{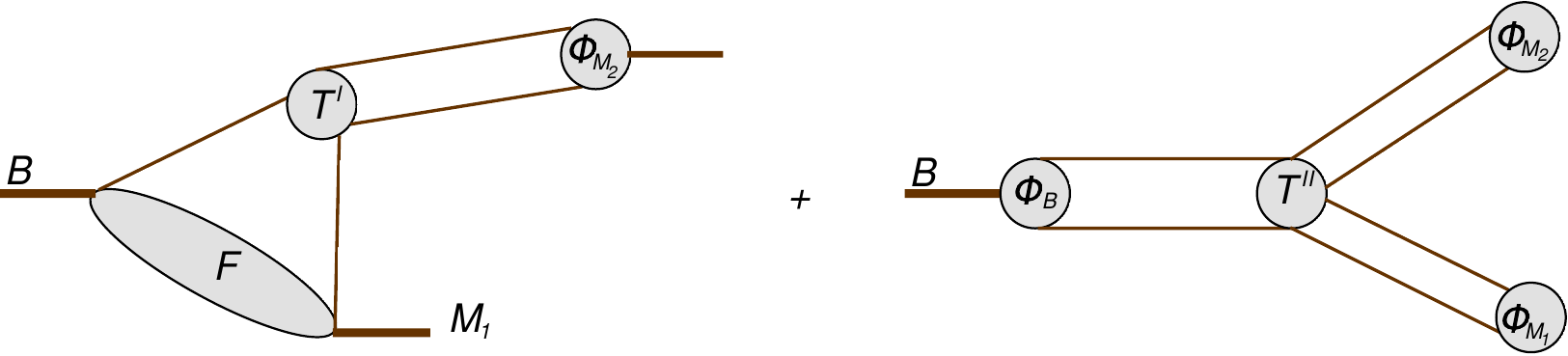}
\caption{Factorization of matrix elements for $B$ meson decays into ``light''-``light'' mesons (both diagrams included)
and ``heavy''-``light'' (only left diagram) in QCDF.}
\label{Fig:QCDFactMatrix}
\end{figure}

\begin{figure}
\centering
\includegraphics[height=2.5cm]{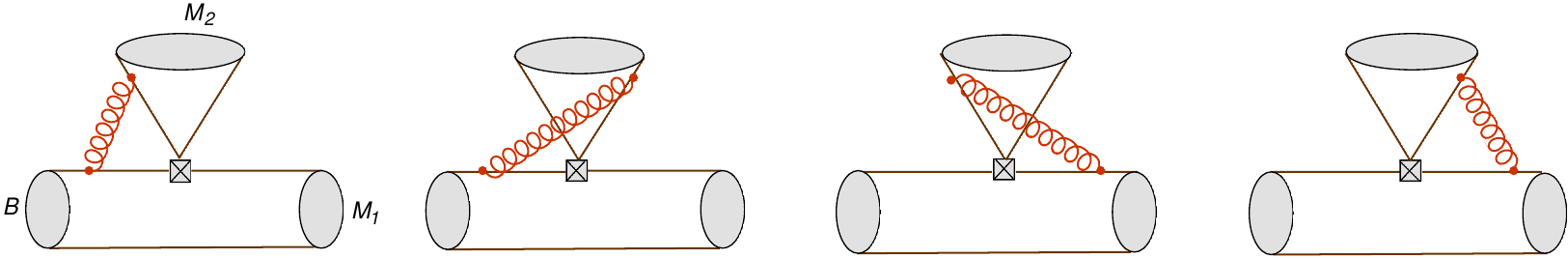}
\caption{NLO Vertex contributions to the process $B\rightarrow M_1 M_2$.}
\label{fig:Vertex}
\end{figure}

\begin{figure}
\centering
\includegraphics[height=2.5cm]{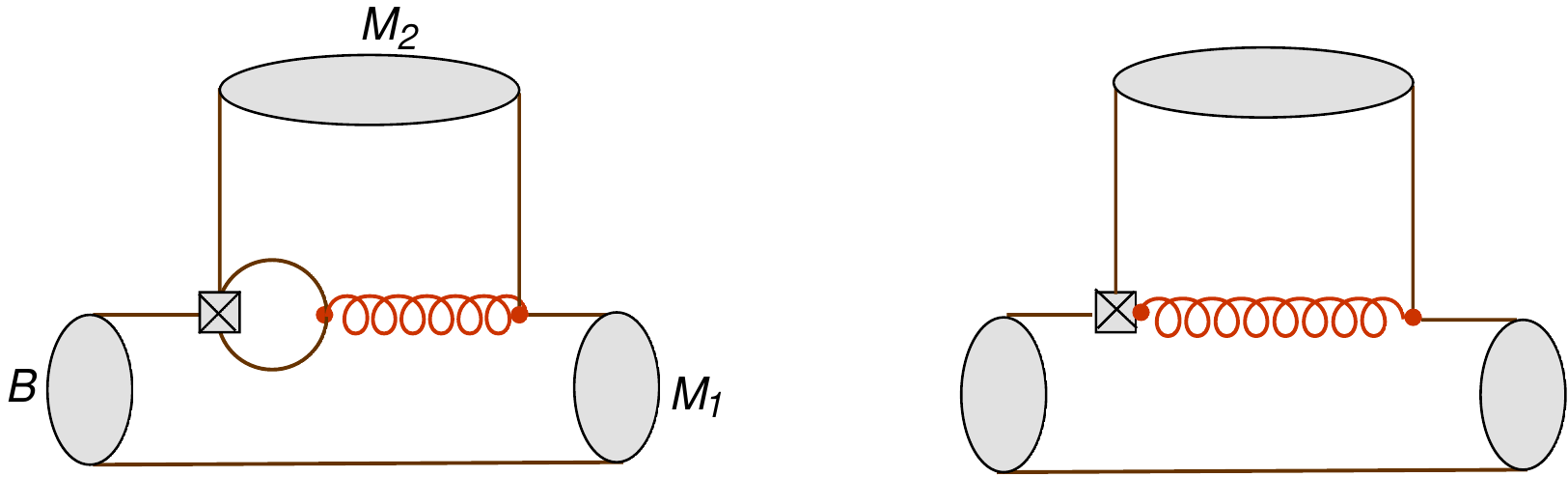}
\caption{NLO penguin contributions to the process $B\rightarrow M_1 M_2$.}
\label{fig:Penguin}
\end{figure}

\begin{figure}
\centering
\includegraphics[height=2.5cm]{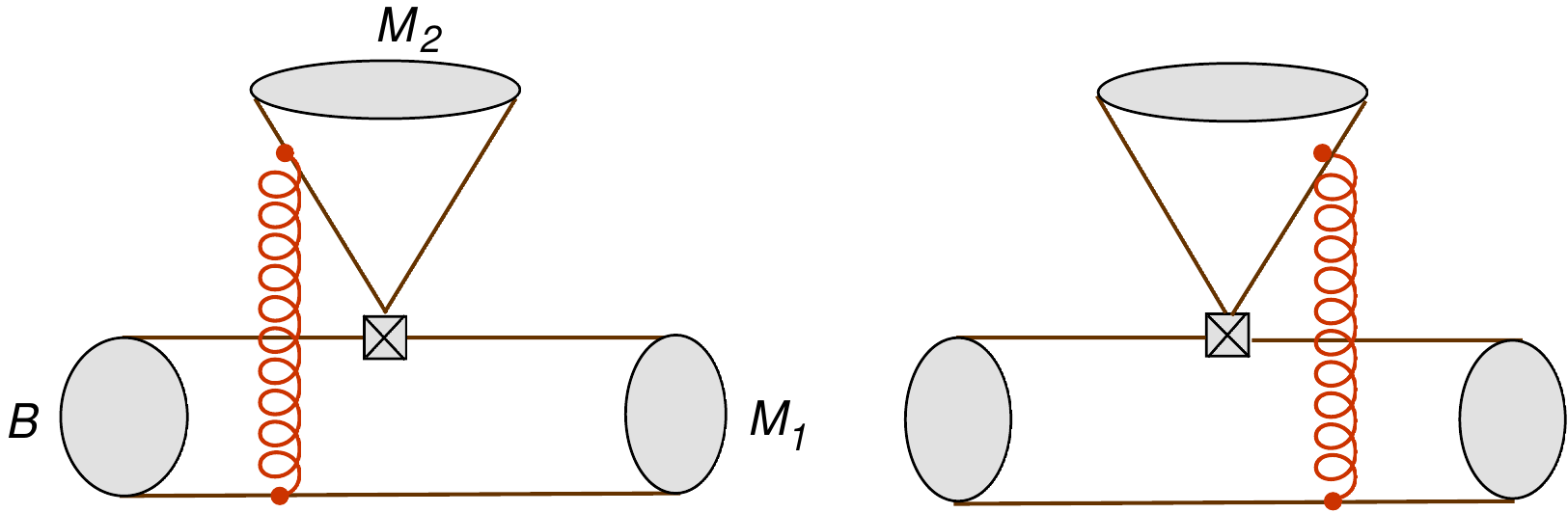}
\caption{Hard spectator-scattering contributions to the decay $B\rightarrow M_1 M_2$.}
\label{fig:Hard_scattering}
\end{figure}

\begin{figure}
\centering
\includegraphics[height=1.5cm]{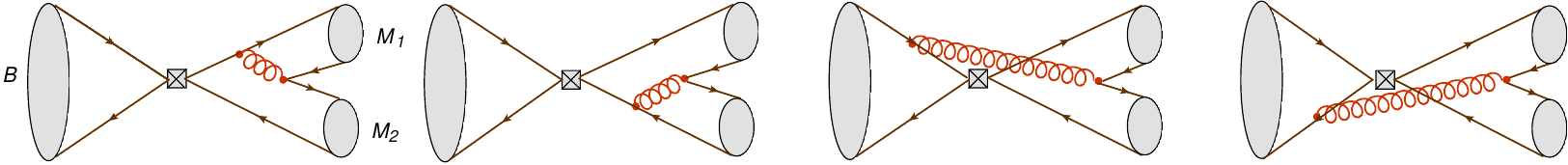}
\caption{Annihilation topologies contributing to the decay process $B\rightarrow M_1 M_2$.}
\label{fig:Annihilation}
\end{figure}

In our analysis we included different observables based on non-leptonic $B$ meson decays 
such as: $B\rightarrow D \pi$, $B\rightarrow \pi\pi$,  $B\rightarrow \pi\rho$ and $B\rightarrow \rho\rho$.
To calculate the corresponding
amplitudes we used the expressions available in the literature
obtained within the QCD Factorization (QCDF) framework  \cite{Beneke:1999br, Beneke:2000ry, Beneke:2001ev, Beneke:2003zv}.
In this section we briefly summarise the QCDF results relevant for the evaluation of some of our flavour constraints.
%
Consider the process $B\rightarrow M_1 M_2$, in which a $B$ meson decays into the final states $M_1$ and $M_2$,
where either $M_1$  and $M_2$ are two ``light'' mesons or $M_1$ is ``heavy'' and $M_2$ is ``light''
\footnote{ A meson
with mass $m$ is considered ``heavy" if $m$
scales with $m_b$ in the heavy quark limit such that $m/m_b$ remains fixed. On the other hand a meson is regarded as ``light''
if its mass remains finite in the heavy quark limit, for a light meson $m\sim\mathcal{O}(\Lambda_{QCD})$ \cite{Beneke:2000ry}.}.
\\
If both $M_1$ and  $M_2$ are light, then the matrix element $\langle M_1 M_2 |\hat{Q}_{i}| B \rangle$ 
of the dimension six effective operators in Eq.~(\ref{eq:mainbasis}) can be written as 
\vspace{-0.2cm}
\begin{eqnarray}
\langle M_1 M_2|\hat{Q}_{i}| B \rangle &=& \sum_{j}F^{B\rightarrow M_{1}}_{j}(0) \int_{0}^{1}du T^{I}_{ij}(u) \Phi_{M_2}(u) + (M_1 \leftrightarrow M_2)\nonumber\\
                                    && + \int_{0}^{1}d\xi du dv T^{II}_{i}(\xi, u, v)\Phi_{B}(\xi)\Phi_{M_1}(v)\Phi_{M_2}(u).
\label{eq:fact1}
\end{eqnarray}
In the right hand side of Eq.~(\ref{eq:fact1}) $F^{B\rightarrow M_{1,2}}_{j}(m^2_{2,1})$ represents 
the relevant form factor to account for the transition $B\rightarrow M_{1}$ (and correspondingly for
$B\rightarrow M_{2}$) and $\Phi_M(u)$ is the non-perturbative Light-Cone Distribution Amplitude (LCDA)
for the meson $M$, see Fig. \ref{Fig:QCDFactMatrix}.
\\
Notice that Eq.~(\ref{eq:fact1}) is written in such a way that it can be applied to situations where the spectator
quark can end in any of the two final state light mesons. If the spectator can go into only one of the final
mesons, this one will be labelled as $M_1$ and just the first and the third terms on the right hand side of Eq.~(\ref{eq:fact1})
should be included.  The functions $T^{I, II}$ are called hard-scattering kernels and can be calculated perturbatively.
The kernel $T^{I}$ contains, at higher order in $\alpha_s$, nonfactorizable contributions from hard gluon exchange or penguin topologies. 
On the other hand, nonfactorizable  hard interactions involving the spectator quark are part of $T^{II}$.
\\
When in the final state the mesons $M_1$ is ``heavy'' and  $M_2$ is ``light'', then the corresponding QCDF formula for the matrix element
$\langle  M_1 M_2|\hat{Q}_{i}| B  \rangle$ becomes
\vspace{-0.2cm}
\begin{eqnarray}
\langle  M_1 M_2|\hat{Q}_{i}| B  \rangle &=&\sum_{j}F^{B\rightarrow M_{1}}_{j}(m^2_{2}) \int_{0}^{1}du T^{I}_{ij}(u) \Phi_{M_2}(u),
\label{eq:fact2}
\end{eqnarray}
where the meaning of the different terms in Eq.~(\ref{eq:fact2}) are analogous to those given for Eq.~(\ref{eq:fact1}).
\\
\noindent
To determine the decay amplitude $\mathcal{A}(B\rightarrow M_1 M_2)$, the matrix element 
$\langle M_1 M_2 |\hat{\mathcal{H}}_{eff}| B \rangle$ should be calculated, with $\hat{\mathcal{H}}_{eff}$  being the effective Hamiltonian
introduced in Eq.~(\ref{eq:Hamiltonian}).  In QCDF the final expression for $\mathcal{A}(B\rightarrow M_1 M_2)$ is written as a linear 
combination of sub-amplitudes $\alpha^{p, M_1 M2}_i$ and $\beta^{p, M_1 M2}_i$, which for the purposes of our discussion will be 
termed ``Topological Amplitudes'' (TA).
The TA $\alpha_i^{p}(M_1 M_2)$, for $p=u,c$, have the following generic structure at NLO in $\alpha_s$ \cite{Beneke:2003zv}
\vspace{-0.2cm}
\begin{eqnarray}
  \alpha_i^{p, M_1 M_2}&=&
  \left[ C_{i}(\mu_b) + \frac{C_{i\pm 1}(\mu_b)}{N_c}\right] N_{i}(M_2)
  \nonumber \\
  && + 
  \frac{\alpha_s(\mu_b)}{4\pi} \  \frac{C_F}{N_c} C_{i\pm 1}(\mu_b) V_{i}(M_2)
  + P^{p}_{i}(M_2)
  \nonumber\\ 
  &&
  + \frac{\alpha_s(\mu_h)}{4\pi} \frac{4 \pi^2 C_F}{N^2_c}  C_{i\pm 1}(\mu_h) H_{i}(M_1 M_2) \, ,
\label{eq:alphaGen0}
\end{eqnarray}

where $C_i$ are the Wilson coefficients calculated at the scale $\mu\sim m_b$,
and the subindex in the coefficient $C_{i\pm 1}$ is assigned following the rule
\vspace{-0.2cm}
\[C_{i\pm1} = \left\{
  \begin{array}{lr}
    C_{i + 1}  : & \hbox{if } i \hbox{ is odd }, \\
    C_{i - 1}  : & \hbox{if } i \hbox{ is even}.
  \end{array}
\right.
\]\\
The Wilson coefficients inside the squared bracket in Eq.~(\ref{eq:alphaGen0}) will be modified to 
allow for NP contributions as discussed below, see Section \ref{sec:strategy}, and $N_c$ denotes the number of 
colours under consideration and will be taken as $N_c=3$.
The global factor $N_{i}(M_2)$ multiplying the square bracket corresponds to the normalisation of the light cone distribution 
for the  meson $M_2$, and is evaluated according to the following rule
\vspace{-0.2cm}
\[N_i(M_2) = \left\{
  \begin{array}{lr}
    0  : & \hbox{if $i=6, 8$ and $M_2$ is a vector meson,}\\
    1  : & \hbox{in any other case}.
  \end{array}
\right.
\]\\
The symbol $V_{i}(M_2)$ in Eq.~(\ref{eq:alphaGen0}) stands for the one loop vertex corrections illustrated in Fig.~\ref{fig:Vertex}. 
Additionally,  the contributions from Penguin diagrams such as those shown in Fig.~\ref{fig:Penguin} are included in
$P_i^{p}(M_2)$, with $p=u,c$.  Finally the hard spectator interactions shown in Fig.~\ref{fig:Hard_scattering} are accounted
for by the term $H_i(M_1 M_2)$.
If $M_1$ and $M_2$ are both pseudoscalar mesons or if one of them is a pseudoscalar and the other is a vector meson, then
the hard spectator function $H_i(M_1 M_2)$ can be written in terms of the leading twist LCDAs 
of $M_1$ and $M_2$, $\Phi_{M_1}$ and $\Phi_{M_2}$ respectively, and the twist-3 LCDA of  $M_1$,  $\Phi_{m_1}$, as \cite{Beneke:2003zv}:
\vspace{-0.2cm}

\begin{eqnarray}
H_{i}(M_1 M_2)&=&\frac{B_{M_1 M_2}}{A_{M_1 M_2}} \int^{1}_{0} d\xi \frac{\Phi_{B}(\xi)}{\xi}\int^{1}_{0} dx
\int^{1}_{0} dy \Bigl[ \frac{ \Phi_{M_2}(x)   \Phi_{M_1}(y)}{\bar{x} \bar{y}} \nonumber\\
&& + r^{M_1}_{\chi} \frac{ \Phi_{M_2}(x)   \Phi_{m_1}(y)}{x \bar{y}}  \Bigl],~\hbox{(for $i=1,...,4,9,10$)} \, ,\nonumber\\
H_{i}(M_1 M_2)&=&-\frac{B_{M_1 M_2}}{A_{M_1 M_2}} \int^{1}_{0} d\xi \frac{\Phi_{B}(\xi)}{\xi}\int^{1}_{0} dx
\int^{1}_{0} dy \Bigl[ \frac{ \Phi_{M_2}(x)   \Phi_{M_1}(y)}{x \bar{y}} \nonumber\\
&& + r^{M_1}_{\chi} \frac{ \Phi_{M_2}(x)   \Phi_{m_1}(y)}{\bar{x} \bar{y}}  \Bigl],~\hbox{(for $i=5,7$)} \, ,\nonumber\\
H_{i}(M_1 M_2)&=&0,~\hbox{(for $i=6, 8$)} \, .
\label{eq:HS1andHS2}
\end{eqnarray}
The analogous expressions for $H_i(M_1 M_2)$ when $M_1$ and $M_2$ are two longitudinally polarised light vector mesons can be found 
in \cite{Beneke:2006hg, Bartsch:2008ps}.
We provide the functions $H_i(M_1 M_2)$  for the processes relevant to this project 
in Appendix \ref{Sec:QCDFact}.  The global coefficients $A_{M_1 M_2}$ and $B_{M_1 M_2}$
presented in Eqs.~(\ref{eq:HS1andHS2}) depend on form factors and decay constants and are given
in Eq.~(\ref{eq:GenPar}) also in Appendix \ref{Sec:QCDFact}.
\\
We want to highlight two sources of uncertainty arising in Eq.~(\ref{eq:HS1andHS2}). The first one
stems from the contribution of the twist-3 LCDA $\Phi_{m_1}(y)$. Since this function does not vanish at $y=1$, the integral 
$\int^1_0 dy \Phi_{m_1}(y)/\bar{y}$ is divergent. To isolate the divergence we follow the prescription
given in \cite{Beneke:2003zv} and write
\vspace{-0.2cm}
\begin{eqnarray}
\label{eq:Twist3}
\int^1_0 \frac{dy}{\bar{y}} \Phi_{m_1}(y)&=& \Phi_{m_1}(1)\int^1_0 \frac{dy}{\bar{y}} +
 \int^1_0 \frac{dy}{\bar{y}} \Bigl[\Phi_{m_1}(y) - \Phi_{m_1}(1) \Bigl]\nonumber\\
&=&\Phi_{m_1}(1) X_H + \int^1_0 \frac{dy}{[\bar{y}]_+}\Phi_{m_1}(y).
\end{eqnarray}
The divergent piece of Eq.~(\ref{eq:Twist3}) is contained in $X_H$. The remaining integral 
$\int^1_0 dy/[\bar{y}]_+\Phi_{m_1}(y)$ is finite (for instance for a pseudo scalar meson $\Phi_{m_1}(y)=1$ and trivially 
$\int^1_0 dy/[\bar{y}]_+\Phi_{m_1}(y)=0$). Physically $X_H$ represents a soft gluon interaction 
with the spectator quark.  It is expected that $X_H\approx \hbox{ln}(m_b/\Lambda_{QCD})$ because the divergence
appearing is regulated by a physical scale of the order
$\Lambda_{QCD}$. A complex coefficient cannot be excluded since multiple soft scattering can introduce a strong interaction phase.
Here we use the standard parameterisation for $X_H$ introduced by Beneke-Buchalla-Neubert-Sachrajda (BBNS) 
\cite{Beneke:2000ry} 
\vspace{-0.2cm}
\begin{eqnarray}
X_H &=& \Bigl( 1 + \rho_H e^{i \phi_H}  \Bigl) \hbox{ln}\frac{m_B}{\Lambda_{h}},
\label{eq:XH}
\end{eqnarray}
where $\Lambda_h \approx \mathcal{O}(\Lambda_{QCD})$ and $\rho_H \approx \mathcal{O}(1)$.
\\ 
The second source of theoretical uncertainty in Eqs.~(\ref{eq:HS1andHS2}) that deserves special attention
is the inverse moment of the LCDA $\Phi_B$ corresponding to the $B$ meson. Following \cite{Beneke:1999br} we write
\vspace{-0.2cm}
\begin{eqnarray}
\label{eq:Mom1LCD}
\int_0^1 d\xi \frac{\Phi_B(\xi)}{\xi}&\equiv&\frac{m_B}{\lambda_B},
\end{eqnarray}
where $\lambda_B$ is expected to be of $\mathcal{O}(\Lambda_{QCD})$. We provide more details about the values for $X_H$ and $\lambda_B$
used in this work at the end of this subsection.
\\
Next we address the contributions from weak annihilation topologies, see Fig.~\ref{fig:Annihilation}, which are power suppressed in 
the $\Lambda_{QCD}/m_b$ expansion with respect to the factorizable amplitudes. Although they do not appear in Eq.~(\ref{eq:fact1}), 
they are included in terms of subamplitudes denoted as $\beta^{p, M_1 M_2}_k$. The numerical subscript $k$ describes the Dirac structure
under consideration: $k=1$  for $(V-A)\otimes(V-A)$, $k=2$ for $(V-A)\otimes (V+A)$ and $k=3$ for $(-2)(S-P)\otimes(S+P)$. 
The annihilation coefficients are expressed in terms of a set of basic ``building blocks'' denoted by $A^{i,f}_k$. Where the subindex 
$k$ also denotes the Dirac structure being considered as previously explained, and the superindices  $i$ and $f$ denote the emission 
of a gluon by an initial or a final state quark as shown in Fig.~\ref{fig:Annihilation}. The coefficients $A^{i,f}_k$ relevant for this 
work can be found in Appendix \ref{Sec:QCDFact}.
The final expressions for annihilation are the result of the convolution of twist-2 and twist-3 LCDA with the corresponding 
hard scattering kernels; as in the case of hard spectator scattering, there are also endpoint singularities that are treated in a
model dependent fashion. To parameterize these divergences, we follow once more the approach of BBNS. Thus, in analogy with hard spectator scattering we introduce \cite{Beneke:2000ry}
\vspace{-0.2cm}
\begin{eqnarray}
X_A &=& \Bigl(1 + \rho_A e^{i \phi_A} \Bigl) \hbox{ln} \frac{m_B}{\Lambda_h}.
\label{eq:XA}
\end{eqnarray}

To finalize this subsection we discuss the numerical inputs used in our evaluations of $\lambda_B$, $X_H$ and $X_A$.
As indicated in Eq.~(\ref{eq:Mom1LCD}), the inverse moment of the LCDA of the $B$ meson introduces the parameter $\lambda_B$.
The description of non-leptonic $B$ decays 
based on QCDF
requires $\lambda_B\sim 200~\hbox{MeV}$~\cite{Beneke:2003zv, Beneke:2009ek}. In contrast, 
QCD sum rules calculations give a higher value. For instance, in ~\cite{Braun:2003wx} the result 
$\lambda_B=(460 \pm 110)~\hbox{MeV}$  was found. In~\cite{Beneke:2011nf}  the usage of the channel $B\rightarrow   \gamma \ell \nu_{\ell}$
was proposed in order to extract $\lambda_B$ experimentally. 
This study was updated in~\cite{ Wang:2016qii, Beneke:2018wjp, Wang:2018wfj} where further effects, including subleading power corrections in 
$1/E_{\gamma}$ and $1/m_b$, were accounted for. 
Based on this idea, the Belle collaboration found  \cite{Heller:2015vvm}
\begin{eqnarray}
\label{eq:bellebound}
\lambda_B \Bigl|_{\rm Belle}> 238~\hbox{MeV},
\end{eqnarray}
at the $90\%$ C.L. and it is expected that the Belle II experiment improves this result ~\cite{Beneke:2018wjp}.
Interestingly the experimental bound in Eq.~(\ref{eq:bellebound}) is compatible
with the QCD sum rules value quoted above and other theoretical approaches,  including the one in \cite{Lee:2005gza} where
the value $\lambda_B=(476.19\pm 113.38)\hbox{ MeV}$ was obtained. For the purposes of our analysis, we consider the following result calculated in
\cite{Bell:2009fm}
with QCD sum rules:
\begin{eqnarray}
\lambda_B= (400 \pm 150) \hbox{  MeV}.
\end{eqnarray}
As discussed above, the calculation of hard spectator interactions and the evaluation of annihilation topologies, 
leads to extra sources of uncertainty associated with endpoint singularities that are power suppressed. 
As indicated in  
Eqs.~(\ref{eq:XH}) and~(\ref{eq:XA}) they can be parameterized through the functions $X_H(\rho_{H},\phi_{H})$ and 
$X_A(\rho_{A},\phi_{A})$  respectively. Using
these models,  we account for the  hard spectator scattering power suppressed singularities through the parameters 
$\rho_H$ and $\phi_H$. Correspondingly,
we introduce $\rho_A$ and $\phi_A$ to address the analogous effects from  annihilation topologies. 
Based on phenomenological considerations we will take into account the intervals 
\cite{Bobeth:2014rra, Hofer:2010ee}
\begin{eqnarray}
 0<\rho_{H,A}<2, &  \hbox{   } 0 <\phi_{H,A}<2\pi, \label{set1}
\end{eqnarray}
which correspond to a $200\%$ uncertainty on $|X_H|$ and $|X_A|$.
\\ 
To evaluate the central values of our observables we take $\rho_{H,A}=0$, or equivalently $X_H=X_A=\hbox{ln}~ m_B/\Lambda_h$.  
Finally, we calculate the percentual error from $X_A$ and $X_H$, by estimating the difference between the maximum and the 
minimum values reached by the hadronic observables when considering the intervals in Eq.~(\ref{set1}), and then we normalize 
by two times the corresponding central values.
%
%
%
%
%
%
%
%
%
%
%
%
%
%
%
%
%
%
%
%
\section{Strategy}
\label{sec:strategy}
Consider the effective Hamiltonian in Eq.~(\ref{eq:Hamiltonian}) written in terms of the basis in Eq.~(\ref{eq:mainbasis}). We
introduce ``new physics'' in the Wilson coefficients $\{C_1, C_2\}$ of the operators $\hat{Q}_1$ 
and $\hat{Q}_2$ following the prescription
\begin{eqnarray}
C_{1}(M_W) := C^{\rm SM}_{1}(M_W) + \Delta C_{1}(M_W),\nonumber\\
C_{2}(M_W) := C^{\rm SM}_{2}(M_W) + \Delta C_{2}(M_W),
\label{eq:NPC12}
\end{eqnarray}
where in the SM
\begin{eqnarray}
\Delta C_{1}(M_W)&=&0,\nonumber\\
\Delta C_{2}(M_W)&=&0.
\end{eqnarray}	
In this paper we present possible bounds on $\Delta C_1$ and $\Delta C_2$ at the matching scale 
$\mu=M_W$ and consider 
changes to each Wilson coefficient independently, e.g. to establish constraints on 
$\Delta C_1(M_W)$ we fix $\Delta C_2(M_W)=0$ and vice versa.
This is a conservative approach, if we allow both parameters 
to change simultaneously this can result into partial cancellations leading to potentially bigger NP allowed regions for 
$\{\Delta C_{1}(M_W), \Delta C_{2}(M_W)\}$.
Since  the theoretical formulae for our observables are calculated at the scale $\mu=m_b$, we evolve down the modified Wilson
coefficients $C_{1}(M_W)$ and $C_{2}(M_W)$ up to this scale using the renormalisation group formalism
described in Section \ref{sec:Heff}.  We consider NP to be leading order only, therefore we treat the SM
contribution $\{C^{\rm SM}_1(M_W), C^{\rm SM}_2(M_W)\}$ and the NP components  $\{\Delta C_{1}(M_W), \Delta C_{2}(M_W)\}$ 
differently under the renormalisation group equations. For instance the evolution of $\{C^{\rm SM}_{1}(M_W), C^{\rm SM}_{2}(M_W)\}$ 
is done using the full NLO expressions 
in Eq.~(\ref{eq:fullevoMat}), on the other hand $\{\Delta C_{1}(M_W), \Delta C_{2}(M_W)\}$ are evolved down using only the LO  
version  shown in Eq.~(\ref{Eq:LOevo}). 
Notice that, even though at the scale $\mu=M_W$ the only modified Wilson coefficients are $C_1(M_W)$ and $C_2(M_W)$,
the non diagonal nature of the evolution matrices propagates these effects
to all the other Wilson coefficients undergoing mixing at $\mu=m_b$. Hence, when writing 
expressions for the different physical observables, it makes sense to consider
NP effects in $C_{i}(m_b)$ even for $i\neq 1,2$.
\subsection{Statistical analysis}
\label{sec:stat}
The values of $\Delta C_1(M_W)$ and $\Delta C_2(M_W)$ compatible with experimental data
are evaluated using the program MyFitter \cite{Wiebusch:2012en}. The full statistical procedure is based on a likelihood ratio 
test. The basic ingredient is the $\chi^2$ function
\begin{eqnarray}
\chi^2(\vec{\omega}) &=&\sum_i \Bigl(\frac{\tilde{O}_{i,\rm exp} - 
                                              \tilde{O}_{i,\rm theo}(\vec{\omega})  }{\sigma_{i,\rm exp}} \Bigl)^2,
\label{eq:chi2}
\end{eqnarray}
where $\tilde{O}_{i, \rm exp}$ and $\tilde{O}_{i,\rm theo}$ are the experimental and theoretical values of the $i-$th observable respectively
and $\sigma_{i,\rm exp}$ is the corresponding experimental uncertainty.
The vector $\vec{\omega}$ contains all the inputs necessary for the evaluation of $\tilde{O}_{i,\rm theo}$ and will be written as
\begin{eqnarray}
\label{eq:omegavector}
\vec{\omega}&=&\Bigl(\Delta C_1(M_W), \Delta C_2(M_W), \vec{\lambda}\Bigl).
\end{eqnarray}
In Eq.~(\ref{eq:omegavector}) we are making a distinction between $\{\Delta C_{1}(M_W), \Delta C_{2}(M_W)\}$ and the rest of the theoretical 
inputs, which have been included in the subvector $\vec{\lambda}$. Examples of the entries inside $\vec{\lambda}$ are masses,
decay constants, form-factors, etc.
Notice that our main target is the determination of $\Delta C_{1}(M_W)$ and $\Delta C_{2}(M_W)$, however, the components 
entering $\vec{\lambda}$ are crucial in defining the uncertainty of our observables and hence in
establishing the potential values of $\Delta C_{1}(M_W)$ and $\Delta C_{2}(M_W)$.
In this respect, we will say that the elements inside $\vec{\lambda}$ are our nuisance parameters, 
and that the determination of the possible NP values compatible with data are obtained by profiling the likelihood with respect to 
$\{\Delta C_{1}(M_W), \Delta C_{2}(M_W)\}$.
During our analysis the elements of $\{\Delta C_1(M_W), \Delta C_2(M_W)\}$ are assumed to be complex and, as indicated in the argument,
the initial evaluation is done at the scale $\mu=M_W$.
The statistical theory behind the $\chi^2$-fit software used, e.g. MyFitter \cite{Wiebusch:2012en}, can be found in the documentation
of the computer program.  Here we only summarize the key steps involved in our analysis:
\begin{enumerate}
\item We first define the Confidence Level $CL$ for the $\chi^2$-fit.  Following the criteria established in 
      \cite{Bobeth:2014rda, Brod:2014bfa} for our study we take
      \begin{eqnarray}
      CL=90\%,	
      \end{eqnarray}
      which is equivalent to $1.64$ standard deviations approximately.
\item \label{step2} Then, we establish a sampling region on the plane defined by the real 
      and the imaginary components of $\{\Delta C_1(M_W), \Delta C_2(M_W)\}$. The sampling region is observable dependent.
      In our case we opt for rectangular grids around the origin of the complex plane defined by $\Delta C_1(M_W)$ and $\Delta C_2(M_W)$.
      Notice that the origin of our complex plane corresponds to the SM value. The number of points in our test grid depends on three factors:
      the numerical stability of our algorithms,
      on the time required to compute a particular combination of observables
      and the size of the NP regions determined by them.
\item Each one of the points inside the sampling grid described in the previous step corresponds to a null-hypothesis for the 
	components of $\Delta C_1(M_W)$ and $\Delta C_2(M_W)$. We test our null-hypothesis values using a likelihood ratio test considering 
	the confidence level established in the first step.
 	For a combination of multiple observables several nuisance parameters are involved and the full statistical procedure becomes time and 
resource consuming. Hence, the parallelization of our calculations using a computer cluster became necessary. We did our first numerical 
evaluations partially at the Institute for Particle Physics and Phenomenology (IPPP, Durham University). The results presented in this work 
were obtained in full using the computing facilities available at the Dutch National Institute for Subatomic Physics (Nikhef).

\end{enumerate}

%
%
%
%
%
%
%
%
%
%
%
%
%
%
%
%
%
%
%
%
%
%
%
%
%
%
%
%
%
%
\section{Individual Constraints}
\label{sec:constraints}
In this section we present the different observables considered during the analysis. From Sections~\ref{sec:bcud} 
to~\ref{sec:bccs} we focus exclusively on observables that constrain individual $b$ decay channels, in our case: 
 $b\rightarrow c\bar{u}d$,
$b\rightarrow u \bar{u}d$,
$b\rightarrow c\bar{c}s$ and $b\rightarrow c\bar{c}d$. In Section~\ref{sec:multiple_channels} we will study observables that affect
multiple $b$ decay channels. In what follows and unless stated otherwise, the SM predictions as well as the experimental determinations
are given at $1~\sigma$, i.e. $68\%~\rm{C. L.}$. However the allowed NP regions  for  $C_1$ and $C_2$ are presented at
$1.64~\sigma$, i.e. $90\%~\rm{C. L.}$.
\\
Following the notation introduced in Eqs.~(\ref{eq:Hamiltonian}) and 
(\ref{eq:mainbasis}) we will denote
the NP effects in the Wilson coefficient of the operator $\hat{Q}^{q, \, pp'}_i$
as $\Delta C^{q, \, pp'}_i$ for $i=1,2$ and $q=d,s$. Then for example, 
$\{\Delta C^{d, \, cu}_{1}(M_W), \Delta C^{d, \, cu}_{2}(M_W)\}$ will quantify the potential deviations from the SM values in
the coefficients of $\{\hat{Q}^{d, \, cu}_{1}, \hat{Q}^{d, \, cu}_{2}\}$ which describe the tree level process
$b\rightarrow c \bar{u} d$.
\\
In this work NP is supposed to be leading order in $\alpha_s$ and $\alpha$ only.
Since all the vertex corrections $V^{M}_{i}$,
penguins $P^{p, M}_{i}$ and
hard scattering spectator interactions $H^{M_1 M_2}_{i}$ inside Eq.~(\ref{eq:alphaGen0})  
are already suppressed by factors
of $\mathcal{O}(\alpha_s)$ and $\mathcal{O}(\alpha)$,
we will consistently drop the extra contributions $\Delta C_1^{d, \, uu}(M_W)$ and $\Delta C_2^{d, \, uu}(M_W)$ affecting any of these terms
for all observables that are described by QCDF.

%
%
%
%
%
%
%
%
\subsection{Observables constraining $b\rightarrow c\bar{u}d$ transitions}
\label{sec:bcud}
We start with the dominant quark level decay $ b \to c \bar{u} d$ and describe our analysis of the potential NP
regions for $\Delta C^{d, \, cu}_{1}(M_W)$ and 
$\Delta C^{d, \, cu}_{2}(M_W)$. The decay $\bar{B}^{0}\rightarrow D^{*+}\pi^{-}$ will exclude large positive values of
$\Delta C^{d, \, cu}_{1}(M_W)$ and it will significantly
constrain $\Delta C^{d, \, cu}_{2}(M_W)$.
\subsubsection{$\bar{B}_d^{0}\rightarrow D^{*+}\pi^{-}$}
\label{sec:RDpi}
Our bounds will be established using the ratio between the decay width for the non-leptonic decay $\bar{B}^0_d\rightarrow D^{*+}\pi^{-}$ 
and the differential rate 
for the semi-leptonic process
$\bar{B}^0_d\rightarrow D^{*+}l^-\bar{\nu}_l$ evaluated at $q^2=m^2_{\pi}$ for $l=e, \mu$
\vspace{-0.2cm}
\begin{eqnarray}
\label{eq:RDpi}	
  R_{D^{*}\pi}&=&\frac{\Gamma(\bar{B}^0\rightarrow D^{*+}\pi^{-})}
  {d\Gamma(\bar{B}^0\rightarrow D^{*+} l^{-}\bar{\nu}_{l})/dq^2|_{q^2=m^2_{\pi}}}
  \simeq 6\pi^2f^2_{\pi}|V_{ud}|^2|\alpha_{2}^{D^{*}\pi}+\beta^{D^* \pi }_2|^2.\nonumber\\
\end{eqnarray}
This observable was proposed by Bjorken to test the factorization hypothesis \cite{Bjorken:1988kk},
it is free from the uncertainties associated with the required form factor to describe the transition $B\rightarrow D^{*}$
and offers the possibility of comparing directly the coefficient $\alpha^{D^{*}\pi}_{2}$ calculated using QCDF 
against experimental observations.
At NLO the TA $\alpha_{2}^{D^*\pi}$ \cite{Beneke:2000ry} is given by
\vspace{-0.2cm}
\begin{align}
	\alpha^{{\rm NLO},  D^{*}\pi}_{2}&=& C^{d, \, cu}_2(\mu_b) + \frac{C^{d, \, cu}_{1}(\mu_b)}{3} + \frac{\alpha_{s}(\mu_b)}{4\pi}\frac{C_F}{N_c}C^{d, \, cu}_{1}(\mu_b)
			\Bigl[-\tilde{B} -6 \hbox{ln}\frac{\mu^2}{m^2_{b}}\nonumber\\
                          &&
                          + \int^1_{0}du F(u, -x_c)\Phi_{\pi}(u) \Bigl]
                        \approx 1.057 \pm 0.040 \, ,
\label{eq:BDpi}
\end{align}
where the term $\tilde{B}$ inside the square bracket cancels the renormalisation scheme dependence of the Wilson coefficients $C^{d, \, cu}_1$ and 
$C^{d, \, cu}_2$, which in naive dimensional regularisation requires $\tilde{B}=11$. The kernel $F(u,-x_c)$ includes QCD vertex corrections
arising in the decay
$b\rightarrow c \bar{u} d$ and has to be evaluated at $x_c=\bar{m}_c(\bar{m}_b) /\bar{m}_b$
before being convoluted with the light-cone distribution $\Phi_{\pi}$
associated with the $\pi^-$ meson in the final state. For the explicit evaluation of Eq.~(\ref{eq:RDpi})
we use the updated determination of the TA $\alpha^{D^{*}\pi}_{2}$  at NNLO calculated in \cite{Huber:2016xod}

\begin{eqnarray}
|\alpha^{{\rm NNLO}, D^{*}\pi}_{2}|&=&1.071^{+0.013}_{-0.014}.	
\end{eqnarray}

The annihilation topologies contributions are taken into account through 
\begin{eqnarray}
\beta^{D^* \pi }_2&=&\frac{C_F}{N^2_c}\frac{B_{D^* \pi}}{A_{D^* \pi}} C_2^{d, \, cu}(\mu_h) A^i_1(\mu_h) \approx 0.014 \pm 0.045 \, ,
\end{eqnarray}
where
\begin{eqnarray}
\frac{B_{D^* \pi}}{A_{D^* \pi}}=\frac{f_B f_{D^*}}{m^2_B A^{B\rightarrow D^*}_0(0)},
\end{eqnarray}
and 
\begin{eqnarray}
  A^i_1(\mu_h)&\approx& 6\pi\alpha_s(\mu_h)
  \Biggl[3\Biggl(X_A -4 + \frac{\pi^2}{3}\Biggl) + r^{D^*}_{\chi}(\mu_h) r^{\pi}_{\chi}(\mu_h)\Bigl(X^2_A -2 X_A\Bigl)\Biggl],
  \nonumber
  \\
\end{eqnarray}
with the parameters $X_A$ are given in Eq.~(\ref{eq:XA}) and the factors $r^{\pi}_{\chi}$
and $r^{D^*}_{\chi}$ quoted in Eq.~(\ref{eq:GenPar}).
Using the numerical inputs given in \ref{Sec:Inputs} we find
\begin{eqnarray}
R^{\rm SM}_{D^{*}\pi} &=&\Bigl(1.12\pm 0.15\Bigl)\hbox{GeV}^2,
\label{eq:BDpi_theo}
\end{eqnarray}
corresponding to { $x_c=0.225$}, the partial contributions to the total error are shown in Table \ref{tab:errorBDpi}.

\begin{table}
	\begin{center}
		\begin{tabular}{ |c|c|}
			\hline
			Parameter  & Relative error \\
			\hline
			\hline
			$X_A$                                & $13.05\%$ \\
                        \hline
                        $\mu$ &                            $2.53\%$ \\
                        \hline
			$f_{\pi}$                            & $1.23\%$ \\
			\hline
			$\Lambda^{QCD}_5$                          & $0.09\%$ \\
			\hline
                        $A_0^{B\rightarrow D^{*}}$               &  $0.08\%$\\
                        \hline
			$f_B$                            &  $0.02\%$ \\
                        \hline
			\hline
		    Total                             & $13.35\%$\\
		    \hline
		\end{tabular}
		\caption{
			Error budget for the observable $R_{D^{*}\pi}$.}
		\label{tab:errorBDpi}
	\end{center}
\end{table}

The SM result is dominated by the contribution of $C_2$, thus we will get from $R_{D^*\pi}$ strong constraints on $C_2$ and
relatively weak ones on $C_1$.
To compute the experimental result we use~\cite{Huber:2016xod}
\begin{eqnarray}
  d\Gamma(\bar{B}^0_d \rightarrow D^{*+}l^{-}\bar{\nu}_l)/dq^2\Bigl|_{q^2=m^2_{\pi}}= (2.04 \pm 0.10)\cdot 10^{-3}
  \hbox{GeV}^{-2}\hbox{ps}^{-1}\, ,
\end{eqnarray}
together with \cite{Amhis:2016xyh}
\begin{eqnarray}
	\mathcal{B}r(\bar{B}^0\rightarrow D^{*+}\pi^{-})&=&
 (2.84\pm 0.15)\cdot 10^{-3},
\end{eqnarray}
to obtain
\begin{eqnarray}
R^{\rm Exp}_{D^{*}\pi}&=&
 (0.92\pm 0.07)\hbox{GeV}^2.
\label{eq:BDpiExp}
\end{eqnarray}
\begin{figure}
\centering
\includegraphics[height=5cm]{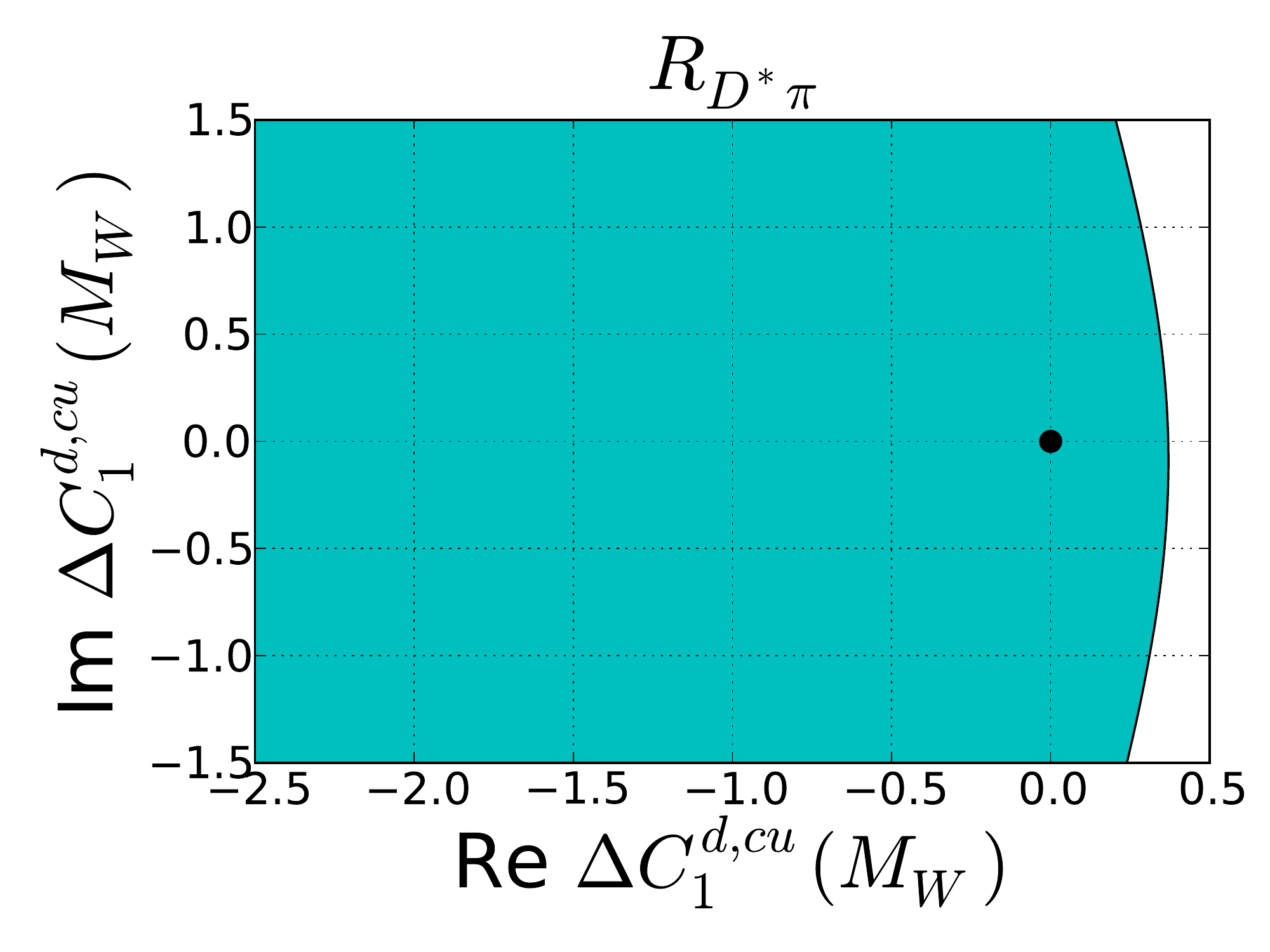}
\includegraphics[height=5cm]{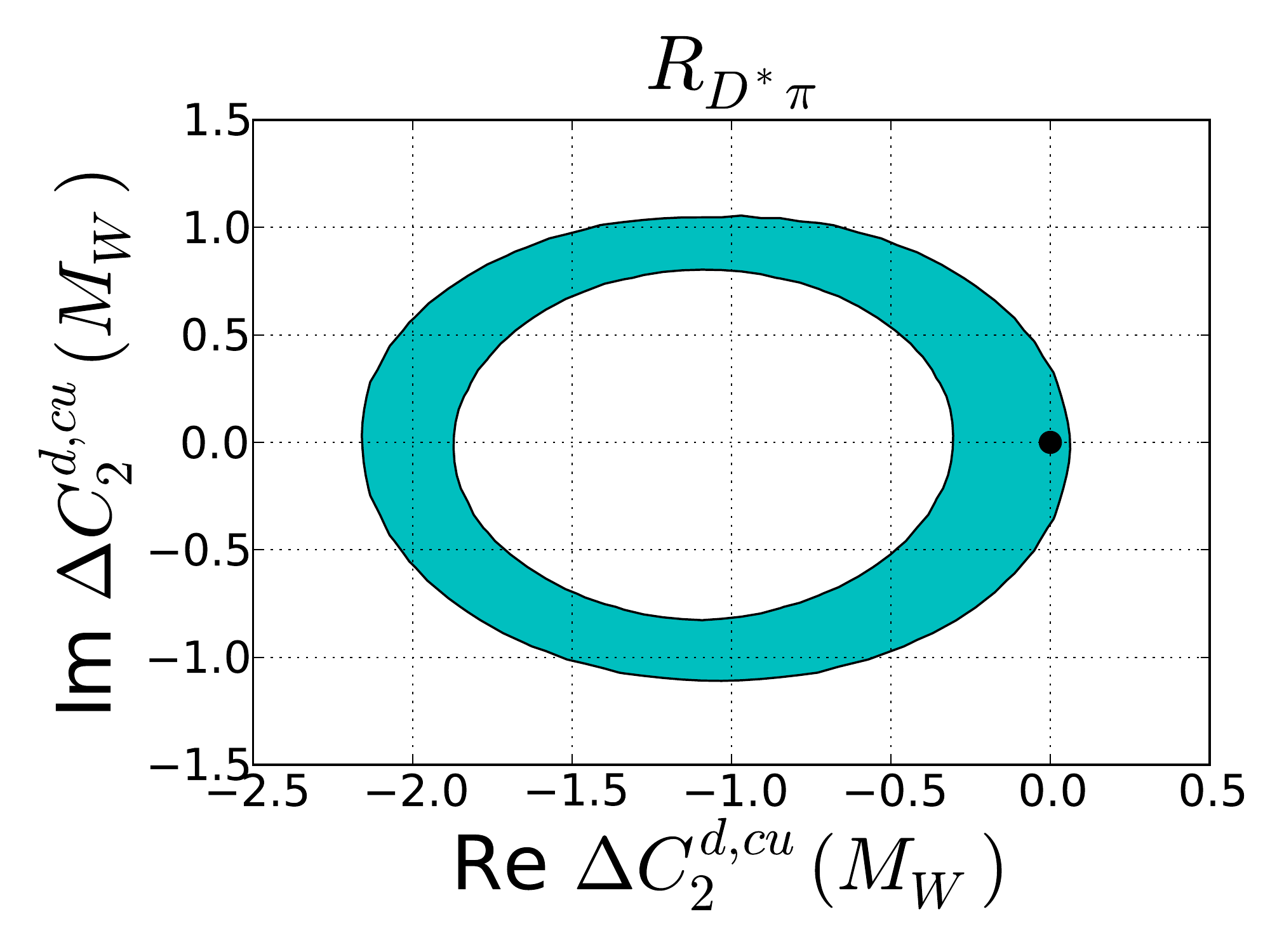}
\caption{
Potential regions for the NP contributions in $\Delta C^{d, cu}_1(M_W)$ and $\Delta C^{d, cu}_2(M_W)$  allowed by the observable 
$R_{D^{*}\pi}$ at $90\%$ C.L.. The black point corresponds to the SM value. Since $R_{D^{*}\pi}$ is dominated by $C_2$, we get strong constraints on $C_2$
and relatively weak ones on $C_1$.}
\label{fig:BDpi}
\end{figure}
Our $\chi^2$-fit provides the $90$ \% confidence level regions allowed by $\Delta C_1^{d, cu}(M_W)$ and $\Delta C_2^{d, cu}(M_W)$ 
displayed in Fig.~\ref{fig:BDpi}, which show that $\Delta C_1^{d, cu}(M_W)$ is quite unconstrained. On the other hand, there are
stronger restrictions on the values that $\Delta C_2^{d, cu}(M_W)$ can assume. This is not surprising considering that $C^{d, cu}_2$ gives the leading 
contribution to $\alpha^{D^*\pi}_2$; this can be seen in the NLO version of the formula for this term in Eq.~(\ref{eq:BDpi}).
%
%
%
%
%
%
%
%
%
%
%
%
%
\subsection{Observables constraining $b\rightarrow u\bar{u}d$ transitions}
\label{sec:buud}
We proceed to describe the constraints to the NP contributions  
$\Delta C^{d, uu}_{1,2}(M_W)$ entering in the CKM suppressed quark level 
transition $b\rightarrow u\bar{u}d$.
Our bounds are obtained taking into account both the branching ratios, but also the
CP asymmetries of the decays $B\to \pi\pi,\, \rho\pi,\, \rho\rho$ and using again QCDF for the theoretical description.
The combination of CP-conserving and CP-violating observables significantly shrinks the allowed region for
$\Delta C^{d, uu}_{2}(M_W)$. 
\subsubsection{$R_{\pi\pi}$}
\label{sec:Rpipi}
\begin{figure}
\centering
\includegraphics[height=5cm]{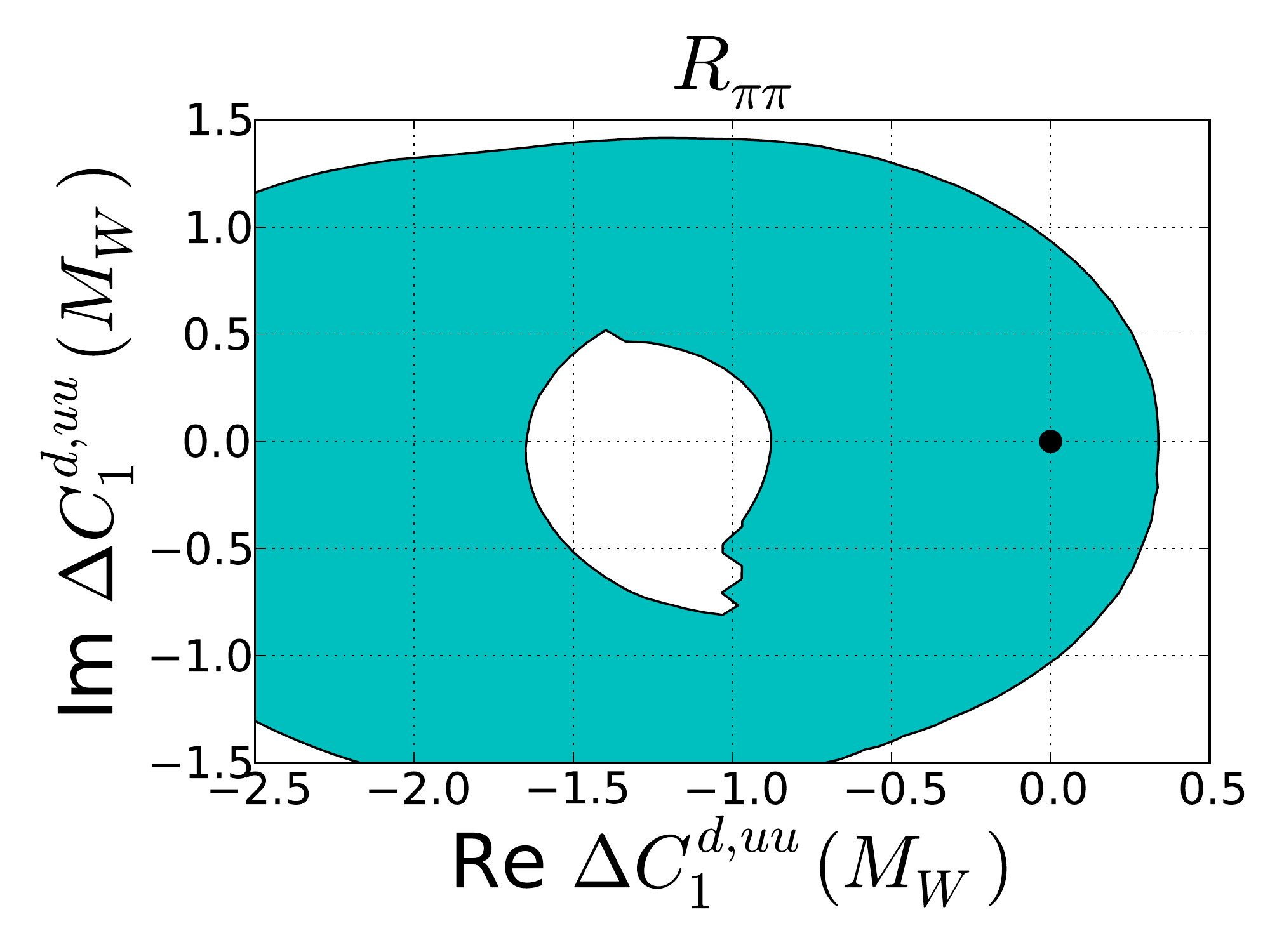}
\includegraphics[height=5cm]{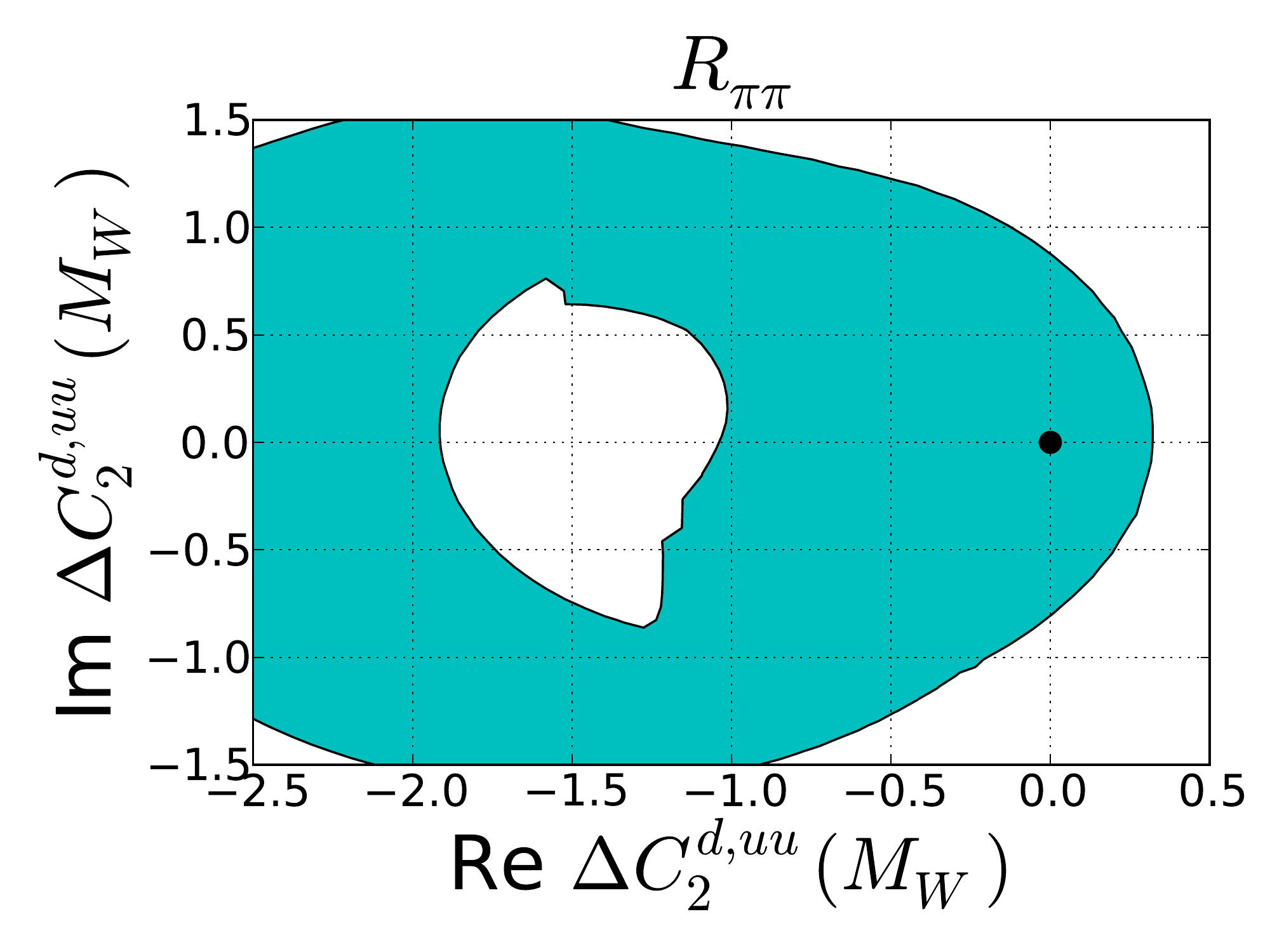}
\caption{Potential regions for the NP contributions $\Delta C^{d, uu}_1(M_W)$ and $\Delta C^{d, uu}_2(M_W)$ allowed by the observable 
  $R_{\pi\pi}$ at $90\%$ C.L.. The black point corresponds to the SM value. The dependence of $ R_{\pi\pi}$ is  symmetric in $C_1$ and $C_2$, therefore
  both Wilson coefficients are  constrained in an almost identical way.}
\label{fig:Rpipi}
\end{figure}

Our first observable is the theoretical clean ratio \cite{Bjorken:1988kk}
\begin{eqnarray}
  R_{\pi\pi}&=&\frac{\Gamma(B^+\rightarrow \pi^+\pi^0)}{d\Gamma(\bar{B}^0_d\rightarrow \pi^{+} \ell^{-}\bar{\nu}_{\ell})/dq^2|_{q^2=0}}
  \simeq
  3\pi^2f^2_{\pi}|V_{ud}|^2|\alpha_1^{\pi\pi} + \alpha_2^{\pi\pi}|^2,
\label{eq:Rpipi}
\end{eqnarray}
where $\ell^{-}=\mu^{-},~e^{-}$ and $\alpha_1^{\pi\pi}$, $\alpha_2^{\pi\pi}$ are the TA associated with the decays
$B\rightarrow \pi\pi$ which were introduced in a generic way in Eq.~(\ref{eq:alphaGen0}).
The dependence of $ R_{\pi\pi}$ is now symmetric in $C_1$ and $C_2$, so both Wilson coefficients will be constrained in an almost identical way.
Notice that the denominator
in Eq.~(\ref{eq:Rpipi}) refers to the differential distribution $d\Gamma(\bar{B}^0_d\rightarrow \pi^{+} \ell^{-}\bar{\nu}_{\ell})/dq^2$
evaluated at $q^2=0$, where $q^2$ is the four momentum transferred to the system composed by the $\ell^{-}$ and $\bar{\nu}_{\ell}$.
In Eq.~(\ref{eq:Rpipi}), our sensitivity to NP enters through the decay $B^+\rightarrow \pi^+\pi^0$ 
which is to a good degree of precision a pure tree level channel. 
We neglect hypothetical BSM effects in $\bar{B}^0_d\rightarrow \pi^{+} \ell^{-}\bar{\nu}_{\ell}$ for $\ell=e, \mu$, see
e.g. \cite{Banelli:2018fnx} for a recent investigation of such a possibility.
The observable  $R_{\pi\pi}$ is theoretically clean since it does not depend on the CKM matrix element $|V_{ub}|$, which cancels in the ratio.
Moreover, at leading order in $\alpha_s$ it is independent of the form factors $F^{B \rightarrow \pi}_+(0)= F^{B \rightarrow \pi}_0(0)$ which account for the hadronic transition
$B\rightarrow \pi$. However, these parameters
enter in the coefficients $\alpha_{1, 2}^{\pi\pi}$ once the spectator interaction contributions $H_{\pi \pi}$ are taken into 
account. More precisely,  they appear in the ratio $B_{\pi\pi}/A_{\pi\pi}$ inside $H_{\pi\pi}$, see Eqs.~(\ref{eq:GenPar}) 
and (\ref{eq:HardScattering}).
Currently, the coefficients $\alpha_{1,2}^{\pi\pi}$ in Eq.~(\ref{eq:Rpipi}) are available  up to NNLO in QCDF
 \cite{Beneke:2009ek, Beneke:2005vv, Bell:2007tv, Bell:2009nk}. In order
to optimize the computation time  of our $\chi^2$-fit,  we have accounted for the  NNLO effects using the following formula
\vspace{-0.2cm}
\begin{eqnarray}
\frac{\alpha_{1,2}^{\pi\pi}}{\alpha_{1,2}^{\text{NNLO}, \pi\pi}}&=&
\frac{\alpha_{1,2}^{\text{NLO}, \pi\pi}(\mu_0)}{\alpha_{1,2}^{(0)~\text{NLO}, \pi\pi}}.
\label{eq:repipi}
\end{eqnarray}
Where in Eq.~(\ref{eq:repipi}):
\begin{itemize}
\item $\alpha_{1,2}^{\text{NLO}, \pi\pi}(\mu_0)$ corresponds to the fully programmed NLO expression for the amplitude $\alpha^{\pi\pi}_{1,2}$.
      For this term, the renormalization scale is kept fixed to the value $\mu_0=m_b$ whereas the rest of the input parameters are allowed to float.
\item $\alpha_{1,2}^{\text{(0)~NLO}, \pi\pi}$ are the  NLO version of the amplitudes $\alpha^{\pi\pi}_{1,2}$  evaluated at the central value
      of all the input parameters and kept constant during the $\chi^2$-fit.
\item $\alpha^{\rm NNLO, \pi\pi}_{1, 2}$  are the NNLO version of the amplitude $\alpha^{\pi\pi}_{1,2}$. 
      We are interested in the NNLO results because of the reduction in the renormalisation scale dependency with respect to the NLO determination. 
      Therefore during the $\chi^2$-fit we have treated the  coefficients
      $\alpha^{\text{NNLO}, \pi\pi}_{1,2}$ as nuisance parameters given by \cite{Bell:2009fm}
\vspace{-0.2cm}
\begin{eqnarray}
\begin{split}
\alpha^{\text{NNLO}, \pi\pi}_{1}&=0.195^{ + 0.025 }_{ - 0.025} - \Bigl(0.101^{ + 0.021}_{ - 0.029}\Bigl)i,\\
\alpha^{\text{NNLO}, \pi\pi}_{2}&=1.013^{ + 0.008 }_{ - 0.011} + \Bigl( 0.027^{+ 0.020}_{ - 0.013}\Bigl)i,
\end{split}
\label{eq:NNLOa1a2_pipi}
\end{eqnarray}
where the error indicated arises only from the renormalization scale uncertainty. Alternatively, we also tested the numerical values provided in 
~\cite{Beneke:2009ek} which give consistent results once the uncertainties arising by varying $\mu$ and $\mu_{h}$, \footnote{T. Huber, private communication.}
are taken into account. 
\end{itemize}
We predict the SM value of $ R_{\pi\pi}$  to be
\vspace{-0.2cm}
\begin{eqnarray}
\label{eq:Rpipi_val}
R^{\rm SM}_{\pi\pi}=\Bigl(0.70 \pm 0.14\Bigl),
\end{eqnarray}
with the partial contributions to the total error shown in Table~\ref{table:Rpipi}.
\begin{table}
\begin{center}
\begin{tabular}{ |c|c|c|}
\hline
Parameter  & Relative Error \\
\hline
\hline
$X_{H}$     & $16.86\%$ \\
\hline
$\lambda_B$ & $8.85\%$  \\ 
\hline
$\mu$ & $4.42\%$  \\
\hline
$a^{\pi}_{2}$ & $2.57\%$ \\
\hline
$F^{B\rightarrow\pi}_+(0)$& $1.77\%$ \\
\hline
$f_{\pi}$ & $1.35\%$ \\
\hline
$m_{s}$ & $0.68\%$ \\
\hline
$\Lambda^{QCD}_5$ & $0.25\%$ \\
\hline
$f_{B}$ & $0.14\%$ \\
\hline
$m_b$ & $0.04\%$ \\
\hline
$V_{us}$ & $0.01\%$ \\
\hline
\hline
Total & $19.86\%$ \\
\hline
\end{tabular}
\caption{Error budget for the observable $R_{\pi\pi}$. Here $X_H$ accounts for the endpoint singularities from 
	hard scattering spectator interactions. $F^{B\rightarrow \pi}_{+}(0)$
is the relevant form factor for the transitions $B\rightarrow \pi$. The parameter $\lambda_B$ is the inverse moment of the
LCDA of the $B$ meson and $a^{\pi}_{2}$  is the second Gegenbauer moment for the $\pi$ meson. 
}
\label{table:Rpipi}
\end{center}
\end{table}
To calculate the experimental result, we consider the following updated value for the branching fraction
for the process $B^{+}\rightarrow \pi^{+}\pi^{0}$ \cite{Tanabashi:2018oca}
\begin{eqnarray}
\label{eq:Bpipi}
\mathcal{B}r(B^{+}\rightarrow \pi^{+}\pi^{0})&=&(5.5\pm 0.4) \cdot 10^{-6},
\end{eqnarray}
together with the product \cite{Gonzalez-Solis:2018ooo}
\begin{eqnarray}
\label{eq:VubFBpi}
|V_{ub}F^{B\rightarrow \pi}_+(0)|&=&(9.25 \pm 0.31)\cdot 10^{-4},
\end{eqnarray}
 which was extracted via a fit  to data including experimental results from BaBar, Belle and CLEO  
\cite{delAmoSanchez:2010af, Lees:2012vv, Ha:2010rf, Sibidanov:2013rkk, Adam:2007pv} under the assumption of the
SM, neglecting the mass of the light leptons  and keeping the mass of the $B^*$ meson fixed.
Using the inputs indicated in Eqs.~(\ref{eq:Bpipi}) and (\ref{eq:VubFBpi}) we obtain the following result for the experimental
value of $R_{\pi\pi}$
\vspace{-0.2cm}
\begin{eqnarray}
R^{\rm Exp}_{\pi\pi}&=&\Bigl(0.83\pm 0.08\Bigl).
\end{eqnarray}
This determination is in agreement with the result given in \cite{Beneke:2009ek}, however, the uncertainty is reduced by nearly
$50\%$ due to the update on the product $|V_{ub}F^{B\rightarrow \pi}_+(0)|$ shown in Eq.~(\ref{eq:VubFBpi}).
The allowed regions for $\Delta C^{d, \, uu}_1(M_W)$ and $\Delta C^{d, \, uu}_2(M_W)$ are shown in Fig.~\ref{fig:Rpipi} - we note here rather stringent
constraints on positive and real values of  $\Delta C^{d, \, uu}_1(M_W)$ and $\Delta C^{d, \, uu}_2(M_W)$. 
\subsubsection{$S_{\pi\pi}$}
\label{sec:Spipi}
Since our NP contributions are allowed to be complex, we are exploring the possibility of
having new CP violating phases. We can constrain these effects through the time-dependent  asymmetries 
\begin{eqnarray}
\mathcal{A}^{CP}_{f}(t)&=& \frac{d\Gamma [\bar{B}^0_q\rightarrow f](t)/dt - d\Gamma[B^0_q\rightarrow f](t)/dt}
{d\Gamma [\bar{B}^0_q\rightarrow f](t)/dt + d\Gamma[B^0_q\rightarrow f](t)/dt}\nonumber\\
&\simeq&S_f \sin \Delta M_q t - C_f \cos \Delta M_q t,
\label{eq:CPint}
\end{eqnarray}
where we have neglected the effects of the observable $\Delta \Gamma_q$ entering in the denominator - this is only justified for the case
of $B_d$-mesons. 
The symbol $f$ in Eq.~(\ref{eq:CPint}) denotes a final state to which both, the $B_q^0$ and the $\bar{B}_q^0$ meson can decay, for $q=d, s$. 
The mixing induced ($S_f$) and direct CP asymmetries ($C_f$) are defined as
\begin{eqnarray}
  S_f\equiv \frac{2~\rm Im (\lambda^q_f)}{1 + |\lambda^q_f|^2},
  &&
  C_f\equiv\frac{1-|\lambda^q_f|^2}{1 + |\lambda^q_f|^2}.
\label{eq:Sf}
\end{eqnarray}
with the parameter $\lambda^q_f$ given by
\begin{eqnarray}
\lambda^q_f:= \frac{q}{p}\Bigl|_{B_q}  \frac{\bar{A}^q_f}{A^q_f}.
\label{eq:lambdaf}
\end{eqnarray}
In Eq.~(\ref{eq:lambdaf}) the amplitude  for the process $B^0_q\rightarrow f$ has been denoted as $A^q_f$ and the one for 
$\bar{B}^0_q\rightarrow f$ as $\bar{A}^q_f$ .
Finally,
\begin{eqnarray}
\frac{q}{p}\Bigl|_{B_q}=\frac{M^{q*}_{12}}{|M^{q}_{12}|},
\label{eq:qop}
\end{eqnarray}
where $M^{d}_{12}$ is the contribution from virtual internal particles to the $B^0_q-\bar{B}^0_q$ mixing diagrams.
For instance in the case of $B_d$ mesons we get
\begin{eqnarray}
\frac{q}{p}\Bigl|_{B_d}=\left[\frac{V_{td} V^*_{tb} }{|V_{td}V^*_{tb}|}\right]^2.
\label{eq:qopBeta}
\end{eqnarray}
Notice that the observable $S_f$, in Eq.~(\ref{eq:Sf}), is particularly sensitive to the imaginary components 
of $\Delta C_1(M_W)$ and $\Delta C_2(M_W)$.
\\ 
For the decays $\bar{B}_d^0\rightarrow \pi^+\pi^-$  and $B_d^0\rightarrow \pi^+\pi^-$ we get
\begin{align}
\label{eq:Spipi_def}
S_{\pi\pi}=\frac{2~\rm Im \Bigl( \lambda^d_{\pi\pi}  \Bigl)}
      {1+| \lambda^d_{\pi\pi} |^2} \, , & \hspace{1cm}
\lambda^d_{\pi\pi}=\left[\frac{V_{td} V^*_{tb} }{|V_{td}V^*_{tb}|}\right]^2\frac{\bar{\mathcal{A}}_{\pi^+\pi^-}}{\mathcal{A}_{\pi^+\pi^-}}.
\end{align}
Here $\bar{\mathcal{A}}_{\pi^+\pi^-}$  and $\mathcal{A}_{\pi^+\pi^-}$ denote the transition amplitudes for the processes
$\bar{B}_d^0\rightarrow \pi^+\pi^-$  and $B_d^0\rightarrow \pi^+\pi^-$ respectively.  They have been calculated in \cite{Beneke:2003zv}
using the QCDF formalism  briefly described in Section~\ref{sec:QCDF}.  The explicit expression for $\bar{\mathcal{A}}_{\pi^+\pi^-}$ is 
\begin{align}
  \bar{\mathcal{A}}_{\pi^+\pi^-}&=
  A_{\pi \pi}\Bigl(\lambda^{(d)}_u \alpha_2^{\pi\pi} + \lambda^{(d)}_u \beta_2^{\pi\pi} +
  \sum_{p=u,c}\lambda^{(d)}_p\Bigl[\tilde{\alpha}_4^{p,\pi\pi}+\tilde{\alpha}^{p,\pi\pi}_{4,EW}\nonumber\\
                & + \beta_3^{p,\pi\pi}  -1/2\beta^{p, \pi\pi}_{3,EW} + 2 \beta^{p, \pi\pi}_4 + 1/2\beta^{p, \pi\pi}_{4,EW}\Bigl]\Bigl).
\label{eq:ABpipi}
\end{align}
To determine the remaining amplitude $\mathcal{A}_{\pi^+\pi^-}$, the CP conjugate of the expression in Eq.~(\ref{eq:ABpipi}) has to be obtained.
The parameters  $\lambda^{(d)}_{u,c}$ in Eq.~(\ref{eq:ABpipi}) correspond to products of CKM matrix elements as defined in 
Eq.~(\ref{eq:lambmdadef}).
Notice that our sensitivity towards NP in tree level enters mainly through  $\alpha_2^{\pi\pi}$, which according to
Eq.~(\ref{eq:alphaGen0}) has a leading dependency on  $\Delta C^{d, \, uu}_2(M_W)$. Therefore, the observable $S_{\pi\pi}$ yields to strong
constraints on  $\Delta C^{d, \, uu}_2(M_W)$, while giving weak ones in $\Delta C^{d, \, uu}_1(M_W)$.
Besides the TA $\alpha_2^{\pi\pi}$, which is introduced in our analysis at NNLO following the prescription 
shown in Eq.~(\ref{eq:repipi}),
there are now also contributions from  QCD and electroweak penguins given by $\tilde{\alpha_4}^{p,\pi\pi}$
and $\tilde{\alpha_4}^{p,\pi\pi}_{EW}$ respectively. Finally $\beta^{p, \pi\pi}_4$ accounts for QCD penguin annihilation and 
$\beta^{p, \pi\pi}_{4, EW}$ for electroweak penguin annihilation. All the TA can be calculated using Eq.~(\ref{eq:alphaGen0}) together with the
information presented in Appendix \ref{Sec:QCDFact}.
At leading order in $\alpha_s$, the  normalization factor $A_{\pi\pi}$ introduced in Eq.~(\ref{eq:GenPar}),
which depends on the form factor $F^{B\rightarrow \pi}_+(0)$ and the decay constant $f_{\pi}$, cancels in the ratio
given in Eq.~(\ref{eq:Spipi_def}). However it appears again once interactions with the spectator are taken into account. This leads
to small effects in the error budget of $\mathcal{O}(1~\%)$ and $\mathcal{O}(0.1~\%)$ from $F^{B\rightarrow \pi}_+(0)$ and  $f_{\pi}$ 
respectively, see Table \ref{tab:tableSpipi}. Our theoretical prediction for the SM value of the asymmetry $S_{\pi\pi}$ is
\begin{eqnarray}
S^{\rm SM}_{\pi\pi} = -0.59 \pm  0.25.
\label{eq:Spipi}
\end{eqnarray}

\begin{figure}
\centering
\includegraphics[height=5cm]{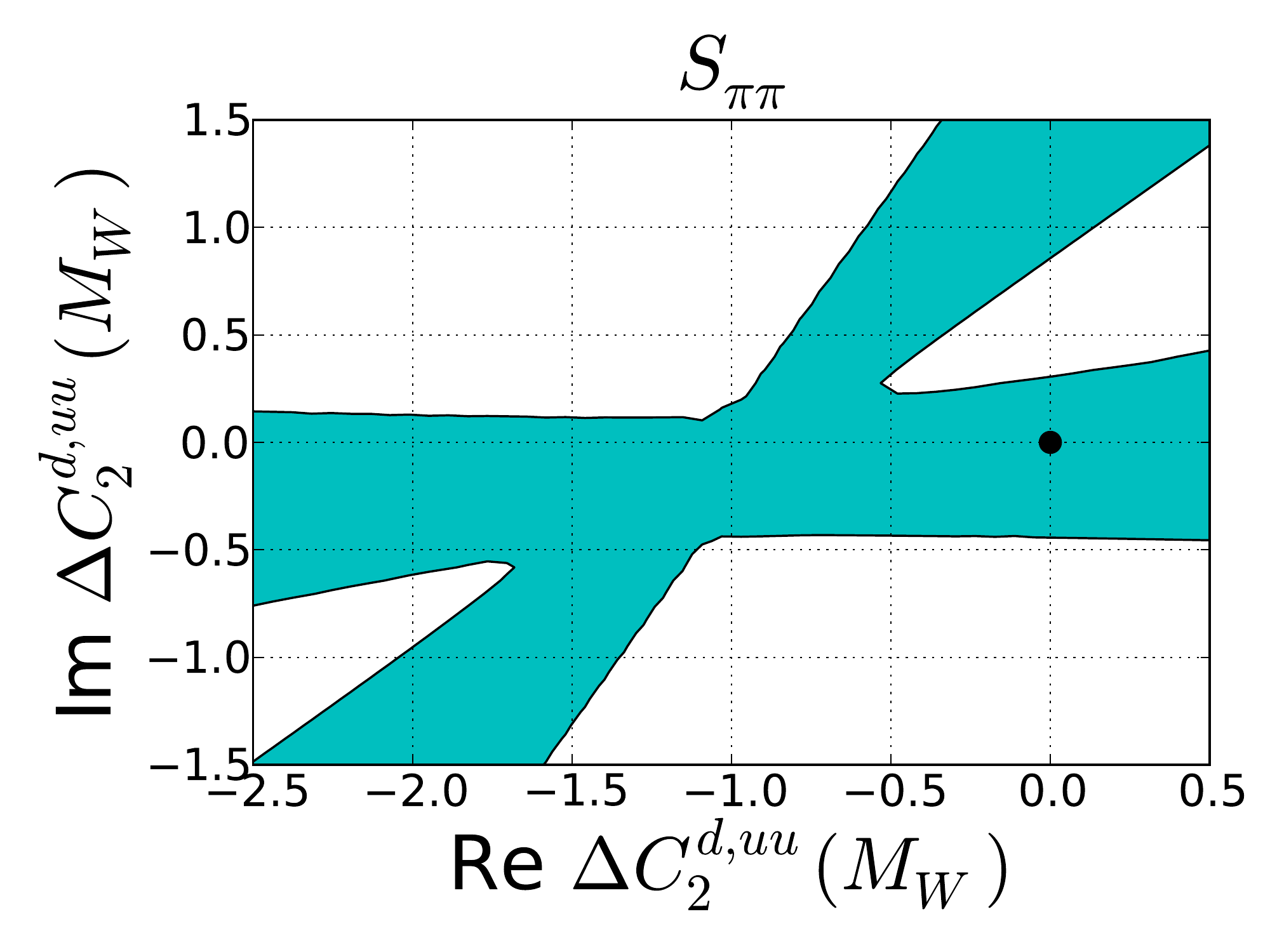}
\caption{
Potential regions for the NP contributions in $\Delta C^{d, uu}_2(M_W)$ allowed by the observable 
$S_{\pi\pi}$ at $90\%$ C.L., the shift in the Wilson coefficient $\Delta C^{d, uu}_1(M_W)$ is only weakly constrained
and therefore not shown.
The black point corresponds to the SM value.}
\label{fig:Spipi}
\end{figure}

\begin{table}
\begin{center}
\begin{tabular}{ |c|c|}
\hline
Parameter  & Relative Error \\
\hline
$X_{A}$     &  $41.76\%$ \\
\hline
$\gamma$                & $6.24\%$  \\
\hline
$m_s$                  & $4.43\%$ \\
\hline
$|V_{ub}/V_{cb}|$              & $4.31\%$ \\
\hline
$X_{H}$   &$3.08\%$ \\
\hline
$\mu$                        & $2.79\%$   \\
\hline
$\Lambda^{QCD}_5$            & $2.25\%$ \\
\hline
$\lambda_B$ & $1.55\%$ \\
\hline
$F^{B\rightarrow\pi}_+$& $0.89\%$  \\
\hline
$m_b$                  & $0.76\%$  \\
\hline
$|V_{us}|$               & $0.13\%$  \\
\hline
$f_B$                  & $0.07\%$  \\
\hline
$m_c$                         & $0.06\%$   \\
\hline
$f_{\pi}$              & $0.06\%$  \\
\hline
$a^{\pi}_{2}$        & $0.03\%$ \\
\hline
\hline
Total & $42.98\%$ \\
\hline
\end{tabular}
\caption{Error budget for the observable $S_{\pi\pi}$. Most of the inputs coincide with those for $R_{\pi\pi}$ described
in Table \ref{table:Rpipi}. Additionally the effects of annihilation topologies are accounted by $X_A$.}
\label{tab:tableSpipi}
\end{center}
\end{table}
For the corresponding experimental value we have \cite{Amhis:2016xyh}
\begin{align}
S^{\rm Exp}_{\pi\pi}&=-0.63 \pm 0.04,
\end{align}
showing consistency with the SM estimation in Eq.~(\ref{eq:Spipi}). The relevant constraints on $\Delta C_2^{d, \, uu} (M_W)$  derived from $S_{\pi\pi}$ are 
presented in Fig.~\ref{fig:Spipi} - constraints on $\Delta C_1^{d, \, uu} (M_W)$ are very weak and will thus not be shown.
\subsubsection{$S_{\rho\pi}$}
\label{sec:Srhopi}
\begin{figure}
\centering
\includegraphics[height=5cm]{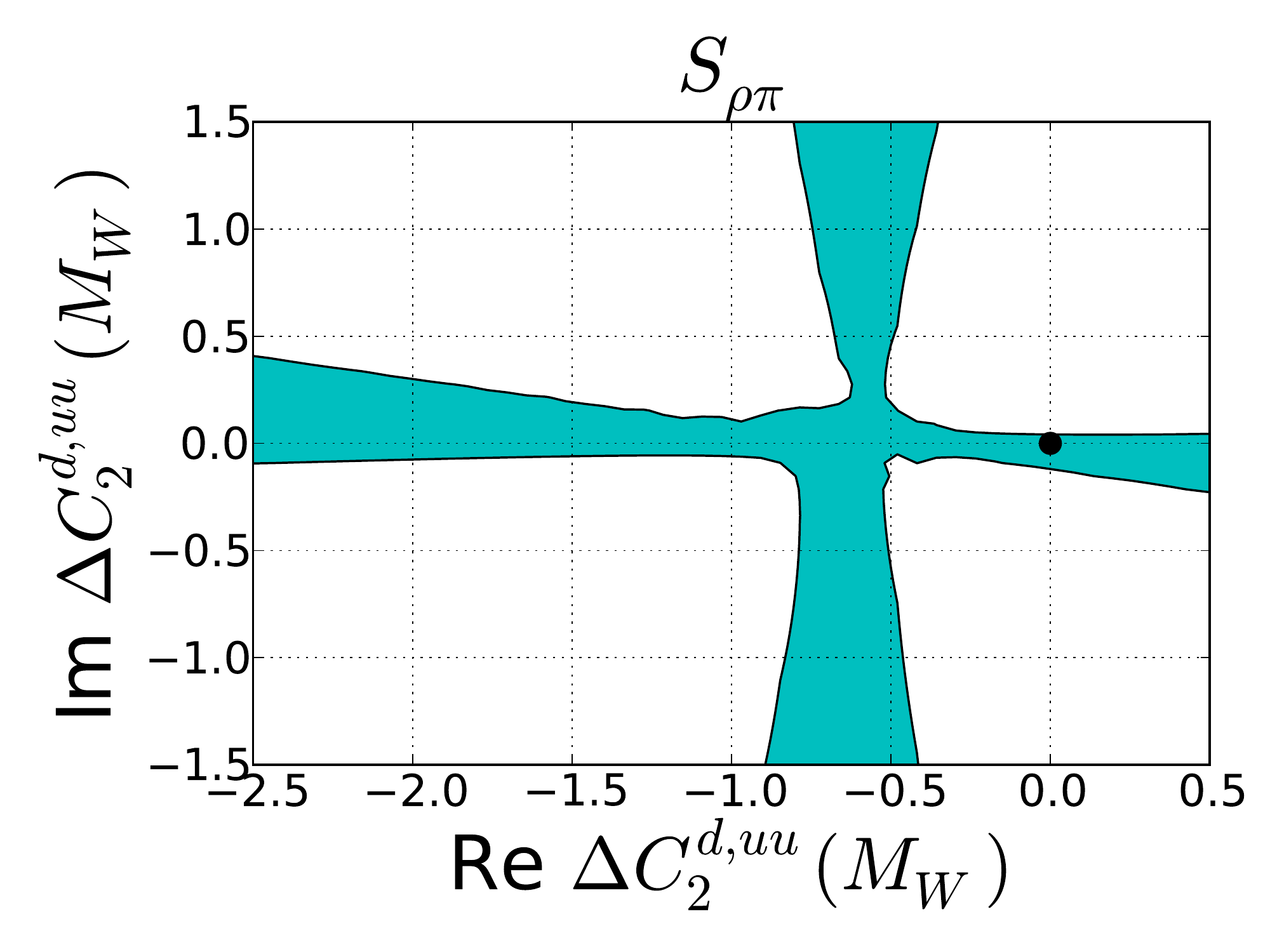}
\caption{
Potential regions for the NP contributions in $\Delta C^{d, uu}_2(M_W)$ allowed by the observable 
$S_{\rho\pi}$ at $90\%$ C.L., the shift in the Wilson coefficient $\Delta C^{d, uu}_1(M_W)$ is only weakly constrained
and therefore not shown.
The black point corresponds to the SM value.}
\label{fig:Srhopi}
\end{figure}
We also included the mixing induced CP asymmetry associated with the decays $B_d, \bar{B}_d \rightarrow \rho\pi$. Our evaluation
is based in the following definition
\begin{align}
S_{\pi\rho}&=\frac{1}{2}\Bigl( \tilde{S}_{\pi\rho}  +  \tilde{S}_{\rho\pi} \Bigl),
\end{align}
with the partial contributions given by
\begin{eqnarray}
\tilde{S}_{\pi\rho}=\frac{2~\rm Im \Bigl( \lambda^{d}_{\pi\rho} \Bigl)}{1 + |\lambda^d_{\pi\rho}|^2},
&&
\tilde{S}_{\rho\pi}=\frac{2~\rm Im \Bigl( \lambda^{d}_{\rho\pi} \Bigl)}{1 + |\lambda^d_{\rho\pi}|^2},
\end{eqnarray}
with
\begin{eqnarray}
\lambda^d_{\pi\rho}= \left[\frac{V_{td} V^*_{tb} }{|V_{td}V^*_{tb}|}\right]^2 \frac{\bar{\mathcal{A}}_{\pi^+\rho^-}}{\mathcal{A}_{\rho^+ \pi^-}},
&&
\lambda^d_{\rho\pi}= \left[\frac{V_{td} V^*_{tb} }{|V_{td}V^*_{tb}|}\right]^2  \frac{\bar{\mathcal{A}}_{\rho^+ \pi^-}}{\mathcal{A}_{ \pi^+ \rho^-}}.
\label{eq:ratiosArhopi}
\end{eqnarray}
The individual amplitudes  $\bar{\mathcal{A}}_{\pi^+\rho^-}$ and $\bar{\mathcal{A}}_{\rho^+ \pi^-}$ for the processes $\bar{B}^0_d\rightarrow \pi^+ \rho^-$   and $\bar{B}^0_d\rightarrow  \rho^+ \pi^-$
are respectively
\begin{align}
\bar{\mathcal{A}}_{\pi^+\rho^-}&= A_{\pi\rho}\Bigl(\lambda^{(d)}_u \alpha^{\pi\rho}_2 + 
\sum_{p=u,c}\lambda^{(d)}_p\Bigl[ \tilde{\alpha}_4^{p, \pi\rho } + \tilde{\alpha}_{4,EW}^{p, \pi\rho}\nonumber\\
&+ \beta^{p, \pi\rho}_3  + \beta^{p, \pi\rho}_4 - \frac{1}{2}\beta^{p, \pi\rho}_{3,EW} -\frac{1}{2}\beta^{p, \pi\rho}_{4,EW} \Bigl] \Bigl)\nonumber\\
&+ A_{\rho\pi}\Bigl(\lambda^{(d)}_u\beta^{\rho\pi}_1 + 
\sum_{p=u,c}\lambda^{(d)}_p\Bigl[\beta^{p, \rho\pi}_4  + \beta^{p, \rho\pi}_{4,EW} \Bigl]\Bigl),\nonumber\\
\bar{\mathcal{A}}_{\rho^+ \pi^-}&= A_{\rho\pi}\Bigl(\lambda^{(d)}_u \alpha^{\rho\pi}_2 + 
\sum_{p=u,c}\lambda^{(d)}_p\Bigl[ \tilde{\alpha}_4^{p, \rho\pi} + \tilde{\alpha}_{4,EW}^{p, \rho\pi} + \beta^{p, \rho\pi}_3\nonumber\\
& + \beta_4^{p,\rho\pi} - \frac{1}{2}\beta^{p, \rho\pi}_{3,EW} -\frac{1}{2}\beta^{p, \rho\pi}_{4,EW} \Bigl] \Bigl) \nonumber\\
& + A_{\pi\rho}\Bigl(\lambda^{(d)}_u\beta^{\pi\rho}_1 +  \sum_{p=u,c} \lambda^{(d)}_p\Bigl[\beta^{p, \pi\rho}_4 + \beta^{p, \pi\rho}_{4,EW}\Bigl]\Bigl),
\label{eq:pirho}
\end{align}
with $\lambda^{(d)}_{u,c}$ given by Eq. (\ref{eq:lambmdadef}).
In analogy with  $S_{\pi\pi}$, there are also tree level amplitudes given by  $\{\alpha_2^{\pi\rho},~\alpha_2^{\rho\pi}\}$, together with
QCD and electroweak penguin contributions introduced through $\{\alpha^{\pi\rho}_4,~\alpha^{\rho\pi}_4\}$ and 
$\{\tilde{\alpha}^{\pi\rho}_4,~\tilde{\alpha}^{\rho\pi}_4\}$ respectively. Moreover, the coefficients  
$\{\beta_{1}^{p, \pi\rho},~\beta_{1}^{p, \rho\pi}\}$ correspond to current-current annihilation,
$\{\beta_{3,4}^{p, \pi\rho},~\beta_{3,4}^{p, \rho\pi}\}$ to QCD penguin annihilation and 
$\{\beta^{p, \pi\rho}_{4, EW},~\beta^{p, \rho \pi}_{4, EW}\}$ to electroweak penguin annihilation.  The TA can be 
obtained using  Eq.~(\ref{eq:alphaGen0}) and the information provided in Appendix \ref{Sec:QCDFact}. 
Our SM determination of the mixing induced CP asymmetry reads
\vspace{-0.2cm}
\begin{align}
S^{\rm SM}_{\pi\rho}= -0.04 \pm 0.08,
\label{eq:Srhopi_value}
\end{align}
which is compatible with the current experimental average \cite{Amhis:2016xyh}
\vspace{-0.2cm}
\begin{align}
S^{\rm Exp}_{\pi\rho}&=0.06 \pm 0.07.
\end{align}

\begin{table}
\begin{center}
\begin{tabular}{ |c|c|}
\hline
Parameter  & Relative Error\\
\hline
\hline
$\gamma$                & $142.75\%$\\
\hline
$X_A$ & $96.41\%$\\
\hline
$X_H$& $58.85\%$\\
\hline
$|V_{ub}/V_{cb}|$              & $46.96\%$\\
\hline
$m_s$                  & $37.31\%$\\
\hline
$\mu$                  & $20.58\%$\\
\hline
$a^{\rho}_{2}$        & $18.34\%$\\
\hline
$\Lambda^{QCD}_5$ & $13.16\%$\\
\hline
$\lambda_B$       & $8.27\%$\\
\hline
$A^{B\rightarrow\rho}_0$& $7.06\%$\\
\hline
$a^{\pi}_{2}$        & $6.26\%$\\
\hline
$m_b$                  & $5.22\%$\\
\hline
$F^{B\rightarrow\pi}_+$ & $2.19\%$\\
\hline
$|V_{us}|$ & $1.38\%$\\
\hline
$f_{\rho}$ & $0.93\%$\\
\hline
\end{tabular}
	\caption{Error budget for the observable $S_{\pi\rho}$ (Part I). Here $A^{B\rightarrow \rho}_0$ is the form factor for the transition 
$B \rightarrow \rho$, $a^{\rho}_{2}$ is the Gegenbauer moment for the leading twist LCDA for the $\rho$ meson.
}
\label{tab:tableSrhopi1}
\end{center}
\end{table}

\begin{table}
\begin{center}
\begin{tabular}{ |c|c|}
\hline
Parameter  & Relative Error\\
\hline
\hline
$f_{\pi}$ & $0.51\%$\\
\hline
$f_B$ & $0.26\%$\\
\hline
$f^{\perp}_{\rho}$ & $0.23\%$\\
\hline
$|V_{cb}|$        & $0.06\%$\\
\hline
$m_c$ & $0.02\%$\\
\hline
\hline
Total & $194.57\%$\\
\hline
\end{tabular}
\caption{Error budget for the observable $S_{\pi\rho}$ (Part II).}
\label{tab:tableSrhopi2}
\end{center}
\end{table}
The relative errors from each one of the inputs for $S_{\pi\rho}$ are presented in Tables \ref{tab:tableSrhopi1} and \ref{tab:tableSrhopi2}, 
it can be seen that 
this observable is highly sensitive to the CKM input $\gamma$ leading to a relative uncertainty of $\mathcal{O}(100 \%)$.  
This is related to the fact
that in the ratio $\lambda_{\rho\pi}$ given in Eq.~(\ref{eq:ratiosArhopi}) we have:
\begin{eqnarray}
  \rm Re \left( \frac{\mathcal{A}_{\rho^+ \pi^-}}{A_{\pi^+\rho^-}} \right)
  \approx
  \rm Im \left( \frac{\mathcal{A}_{\rho^+ \pi^-}}{A_{\pi^+\rho^-}} \right),
\end{eqnarray}
and
\begin{eqnarray}
  \rm Re \left(\left[\frac{V_{td} V^*_{tb} }{|V_{td}V^*_{tb}|}\right]^2\right)
  \approx 
- \rm Im \left(\left[\frac{V_{td} V^*_{tb} }{|V_{td}V^*_{tb}|}\right]^2\right) \, ,
\end{eqnarray}
which lead to a very strong cancellation on the resulting imaginary component.
The allowed NP regions for $\Delta C^{d, uu}_2(M_W)$
are displayed in Fig.~\ref{fig:Srhopi}. Here we can see how, in spite of having an uncertainty of $\mathcal{O}(100 \%)$, the observable
$S_{\pi \rho}$ rules out large sections in the complex plane of $\Delta C^{d, uu}_2(M_W)$ and consequently
deserves to be included in the analysis of $C^{d, uu}_2$. In contrast
we find weak bounds for $\Delta C^{d, uu}_1(M_W)$ that are not strong enough to be taken into account.  This is explained
by the strong dependence of the amplitudes in Eqs.~(\ref{eq:pirho}) on $C^{d, uu}_2(M_W)$, which enters through
 $\alpha^{\pi \rho}_2$ and $\alpha^{\rho \pi}_2$ as shown in Eq.~(\ref{eq:alphaGen0}).
\subsubsection{$R_{\rho\rho}$}
\label{sec:Rrhorho}
\begin{figure}
\centering
\includegraphics[height=5cm]{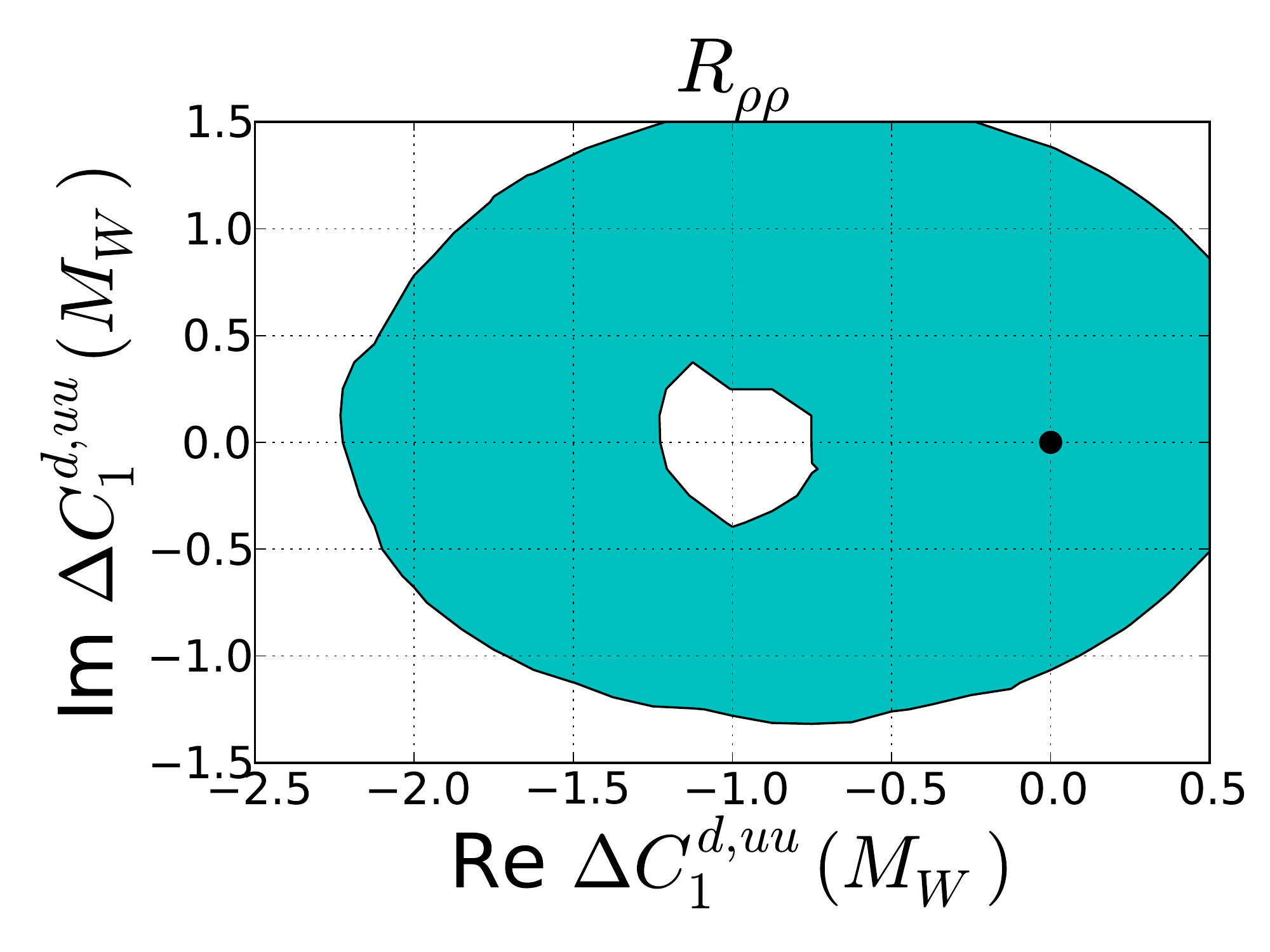}
\caption{
Potential regions for the NP contributions in $\Delta C^{d, uu}_1(M_W)$ allowed by the observable 
$R_{\rho\rho}$ at $90\%$ C.L.. The bounds on  $\Delta C^{d, uu}_2(M_W)$ are very weak and hence not shown.
The black point corresponds to the SM value.}
\label{fig:Rrhorho}
\end{figure}

To obtain extra constraints on NP contributions to the tree level Wilson coefficients for the transition $b\rightarrow u\bar{u}d$ 
we include the ratio
\begin{eqnarray}
  R_{\rho\rho} &= &
  \frac{\mathcal{B}r\left(B^{-}\rightarrow \rho_L^{-}\rho_L^{0}\right)}
       {\mathcal{B}r\left(\bar{B}_d^0\rightarrow\rho_L^{+}\rho_L^{-}\right)}
  =\frac{\left|\mathcal{A}_{\rho^-\rho^0}\right|^2}{\left|\mathcal{A}_{\rho^+\rho-}\right|^2} \, ,
\label{eq:Rrhorhodef}
\end{eqnarray}
where $\mathcal{A}_{\rho^-\rho^0} $ and $\mathcal{A}_{\rho^+\rho-}$ are the amplitudes for the processes $B^-\rightarrow \rho^-_L\rho^0_L$ and
$\bar{B}_d^0\rightarrow \rho^+_L\rho^-_L$ respectively. In terms of TAs they can be written
as \cite{Bartsch:2008ps, Beneke:2006hg}
\vspace{-0.2cm}
\begin{align}
\label{eq:A_rho_rho}
\mathcal{A}_{\rho^-\rho^0}&= \frac{A_{\rho\rho}}{\sqrt{2}}  \Bigl[ \lambda^{(d)}_u\Bigl(\alpha^{\rho\rho}_{1}+\alpha^{\rho\rho}_{2}\Bigl)
                                     +\frac{3}{2}\sum_{p=u,c}\lambda^{(d)}_{p}\Bigl(\alpha^{p, \rho\rho}_{7}+
                                      \alpha^{p,\rho\rho}_{9}+\alpha^{p,\rho\rho}_{10}\Bigl)\Bigl],\nonumber\\
\mathcal{A}_{\rho^+\rho^-}&=A_{\rho\rho}\Bigl[\lambda^{(d)}_u\Bigl(\alpha^{\rho\rho}_{2} +\beta^{\rho\rho}_{2}\Bigl)+
                                        \sum_{p=u,c}\lambda^{(d)}_p\Bigl(\alpha^{p,\rho\rho}_4 + \alpha^{p,\rho\rho}_{10}\nonumber\\
                                        &+\beta^{p,\rho\rho}_3+2\beta^{p,\rho\rho}_{4}-\frac{1}{2}\beta^{p,\rho\rho}_{3,EW}
                                          +\frac{1}{2}\beta^{p,\rho\rho}_{4,EW} \Bigl)\Bigl].
\end{align}
Here we expect a stronger dependence on $C_1$ compared to $C_2$.
As indicated in Eq.~(\ref{eq:A_rho_rho}), in addition to the tree level contributions $\alpha^{\rho\rho}_{1,2}$, we can also identify
QCD $\alpha^{\rho\rho}_4$ and electroweak penguins $\alpha^{\rho\rho}_{7, 9, 10}$. Moreover 
 QCD  penguin annihilation topologies enter through $\beta^{p, \rho\rho}_{3,4}$. On the other hand electroweak penguin annihilation is given by
$\beta^{p, \rho\rho}_{3,4, EW}$. The expressions for the topological amplitudes obey the structure indicated in Eq.~(\ref{eq:alphaGen0}) 
and can be calculated explicitly using the information provided in Appendix \ref{Sec:QCDFact}.
Currently  $\alpha^{\rho\rho}_{1,2}$ are available up to NNLO, we introduce these effects following the same procedure used
for the determination of $\alpha^{\pi\pi}_{1,2}$. Thus, we apply Eq.~(\ref{eq:repipi}) under the replacements
$\alpha^{\rm NNLO, \pi \pi}_i \rightarrow \alpha^{\rm NNLO, \rho_L \rho_L}_i$, 
$\alpha^{\rm NLO, \pi\pi}_i \rightarrow \alpha^{\rm NLO, \rho_L\rho_L}_i$ and $\alpha^{\rm NLO, \pi\pi}_{i} 
\rightarrow \alpha_{0,i}^{\rm NLO, \rho_L\rho_L}$, with $i=1,2$. 
For the corresponding NNLO components we use \cite{Bell:2009fm}
\begin{align}
\alpha^{\text{NNLO},\rho_L\rho_L}_{1} &=0.177^{+ 0.025}_{- 0.029} - \Bigl( 0.097^{+ 0.021}_{-0.029}\Bigl)i,\nonumber\\
\alpha^{\text{NNLO},\rho_L\rho_L}_{2} &=1.017^{+ 0.010}_{- 0.011}+\Bigl(0.025^{+0.019}_{-0.013}\Bigl)i.
\label{eq:rhorhoNNLO}
\end{align}
The uncertainty shown in Eq.~(\ref{eq:rhorhoNNLO}) has its origin in higher order perturbative corrections, we have taken this as the 
corresponding renormalization scale uncertainty when treating $\alpha^{\rm NNLO, \rho_L\rho_L}_{1,2}$
as nuisance parameters. Our SM determination for $R_{\rho\rho}$ is
\begin{eqnarray}
R^{\rm SM}_{\rho\rho}=\Bigl( 67.5\pm 25.7\Bigl)\cdot 10^{-2}.
\end{eqnarray}
\begin{table}
\begin{center}
\begin{tabular}{ |c|c|}
\hline
Parameter  & Relative Error\\
\hline
\hline
$X_A$                & $26.40\%$\\
\hline
$X_H$ & $23.33\%$\\
\hline
$\lambda_B$              & $12.32\%$\\
\hline
$\mu$                  & $6.78\%$\\
\hline
$A^{B\rightarrow \rho}_0$                  & $2.54\%$\\
\hline
$a^{\rho}_{2}$        & $2.24\%$\\
\hline
$f_{\rho}$ & $0.46\%$\\
\hline
$\Lambda^{QCD}_5$            & $0.45\%$\\
\hline
$\gamma$                  & $0.38\%$\\
\hline
$m_b$        & $0.27\%$\\
\hline
$f_B$ & $0.15\%$\\
\hline
$f^{\perp}_{\rho}$ & $0.15\%$\\
\hline
$m_c$        & $0.12\%$\\
\hline
$f^{\perp}_{\rho}$ & $0.07\%$\\
\hline
$|V_{ub}/V_{cb}|$ & $0.02\%$\\
\hline
\hline
Total & $38.09\%$\\
\hline
\end{tabular}
\end{center}
\caption{Error budget for the observable $R_{\rho\rho}$.}
\label{tab:Rrhorho}
\end{table}
The experimental result for $R_{\rho\rho}$ is obtained by calculating the ratio
of $\mathcal{B}r(B^{-}\rightarrow \rho_L^{-}\rho_L^{0})$ and $\mathcal{B}r(\bar{B}_d^0\rightarrow\rho_L^{+}\rho_L^{-})$ 
weighted by the corresponding longitudinal polarization fractions $f^{-0}_L$ and  $f^{+-}_L$. Using the numerical values 
available in the PDG  \cite{Tanabashi:2018oca} we obtain
\vspace{-0.2cm}
\begin{eqnarray}
R_{\rho\rho}^{\rm Exp}&=&\Bigl( 83.14\pm 8.98\Bigl)\cdot 10^{-2}.
\end{eqnarray}
The partial contributions to the error budget are presented in Table~\ref{tab:Rrhorho} and the constraints derived for $\Delta C^{d, uu}_1(M_W)$ 
in Fig.~\ref{fig:Rrhorho}. We do not show the associated regions for $\Delta C^{d, uu}_2(M_W)$ because, for $R_{\rho\rho}$, 
the results are weaker than those derived from other observables in our study.

%
%
%
%
%
%
%
%
%
%
%


\subsection{Observables constraining $b\rightarrow c\bar{c} s$ transitions}
\label{sec:bccs}
In this section we study bounds for $\Delta C_{1,2}^{s, cc}(M_W)$ stemming from 
$\mathcal{B}r(\bar{B}\rightarrow X_s \gamma)$, the mixing observable $\Delta \Gamma_s$,
the CKM angle $\sin(2\beta_s)$ and the lifetime ratio $\tau_{B_s}/\tau_{B_d}$.
These observables give very constrained regions for $\Delta C_{1,2}^{s, cc}(M_W)$.
\subsubsection{$\bar{B}\rightarrow X_s \gamma$}
\label{sec:Bsgamma}
The process $\bar{B}\rightarrow X_s \gamma$ is of mayor interest for BSM studies for several reasons. 
To begin with, within 
the SM  it is generated mainly at the loop level (its branching fraction actually receives contributions below $0.4\%$
from the tree-level 
CKM-suppressed transitions $b\rightarrow u\bar{u} s\gamma$ when the energy of the photon is within the phenomenologically
relevant range 
$E_{\gamma}\geq 1.6~\hbox{GeV}$   \cite{Kaminski:2012eb}).
In the HQET, it corresponds to a flavour changing
neutral current sensitive to new particles. Additionally, the experimental and theoretical precision achieved
on its determination have an 
accuracy of the same order. Moreover, this transition
is useful to constrain CKM elements involving the top quark.
\\
The experimental world average for $\mathcal{B}r(\bar{B}\rightarrow X_s \gamma)$ up to
date combines measurements from CLEO, Belle and BaBar 
leading to \cite{Amhis:2016xyh}
\vspace{-0.2cm}
\begin{eqnarray}
\mathcal{B}r^{\rm Exp}(\bar{B}\rightarrow X_s \gamma)&=&\Bigl(3.32 \pm 0.15 \Bigl)\cdot 10^{-4}.
\end{eqnarray}
On the theoretical side there has been  a huge effort on the determination of this observable;
the most precise results available are 
obtained at NNLO. Here we consider~\cite{Czakon:2015exa}
\begin{eqnarray}
\mathcal{B}r^{\rm SM}(\bar{B}\rightarrow X_s \gamma)&=&\Bigl(3.36 \pm 0.22 \Bigl)\cdot 10^{-4},
\label{eq:B_SM_NNLO}
\end{eqnarray}
where the energy of the photon satisfies the cut
\begin{eqnarray}
E_{\gamma}> E_0 = 1.6 \hbox{ GeV}.
\label{eq:PhotonE}
\end{eqnarray}
The calculation of the branching ratio for the process $\bar{B}\rightarrow X_s \gamma$
can be written as \cite{Misiak:2006ab}
\vspace{-0.2cm}
\begin{eqnarray}
\label{eq:BXsgammafull}
\mathcal{B}r(\bar{B}\rightarrow X_s\gamma)_{E_{\gamma}>E_0}
&=&
\mathcal{B}r(\bar{B}\rightarrow X_c e \bar{\nu})_{\rm exp}
\Bigl|\frac{V^{*}_{ts} V_{tb}}{V_{cb}}\Bigl|^2
\frac{6\alpha_{\rm em}}{\pi C}
\left[P(E_0) + N(E_0)\right].
\nonumber
\\
\end{eqnarray}
In Eq.~(\ref{eq:BXsgammafull}), $P(E_0)$ and $N(E_0)$ denote the perturbative and the non-perturbative 
contributions to the decay probability
respectively. They depend on the lower
cut for the energy of the photon in the Bremsstrahlung correction $E_0$ shown in Eq.~(\ref{eq:PhotonE}).
Using the parameterisation given 
in Ref.~\cite{Chetyrkin:1996vx}
we write $E_0=m^{1S}_b/2\Bigl(1-\delta'\Bigl)$ and choose $\delta'$ such that the lower bound in
Eq.~(\ref{eq:PhotonE}) is saturated.
The perturbative contribution $P(E_0)$ is given by \cite{Misiak:2006ab}
\vspace{-0.2cm}
\begin{eqnarray}
P(E_0)=\sum^8_{i,j=1} C^{eff}_i(\mu_b) C^{eff*}_j(\mu_b) K_{ij}(E_0, \mu_b)
\label{eq:perturbative}
\end{eqnarray}
with $K_{ij}=\delta_{i7}\delta_{j7} + \mathcal{O}(\alpha_s)$. The effective Wilson coefficients $C^{eff}_i$ are expressed
in terms of linear combinations of the coefficients for the operators  $\hat{Q^s_i}$
($i=1,..,6$), $\hat{Q}^s_{7\gamma}$ and $\hat{Q}^s_{8g}$
introduced in Section \ref{sec:Heff}.
For the denominator of Eq.~(\ref{eq:BXsgammafull}) we have \cite{Misiak:2006ab}
\vspace{-0.2cm}
\begin{eqnarray}
C=\Bigl|\frac{V_{ub}}{V_{cb}}\Bigl|^2\frac{\Gamma(\bar{B}\rightarrow X_c e \bar{\nu})}{\Gamma(\bar{B}\rightarrow X_u e \bar{\nu})}.
\end{eqnarray}
In order to account for the NNLO result in Eq.~(\ref{eq:B_SM_NNLO}) we write
\vspace{-0.2cm}
\begin{eqnarray}
\mathcal{B}r(\bar{B}\rightarrow X_s \gamma) &=&
\mathcal{B}r^{\rm SM, ~ NNLO}(\bar{B} \rightarrow X_s \gamma) \cdot
	\frac{\mathcal{B}r^{\rm NLO}(\bar{B}\rightarrow X_s \gamma)(\mu_0)}{\mathcal{B}r^{(0)~\rm SM,~ NLO}_{0}(\bar{B} \rightarrow X_s \gamma  )}
\, .
\nonumber
\\
\label{eq:BXsgammaNNLO}
\end{eqnarray}

Where

\begin{figure}
\centering
\includegraphics[height=5cm]{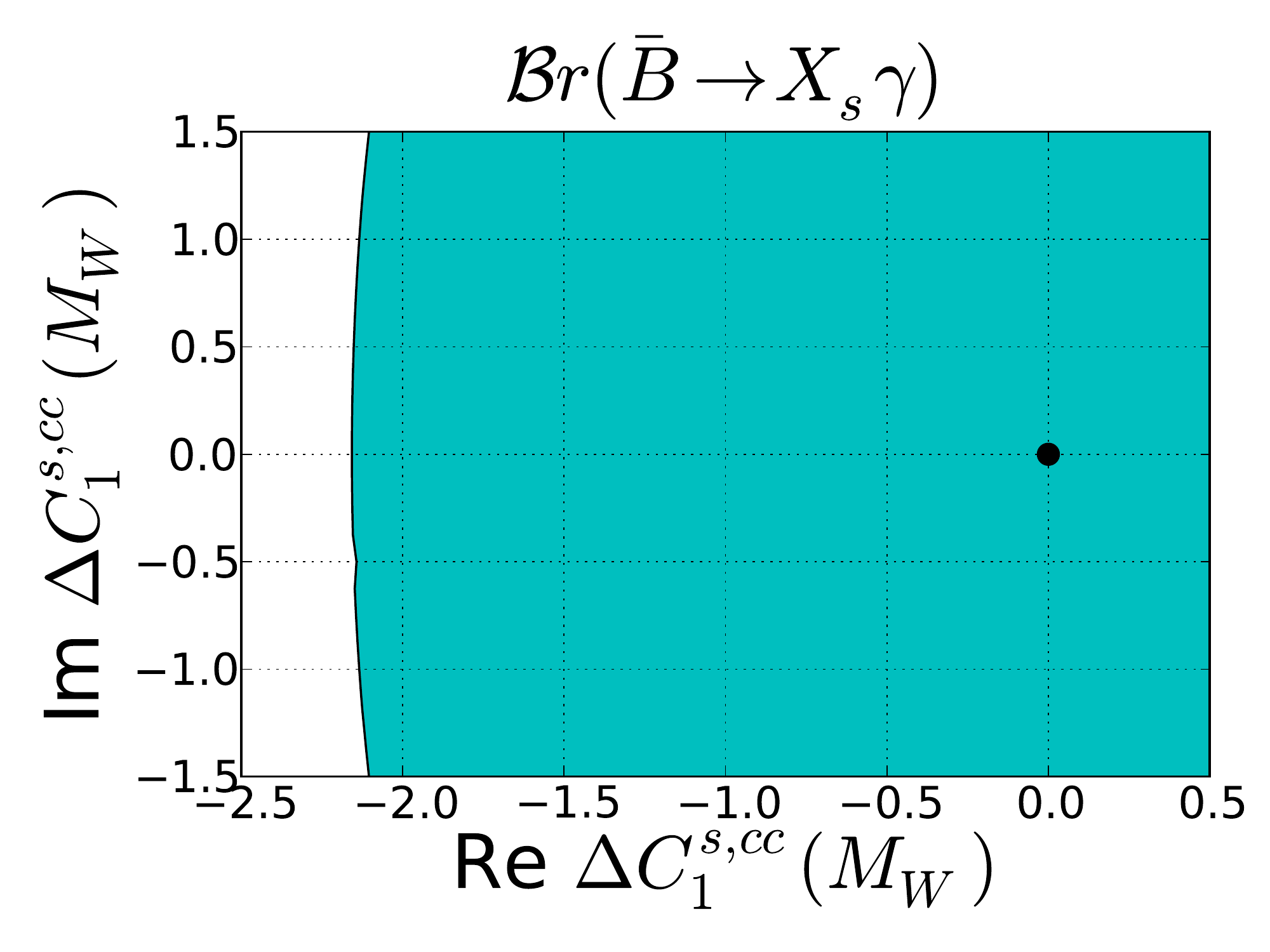}
\includegraphics[height=5cm]{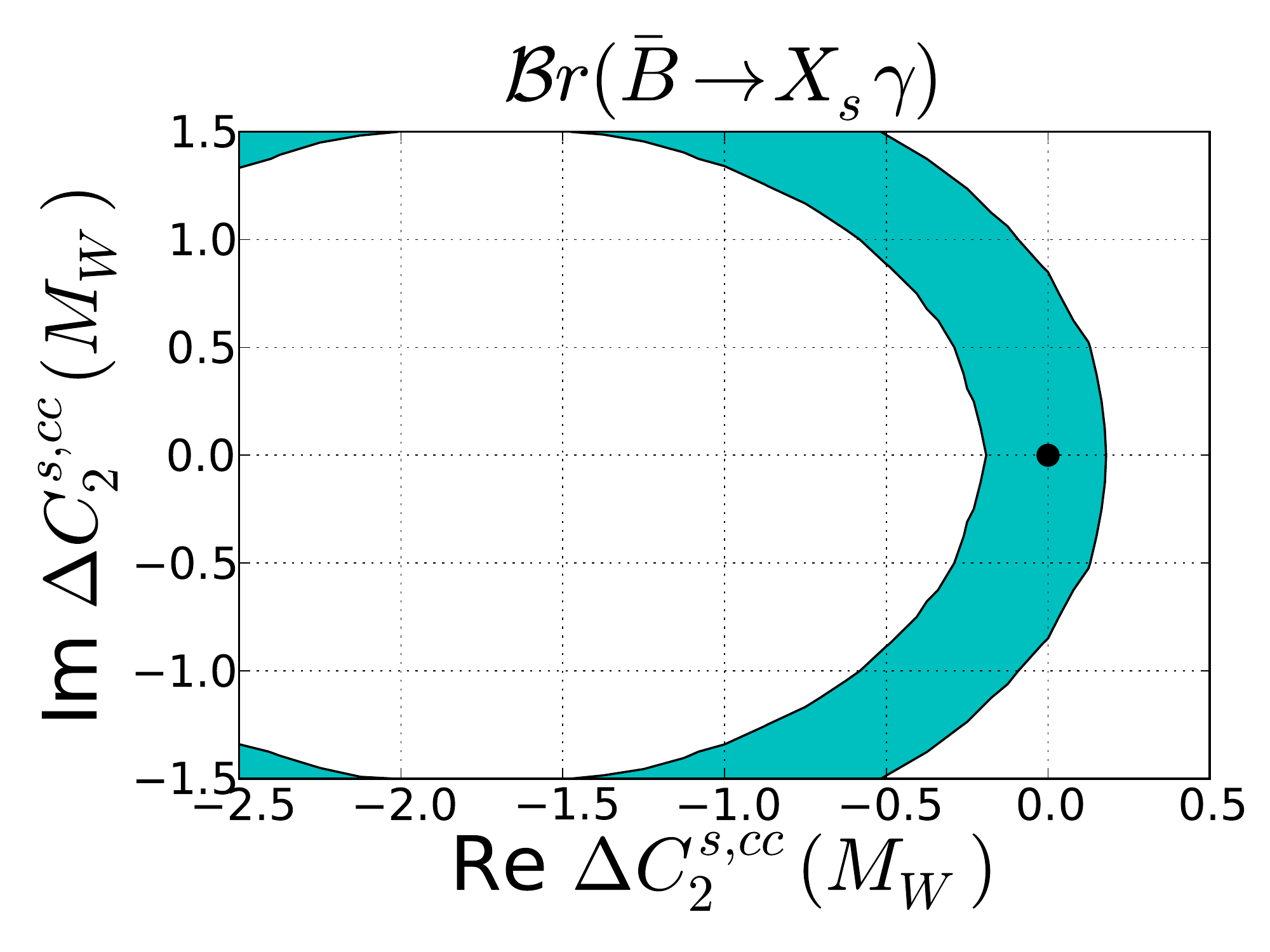}
\caption{
Potential regions for the NP contributions in $\Delta C^{s, cc}_1(M_W)$ and $\Delta C^{s, cc}_2(M_W)$  allowed by the observable 
$\mathcal{B}r(\bar{B}\rightarrow X_s \gamma)$ at $90\%$ C.L.. The black point corresponds to the SM value.}
\label{fig:Bs_gamma}
\end{figure}

\begin{itemize}

\item $\mathcal{B}r^{\hbox{NLO}}(\bar{B}\rightarrow X_s \gamma)$ is the branching ratio for the process
      $\bar{B}\rightarrow X_s \gamma$
		calculated at NLO including NP effects from $\Delta C^{s, cc}_{1,2}(M_W)$. All inputs are allowed to float except
      the renormalisation scale, which is fixed at $\mu_0=m_b$. Our calculations
      are determined using the anomalous dimension matrices provided in~\cite{Chetyrkin:1996vx}.
      NP contributions are introduced according to Eq.~(\ref{eq:NPC12}). They propagate to the rest
      of the Wilson coefficients $C_{i}$ after applying the renormalisation group equations, 
      described in Section 2 of Ref.~\cite{Chetyrkin:1996vx}.
\item $\mathcal{B}r^{\hbox{SM, NLO}}_{0}(\bar{B}\rightarrow X_s \gamma)$ is the SM branching ratio for the process
      $\bar{B}\rightarrow X_s \gamma$ calculated at NLO and  evaluated at the central values of all the input
      parameters and then kept constant during the $\chi^2$-fit.
\item $\mathcal{B}r^{\hbox{SM, NNLO}}(\bar{B}\rightarrow X_s \gamma)$ is the SM branching ratio for the process
      $\bar{B}\rightarrow X_s \gamma$ calculated at NNLO and allowed to float within the uncertainty associated with the renormalisation
      scale. In the case of the theoretical result given in  Eq.(\ref{eq:B_SM_NNLO}) this corresponds to
$3\%$ of the central value~\cite{Czakon:2015exa}\footnote{
In the NNLO determination in \cite{Czakon:2015exa} two scales $\mu_b$
and $\mu_c$ are introduced.  The $3\%$ variation indicated in the error 
budget 
is derived from considering the variation $1.25~\rm{GeV}\leq\mu_{b,c}\leq 5~\rm{GeV}$ which 
accounts for about $2.4\%$. However a more conservative value is taken due to the lack of 
certainty on extra contributions  to the perturbation series involved, see more details
in \cite{Czakon:2015exa}. 
}
.
\end{itemize}
The partial contributions to the final error are described in Table~\ref{tab:errorBXsgamma}.
The allowed regions for $\Delta C^{s, cc}_1(M_W)$ and $\Delta C^{s, cc}_2(M_W)$ are shown in Fig.~\ref{fig:Bs_gamma},
where it can be seen 
how this observable
imposes strong constraints on  $\Delta C^{s, cc}_2(M_W)$. The bounds in Fig.~\ref{fig:Bs_gamma} are consistent with those reported in 
\cite{Jager:2019bgk} once a $68\%$ C.L. is taken into account.
\begin{table}
\begin{center}
\begin{tabular}{ |c|c|}
 \hline
Parameter  & Relative error\\
\hline
\hline
$ N(E_0)$& $5.00\%$\\
\hline
$\mu$ & $3.00\%$\\
\hline
$\mathcal{B}r(\bar{B}\rightarrow X_c e\bar{\nu}_e)$                                & $2.68\%$ \\
\hline
$m_c(m_c)$& $1.10\%$\\
\hline
$m^{1S}_b$& $0.61\%$\\
\hline
$\Lambda^{QCD}_5$& $0.26\%$\\
\hline
$\gamma$& $0.10\%$\\
\hline
$|V_{ub}/V_{cb}|$ & $0.04\%$\\
\hline
$|V_{us}|$ & $0.01\%$ \\
\hline
\hline
Total & $6.55\%$\\
\hline
\end{tabular}
\caption{Error budget for the observable $\mathcal{B}r(\bar{B}\rightarrow X_s \gamma)$. Here $N(E_0)$ determines the uncertainty 
arising from non-perturbative contributions.
}
\label{tab:errorBXsgamma}
\end{center}
\end{table}

%
%
%
%
%
%
%
%
%
\subsubsection{$\Delta \Gamma_s$: Bounds and SM update}
\label{subsec:DGs}
The decay rate differences $\Delta \Gamma_{q}$ and the semileptonic asymmetries $a^{q}_{sl}$ arising from neutral $B_q$ meson mixing 
are sensitive to the tree-level transitions $b\rightarrow u\bar{u} q$, $b\rightarrow u\bar{c} q$ ,  $b\rightarrow c\bar{u} q$ and 
$b\rightarrow c\bar{c} q$ for $q=s,d$.
We will, however, show below that for the decay rate difference of $B_s$-mesons our BSM study is completely dominated by the $b \to c \bar{c} s$
transition, yielding therefore
strong constraints to  $\Delta C^{s, cc}_1(M_W)$ and $\Delta C^{s, cc}_2(M_W)$. 
\\
The definitions of the observables $\Delta \Gamma_{q}$ and $a^{q}_{sl}$
in terms of $\Gamma^q_{12}/M^q_{12}$ were introduced  in 
Eqs.~(\ref{eq:dGammaq}) and (\ref{eq:aslq}). Since, as explained in Section \ref{sec:HQE}, the elements $\Gamma^q_{12}$ are determined 
from the double insertion of $\mathcal{\hat{H}}_{eff}^{|\Delta B|=1}$ Hamiltonians, there are leading order contributions originating from
the insertion of two current-current operators $\hat{Q}^{q, ab}_{j}$  for $ab= uu, uc, cc$ and $j=1, 2$, 
see Eq.~(\ref{eq:mainbasis}). Additionally, there are also double insertions from a single current-current $\hat{Q}^{q, ab}_{1,2}$ and a
penguin operator $\hat{Q}^q_{3,4,5,6} $. In this section, we will only include NP effects to $\Gamma^q_{12}$, while
we neglect tree level NP contributions to $M^q_{12}$ (these contribution are discussed in Section \ref{sec:sin2betaM12}
and they yield considerably weaker bounds
for the observables $\Delta \Gamma_q$ and $a^q_{sl}$).
To show the dominance of the $b \to c \bar{c} s$ contribution for $B_s$-mixing, we decompose $\Gamma^q_{12}$ into
partial contributions  $\Gamma^{q, ab}_{12}$, 
where the indices $ab=uu, uc, cc$ indicate which``up'' type quarks are included inside the corresponding effective
fermionic loops. Thus, the
expression for $\Gamma^q_{12}/M_{12}^q$ becomes
\begin{eqnarray}
  \frac{  \Gamma^q_{12}}{M_{12}^q}
  &=&
- \frac{ \left(\lambda^{(q)}_c\right)^2\Gamma^{q, cc}_{12} + 2 \lambda^{(q)}_u \lambda^{(q)}_c \Gamma^{q, uc}_{12}
    + \left(\lambda^{(q)}_u\right)^2 \Gamma^{q, uu}_{12} }{{M}_{12}^q}
  \nonumber
  \\
&=&
  - \frac{
	  (\lambda^{(q)}_t)^2 \Gamma^{q, cc}_{12} + 2 \lambda^{(q)}_t \lambda^{(q)}_u \Bigl[\Gamma^{q, cc}_{12} - \Gamma^{q, uc}_{12}\Bigl] 
  +(\lambda^{(q)}_u)^2 \Bigl[ \Gamma_{12}^{q, cc} - 2 \Gamma_{12}^{q, uc} + \Gamma^{q, uu}_{12} \Bigl]
  }
  {(\lambda_t^{(q)})^2 \tilde{M}_{12}^q}
  \nonumber
  \\
&=& - 10^{-4} \left[ c^q + a^q \frac{\lambda_u^{(q)}}{\lambda_t^{(q)}} + b^q \left( \frac{\lambda_u^{(q)}}{\lambda_t^{(q)}} \right)^2 \right]
    .\nonumber\\
          \label{Eq:Gamma_d_cont}
\end{eqnarray}
We have used here the unitarity of the CKM matrix: $\lambda_u^{(q)} + \lambda_c^{(q)} + \lambda_t^{(q)} = 0$ and we have split off the CKM dependence
from ${M}_{12}^q$ by introducing the quantity  $\tilde{M}_{12}^q$.
The GIM suppressed \cite{Glashow:1970gm} terms $a$ and $b$ vanish in the limit $m_c \to m_u$ and the
numerical values show a clear hierarchy
\begin{eqnarray}
  c^q \approx  - 48\, , && a^q \approx 11 \, , \hspace{1cm} b^q \approx 0.23 \, .
  \end{eqnarray}
For the ratio of CKM elements we obtain
\begin{eqnarray}
  \frac{\lambda_u^{(q)}}{\lambda_t^{(q)}} & \approx & \left\{
                                              \begin{array}{cc}
						      1.7 \cdot 10^{-2} - 4.2 \cdot 10^{-1}~ i & \mbox{for} \, \, \,q=d
                                                \\
                                               -8.8 \cdot 10^{-3} + 1.8 \cdot 10^{-2}~ i & \mbox{for} \, \, \, q=s
                                         \end{array}
                                          \right.
                                          \\
  \left(\frac{\lambda_u^{(q)}}{\lambda_t^{(q)}}\right)^2 & = & \left\{
                                              \begin{array}{cc}
                                                  -1.8 \cdot 10^{-1} - 1.5 \cdot 10^{-2}~i & \mbox{for} \, \, \, q=d
                                                \\
                                                  -2.5 \cdot 10^{-4} - 3.2 \cdot 10^{-4}~i & \mbox{for} \, \, \, q=s
                                         \end{array}
                                          \right.
\end{eqnarray}
Within the SM we find a very strong hierarchy of the three contributions in Eq. (\ref{Eq:Gamma_d_cont}).
The by far largest term is given by $c^q$ and it is real. The second term proportional to $a^q$ is GIM 
and CKM suppressed - slightly for the case of $B_d$ mesons and more pronounced for $B_s$. Since $\lambda_u^{(q)}/\lambda_t^{(q)}$
is complex, this contribution gives rise to an imaginary part of  $\Gamma^q_{12}/M_{12}^q$.
Finally $b^q$ is even further GIM suppressed and again slightly/strongly CKM suppressed for $B_d$/$B_s$  mesons 
- this contribution has also both a  real and an imaginary part.
According to Eqs.~(\ref{eq:dGammaq}) the decay rate difference $\Delta \Gamma_q$, given by the real part of
$\Gamma^q_{12}/M_{12}^q$,
is dominated by the coefficient $c^q$ - stemming from $b \to c \bar{c} q$ transitions - and the
coefficients  $a^q$ and  $b^q$ yield corrections of the order of 2 per mille.
The semi-leptonic asymmetries are given by the imaginary part of  $\Gamma^q_{12}/M_{12}^q$   (c.f. Eq. (\ref{eq:aslq})),
which in turn
is dominated by the coefficient $a^q$, with $b^q$ giving sub-per mille corrections and no contributions from $c^q$.
\\
Allowing new, complex contributions to $C_1$ and $C_2$ for individual quark level contributions we get the following effects:
\begin{enumerate}
\item The numerically leading coefficient $c^q$ can now also obtain an imaginary part.
\item The GIM cancellations in the coefficients $a^q$ and $b^q$ can be broken, if $b \to c \bar{c} q$, $b \to c \bar{u} q$,
  $b \to u \bar{c} q$
  and  $b \to u \bar{u} q$ are differently affected by NP. If there is a universal BSM contribution then the
  GIM cancellation will stay. 
\item The CKM suppression will not be affected by our BSM modifications.
\end{enumerate}
For the real part of  $\Gamma^s_{12}/M_{12}^s$, we  expect at most a correction of 2 per cent due to $a^s$ and $b^s$,
even if the corresponding
GIM suppression is completely lifted - thus $\Delta \Gamma_s$ is even in our BSM approach, completely dominated by $c^s$
and gives therefore only bounds on
$b \to c \bar{c} s$. In the case of $B_d$ mesons, the corrections due to  $a^d$ and $b^d$ could be as large as 40 per cent -
here all possible decay channels have
to be taken into account - except we are considering universal BSM contributions to all decay channels. Since $\Delta \Gamma_d$
is not yet measured,
we will revert our strategy and use the obtained bounds on the
Wilson coefficients $C_1$ and $C_2$ to obtain potential enhancements or reductions of $\Delta \Gamma_d$  due to BSM effects
in non-leptonic tree-level
decays.
Considering the imaginary part of $\Gamma^s_{12}/M_{12}^s$, we can get dramatically enhanced  values for the semi-leptonic CP asymmetries,
if $C_1$ or $C_2$ are complex, which will result in an  imaginary part of the GIM-unsuppressed coefficient $c^q$.
On the other hand new contributions to e.g. only $b \to c \bar{u} q$ or  $b \to u \bar{c} q$
would have no effect on $c^q$,
but they could lift the
GIM suppression of the coefficient $a^q$ and thus lead to also large effects. Therefore the semileptonic CP asymmetries are not completely dominated by the
$b \to c \bar{c} q$ transitions.
\\
Next we explain in detail how to implement BSM contributions to $C_1$ and $C_2$ in the theoretical description of $\Gamma_{12}^q$.
Each one of the functions $\Gamma^{q, ab}_{12}$ in Eq.~(\ref{Eq:Gamma_d_cont}) are given by \cite{Lenz:2006hd}
\begin{eqnarray}
\label{eq:Gamma12}
\Gamma^{q, ab}_{12}
&=& \frac{G^2_F m^2_b}{24 \pi M_{B_q}} \Bigl[ \Bigl(G^{q, ab} + \frac{1}{2}\alpha_2 G^{q, ab}_{S}  \Bigl)
  \langle B_q |\hat{Q}_1 |\bar{B}_q \rangle + 
    \alpha_1  G^{q, ab}_S \langle B_q| \hat{{Q}}_3 |\bar{B}_q\rangle\Bigl] + \tilde{\Gamma}^{q, ab}_{12, 1/m_b}.\nonumber\\ 
\end{eqnarray}
The coefficients  $\alpha_1$ and $\alpha_2$ in Eq.~(\ref{eq:Gamma12}) include NLO corrections and are written in the $\overline{\hbox{MS}}$ scheme as
\begin{eqnarray}
\alpha_1=1+\frac{\alpha_s(\mu)}{4\pi}C_F\Bigl(12\ln\frac{\mu}{m_b} + 6 \Bigl),&&
\alpha_2=1+\frac{\alpha_s(\mu)}{4\pi}C_F\Bigl(12\ln\frac{\mu}{m_b} + \frac{13}{2}\Bigl).\nonumber\\
\end{eqnarray}
Furthermore, the expressions for $G^{q, ab}$ and $G^{q, ab}_S$ in Eq.~(\ref{eq:Gamma12}) are decomposed as
\begin{eqnarray}
G^{q, ab}=  F^{q, ab} + P^{q, ab}, && G_S^{q, ab}= - F_S^{q, ab} - P_S^{q, ab},
\end{eqnarray}
with $F^{q, ab}$ and $F^{q, ab}_S$ encoding the perturbative contributions resulting from the double insertion of current-current operators. Finally,
 $P^{q, ab}$ and  $P^{q, ab}_S$  contain the perturbative effects from the combined insertion of a current-current and a penguin operators.
In terms of the tree-level Wilson coefficients $C^{q, ab}_1$ and $C^{q, ab}_2$, the equations for $F^{q, ab}$  and $F^{q, ab}_S$ have the following generic structure
\begin{eqnarray}
F^{q, ab}=F^{q, ab}_{11}\left[ C^{q, ab}_1(\mu) \right]^2 + F^{q, ab}_{12}C^{q, ab}_{1}(\mu)C^{q, ab}_{2}(\mu) + F^{q, ab}_{22} \left[ C^{q, ab}_2(\mu) \right]^2,
\label{eq:GenF}
\end{eqnarray}
where the individual factors $F^{q, ab}_{11, 12, 22}$ are available in the literature up to NLO
\vspace{-0.2cm}
\begin{eqnarray}
F^{q, ab}_{ij}=F^{q, (0)}_{ij} + \frac{\alpha_s(\mu)}{4\pi}F^{q, (1)}_{ij}.
\end{eqnarray}
To account for NP effects, the Wilson coefficients inside Eq.~(\ref{eq:GenF}) should be determined using Eq.~(\ref{eq:NPC12})
and applying the renormalization group equations introduced in Sec.~\ref{sec:basic}.
Notice that Eq.~(\ref{eq:GenF})  is sensitive to the different transitions $b\rightarrow c\bar{c} q$, $b\rightarrow u\bar{c} q$, $b\rightarrow c\bar{u} q$ and
$b\rightarrow u\bar{u} q$. To be consistent with the inclusion of NP effects $\Delta C^{q, ab}_1(M_W)$ and $\Delta C^{q, ab}_2(M_W)$  at LO 
only,
we omit all the terms involving products between $\alpha_s(\mu)$ and the NP factors $\Delta C^{q, ab}_{1,2}(M_W)$  inside Eq.~(\ref{eq:GenF}).
The penguin functions $P^{q,ab}$ and $P^{q,ab}_S$ also contain LO contributions from $C^{q, ab}_{1,2}$. For the purposes
of illustration we will show the explicit expressions for the functions $P^{s, cc}$ and $P^{s, cc}_S$ corresponding to the 
$B^0_s-\bar{B}^0_s$ system. At NLO we have \cite{Beneke:1998sy}
\begin{eqnarray}
\label{eq:FPFPS}
	P^{s, cc}&=&\sqrt{1-4\bar{z}}\Bigl[(1-\bar{z})K^{' cc}_1(\mu) + \frac{1}{2}(1-4\bar{z})K^{' cc}_2(\mu) + 3 \bar{z} K^{' cc}_3(\mu)  \Bigl] \nonumber\\
&&+ \frac{\alpha_s(\mu)}{4\pi}F^{cc}_p(\bar{z})\Bigl[C^{s, cc}_2(\mu)\Bigl]^2,\nonumber\\
P^{s, cc}_S&=&\sqrt{1-4\bar{z}}\Bigl[1+ 2 \bar{z}\Bigl]\Bigl[K^{' cc}_1(\mu) - K^{' cc}_2(\mu)\Bigl]-\frac{\alpha_s(\mu)}{4\pi}8F_p(\bar{z})\Bigl[C^{s, cc}_2(\mu)\Bigl]^2. \nonumber\\
\end{eqnarray}
Where the following definition for the ratio of the masses of the bottom and charm quarks, evaluated in the $\overline{\hbox{MS}}$ scheme \cite{Lenz:2006hd}, has been used
\begin{eqnarray}
\bar{z}&=& \Bigl[\overline{m}_c(\overline{m}_b)/\overline{m}_b(\overline{m}_b)\Bigl]^2.
\end{eqnarray}
The functions $K^{' cc}_{1,2,3}$ inside
Eq.~(\ref{eq:FPFPS}) are given by
\begin{eqnarray}
\label{eq:Kfunctions}
K^{' cc}_1(\mu)&=&2\Bigl[3 C^{s, cc}_1(\mu) C^{s}_3(\mu) + C^{s, cc}_1(\mu) C^{s}_4(\mu) + C^{s, cc}_2(\mu) C^{s}_3(\mu)\Bigl],\nonumber\\
K^{' cc}_2(\mu)&=&2C^{s, cc}_2(\mu) C^{s}_4(\mu),\nonumber\\
K^{' cc}_3(\mu)&=&2\Bigl[3 C^{s, cc}_1(\mu) C^{s}_5(\mu) + C^{s, cc}_1(\mu) C_6(\mu) + C^{s, cc}_2(\mu) C^{s}_5(\mu) +C^{s, cc}_2(\mu) C^{s}_6(\mu)\Bigl],\nonumber\\
\end{eqnarray}
and the expression for the NLO correction function $F^{cc}_p(z) $ is
\begin{eqnarray}
F^{cc}_p(z)&=&-\frac{1}{9}\sqrt{1-4\bar{z}}\Bigl(1+2\bar{z}\Bigl)\Bigl[2\hbox{ln}\frac{\mu}{m_b} + \frac{2}{3} + 4\bar{z} - \hbox{ln}\bar{z} \nonumber\\
	&&+ \sqrt{1-4\bar{z}}\Bigl(1 + 2\bar{z} \Bigl)\hbox{ln}\frac{1-\sqrt{1-4\bar{z}}}{1+\sqrt{1+4\bar{z}}} + \frac{3C^s_{8g}(\mu)}{C^{s, cc}_2(\mu)} \Bigl].
\end{eqnarray}
The Wilson coefficients inside Eqs.~(\ref{eq:Kfunctions}) should be calculated by introducing NP deviations at the
scale $\mu=M_W$ and then running down their corresponding values to the scale $\mu\sim m_b$ through the renormalization 
group equations, for details see the discussion in Sec.~\ref{sec:basic}.
In Appendix \ref{Sec:Inputs}, we provide details on the numerical inputs used.
Since there was tremendous progress \cite{King:2019rvk,DiLuzio:2019jyq}
in the theoretical precision of the mixing observables we will present in this work numerical updates
of all mixing observables:
$\Delta \Gamma_q$ below,
$\Delta M_q$ in Section \ref{sec:sin2betaM12}
and the semi-leptonic CP asymmetries
$a_{sl}^q$ and $\phi_q$ in Section \ref{sec:multiple_channels}.
For our numerical analysis we use results for $\Gamma_{12,3}^{q,(0)}$, $\Gamma_{12,3}^{q,(1)}$ and $\Gamma_{12,4}^{q,(0)}$,
from 
\cite{Beneke:1998sy,Beneke:2002rj,Beneke:1996gn,Dighe:2001gc,Ciuchini:2003ww,Beneke:2003az,Lenz:2006hd} and for
the hadronic matrix elements the averages presented in
\cite{DiLuzio:2019jyq}
based on \cite{Grozin:2016uqy,Kirk:2017juj,King:2019lal} and \cite{Christ:2014uea,Bussone:2016iua,Hughes:2017spc,Bazavov:2017lyh}, as
well as the dimension seven matrix elements from \cite{Davies:2019gnp}.
The new SM determinations for $\Delta \Gamma_s$ and $\Delta \Gamma_d$ are
\begin{eqnarray}
\label{eq:DeltaGamma}
  \Delta \Gamma^{\rm SM}_s
    &=&  \Bigl(9.1 \pm 1.3 \Bigl)\cdot 10^{-2}~\hbox{ps}^{-1},
  \\
  \Delta \Gamma^{\rm SM}_d
  &=& \Bigl( 2.6 \pm 0.4\Bigl)\cdot 10^{-3}~\hbox{ps}^{-1}.
\end{eqnarray}
The error budgets
of the mixing observables  $\Delta \Gamma_s$ and $\Delta \Gamma_d$
are presented in Tabs. \ref{error:DGs} and \ref{error:DGd} respectively. Compared to the SM estimates for $\Delta \Gamma_s$ stemming from
2006 \cite{Lenz:2006hd},
2011 \cite{Lenz:2011ti}
and
2015 \cite{Artuso:2015swg}
we
find a huge improvement in the SM precision. Moreover,   
 the value of $\Delta \Gamma_s$ in Eq.~(\ref{eq:DeltaGamma}) is in good agreement with the corresponding result of $\Delta \Gamma_s= \Bigl(9.2 \pm 1.4 \Bigl)\cdot 10^{-2}~\hbox{ps}^{-1}$ obtained in \cite{Davies:2019gnp}.

\begin{figure}
\centering
\includegraphics[height=5cm]{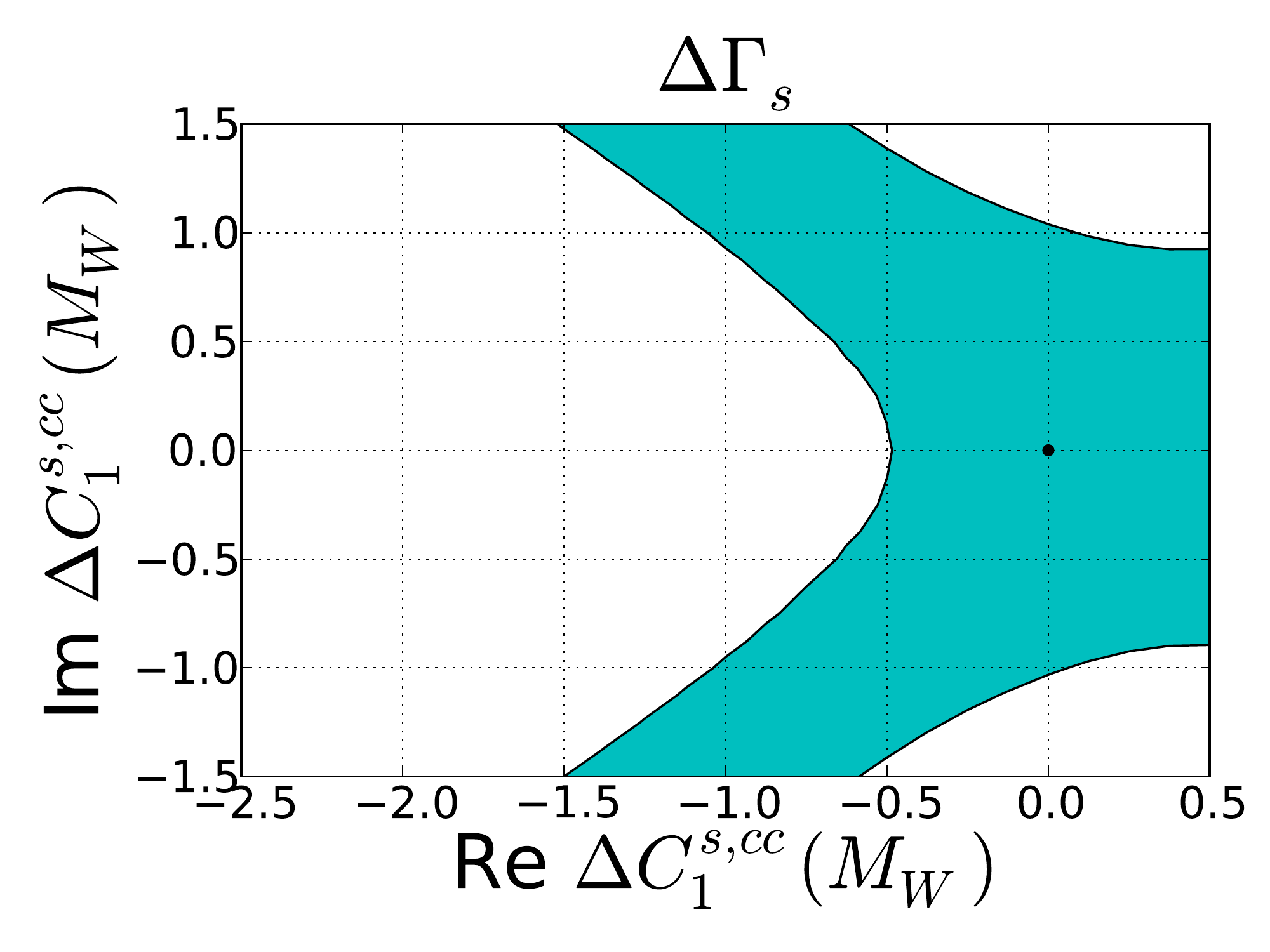}
\includegraphics[height=5cm]{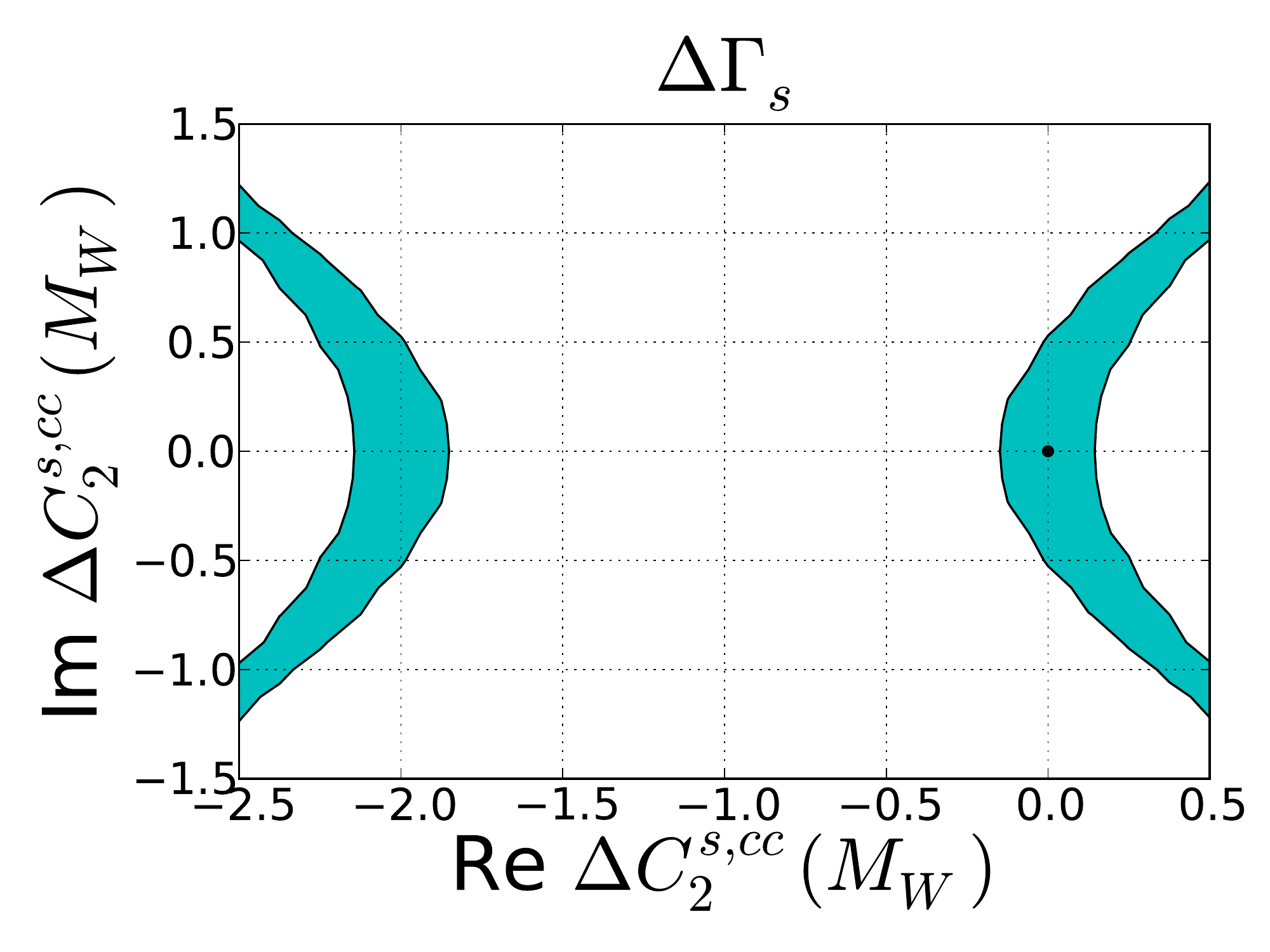}
\caption{
Potential regions for the NP contributions in $\Delta C^{s, cc}_1(M_W)$ and $\Delta C^{s, cc}_2(M_W)$  allowed by 
the observable $\Delta \Gamma_s$ at $90\%$ C.L.. The black point corresponds to the SM value.
}
\label{fig:dGammas}
\end{figure}

\begin{table}
  \begin{center}
\begin{tabular}{|c||c|c|c|c|}
\hline
$\Delta \Gamma_s^{\rm SM} $&$  \mbox{this work} $&$\mbox{ABL 2015}  $&$\mbox{LN 2011}  $&${\mbox{LN 2006}}$
\\
\hline
\hline
$\mbox{Central Value}   $&$  0.091 \, \mbox{ps}^{-1} $&$  0.088 \, \mbox{ps}^{-1} $&$0.087 \, \mbox{ps}^{-1} $&$ 0.096 \, \mbox{ps}^{-1}$
\\
\hline
\hline
$B^s_{\widetilde R_2}$&$ 10.9 \% $&$ 14.8 \% $&$17.2 \% $&$15.7 \%$
\\
\hline
$\mu            $  &$ 6.6 \% $&$  8.4 \% $&$ 7.8 \% $&$13.7 \%$
\\
\hline
$V_{cb}        $  &$ 3.4 \% $ &$  4.9 \% $&$ 3.4 \% $&$ 4.9 \%$
\\
\hline
$B^s_{R_0}      $   &$3.2 \%$ &$  2.1 \% $&$ 3.4 \% $&$ 3.0 \%$
\\
\hline
$f_{B_s} \sqrt{B^s_1}$ &$ 3.1 \% $&$ 13.9 \% $&$13.5 \% $&$34.0 \%$
\\
\hline
$B^s_3$  &$  2.2 \% $ &$  2.1 \% $&$  4.8 \% $&$ 3.1 \%$
\\
\hline
$\bar z        $   &$0.9 \%$ &$  1.1 \% $&$ 1.5 \% $&$ 1.9 \%$
\\
\hline
$m_b            $  &$0.9 \%$&$  0.8 \% $&$ 0.1 \% $&$  1.0 \%$
\\
\hline
$B^s_{R_3}          $&$0.5 \%$&$ 0.2 \% $&$ 0.2 \% $&$ ---$
\\
\hline
$B^s_{\tilde{R}_3} $  & - &$  0.6 \% $&$ 0.5 \% $&$  ----$
\\
\hline
$m_s              $&$0.3 \%$&$ 0.1 \% $&$ 1.0 \%  $&$  1.0 \%$
\\
\hline
$B^s_{\tilde{R}_1}$  &$0.2 \%$ &$  0.7\%  $&$ 1.9 \% $&$  ---$
\\
\hline
$\Lambda_5^{\rm QCD}    $&$ 0.1 \% $&$ 0.1 \% $&$ 0.4 \%  $&$ 0.1 \%$
\\
\hline
$\gamma           $&$ 0.1 \% $&$ 0.1 \% $&$ 0.3 \%  $&$ 1.0 \%$
\\
\hline
$B^s_{R_1}          $&$ 0.1 \% $&$ 0.5 \% $&$ 0.8 \% $&$  ---$
\\
\hline
$|V_{ub}/V_{cb}|  $&$ 0.1 \% $&$ 0.1 \% $&$  0.2 \% $&$ 0.5 \%$
\\
\hline
$\bar{m}_t(\bar{m}_t) $&$ 0.0 \% $&$ 0.0 \% $&$ 0.0 \%  $&$ 0.0 \%$
\\
\hline
\hline
Total &$14.1 \% $&$22.8 \% $&$ 24.5 \%                $&$40.5 \%$
\\
\hline
\end{tabular}
\end{center}
\caption{List of the individual contributions to the theoretical error of the decay rate difference 
$\Delta \Gamma_s$ within the Standard Model and comparison with the values obtained in 2015 \cite{Artuso:2015swg}, in 2011 \cite{Lenz:2011ti}
and in 2006 \cite{Lenz:2006hd}. We have used equations of motion in the current analysis to get rid of the operator $\tilde{R}_3$.}
\label{error:DGs}
\end{table}
In addition, the current SM predictions are based for the first time on a non-perturbative determination  \cite{Davies:2019gnp} of the
leading uncertainty due to dimension seven operators. All previous predictions had to rely on vacuum insertion approximation for
the corresponding matrix elements.
To further reduce the theory uncertainties, improvements in the lattice determination would be very welcome
or a corresponding sum rule calculation. The next important uncertainty stems from the renormalisation scale dependence, to reduce this
a NNLO calculation is necessary. First steps in that direction have been done in \cite{Asatrian:2017qaz}.
\begin{table}
  \begin{center}
    \begin{tabular}{|c||c|c|}
\hline
$ \Delta \Gamma_d^{\rm SM}  $& This work & ABL 2015
\\
\hline
\hline
$\mbox{Central Value}$      &$2.61 \cdot 10^{-3} \, \mbox{ps}^{-1} $ &$2.61 \cdot 10^{-3} \, \mbox{ps}^{-1} $
\\
\hline
\hline
$B^d_{\widetilde R_2} $ & $11.1\% $& $14.4\% $
\\
\hline
$f_{B_d} \sqrt{B_1^d}   $&$3.6\%$&$13.7\%$
\\
\hline
$\mu                $  &$ 6.7\% $&$ 7.9\% $
\\
\hline
$V_{cb}             $  &$ 3.4\%$&$ 4.9\%$
\\
\hline
$B_3^d    $            &$ 2.4\% $ &$ 4.0\% $
\\
\hline
$B_{R_0}^d             $&$ 3.3\% $&$ 2.5\% $
\\
\hline
$\bar z             $&$ 0.9\% $&$ 1.1\% $
\\
\hline
$m_b                $&$ 0.9\% $&$ 0.8\% $
\\
\hline
$\tilde{B}_{{R}_3}^d    $& - &$ 0.5\% $
\\
\hline
$B_{{R}_3}^d    $& $ 0.5\% $&$ 0.2\% $
\\
\hline
$\gamma            $&$ 2.2 \%$&$ 2.5\%$
\\
\hline
$\Lambda_5^{\rm QCD}       $&$ 0.1 \%  $&$ 0.1 \%  $
\\
\hline
$|V_{ub}/V_{cb}|    $&$  0.0 \% $&$  0.1 \% $
\\
\hline
$\bar{m}_t(\bar{m}_t)$&$ 0.0 \%  $&$ 0.0 \%  $
\\
\hline
\hline
Total               &$14.7\%$ &$22.7\%$
\\
\hline
    \end{tabular}
    \end{center}
\caption{List of the individual contributions to the theoretical 
  error of the mixing quantity $\Delta \Gamma_d$ and comparison with the values obtained in 2015 \cite{Artuso:2015swg}.
 We have used equations of motion in the current analysis to get rid of the operator $\tilde{R}_3$.}
\label{error:DGd}
\end{table}
In the ratio $\Delta \Gamma_q/\Delta M_q$ uncertainties due to the matrix elements of dimension six are cancelling -
so for a long time this ratio was considerably better known
than the individual value of $\Delta \Gamma_s$. Due to the huge progress in determining precise values for these
non-perturbative parameter, this advantage is now considerably less pronounced, see Table \ref{error4}.
\begin{table}
  \begin{center}
    \begin{tabular}{|c||c|c|c|c|}
\hline
$\Delta  \Gamma_s^{\rm SM} / \Delta M_s^{\rm SM}  $&$  \mbox{this work} $&$ \mbox{ABL 2015}$&$ \mbox{LN 2011}
        $&$ {\mbox{LN 2006}}$
\\
\hline
\hline
$\mbox{Central Value}    $&$48.2 \cdot 10^{-4}$&$48.1 \cdot 10^{-4}$&$50.4 \cdot 10^{-4}$&$  49.7 \cdot 10^{-4}$
\\
\hline
\hline
$B_{R_2}^s       $&$ 10.9 \%          $&$ 14.8 \%          $&$   17.2 \%        $&$  15.7 \%$
\\
\hline
$\mu            $&$ 6.6 \%          $&$  8.4 \%          $&$   7.8 \%         $&$   9.1 \%$
\\
\hline
$B_{R_0}^s        $&$ 3.2 \%          $&$  2.1 \%          $&$   3.4 \%         $&$   3.0 \%$
\\
\hline
$B_3^s$&$ 2.2 \%          $&$  2.1 \%          $&$   4.8 \%         $&$  3.1 \%$
\\
\hline
$\bar z         $&$ 0.9 \%          $&$  1.1 \%          $&$   1.5 \%         $&$   1.9 \%$
\\
\hline
$m_b            $&$ 0.9 \%          $&$   0.8 \%         $&$  1.4 \%          $&$   1.0 \%$
\\
\hline
$B_{R_3}^s        $&$ 0.5 \%          $&$   0.2 \%         $&$  0.2 \%          $&$  ---$
\\
\hline
$B_{\tilde{R}_3}^s$&$  -         $&$   0.6 \%         $&$  0.5 \%          $&$   ----$
\\
\hline
$\bar m_t (\bar m_t)            $&$ 0.3 \%          $&$   0.7 \%         $&$  1.1 \%          $&$   1.8 \%$
\\
\hline
$m_s            $&$ 0.3 \%          $&$   0.1 \%         $&$ 1.0 \%           $&$   0.1 \%$
\\
\hline
$\Lambda_5^{\rm QCD}       $&$ 0.2 \%          $&$   0.2 \%         $&$  0.8 \%          $&$  0.1 \%$
\\
\hline
$B_{\tilde{R}_1}^s$&$ 0.2 \%          $&$   0.7 \%          $&$   1.9 \%         $&$   ---$
\\
\hline
$B_{R_1}^s        $&$ 0.1 \%          $&$   0.5 \%         $&$  0.8 \%          $&$   ---$
\\
\hline
$\gamma         $&$ < 0.1 \%          $&$   0.0 \%         $&$ 0.0 \%           $&$  0.1 \%$
\\
\hline
$|V_{ub}/V_{cb}|$&$ < 0.1 \%          $&$   0.0 \%         $&$ 0.0 \%           $&$  0.1 \%$
\\
\hline
$V_{cb}         $&$ < 0.1 \%          $&$   0.0 \%         $&$  0.0 \%          $&$ 0.0 \%$
\\
\hline
\hline
Total          &$ 13.4 \%          $&$  17.3 \%         $&$ 20.1 \%          $&$ 18.9 \%$
\\
\hline
\end{tabular}
  \end{center}
  \caption{List of the individual contributions to the theoretical error of the ratio
  $\Delta \Gamma_s$/$\Delta M_s$ within the Standard Model and comparison with the values obtained in 2015 \cite{Artuso:2015swg},
  in 2011 \cite{Lenz:2011ti}
and in 2006 \cite{Lenz:2006hd}.  We have used equations of motion in the current analysis to get rid of the operator $\tilde{R}_3$.}
\label{error4}
\end{table}
For the corresponding experimental values we use the HFLAV averages
\begin{eqnarray}
\Delta \Gamma^{\rm Exp}_s&=&\Bigl(8.8 \pm 0.6\Bigl)\cdot 10^{-2}~\hbox{ps}^{-1}, \hbox{\cite{Amhis:2016xyh}}\nonumber\\
\Delta \Gamma^{\rm Exp}_d&=&\Bigl(-1.3 \pm 6.6\Bigl)\cdot 10^{-3}\hbox{~ps}^{-1},
\label{eq:DGammaExp}
\end{eqnarray}
where $\Delta \Gamma^{Exp}_d$ was obtained using \cite{Amhis:2016xyh}
\vspace{-0.2cm}
\begin{eqnarray}
\Bigl(\Delta \Gamma_d/\Gamma_d\Bigl)^{\rm Exp}=-0.002\pm 0.010, &&
\tau^{\rm Exp}_{B^0_d}=\Bigl(1.520 \pm 0.004\Bigl) \hbox{~ps}.
\end{eqnarray}
The resulting regions for $\Delta C^{s, cc}_1(M_W)$ and $\Delta C^{s, cc}_2(M_W)$ allowed by $\Delta \Gamma_s$ 
are presented in Fig.\ref{fig:dGammas}. 
%
%
%
%
%
%
%
%
%
%
%
%
%
\subsubsection{$S_{J/\psi \phi}$}
\label{subsub:SJPsiPhi}
The mixing induced CP asymmetry for the decay $\bar{B}_s\rightarrow J/\psi \phi$, given as
\begin{eqnarray}
\label{eq:SJPsiPhi}	
S_{J/\psi \phi}&=&\frac{2~\mathcal{I}m\Bigl(\lambda^s_{J/\psi \phi}\Bigl)}{1+\Bigl|\lambda^s_{J/\psi \phi}\Bigl|^2}
	=\sin(2\beta_s),
\end{eqnarray}
can be used to constrain $\Delta C^{s, cc}_{1}(M_W)$.
In Eq.~(\ref{eq:SJPsiPhi}), $\lambda^s_{J/\psi \phi}$ is determined according 
to Eq.~(\ref{eq:lambdaf}) considering the amplitudes $\bar{\mathcal{A}}_{J/\psi \phi}$ and $\mathcal{A}_{J/\psi \phi}$ for the decays
$\bar{B}^0_s\rightarrow J/\psi \phi$ and $B^0_s\rightarrow J/\psi \phi$ respectively. The required theoretical expressions have been
calculated explicitly within the QCDF formalism in \cite{Cheng:2001ez}. 
The equation for the decay amplitude obeys the structure 
\begin{eqnarray}
\mathcal{A}^{h}_{J/\psi \phi}&\propto& \alpha^{J/\psi \phi, h}_{1} + \alpha^{J/\psi \phi, h}_{3} + \alpha^{J/\psi \phi, h}_{5} + 
\alpha^{J/\psi \phi, h}_{7}  + \alpha^{J/\psi \phi, h}_{9},
\label{eq:masterAmpphi}
\end{eqnarray}
where the proportionality constant has been omitted since it cancels in the ratio $\lambda^s_{J/\psi \phi}$. 
The amplitudes $\alpha^{J/\psi \phi}_i$ appearing in Eq.~(\ref{eq:masterAmpphi}) obey the structure given in 
Eq.~(\ref{eq:alphaGen0}).  The required expressions for the vertices and hard-scattering functions can be found in the appendix. 
The index $h=0,\pm$ indicated in Eq.~(\ref{eq:masterAmpphi}) makes reference to helicity of the particles in the final state.
During our 
analysis we average over the different helicity contributions. Therefore we take
\begin{eqnarray}
S_{J/\psi\phi}=\frac{S^0_{J/\psi\phi} + S^+_{J/\psi\phi} + S^-_{J/\psi\phi}}{3},
\end{eqnarray}
where each one of the asymmetries $S^h_{J/\psi\phi}$, are determined individually considering the corresponding
amplitude $\mathcal{A}^h_{J/\psi \phi}$ for $h=0,\pm$.
\\
Neglecting penguin contributions our theoretical evaluation leads to
\begin{eqnarray}
\sin(2\beta^{\rm SM}_s)=0.037 \pm 0.001,
\end{eqnarray}
which numerically coincides with $2\beta^{\rm SM}_s$ within the precision under consideration.
The error budget is shown in Table~\ref{tab:sinBetas}.
On the experimental side we use the average \cite{Amhis:2016xyh}
\begin{eqnarray}
2\beta^{\rm Exp}_s=0.021 \pm 0.031.
\end{eqnarray}
 The effect of $S_{J/\psi\phi}$ on the allowed values for $\Delta C^{s, cc}_{1}(M_W)$ is not as strong as the results derived 
 from other observables. However we included it in our analysis for completeness. For this reason we do not show the individual constraints
 from $S_{J/\psi\phi}$ and present only its effect in the global $\chi^2$-fit described in Section~\ref{sec:Universal_fit}.
\begin{table}
\begin{center}
\begin{tabular}{ |c|c|c|}
 \hline
Parameter  & Relative error \\
\hline
\hline
$|V_{ub}/V_{cb}|$& $2.44\%$ \\
\hline
$\gamma$& $1.39\%$ \\
\hline
$|V_{us}|$& $0.07\%$ \\
\hline
\hline
Total   &  $2.81\%$\\
\hline
\end{tabular}
\caption{Error budget for the observable $\sin(2\beta_s)$.}
\label{tab:sinBetas}
\end{center}
\end{table}


\subsubsection{$\tau_{B_s}/\tau_{B_d}$}
The lifetime ratio $\tau_{B_s}/\tau_{B_d}$ gives us sensitivity to $\Delta C^{s,cc}_1(M_W)$ and  $\Delta C^{s,cc}_2(M_W)$
via the weak exchange diagram contributing to the $B_s$-lifetime as CKM leading part. We assumed here that no new effects are
arising in the $B_d$-lifetime, where the CKM leading part is given by a $b \to c \bar{u} d$ transition. Allowing new effects in both
$b \to c \bar{c} s$ and $b \to c \bar{u} d$ the individually large effects will hugely cancel.
We also neglect the currently unknown contribution of the Darwin term \footnote{Recently the Wilson coefficient of the Darwin operator was found to be large 
\cite{Lenz:2020oce, Mannel:2020fts} for B mesons. Due to the currently unknown size of the matrix element of this operator in between $B_s$ states, the numerical effect of these new contributions 
on the lifetime ratio 
$\tau_{B_s} / \tau_{B_d}$ - being proportional to 
$\langle B_s | \rho_D^3  | B_s \rangle - \langle B_d | \rho_D^3  | B_d \rangle$ -   cannot yet be estimated.	
}.

Using the results presented in \cite{Jager:2017gal} we write
\begin{eqnarray}
\frac{\tau_{B_s}}{\tau_{B_d}}=\left(\frac{\tau_{B_s}}{\tau_{B_d}}\right)^{\rm SM}+\left(\frac{\tau_{B_s}}{\tau_{B_d}}\right)^{\rm NP} \, ,
  \label{eq:TausTauSM+NP}
\end{eqnarray}
for the SM value we take \cite{Kirk:2017juj}
\begin{eqnarray}
  \left(\frac{\tau_{B_s}}{\tau_{B_d}}\right)^{\rm SM}&=&1.0006\pm 0.0020.
  \label{eq:TausTauSM}
\end{eqnarray}
The experimental result for the ratio is \cite{Amhis:2016xyh}
\begin{eqnarray}
\left(\frac{\tau_{B_s}}{\tau_{B_d}}\right)^{\rm Exp}&=& 0.994\pm 0.004.
  \label{eq:TausTauExp}
\end{eqnarray}
To estimate the NP contribution $(\tau_{B_s}/\tau_{B_d})^{\rm NP}$ we consider the following function~\cite{Jager:2017gal}
\begin{eqnarray}
F_{\tau_{B_s}/\tau_{B_d}}(C_1,C_2)
  &=&
  G^2_F |V_{cb} V_{cs}|^2 m^2_b M_{B_s} f^2_{B_s} \tau_{B_s}\frac{\sqrt{1-4 x^2_c}}{144\pi}\Biggl\{ (1-x^2_c)\Biggl[4 |C'|^2 B_1\nonumber\\
&& + 24 |C_{2}|^2 \epsilon_1  \Biggl] - \frac{M^2_{B_s} (1 + 2 x^2_c)}{(m_b+m_s)^2}\Biggl[4 |C'|^2 B_2+ 24 |C_{2}|^2 \epsilon_2  \Biggl]\Biggl\},\nonumber\\
\end{eqnarray}
where $x_c = m_c/m_b$ and  $C'$  denotes the following combination of tree-level Wilson coefficients
\begin{eqnarray}
C'\equiv3 C_1 +C_2.
\end{eqnarray}
The non-perturbative matrix elements of the arising four-quark $\Delta B = 0$ operators are parameterised in
terms of the  decay constant $f_{B_s}$
and the bag parameter
$B_1$, $B_2$, $\epsilon_1$ and $\epsilon_2$, which we take from the recent evaluation in \cite{Kirk:2017juj}.
The numerical values used are listed in Appendix \ref{Sec:Inputs}.
The NP contribution to the lifetime ratio can be written as
\begin{eqnarray}
\Bigl(\frac{\tau_{B_s}}{\tau_{B_d}}\Bigl)^{\rm NP}&=&
F_{\tau_{B_s}/\tau_{B_d}}(C^{s, cc}_1(\mu),C^{s, cc}_2(\mu))\nonumber\\
	&&-F_{\tau_{B_s}/\tau_{B_d}}(C^{s, cc}_1(\mu),C^{s, cc}_2(\mu))\Biggl|_{\rm SM},
\label{eq:LTRatio}	
\end{eqnarray}
where in the second term in Eq.~(\ref{eq:LTRatio}) we have dropped the NP contributions $
\Delta C^{s, cc}_1(\mu)$ and $\Delta C^{s, cc}_2(\mu)$.
\begin{figure}
	\centering
	\includegraphics[height=5cm]{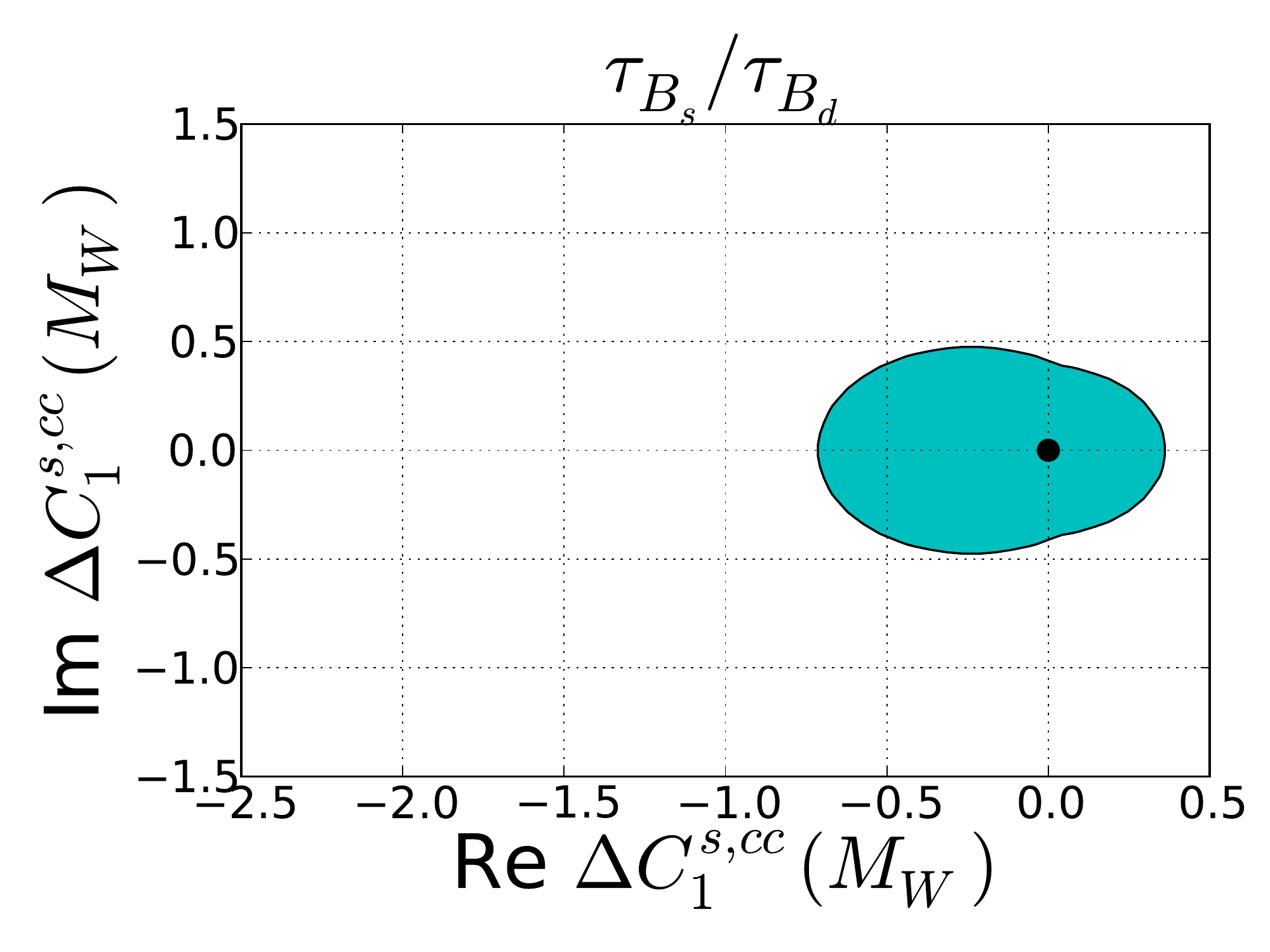}
	\caption{ Potential regions for the NP contributions in $\Delta C^{s, cc}_1(M_W)$ and $\Delta C^{s, cc}_2(M_W)$  allowed by 
	  the life-time ratio $\tau_{B_s}/\tau_{B_d}$ at $90\%$ C.L.. Here we assumed only BSM contributions to the decay channel $b \to c \bar{c}s$,
          but none to $b \to c \bar{u} d$. The black point corresponds to the SM value.}
	\label{Fig:TausTau}
\end{figure}
Our bounds for $\Delta C^{s, cc}_1(M_W)$ are shown in Fig.~\ref{Fig:TausTau}, the corresponding results for $\Delta C^{s, cc}_2(M_W)$ 
turn out to be weak and therefore we do not display them. We would like to highlight the consistency between our regions and those
presented in~\cite{Jager:2019bgk} which were calculated at the $68\%$ C. L..



\subsection{Observables constraining $b\rightarrow c\bar{c}d$ transitions}
\label{sec:bccd}
We devote this section to the derivation of bounds on $\Delta C^{d,cc}_{1}(M_W)$ and $\Delta C^{d,cc}_{2}(M_W)$ from
$\sin(2\beta_d)$ and $B\rightarrow X_d \gamma$.
In our final analysis we also included contributions from $a_{sl}^d$ which will
be described in more detail in Section~\ref{sec:multiple_channels}.
\subsubsection{$\sin(2\beta_d)$ and SM update of $\Delta M_q$}
\label{sec:sin2betaM12}
In our BSM framework mixing induced CP asymmetries can be modified by
changes in the tree-level decay or by changes to the neutral $B$-meson mixing.
The first effect was studied in Section \label{subsub:SJPsiPhi} for the case of
$B_s \to J/ \Psi \phi$ and found to give very weak bounds. Thus we will not consider them here.
The second effect is also expected to give relatively weak bounds, but since the lack of strong bounds on new
contributions to $b \to c \bar{c}d$ we will consider it here - in the $b \to c \bar{c} s$ we neglected it, because
of much stronger constraints from other observables.
\\
We can constrain $\Delta C^{d, cc}_{2}(M_W)$ with the observable
\begin{eqnarray}
\sin(2\beta_d)&=&-S_{J/\psi K_S}
\end{eqnarray}
which can be evaluated by applying the generic definition of the CP asymmetry shown in Eq.~(\ref{eq:Sf}) and using 
\begin{eqnarray}
\lambda^d_{J/\psi K_S}&=&\frac{q}{p}\Bigl|_{B_d} \frac{\bar{\mathcal{A}}_{J/\psi K_S} }{\mathcal{A}_{J/\psi K_S}}.
\label{eq:lambdaJPsi}
\end{eqnarray}
Where in Eq.~(\ref{eq:lambdaJPsi}), $\mathcal{A}_{J/\psi K_S}$ and $\bar{\mathcal{A}}_{J/\psi K_S}$ correspond to the 
amplitudes for the processes $B^0\rightarrow J/\psi K_S$ and $\bar{B}^0\rightarrow J/\psi K_S$ respectively.
\\
We study here  modifications of $q/p|_{B_d}$, while we neglect the change of the  amplitudes $\mathcal{A}_{J/\psi K_S}$
and $\bar{\mathcal{A}}_{J/\psi K_S}$ - since an exploratory study found much weaker bounds.
The definition of $q/p|_{B_d}$ in terms of the $B_d$ matrix element $M^d_{12}$ is given in Eq.~(\ref{eq:qop}). 
\\
In the SM we have
\begin{eqnarray}
M^{d, \rm{SM}}_{12}=\frac{\langle B^0_d| \hat{\mathcal{H}_d}^{|\Delta B|=2,\rm{SM}}|\bar{B}^0_d \rangle}{2M_{B^0_d}},
\label{eq:M12}
\end{eqnarray}
with
\begin{eqnarray}\label{eq:EffDelta2}
\mathcal{\hat{H}}^{|\Delta B| = 2, \rm{SM}}_d
&=&\frac{G^2_F}{16\pi^2}(\lambda_t^{(d)})^2 C^{|\Delta B| =2}(m_t, M_W, \mu) \hat{Q}^d_1+ h.c..
\end{eqnarray}
The dimension six effective  ${|\Delta B| =2}$  operator ${Q}^d_1$ in Eq.~(\ref{eq:EffDelta2}) is given by
\begin{eqnarray}
\hat{Q}^d_1&=& \Bigl(\bar{\hat{d}} \hat{b} \Bigl)_{V-A} \Bigl(  \bar{\hat{d}}\hat{b}\Bigl)_{V-A},
\label{eq:DeltaB2}
\end{eqnarray}
and the Wilson coefficient $C^{|\Delta B| =2}(m_t, M_W, \mu)$ corresponds to
\begin{eqnarray}
C^{|\Delta B| =2}(m_t, M_W, \mu)&=&\tilde{\eta}  M^2_W S_0(x_t),
\end{eqnarray}
where the factor $\tilde{\eta}$ accounts for the renormalization group evolution from 
the scale $m_t$ down to the renormalization scale $\mu \sim m_b$ \cite{Buras:1990fn} and $S_0(x_t)$ is the 
Inami-Lim function \cite{Inami:1980fz}
\begin{eqnarray}
S_0(x_t)=\frac{x_t}{(1-x_t)^2}\Bigl(1-\frac{11}{4} x_t +\frac{x^2_t}{4} -\frac{3 x^2_t \ln x_t}{(1-x_t)}\Bigl).
\end{eqnarray}
Using the new averages presented in
\cite{DiLuzio:2019jyq} for the hadronic matrix elements 
(based on the non-perturbative calculations in
\cite{Grozin:2016uqy,Kirk:2017juj,King:2019lal} and \cite{Christ:2014uea,Bussone:2016iua,Hughes:2017spc,Bazavov:2017lyh})
we get the new updated SM results
\begin{eqnarray}
  \Delta M_s^{\rm SM} & = & \left( 18.77 \pm 0.86 \right) \mbox{ps}^{-1} \, ,
  \\
  \Delta M_d^{\rm SM} & = & \left( 0.543 \pm 0.029 \right) \mbox{ps}^{-1} \, ,
\end{eqnarray}
where we observe a huge reduction of the theoretical uncertainty, see Tables \ref{error1} and \ref{error7}.
Our numbers agree with the ones quoted in \cite{DiLuzio:2019jyq} - a tiny difference stems from a different treatment of the
top quark mass, the CKM input and the symmetrisation of the error we have performed here.
\begin{center}
\begin{table}
  \begin{center}
    \begin{tabular}{|c||c|c|c|c|}
\hline
$\Delta M_s^{\rm SM} $   &   $\mbox{This work}  $ & $\mbox{ABL 2015}$  &  $\mbox{LN 2011}  $      &  $ {\mbox{LN 2006}} $
\\
\hline
\hline
 $\mbox{Central Value} $   &  $18.77 \, \mbox{ps}^{-1} $&  $18.3 \, \mbox{ps}^{-1} $ &  $ 17.3 \, \mbox{ps}^{-1 } $ &  $ 19.3 \, \mbox{ps}^{-1} $
\\
\hline
\hline
 $f_{B_s} \sqrt{B_1^s} $ & $3.1 \%$ &  $ 13.9\%  $          & $  13.5 \%   $              & $  34.1 \% $
\\
\hline
 $V_{cb}    $     &  $3.4 \%$ &$  4.9 \%   $         &  $   3.4 \%   $             &  $ 4.9 \% $
\\
\hline
$\bar{m}_t(\bar{m}_t)$       &  $0.3 \%$ &$ 0.7 \%   $          &  $  1.1 \%    $             & $  1.8 \% $
\\ 
\hline 
$ \Lambda_5^{\rm QCD}   $     & $0.2 \%$  &$ 0.1 \%    $        &  $  0.4 \%      $           &  $ 2.0 \% $
\\
\hline
 $\gamma   $      &  $0.1 \%$ & $0.1 \%  $           &  $  0.3 \%      $           &  $ 1.0 \% $
\\
\hline
 $|V_{ub}/V_{cb}| $ &  $<0.1 \%$ & $0.1 \%   $        &  $  0.2 \%    $             &   $0.5 \% $
\\
\hline
 $\overline{m}_b  $      & $<0.1 \%$ &  $<0.1 \%    $        &  $  0.1 \%   $              &  $ --- $
\\
\hline
\hline
Total           & $ 4.6 \%   $ &  $ 14.8 \%   $        &  $14.0 \%    $             &  $34.6 \% $
\\
\hline
    \end{tabular}
    \end{center}
\caption{List of the individual contributions to the theoretical error of the mass difference 
$\Delta M_s$ within the Standard Model and comparison with the values obtained in  2015 \cite{Artuso:2015swg}, in 2011 \cite{Lenz:2011ti}
and in 2006 \cite{Lenz:2006hd}.}
\label{error1}
\end{table}
\end{center}
\begin{table}
  \begin{center}
    \begin{tabular}{|c||c|c|}
\hline
$  \Delta M_d^{\rm SM}  $  & This work & \mbox{ABL \, 2015}        
\\
\hline
\hline
$\mbox{Central Value}$&$  0.543 \, \mbox{ps}^{-1} $&$  0.528 \, \mbox{ps}^{-1} $
\\
\hline
\hline
$f_{B_d} \sqrt{B_1^d}   $&$3.6\%$&$13.7\% $
\\
\hline
$V_{cb}             $&$ 3.4\%$&$ 4.9\% $
\\
\hline
$m_b                $&$ 0.1\%$&$ 0.1\% $
\\
\hline
$\gamma             $&$ 0.2\%$&$ 0.2 \%  $
\\
\hline
$\Lambda_5^{\rm QCD} $&$ 0.2\%$&$ 0.1 \%  $
\\
\hline
$|V_{ub}/V_{cb}|   $&$ 0.1\%$&$  0.1 \% $
\\
\hline
$\bar{m}_t(\bar{m}_t)$&$ 0.3\%$&$ 0.1 \%  $
\\
\hline
\hline
Total               &$5.3 \%$&$14.8\%$
\\
\hline
    \end{tabular}
    \end{center}
\caption{List of the individual contributions to the theoretical 
error of the mixing quantity $\Delta M_d$ and comparison with the values obtained in  2015 \cite{Artuso:2015swg}.}
\label{error7}
\end{table}
HFLAV \cite{Amhis:2016xyh} gives for the experimental values
\begin{eqnarray}
  \Delta M_s^{\rm Exp} & = & \left( 17.757 \pm 0.021 \right) \mbox{ps}^{-1} \, ,
  \\
  \Delta M_d^{\rm exp} & = & \left( 0.5064 \pm 0.0019 \right) \mbox{ps}^{-1} \, .
\end{eqnarray}
We introduce BSM effects to  Eq.~(\ref{eq:M12}) by adding to the SM expression in Eq.~(\ref{eq:EffDelta2}) the 
double insertion of the effective Hamiltonian
\begin{eqnarray}
\hat{\mathcal{H}}_{eff}^{|\Delta B|=1}&=&\frac{G_F}{\sqrt{2}}\Bigl(\sum_{p,p'=u,c}
\lambda^{(d)}_{pp'} C^{d, pp'}_2 \hat{Q}^{d, pp'}_2 + h.c. \Bigl).
\end{eqnarray}
Following~\cite{Boos:2004xp} we evaluate the full combination at the scale $\mu_c=m_c$, where the extra contribution to the SM
$|\Delta B|=2$ Hamiltonian in Eq.~(\ref{eq:EffDelta2}) is given by
\begin{eqnarray}
\hat{\mathcal{H}}^{|\Delta B|=2}_{extra} &\approx&
\frac{G^2_F}{16\pi^2} \Biggl\{ C'_1(\mu_c) \hat{P}_1 +  C'_2(\mu_c) \hat{P}_2
\nonumber\\ 
	&&+ \Biggl[ \Bigl(2\lambda^{(d)}_c\lambda^{(d)}_t \tilde{C}_3(x_t^2)
+ (\lambda^{(d)}_c)^2\Bigl)
 + C'_3(\mu_c) \Biggl] \hat{P}_3 \Biggl\},\nonumber\\
\label{eq:H2extra}
\end{eqnarray}
with
\begin{eqnarray}
C'_{1}(\mu_c)&=&- \frac{2}{3} \hbox{ ln}\Bigl[\frac{\mu_c^2}{M^2_W}\Bigl] \Bigl\{\frac{(\lambda^{(d)}_c)^2}{2} \Bigl(C^{d, cc}_2\Bigl)^2 -
  (\lambda^{(d)}_c)^2 C^{d, cu}_2 C_2^{d, uc} - \lambda^{(d)}_c \lambda^{(d)}_t C_2^{d, cu} C_2^{d, uc}\nonumber\\
&&+ \frac{(\lambda^{(d)}_c)^2}{2} \Bigl(C_2^{d, u u}\Bigl)^2 + \lambda^{(d)}_c \lambda^{(d)}_t \Bigl(C_2^{d, uu}\Bigl)^2 + \frac{(\lambda^{(d)}_t)^2}{2}
\Bigl(C_2^{d, uu} \Bigl)^2\Bigl\},\nonumber\\
C'_{2}(\mu_c)&=&\frac{2}{3} \hbox{ ln}\Bigl[\frac{\mu_c^2}{M^2_W}\Bigl]
\Bigl\{(\lambda^{(d)}_c)^2 \Bigl(C_2^{d, cc}\Bigl)^2 - 2 (\lambda^{(d)}_c)^2 C_2^{d, cu} C_2^{d, uc} - 2 \lambda^{(d)}_c \lambda^{(d)}_t C_2^{d, cu} C_2^{d, uc} \nonumber\\
&&+ (\lambda^{(d)}_c)^2 \Bigl(C_2^{d, uu}\Bigl)^2 + 2 \lambda_c \lambda^{(d)}_t \Bigl(C_2^{d, uu}\Bigl)^2 + (\lambda^{(d)}_t)^2 \Bigl(C_2^{d, uu}\Bigl)^2\Bigl\},\nonumber\\
C'_{3}(\mu_c)&=&\frac{2}{3} \hbox{ ln}\Bigl[\frac{\mu_c^2}{M^2_W}\Bigl]
\Bigl\{ 3 (\lambda^{(d)}_c)^2 \Bigl(C_2^{d, cc}\Bigl)^2 - 3 (\lambda^{(d)}_c)^2 C_2^{d, c u} C_2^{d, uc} - 3 \lambda^{(d)}_c \lambda^{(d)}_t C_2^{d, cu} C_2^{d, uc}\Bigl\}.
\nonumber\\
\end{eqnarray}

and

\begin{eqnarray}
\tilde{C}_3(x_t)=\ln x_t - \frac{3 x_t}{4(1-x_t)}-
\frac{3 x^2_t\ln x_t}{4(1-x_t)^2}.	
\end{eqnarray}	

\begin{figure}
\centering
\includegraphics[height=5cm]{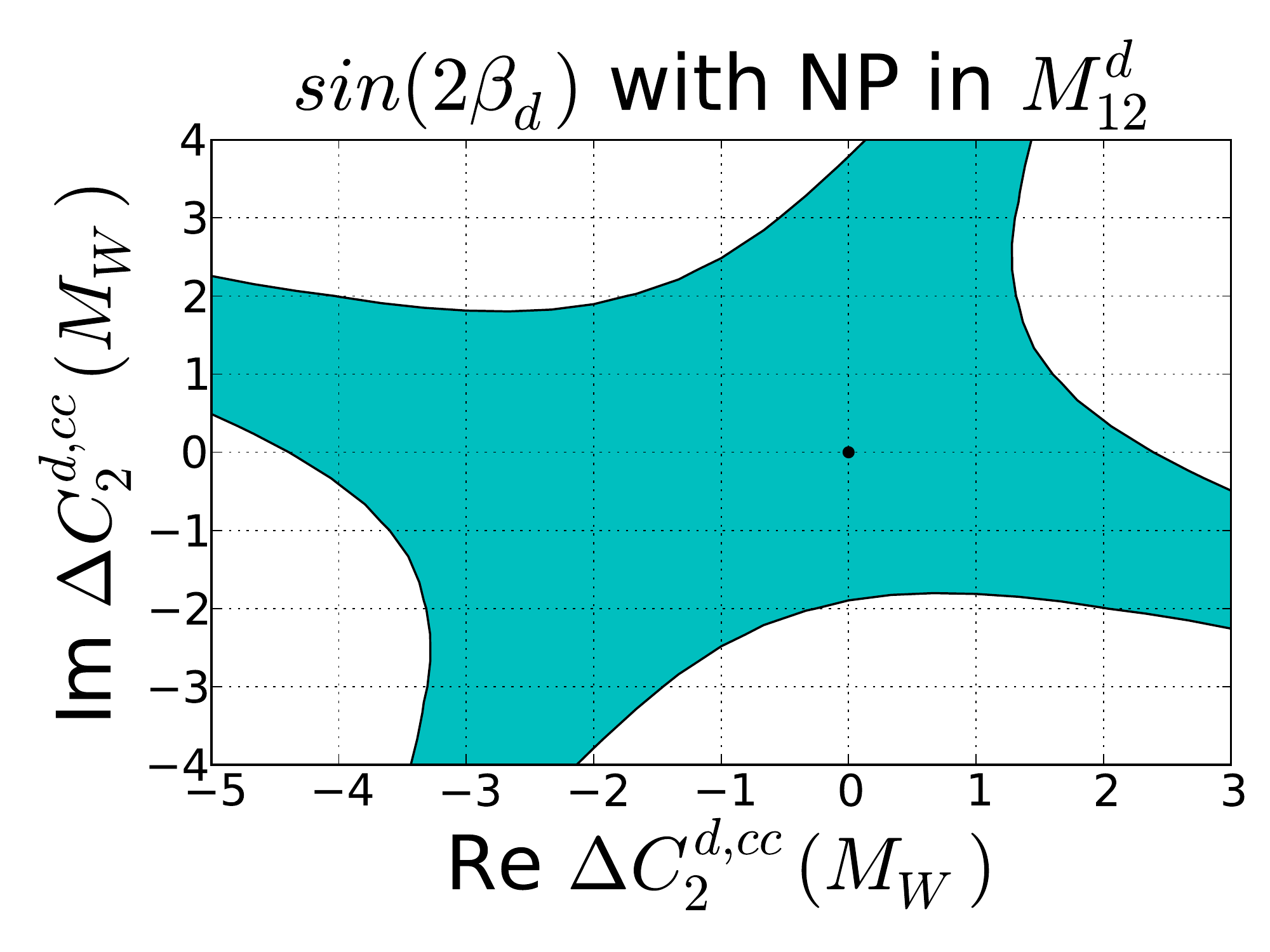}
\caption{
Potential regions for the NP contributions in $\Delta C^{d, cc}_2(M_W)$ allowed by the observable 
$\sin(2\beta_d)$ from modifications in $M^d_{12}$ through double insertions of the $\Delta B=1$ 
effective Hamiltonian at $90\%$ C.L.. Due to the weakness of the current bounds, penguin pollution has been neglected in the analysis.
The black point corresponds to the SM value.
}
\label{fig:M12d}
\end{figure}

The set of HQET operators required in Eq.~(\ref{eq:H2extra}) are 
\begin{eqnarray}
\hat{P}_0=(\bar{\hat{h}}^{(+)}\hat{d})_{V-A}(\bar{\hat{h}}^{(-)} \hat{d})_{V-A},&&
\hat{P}_1=m^2_b \hat{P}_0,\nonumber\\
&&\nonumber\\
\hat{P}_2=m^2_b\Bigl(\bar{\hat{h}}^{(+)}_{v}\Bigl[1-\gamma_5 \Bigl] \hat{d}\Bigl)\Bigl(\bar{\hat{h}}^{(-)}_{v} \Bigl[1-\gamma_5 \Bigl]\hat{d}\Bigl),&&\hat{P}_3= m^2_c\hat{P}_0.
\end{eqnarray}
Thus, our full determination of $M^d_{12}$ is given by
\begin{eqnarray}
M^d_{12}&=&\frac{\langle B^0_d|\hat{\mathcal{H}}^{|\Delta B|=2 ,SM}_d
+\hat{\mathcal{H}}_{extra}^{|\Delta B|=2}
 |\bar{B^0_d}\rangle}{2 M_{B^0_d}},
\end{eqnarray}
where the $|\Delta B|=2$ operator $\hat{Q}^d_1$ is matched at the scale $\mu_c=m_c$ into $\hat{P}_0$ \cite{Boos:2004xp}.
The required matrix elements for the numerical evaluations are \cite{Kirk:2017juj}
\begin{eqnarray}
\label{eq:matP0}
\langle B^0_d| \hat{P}_0 |\bar{B}^0_d \rangle &=&\frac{8}{3}f^2_{B_d}M^2_{B_d}B^d_1(\mu_c),\nonumber\\
\langle B^0_d| \hat{P}_2 |\bar{B}^0_d \rangle&=& -\frac{5}{3} m^2_b  \Bigl(\frac{M_{B_d}}{m_b + m_d}\Bigl)^2f^2_{B_d}M^2_{B_d}B^{d}_2(\mu_c),
\end{eqnarray}
with the values for the Bag parameters as indicated in Appendix \ref{Sec:Inputs}.
Our theoretical result - neglecting contributions from penguins - is
\begin{eqnarray}
\sin(2\beta^{\rm SM}_d)=0.707\pm 0.030,
\label{eq:Sin2betadfitdir}
\end{eqnarray}
the full error budget in the SM can be found in Table \ref{tab:MdMd}. Notice that, the contributions from 
double insertions of the $|\Delta B|=1$ effective Hamiltonian are relevant only when $\Delta C^{d, cc}_{2}(M_W)\neq 0$, hence they do 
not appear in Table~\ref{tab:MdMd}.
On the experimental side we use the average from direct measurements \cite{Amhis:2016xyh}
\begin{eqnarray}
\sin(2\beta^{\rm Exp}_d)=0.699\pm 0.017,
\end{eqnarray}
our results for the allowed regions on $\Delta C^{d, cc}_2(M_W)$ are shown in Fig.~\ref{fig:M12d}.
\begin{table}
\begin{center}
\begin{tabular}{ |c|c|c|}
 \hline
Parameter  & Relative error \\
\hline
\hline
$|V_{ub}/V_{cb}|$& $4.22\%$ \\
\hline
$|V_{us}|$ & $0.20\%$ \\
\hline
$\gamma$ &$0.04\%$\\
\hline
$\mu_c$    & $0.02\%$  \\
\hline
$|V_{cb}|$& $0.01\%$ \\
\hline
\hline
Total   &  $4.22\%$\\
\hline
\end{tabular}
\caption{Error budget for the observable $\sin(2\beta_d)$.}
\label{tab:MdMd}
\end{center}
\end{table}

\subsubsection{$\bar{B}\rightarrow X_d \gamma$}
\label{sec:Bdgamma}
\begin{figure}
\centering
\includegraphics[height=5cm]{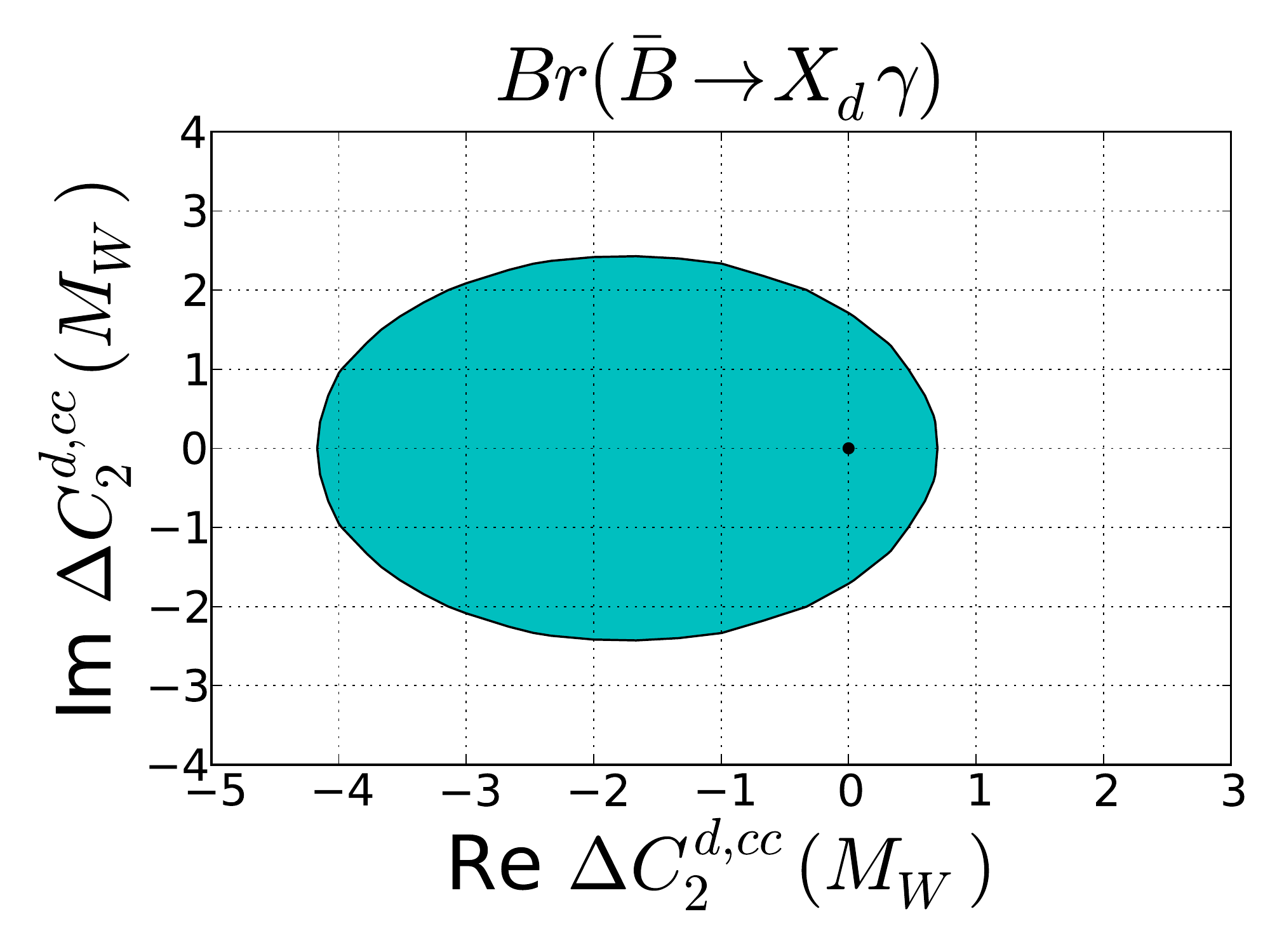}
\caption{
Potential regions for the NP contributions in $\Delta C^{d, cc}_2(M_W)$ allowed by the observable 
$\mathcal{B}r(\bar{B}\rightarrow X_d \gamma)$ at $90\%$ C.L.. The black point corresponds to the SM value.}
\label{fig:Bd_gamma}
\end{figure}

The branching ratio of the process $\bar{B}\rightarrow X_d \gamma$ allows us to impose further constraints on the NP contribution
$\Delta C^{d, cc}_2(M_W)$. For the theoretical determination, we used the NNLO branching ratio for the transition 
$\bar{B}\rightarrow X_d \gamma$ given in \cite{Misiak:2015xwa}
\begin{eqnarray}
\mathcal{B}^{\hbox{NNLO}}_r(\bar{B}\rightarrow X_d \gamma)=(1.73^{+0.12}_{-0.22})\cdot 10^{-5}\quad\hbox{ for }E_{\gamma}>1.6~\hbox{ GeV}.
\end{eqnarray}
On the experimental side we consider~\cite{Crivellin:2011ba, delAmoSanchez:2010ae, Wang:2011sn}
\begin{eqnarray}
\mathcal{B}_{r}^{\rm Exp}(\bar{B}\rightarrow X_d \gamma)&=&\Bigl(1.41 \pm 0.57 \Bigl)\cdot 10^{-5}.
\label{eq:BdgammaExp}
\end{eqnarray}
The NP regions on $\Delta C^{cc, d}_1(M_W)$ derived from $\mathcal{B}^{\hbox{NNLO}}_r(\bar{B}\rightarrow X_d \gamma)$ are 
shown in Fig.~\ref{fig:Bd_gamma}. Our treatment for $\bar{B}\rightarrow X_d \gamma$ is analogous to the one of $\bar{B}\rightarrow X_s \gamma$,
therefore our discussion here is rather short and we refer the reader to the details provided  in Section~\ref{sec:Bsgamma}. 
%
%
%
%
%
%
%
%
%
%
\subsection{Observables constraining multiple channels}
Several observables like $\Delta \Gamma_q$, $\tau (B_s) / \tau (B_d)$ and the semi-leptonic CP asymmetries are affected
by different decay channels. We have shown that $\Delta \Gamma_s$ is by far dominated by the $ b \to c \bar{c} s$ transition,
$\Delta \Gamma_d$ has not yet been measured. In $\tau (B_s) / \tau (B_d)$ a new effect in the $ b \to c \bar{c} s$ transition
roughly cancels a similar size effect in a $ b \to c \bar{u} d$ transition, thus we have assumed for this observable only BSM effects
in the  $ b \to c \bar{c} s$ transition. Below we will study constraints stemming from $a_{sl}^q$, which is affected by the decay channels
$b \to c \bar{c} q$, $b \to c \bar{u} q$, $b \to u \bar{c} q$ and $b \to u \bar{u} q$.
\label{sec:multiple_channels}
\subsubsection{$a^s_{sl}$ and $a^d_{sl}$: Bounds and SM update}
The theoretical description of the semi-leptonic CP asymmetries was already presented in detail
in Section \ref{subsec:DGs}.
Our SM predictions for the semileptonic asymmetries $a^s_{sl}$ and $a^d_{sl}$ are
\begin{eqnarray}
  a^{s,\rm SM}_{sl}  &=&  \Bigl(2.06 \pm 0.18 \Bigl)\cdot 10^{-5},
  \\
  a^{d,\rm SM}_{sl}  &=&    \Bigl(-4.73\pm 0.42 \Bigl)\cdot 10^{-4}.
\label{eq:aslasldSM}
\end{eqnarray}
The error budgets of the mixing observables $a^s_{sl}$ and $a^d_{sl}$  within the SM 
are presented in Tabs.~\ref{tab:asls} and  \ref{tab:asld}  respectively.
\begin{table}
  \begin{center}
\begin{tabular}{|c||c|c|c|c|}
\hline
 $a_{\rm sl}^{s,{\rm SM}} $&$ \mbox{this work}   $&$  \mbox{ABL 2015}     $&$  \mbox{LN 2011}     $&$  {\mbox{LN 2006}} $
\\
\hline
\hline
 $\mbox{Central Value}    $&$2.06 \cdot 10^{-5}  $&$2.22 \cdot 10^{-5}  $&$  1.90\cdot 10^{-5} $&$  2.06 \cdot 10^{-5} $
\\
\hline
\hline
 $\mu            $&$6.7 \%$&$9.5  \%             $&$    8.9 \%           $&$  12.7 \% $
\\
\hline
 $\bar z         $&$4.0 \%$&$4.6 \%              $&$   7.9 \%            $&$  9.3 \% $
\\
\hline
 $|V_{ub}/V_{cb}|$&$2.6 \%$&$5.0 \%              $&$  11.6 \%            $&$  19.5 \% $
\\
\hline
 $B_{R_3}^s       $&$2.3 \%$&$ 1.1 \%            $&$   1.2 \%            $&$  1.1 \% $
\\
\hline
$B_{\tilde{R}_3}^s$& - &$2.6 \%                   $&$  2.8 \%             $&$   2.5 \% $
\\
\hline
$m_b           $&$1.3 \%$&$ 1.0 \%              $&$   2.0 \%            $&$  3.7 \% $
\\
\hline
$\gamma        $&$1.1 \%$&$ 1.3 \%              $&$   3.1 \%            $&$  11.3 \% $
\\
\hline
 ${ B}_{R_2}^s    $&$0.8 \%$&$ 0.1 \%            $&$   0.1 \%            $&$  --- $
\\
\hline
 $\Lambda_5^{\rm QCD} $&$0.6 \%$&$ 0.5 \%        $&$   1.8 \%            $&$  0.7 \% $
\\
\hline
	$\bar m_t (\bar m_t)           $&$0.3 \%$&$ 0.7 \%             $&$   1.1 \%            $&$  1.8 \% $
\\
\hline
 $B_{3}^s         $&$0.3 \%$&$ 0.3 \%           $&$   0.6 \%            $&$  0.4 \% $
\\
\hline 
 ${B}_{R_0}^s $&$0.3 \%$&$ 0.2 \%                $&$    0.3 \%           $&$   --- $
\\
\hline
$m_s           $&$<0.1 \%$&$ 0.1 \%        $&$   0.1 \%                   $&$  0.1 \% $
\\
\hline
 $B_{\tilde{R}_1}^s$&$<0.1 \%$&$0.5 \%              $&$  0.2 \%             $&$   --- $
\\
 \hline
 ${ B}_{R_1}^s $&$<0.1 \%$&$ <0.1 \%       $&$    0.0 \%                  $&$   --- $
\\
\hline
 $V_{cb}        $&$<0.1 \%$&$  0.0 \%       $&$   0.0 \%                  $&$ 0.0 \% $
\\
\hline
\hline
Total           &$8.8 \%$&$  12.2 \%      $&$  17.3 \%                   $&$ 27.9 \% $
\\
\hline
\end{tabular}
\end{center}
\caption{List of the individual contributions to the theoretical error of the semileptonic CP 
asymmetries
$a_{sl}^s$ within the Standard Model and comparison with the values obtained in 2015 \cite{Artuso:2015swg}, in 2011 \cite{Lenz:2011ti}
and in 2006 \cite{Lenz:2006hd}. We have used equations of motion in the current analysis to get rid of the operator $\tilde{R}_3$.}
\label{tab:asls}
\end{table}

\begin{figure}
\centering
\includegraphics[height=5cm]{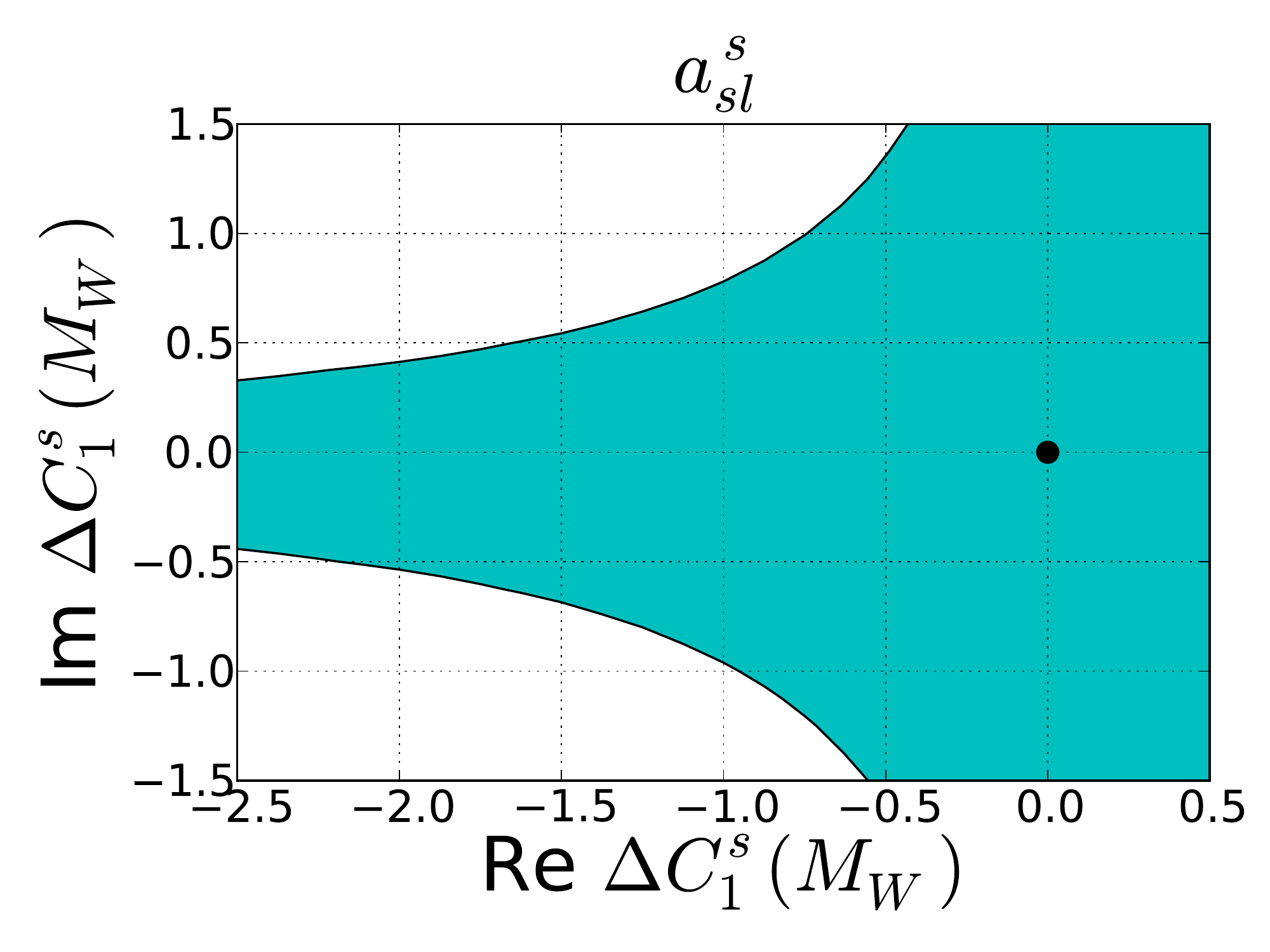}
\includegraphics[height=5cm]{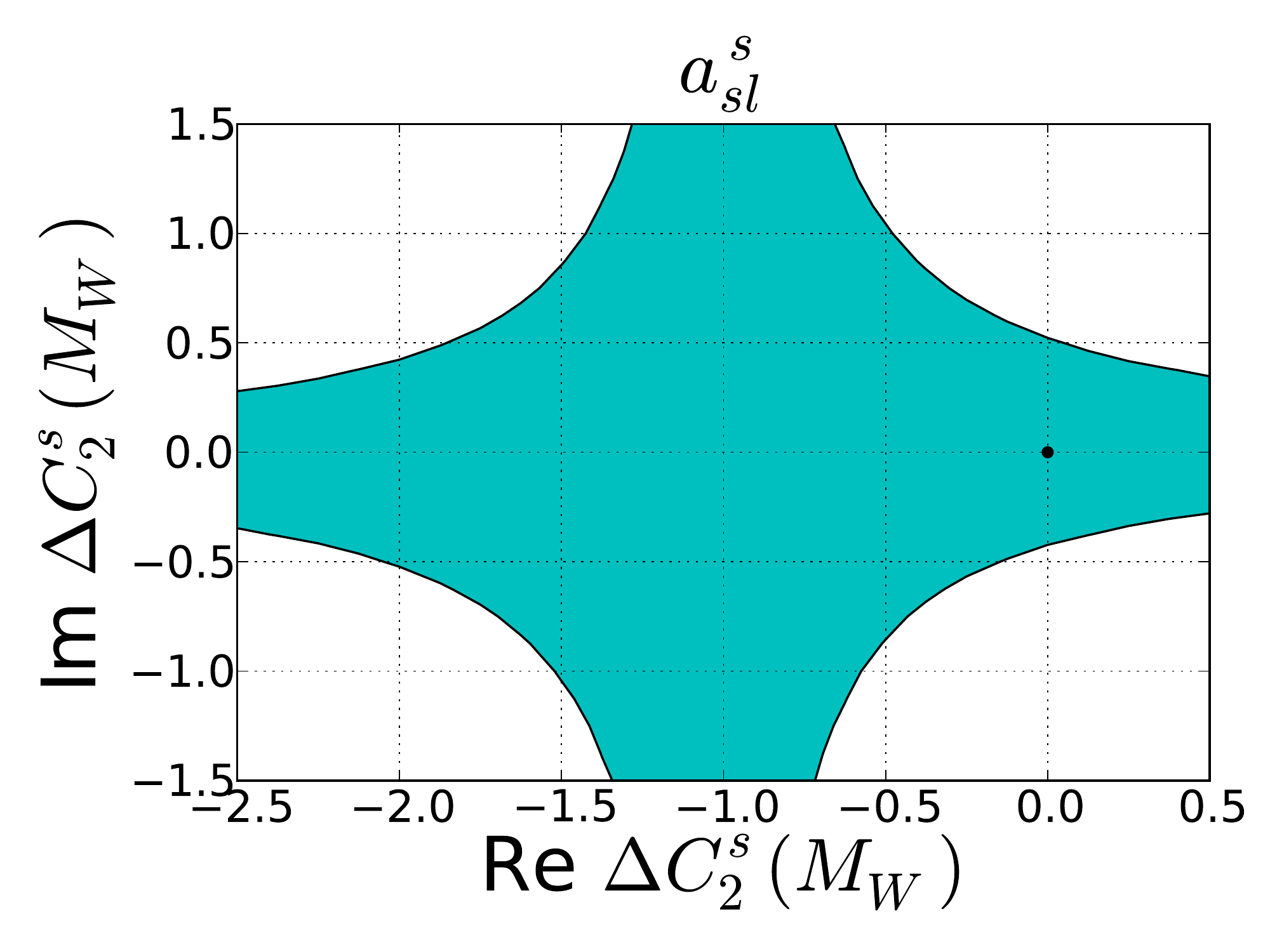}
\caption{
Potential regions for the NP contributions in $\Delta C^{s}_1(M_W)$ and $\Delta C^{s}_2(M_W)$  allowed by 
	the semileptonic asymmetry $a^s_{sl}$ at $90\%$ C.L.. The black point corresponds to the SM value. For the purposes
	of illustration we have made the universality assumptions: $\Delta C^{s, uu}_{1}(M_W)=\Delta C^{s, cu}_{1}(M_W)=\Delta C^{s, uc}_1(M_W)=\Delta C^{s, cc}_{1}(M_W)=\Delta C^s_{1}(M_W)$ and similarly for $\Delta C^s_{2}(M_W)$.}
\label{fig:asls}
\end{figure}

\begin{figure}
\centering
\includegraphics[height=5cm]{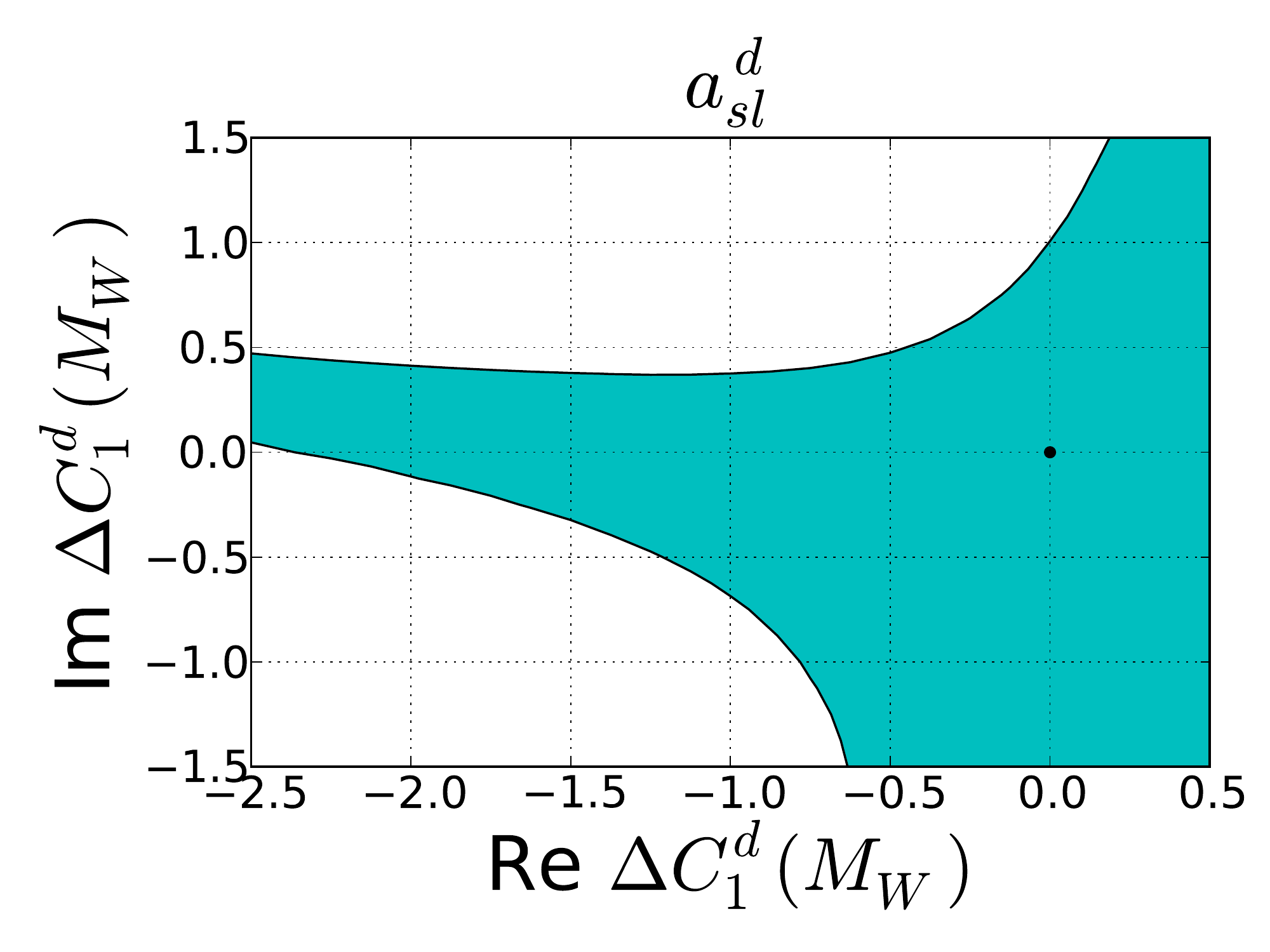}
\includegraphics[height=5cm]{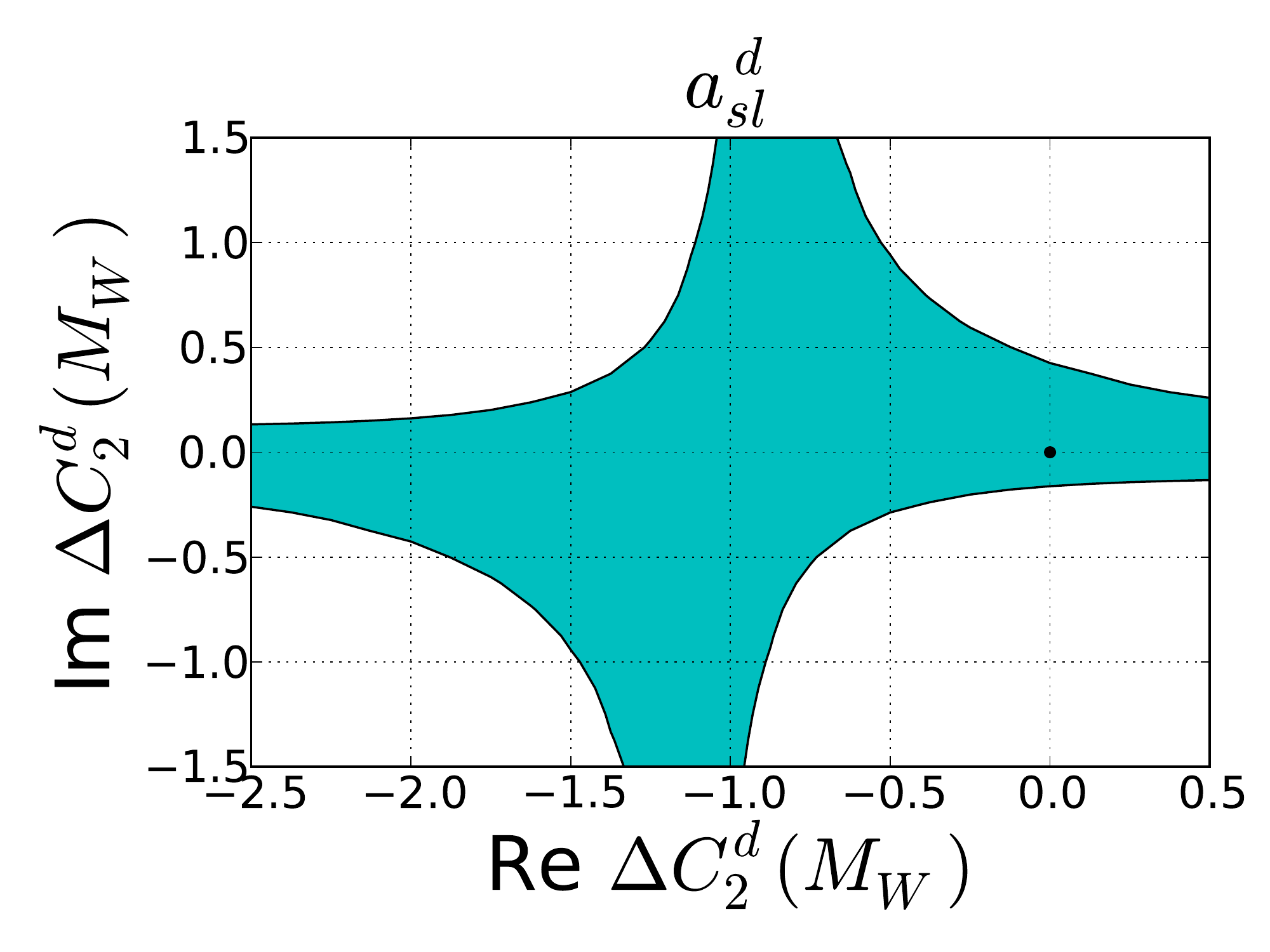}
\caption{
Potential regions for the NP contributions in $\Delta C^{d}_1(M_W)$ and $\Delta C^{d}_2(M_W)$  allowed by 
	the semileptonic asymmetry $a^d_{sl}$ at $90\%$ C.L.. The black point corresponds to the SM value. For the purposes
	of illustration we have made the universality assumptions: $\Delta C^{d, uu}_{1}(M_W)=\Delta C^{d, cu}_{1}(M_W)=\Delta C^{d, uc}_1(M_W)=\Delta C^{d, cc}_{1}(M_W)=\Delta C^d_{1}(M_W)$ and similarly for $\Delta C^d_{2}(M_W)$.}	
\label{fig:asld}
\end{figure}

\begin{table}
  \begin{center}
    \begin{tabular}{|c||c|c|}
\hline
$a_{\rm sl}^{d, \rm SM}$ & This work &   \mbox{ABL \, 2015}       
\\
\hline
\hline
$\mbox{Central Value}$&$ -4.7 \cdot 10^{-4}  $&$ -4.7 \cdot 10^{-4}  $
\\
\hline
\hline
$B_{\widetilde R_2}^d $&$0.8 \%$ &$0.1 \%$
\\
\hline
$\mu                $&$ 6.7 \%$ &$ 9.4 \%$
\\
\hline
$V_{cb}            $ &$ 0.0 \%$ &$ 0.0 \%$
\\
\hline
$B_3^d             $ &$ 0.4 \%$  &$ 0.6 \%$
\\
\hline
$B_{R_0}^d          $ &$ 0.3 \%$ &$ 0.2 \%$
\\
\hline
$\bar z             $&$ 4.1 \%$ &$ 4.9 \%$
\\
\hline
$m_b                $&$  1.3 \%$ &$  1.3 \%$
\\
\hline
$B_{\tilde{R}_3}^d   $ &$  - \%$ &$  2.7 \%$
\\
\hline
$B_{R_3}^d           $&$ 2.3 \%$&$ 1.2 \%$
\\
\hline
$\gamma             $&$ 1.0 \%$&$ 1.1 \%$
\\
\hline
$\Lambda_5^{\rm QCD}  $&$ 0.8 \%$&$ 0.5 \%$
\\
\hline
$|V_{ub}/V_{cb}|    $&$ 2.7 \%$&$ 5.2 \%$
\\
\hline
$\bar{m}_t(\bar{m}_t)$&$ 0.3 \%$&$ 0.7 \%$
\\
\hline
\hline
Total              &$8.8 \%$&$12.3 \%$
\\
\hline
    \end{tabular}
    \end{center}
\caption{List of the individual contributions to the theoretical 
  error of the mixing quantity $a_{\rm sl}^{d, \rm SM}$ in the $B^0$-sector  and comparison with the values obtained in 2015 \cite{Artuso:2015swg}.
We have used equations of motion in the current analysis to get rid of the operator $\tilde{R}_3$.}
\label{tab:asld}
\end{table}
The current experimental bounds \cite{Amhis:2016xyh} are far above the SM predictions
\vspace{-0.2cm}
\begin{eqnarray}
a^{s,\rm Exp}_{sl}&=&\Bigl(60 \pm 280\Bigl)\cdot 10^{-5},\nonumber\\
a^{d,\rm Exp}_{sl}&=&\Bigl(-21 \pm 17 \Bigl)\cdot 10^{-4}.
\end{eqnarray}
Nevertheless, these observables yield already, with the current experimental precision, strong bounds on
$C_1$ and $C_2$ due to the pronounced sensitivity of Im$(\Gamma_{12}^q / M_{12}^q)$ on the imaginary components
of the $\Delta B = 1$ Wilson coefficients.
The regions for $\Delta C_1(M_W)$ and $\Delta C_2(M_W)$ allowed by the observables $a^s_{sl}$ and $a^d_{sl}$ 
are presented in Figs.\ref{fig:asls} and \ref{fig:asld} 
respectively where for simplicity
we have assumed the universal behaviour 
\begin{eqnarray}
  \Delta C^{q, uu}_{j}(M_W)=\Delta C^{q, uc}_{j}(M_W)=\Delta C^{q, cc}_{j}(M_W),
  \label{eq:asl_universal}
\end{eqnarray}
for $j=1,2$.
As discussed in Section \ref{subsec:DGs} different BSM effects in individual decay channels could lift the severe GIM suppression
and lead to large effects, while the scenario given in Eq.(\ref{eq:asl_universal}) is dominated by $b \to c \bar{c} q$ transitions.
However, in Secs. \ref{sec:buudfit}, \ref{sec:bcudfit} and \ref{sec:bccd}  we will also study the effects of $a_{sl}^d$ on the different $b$-quark
decay channels $b\rightarrow u \bar{u} d$,  $b\rightarrow c \bar{u} d$,
and $b\rightarrow c \bar{c} d$ independently. 

%
%
%
%
%
%
%
%
%
%
%
%
%
%
%
%
%
%
%
%

\section{Global $\chi^2$-fit results}
\label{sec:Globalfitresults}
So far, we have limited our discussion to  constraints derived from individual observables.
In this section, we present, as the main result of this work, the resulting regions for $\Delta C_1(M_W)$ and $\Delta C_2(M_W)$
obtained after combining observables for the different exclusive $b$ quark transitions.
We will investigate three consequences of BSM effects in non-leptonic tree-level decays.
\begin{enumerate}
\item The allowed size of BSM contributions to the Wilson coefficients $C_1$ and $C_2$, governing the leading tree-level decays.
\item The impact of these new effects on the possible size of the observable $\Delta \Gamma_d$, which has not been measured yet.
      Notice that, if one sigma deviations are considered, the current experimental uncertainty associated with $\Delta \Gamma_d$, 
      see Eq.~(\ref{eq:DGammaExp}), allows enhancement factors within the interval
      \vspace{-0.2cm}
      \begin{eqnarray}
      -3.40<\Delta \Gamma^{\rm Exp}_d/\Delta \Gamma^{\rm SM}_d<2.27.
      \label{eq:enhacementsonDGammadonesigma}
      \end{eqnarray}
      On the other hand, if the confidence interval is increased up to 1.65 sigmas, i.e. $90\%$ C.L., then the potential effects
      in $\Delta \Gamma_d$ become
      \begin{eqnarray}
      -5.97<\Delta \Gamma^{\rm Exp}_d/\Delta \Gamma^{\rm SM}_d<4.67.
      \label{eq:enhacementsonDGammadonesixsigma}
      \end{eqnarray}
      The measured value of the dimuon asymmetry by the D0-collaboration \cite{Abazov:2010hv,Abazov:2010hj,Abazov:2011yk,Abazov:2013uma}
      seems to be in conflict with the current experimental bounds on $a_{sl}^d$ and $a_{sl}^s$, see e.g. the discussion in \cite{Lenz:2014nka}.
      An enhanced value of $\Delta \Gamma_d$  could solve this experimental discrepancy
      \cite{Borissov:2013wwa}, at the expense of introducing new physics in $\Delta \Gamma_d$ and potentially also in $a_{sl}^s$ and $a_{sl}^d$.
      If all BSM effects in the dimuon asymmetry are due to  $\Delta \Gamma_d$, then an enhancement factor of 6 with respect to its SM value is 
      required. On the other hand, if there are also BSM contributions in  $a_{sl}^s$ and $a_{sl}^d$,  then the BSM enhancement factor in  
      $\Delta \Gamma_d$ can be smaller.
\item The impact of these new effects on the determination of the  CKM angle $\gamma$. Within the SM, this quantity can be extracted with negligible
      uncertainties from $B \to DK$ tree-level decays \cite{Bigi:1981qs,Gronau:1990ra,Gronau:1991dp,Atwood:1996ci,Atwood:2000ck,Giri:2003ty}.
      This quantity is currently extensively tested by experiments, see e.g.\cite{Kenzie:2018oob,Amhis:2019ckw}
      and future measurements will dramatically improve its precision to the one degree level \cite{Bediaga:2018lhg}.
      This observable is particular interesting since direct measurements, e.g. LHCb \cite{Kenzie:2018oob}, seem to be larger than bounds
      from B-mixing \cite{King:2019rvk}\footnote{Similar observations were made in e.g.\cite{Blanke:2016bhf,Blanke:2018cya}.}.
      \begin{eqnarray}
        \gamma^{\rm LHCb}\hspace{0.5cm}  & = & \left( 74.0^{+5.0}_{-5.8} \right)^\circ \, ,
        \\
         \gamma^{\rm B-mixing} & \leq & 66.9^\circ \, .
       \end{eqnarray}
\end{enumerate}
Therefore, in Sections~\ref{sec:buudfit} to \ref{sec:bccdfit} we combine our bounds 
from the  $b\rightarrow u \bar{u} d$, $b\rightarrow c \bar{u} d$ and $b\rightarrow c \bar{c} d$ transitions,
and evaluate the corresponding potential enhancement in $\Delta \Gamma_d$.
We do not present the allowed regions for the NP contributions related to the channel 
$b\rightarrow u\bar{c} d$, since the bounds  are expected to be rather weak considering that our only bound will arise from $a^d_{sl}$.
In Section~\ref{sec:Universal_fit} we report the maximal bounds on $\Delta C_1(M_W)$ and $\Delta C_2(M_W)$, assuming universal
BSM contributions to all different quark level decays. Hence, we combine all our possible bounds regardless of the quark level 
transition and asses the implications on the measurement of the CKM angle $\gamma$.  The target of this part of analysis, is to update 
the investigations reported in \cite{Brod:2014bfa} in the light of a far more detailed study of BSM effects in non-leptonic tree-level
decays. In particular we account here for uncertainties neglected in the former study and we also make a very careful choice of reliable
observables.
%
%
%
%
%
%
%
%
%
%
%
%
%
%
%
%
%
%
%
%
%
%
%
%
%
%
%
%
\subsection{$\chi^2$-fit for the $b\rightarrow u\bar{u}d$ channel and bounds on $\Delta \Gamma_d$}
\label{sec:buudfit}
\begin{figure}
\centering
\includegraphics[height=5cm]{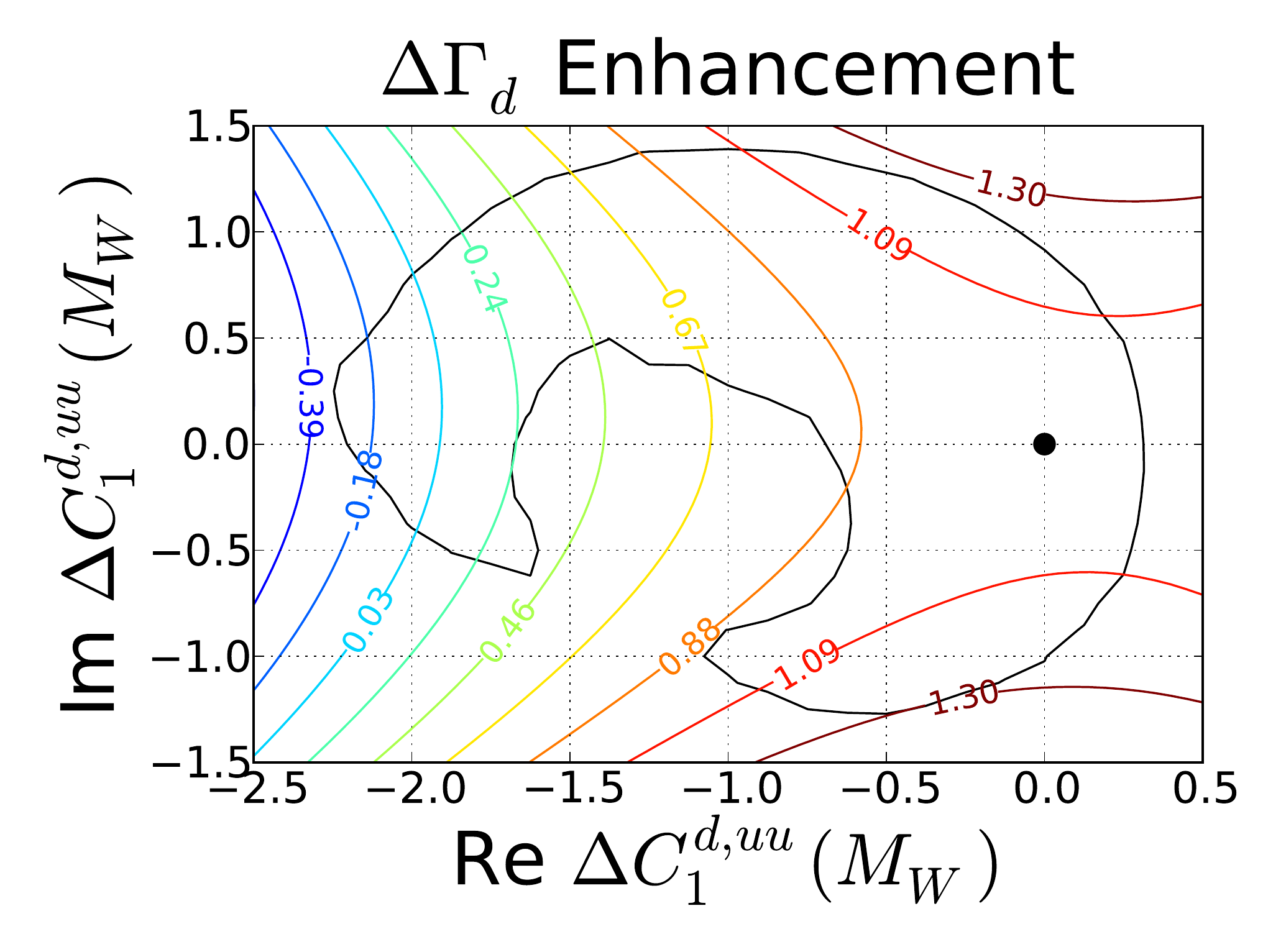}
\includegraphics[height=5cm]{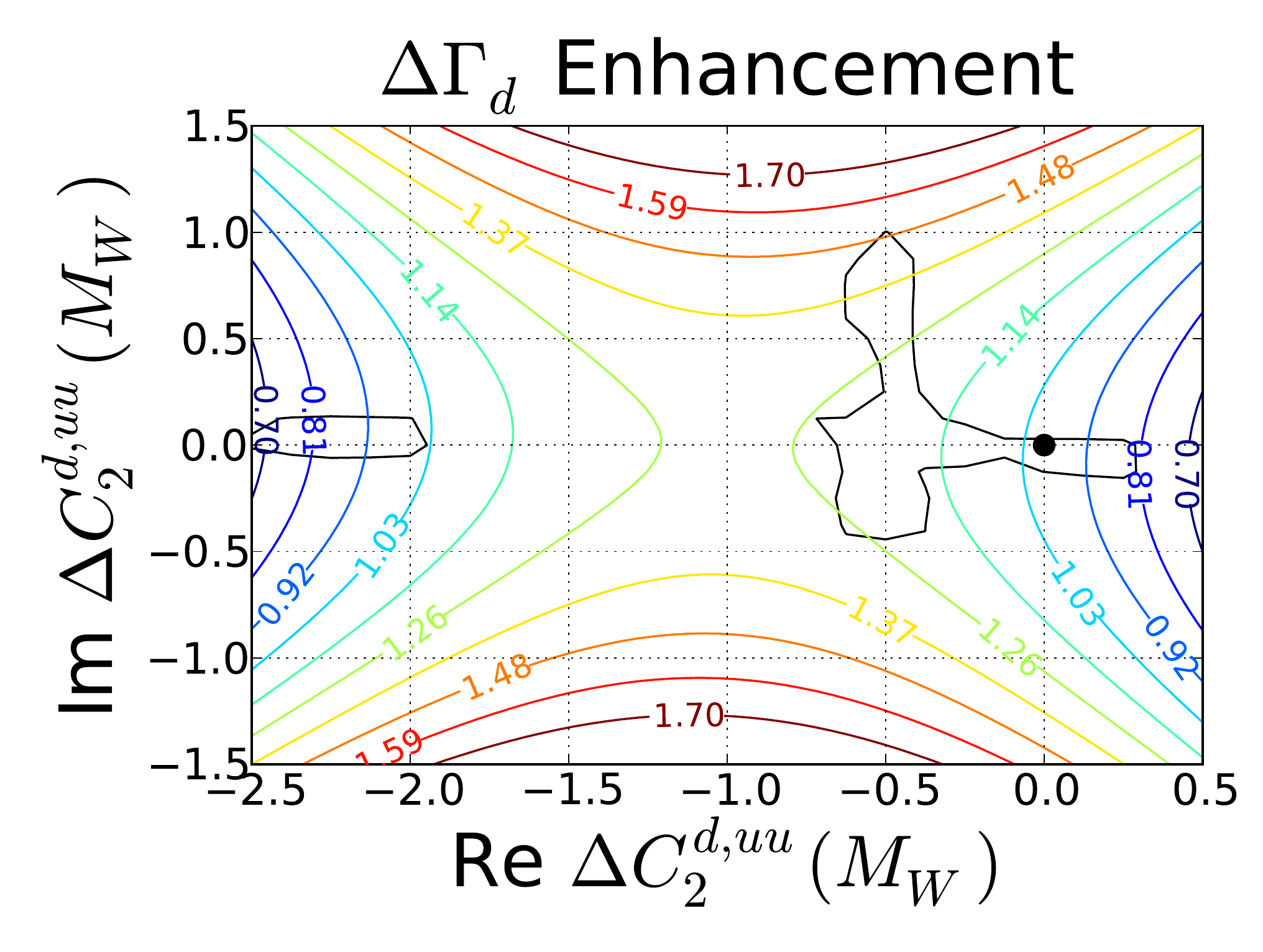}
\caption{Global $\chi^2$-fit including observables constraining the inclusive transition
$b\rightarrow u \bar{u}d$. The $90\%$ C.L. allowed regions correspond to the areas contained within the black contours. 
The colored curves indicate the possible enhancements on $\Delta \Gamma_d$ with respect to the SM value. The black dot corresponds to the SM result.
	}
\label{fig:Global_fit_uu}
\end{figure}

We perform a combined $\chi^2$-fit including  $R_{\pi\pi}$, $S_{\pi\pi}$, $S_{\rho\pi}$, $R_{\rho\rho}$ and $a^d_{sl}$ 
with the aim of constraining $\Delta C^{d, uu}_1(M_W)$ and $\Delta C^{d, uu}_2(M_W)$. The resulting regions are shown in 
Fig.~\ref{fig:Global_fit_uu}.
 $\Delta C^{d, uu}_2(M_W)$ is considerably stronger constrained than $\Delta C^{d, uu}_1(M_W)$, but sizeable deviations can still not be excluded.
Due to the irregularity of the regions for $\Delta C^{d, uu}_1(M_W)$ and $\Delta C^{d, uu}_2(M_W)$, expressing 
the possible NP values for the tree level contributions in terms of simple inequalities is not possible. Instead,
we limit ourselves to quote the minimum and maximum bounds for the real and the imaginary components of our NP
regions.
For $\Delta C^{d, uu}_1(M_W)$ we have
\begin{eqnarray}
{\rm Re}~\Bigl[\Delta C^{d, uu}_1(M_W)\Bigl]\Biggl|_{\rm min}=-2.23,&&
{\rm Im}~\Bigl[\Delta C^{d, uu}_1(M_W)\Bigl]\Biggl|_{\rm min}=-1.27,\nonumber\\
{\rm Re}~\Bigl[\Delta C^{d, uu}_1(M_W)\Bigl]\Biggl|_{\rm max}=~~0.32,&&
{\rm Im}~\Bigl[\Delta C^{d, uu}_1(M_W)\Bigl]\Biggl|_{\rm max}=~1.40.\nonumber\\
\end{eqnarray}
On the other hand for $\Delta C^{d, uu}_2(M_W)$  we get
\begin{eqnarray}
{\rm Re}~\Bigl[\Delta C^{d, uu}_2(M_W)\Bigl]\Biggl|_{\rm min}=-2.5,&&
{\rm Im}~\Bigl[\Delta C^{d, uu}_2(M_W)\Bigl]\Biggl|_{\rm min}=-0.44,\nonumber\\
{\rm Re}~\Bigl[\Delta C^{d, uu}_2(M_W)\Bigl]\Biggl|_{\rm max}=~~0.28,&&
{\rm Im}~\Bigl[\Delta C^{d, uu}_2(M_W)\Bigl]\Biggl|_{\rm max}=~1.00.\nonumber\\
\end{eqnarray}
We have also included the contour lines showing the potential enhancement of the observable
$\Delta \Gamma_d$. Accounting for the uncertainties in theory and experiment we find the following $90\%$ C.L. intervals for
$\Delta \Gamma_d$ due to NP at tree level:
\begin{eqnarray}
\hbox{for $\Delta C^{d, uu}_1(M_W)$:}&&-0.39<\Delta \Gamma_d/\Delta \Gamma^{\rm SM}_d<1.30,\nonumber\\ 
\hbox{for $\Delta C^{d, uu}_2(M_W)$:}&&~~0.70<\Delta \Gamma_d/\Delta \Gamma^{\rm SM}_d<1.48.	
\end{eqnarray}
Thus only moderate enhancements of $\Delta \Gamma_d$ seem to be possible, while a reduction to up to $-39 \%$ of its SM
values is still possible. This scenario could thus not be a solution for the dimuon asymmetry.
%
%
%
%
%
%
%
%
%
%
%
%
%
%
%
%
%
%
%
%
%
%
%
\subsection{$\chi^2$-fit for the $b\rightarrow c\bar{u}d$ channel and bounds on $\Delta \Gamma_d$}
\label{sec:bcudfit}
\begin{figure}
\centering
\includegraphics[height=5cm]{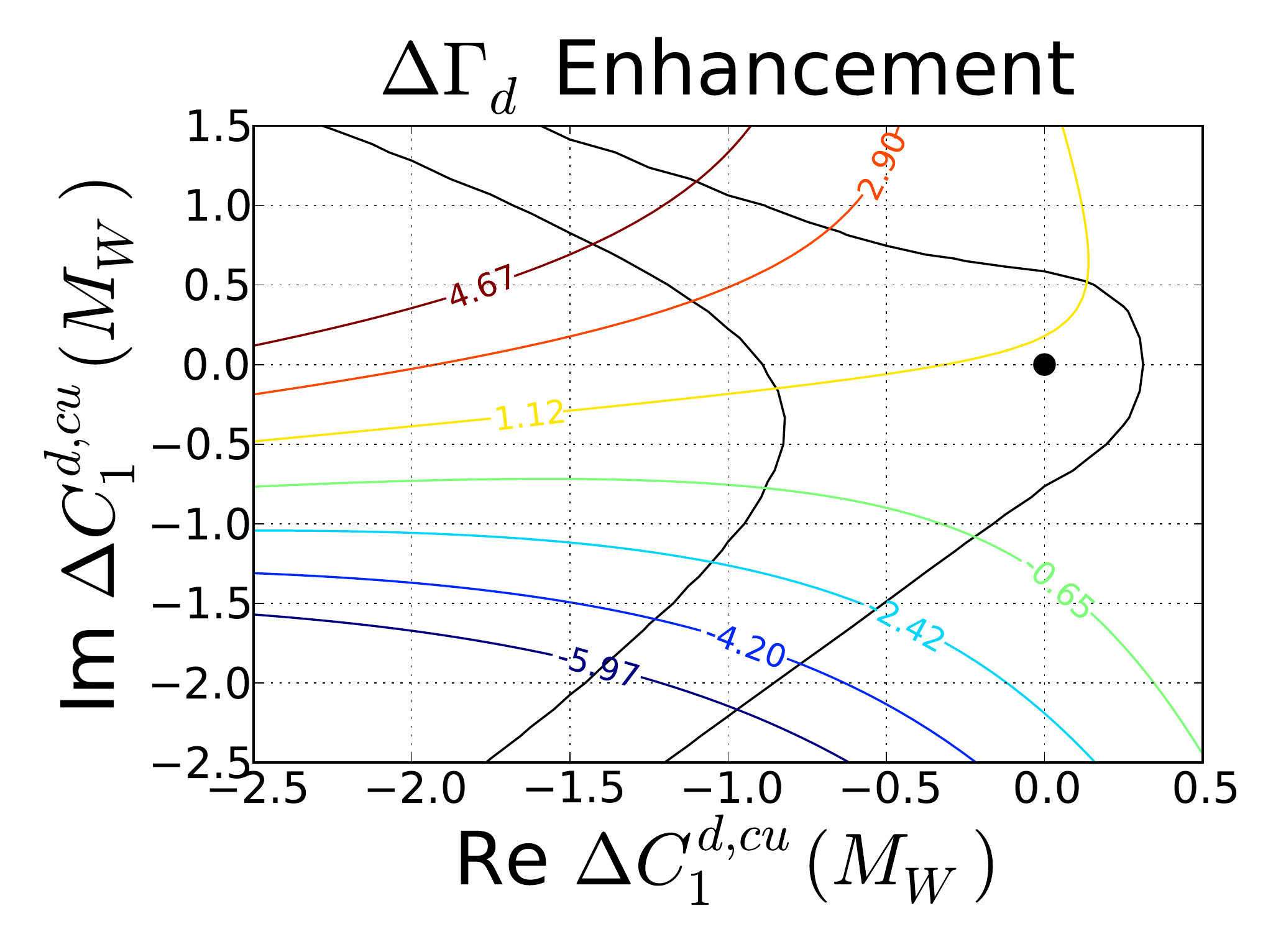}
\includegraphics[height=5cm]{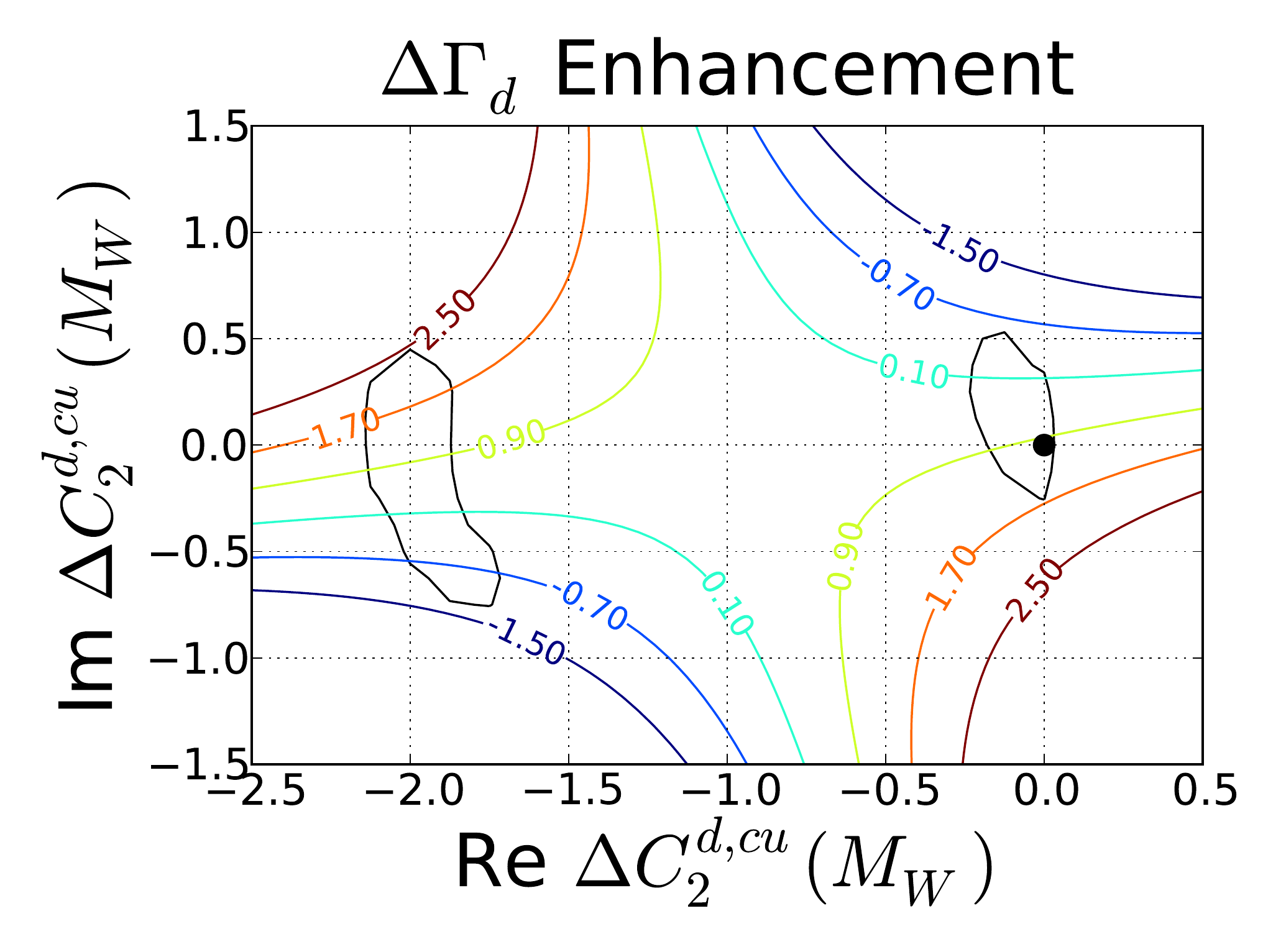}
\caption{Global $\chi^2$-fit including observables constraining the inclusive transition
$b\rightarrow c \bar{u}d$. The $90\%$ C.L. allowed regions correspond to the areas contained within the black contours. The colored
curves indicate the possible enhancements on $\Delta \Gamma_d$ with respect to the SM value. The black dot corresponds to the SM result.}
\label{fig:Global_fit_cu}
\end{figure}

To establish constraints on $\Delta C^{d, cu}_{1}(M_W)$ and $\Delta C^{d, cu}_{2}(M_W)$ we combine $R_{D^{*}\pi}$ together with
$a^d_{sl}$. Our results are presented in Fig.~\ref{fig:Global_fit_cu}.
At the $90\%$ C.L. we find the possibility of huge enhancements/reductions of $\Delta \Gamma_d$:
\begin{eqnarray}
\hbox{for $\Delta C^{d, cu}_1(M_W)$:}&&~   -5.97 <\Delta \Gamma_d/\Delta \Gamma^{\rm SM}_d<4.67,\nonumber\\
\hbox{for $\Delta C^{d, cu}_2(M_W)$:}&&~~-1.5<\Delta \Gamma_d/\Delta \Gamma^{\rm SM}_d<2.50.	
\label{eq:CdcuDeltaGammad}
\end{eqnarray}
Based on the bounds shown in Eq.~(\ref{eq:CdcuDeltaGammad}), we find that this scenario could  solve the dimuon asymmetry.
Since the experimental bounds for $\Delta \Gamma_d$ are saturated in the case of $\Delta C^{d, cu}_1(M_W)$ in Eq.~(\ref{eq:CdcuDeltaGammad}),
it turns out that $\Delta \Gamma_d$ acts as a constraint in itself.
Using this additional information we establish 
the following bounds for $\Delta C^{d, cu}_1(M_W)$ 
\begin{eqnarray}
{\rm Re}~\Bigl[\Delta C^{d, cu}_1(M_W)\Bigl]\Biggl|_{\rm min}=-1.40,&&
{\rm Im}~\Bigl[\Delta C^{d, cu}_1(M_W)\Bigl]\Biggl|_{\rm min}=-2.17,\nonumber\\
{\rm Re}~\Bigl[\Delta C^{d, cu}_1(M_W)\Bigl]\Biggl|_{\rm max}=~~0.32,&&
{\rm Im}~\Bigl[\Delta C^{d, cu}_1(M_W)\Bigl]\Biggl|_{\rm max}=~1.15.\nonumber\\
\end{eqnarray}
The corresponding bounds for $\Delta C^{d, cu}_2(M_W)$  read
\begin{eqnarray}
{\rm Re}~\Bigl[\Delta C^{d, cu}_2(M_W)\Bigl]\Biggl|_{\rm min}=-2.14,&&
{\rm Im}~\Bigl[\Delta C^{d, cu}_2(M_W)\Bigl]\Biggl|_{\rm min}=-0.75,\nonumber\\
{\rm Re}~\Bigl[\Delta C^{d, cu}_2(M_W)\Bigl]\Biggl|_{\rm max}=~~0.04,&&
{\rm Im}~\Bigl[\Delta C^{d, cu}_2(M_W)\Bigl]\Biggl|_{\rm max}=~0.53.\nonumber\\
\end{eqnarray}
%
%
%
%
%
%
%
%
%
%
%
%
%
%
%
%
%
%
%
%
%
%
%
%
%
%
%
%
%
%
%
%
%
%
\subsection{$\chi^2$-fit for the $b\rightarrow c\bar{c}d$ channel and bounds on $\Delta \Gamma_d$}
\label{sec:bccdfit}
\begin{figure}
\centering
\includegraphics[height=5cm]{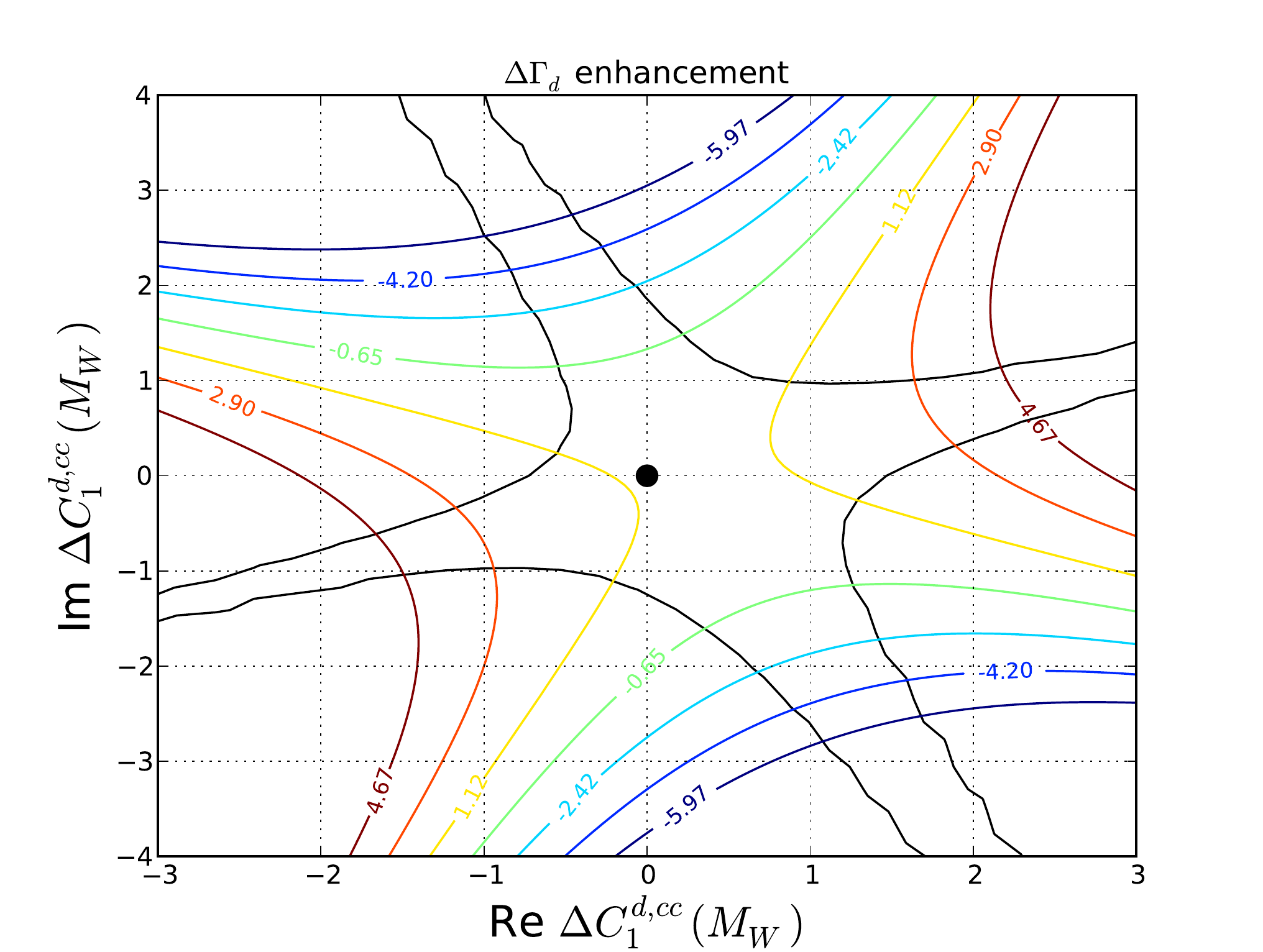}
\includegraphics[height=5cm]{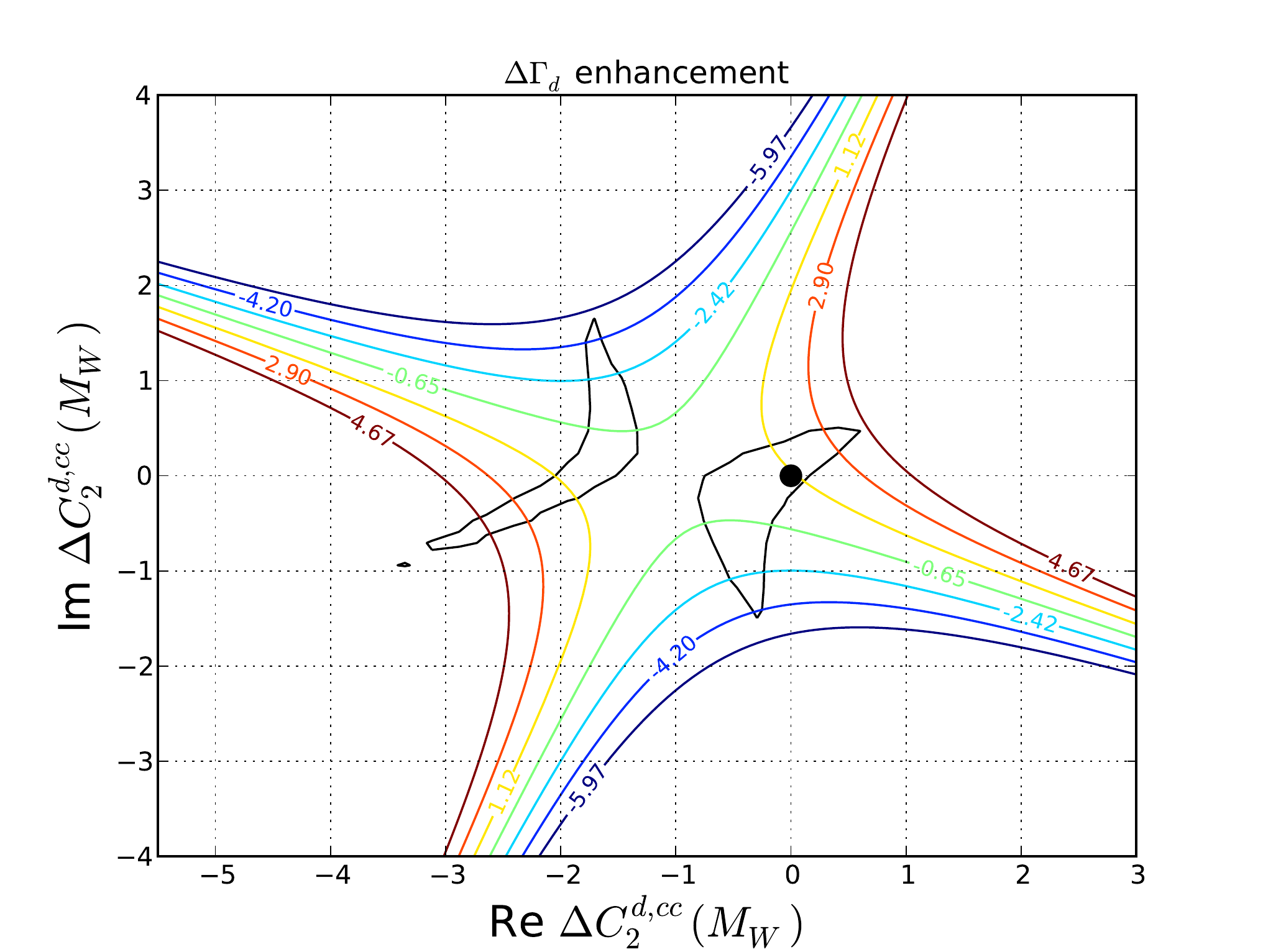}
\caption{Global $\chi^2$-fit including observables constraining the inclusive transition
$b\rightarrow c \bar{c}d$. The $90\%$ C.L. allowed regions correspond to the areas contained within the black contours. The colored
curves indicate the possible enhancements on $\Delta \Gamma_d$ with respect to the SM value.
The black dot corresponds to the SM result.}
\label{fig:Global_fit_cc}
\end{figure}

Next we perform a $\chi^2$-fit including $\mathcal{B}r(B\rightarrow X_d \gamma)$, $a^d_{sl}$ and $\sin(2\beta_d)$. These
observables give strong constraints for $\Delta C^{d, cc}_2(M_W)$ (see Fig.~\ref{fig:Global_fit_cc}),
which turn out to saturate the current experimental bounds on
$\Delta \Gamma_d$.
At the $90\%$ C.L. we find
\begin{eqnarray}
\hbox{for $\Delta C^{d, cc}_1(M_W)$ and $\Delta C^{d, cc}_2(M_W)$:}&&-5.97<\Delta \Gamma_d/\Delta \Gamma^{\rm SM}_d<4.67.\nonumber\\	
\label{eq:CdccDeltaGammad}
\end{eqnarray}
We find again that this scenario could  solve the tension between theory and experiment found in the measurement of the dimuon asymmetry.
Considering the results shown in Fig.~\ref{fig:Global_fit_cc} we see that $\Delta \Gamma_d$ is indeed a powerful constraint for $\Delta C^{d, cc}_1(M_W)$ 
and $\Delta C^{d, cc}_2(M_W)$, which together with $\mathcal{B}r(B\rightarrow X_d \gamma)$, $a^d_{sl}$ and $\sin(2\beta_d)$ 
defines the following limits

\begin{eqnarray}
{\rm Re}~\Bigl[\Delta C^{d, cc}_1(M_W)\Bigl]\Biggl|_{\rm min}=-1.66,&&
{\rm Im}~\Bigl[\Delta C^{d, cc}_1(M_W)\Bigl]\Biggl|_{\rm min}=-2.80,\nonumber\\
{\rm Re}~\Bigl[\Delta C^{d, cc}_1(M_W)\Bigl]\Biggl|_{\rm max}=~~2.36,&&
{\rm Im}~\Bigl[\Delta C^{d, cc}_1(M_W)\Bigl]\Biggl|_{\rm max}=~2.74,\nonumber\\
\end{eqnarray}

and

\begin{eqnarray}
{\rm Re}~\Bigl[\Delta C^{d, cc}_2(M_W)\Bigl]\Biggl|_{\rm min}=-2.70,&&
{\rm Im}~\Bigl[\Delta C^{d, cc}_2(M_W)\Bigl]\Biggl|_{\rm min}=-1.46,\nonumber\\
{\rm Re}~\Bigl[\Delta C^{d, cc}_2(M_W)\Bigl]\Biggl|_{\rm max}=~~0.58,&&
{\rm Im}~\Bigl[\Delta C^{d, cc}_2(M_W)\Bigl]\Biggl|_{\rm max}=~1.65.\nonumber\\
\end{eqnarray}
As can be seen on the l.h.s. of Fig.~\ref{fig:Global_fit_cc} 
$\Delta C^{d, cc}_1(M_W)$ is only weakly constrained by the semi-leptonic CP asymmetries,
here additional information stemming from $\Delta \Gamma_d$ will be important to shrink the allowed regions.
%
%
%
%
%
%
%
%
%
%
%
%
%
%
%
%
%
%
%
%
%
%
%
%
%
%
%
%
%
%
%
%
%
%
\subsection{Universal fit on $\Delta C_1(M_W)$ and $\Delta C_2(M_W)$}
\label{sec:Universal_fit}
\begin{figure}
\centering
\includegraphics[height=5cm]{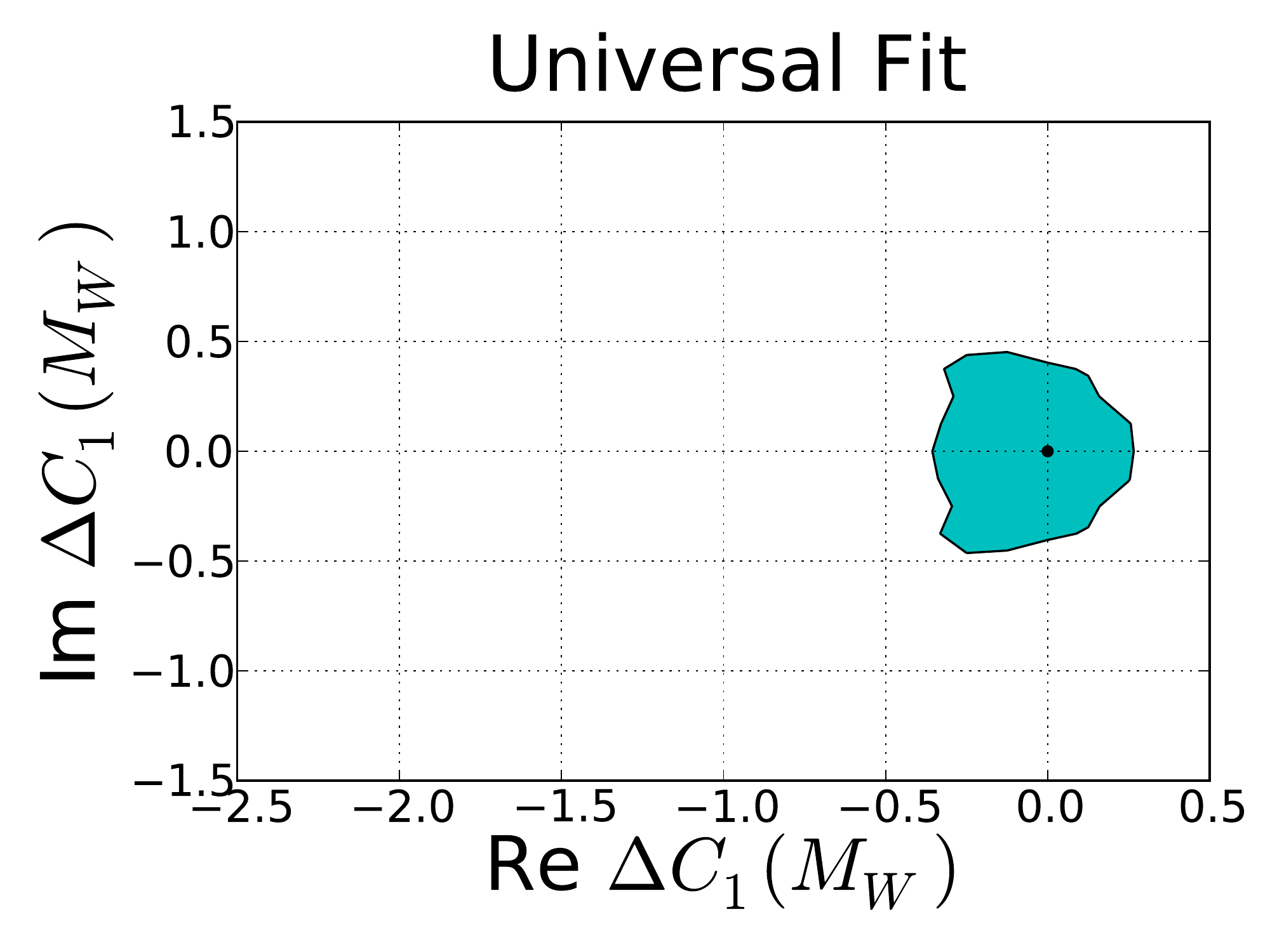}
\includegraphics[height=5cm]{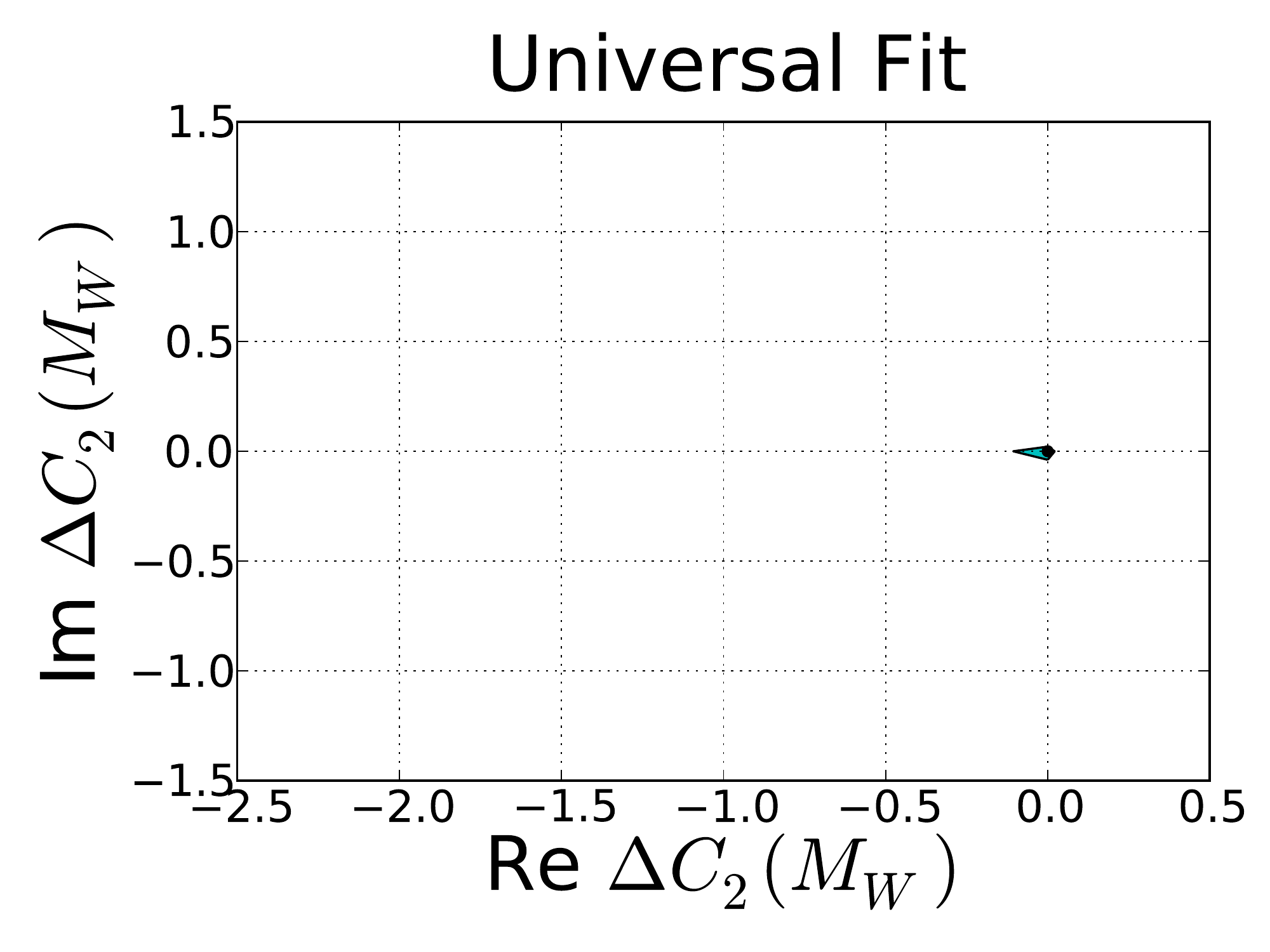}
\caption{
Potential regions for the NP contributions $\Delta C_{1}(M_W)$  and $\Delta C_{2}(M_W)$ 
allowed by the observables used in our analysis at $90\%$ C.L. assuming universal NP contributions. The black dot corresponds to the SM result.}
\label{fig:Global_fit}
\end{figure}

\begin{figure}
\centering
\includegraphics[height=5cm]{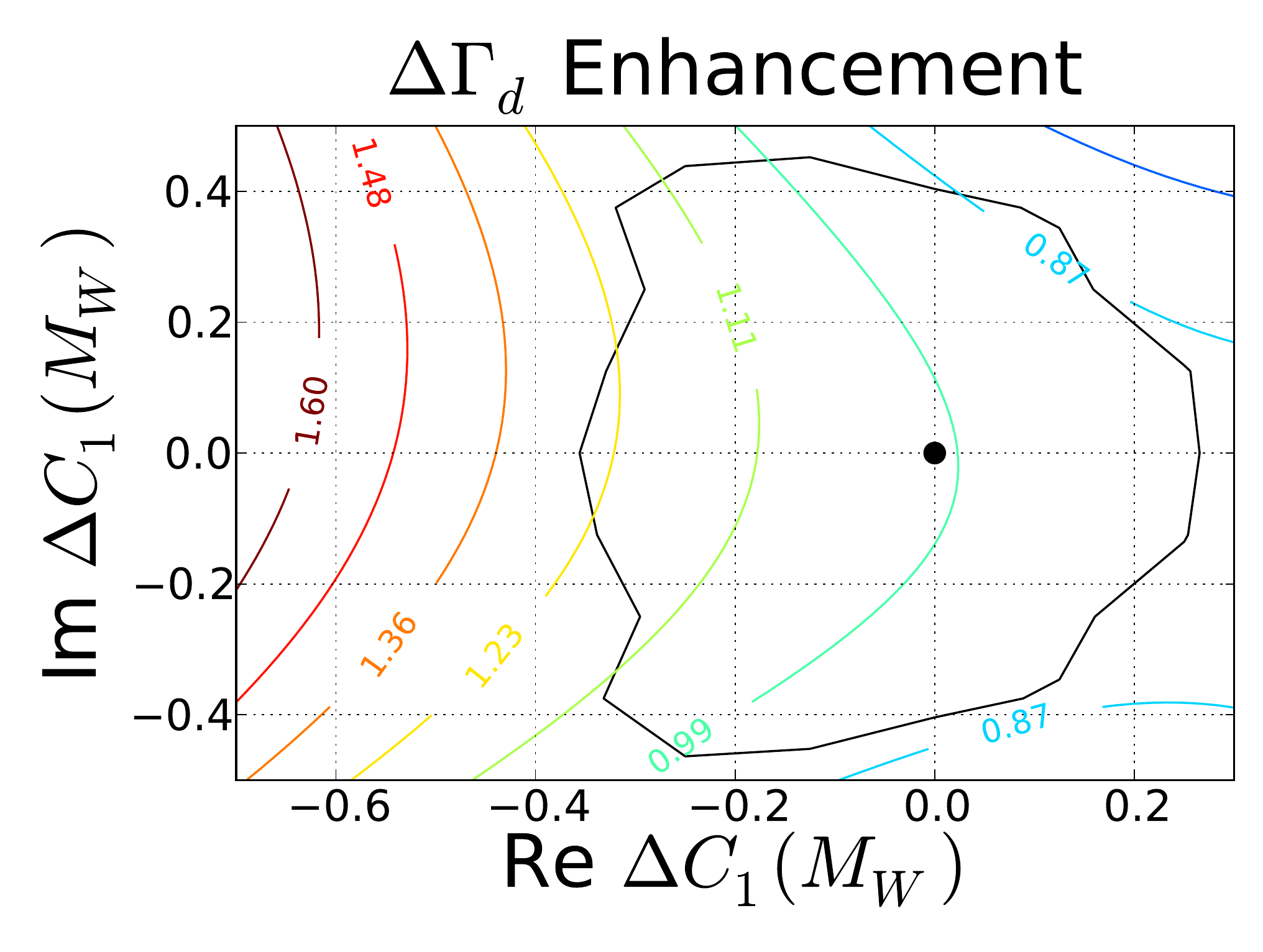}
\caption{Enhancements on $\Delta \Gamma_d$ when assuming universal NP effects in $C_1(M_W)$. The black dot corresponds to the SM result.}
\label{fig:Global_fit_Delta_Gammad}
\end{figure}

In this section we work under the assumptions
\begin{eqnarray}
\Delta C^{s, ab}_{1}(M_W)=\Delta C^{d, ab}_{1}(M_W)=\Delta C_{1}(M_W)\\
\Delta C^{s, ab}_{2}(M_W)=\Delta C^{d, ab}_{2}(M_W)=\Delta C_{2}(M_W)
\end{eqnarray}
for $a=u,~d$ and $b= u,~d$. This procedure allows us to obtain the maximal constraints for our NP contributions.  Making a combined 
$\chi^2$-fit is time and resource consuming, consequently we select the set of observables that give the strongest possible bounds.
For $\Delta C_1(M_W)$ this includes: $R_{D^{*}\pi}$, $S_{\rho\pi}$, $\Delta \Gamma_s$, $\mathcal{B}r(\bar{B} \rightarrow  X_s \gamma)$ and $a^d_{sl}$
and for $\Delta C_2(M_W)$ we use: $R_{D^{*}\pi}$, $R_{\pi\pi}$, $\Delta \Gamma_s$, $S_{J/\psi \phi}$ and $\tau_{B_s}/\tau_{B_d}$.
We show in Fig.~\ref{fig:Global_fit}
our resulting regions
from which we extract
\begin{eqnarray}
{\rm Re}~\Bigl[\Delta C_1(M_W)\Bigl]\Biggl|_{\rm min}=-0.36,&&
{\rm Im}~\Bigl[\Delta C_1(M_W)\Bigl]\Biggl|_{\rm min}=-0.47,\nonumber\\
{\rm Re}~\Bigl[\Delta C_1(M_W)\Bigl]\Biggl|_{\rm max}=~~0.26,&&
{\rm Im}~\Bigl[\Delta C_1(M_W)\Bigl]\Biggl|_{\rm max}=~~~0.45,\nonumber\\
\label{eq:dC1}
\end{eqnarray}
and
\begin{eqnarray}
{\rm Re}~\Bigl[\Delta C_2(M_W)\Bigl]\Biggl|_{\rm min}=-0.11,&&
{\rm Im}~\Bigl[\Delta C_2(M_W)\Bigl]\Biggl|_{\rm min}=-0.04,\nonumber\\
{\rm Re}~\Bigl[\Delta C_2(M_W)\Bigl]\Biggl|_{\rm max}=~~0.02,&&
{\rm Im}~\Bigl[\Delta C_2(M_W)\Bigl]\Biggl|_{\rm max}=0.02.\nonumber\\
\label{eq:dC2}
\end{eqnarray}
We can see from Eqs.~(\ref{eq:dC1}) and ~(\ref{eq:dC2}) how severely constrained is $\Delta C_2(M_W)$ allowing deviations with respect to the
SM point of a few percent at most. This behaviour is clearly in contrast with the results obtained for $\Delta C_1(M_W)$,
where effects of almost up to $\pm 0.5$ are still possible.
For completeness we present  the implications of universal 
NP in $\Delta C_1(M_W)$ on $\Delta \Gamma_d$ in Fig.~\ref{fig:Global_fit_Delta_Gammad}. We find that
at $90\%$ C.L. only $\mathcal{O}(20\%)$ deviations on $\Delta \Gamma_d$ with respect to its SM value can be induced, which is in a similar
ballpark as the SM uncertainties of $\Delta \Gamma_d$ and can clearly  not explain the D0 measurement of the dimuon asymmetry.
%
%
%
%
%
%
%
%
%
%
%
%
%
%
%
%
%
%
%
%
%
%
%
%
%
%
%
%
%
%
%
%
%
\subsection{NP in non-leptonic tree-level decays and its interplay with the CKM angle $\gamma$}
\label{sec:CKMgamma}
\begin{figure}
\centering
\includegraphics[height=5.5cm]{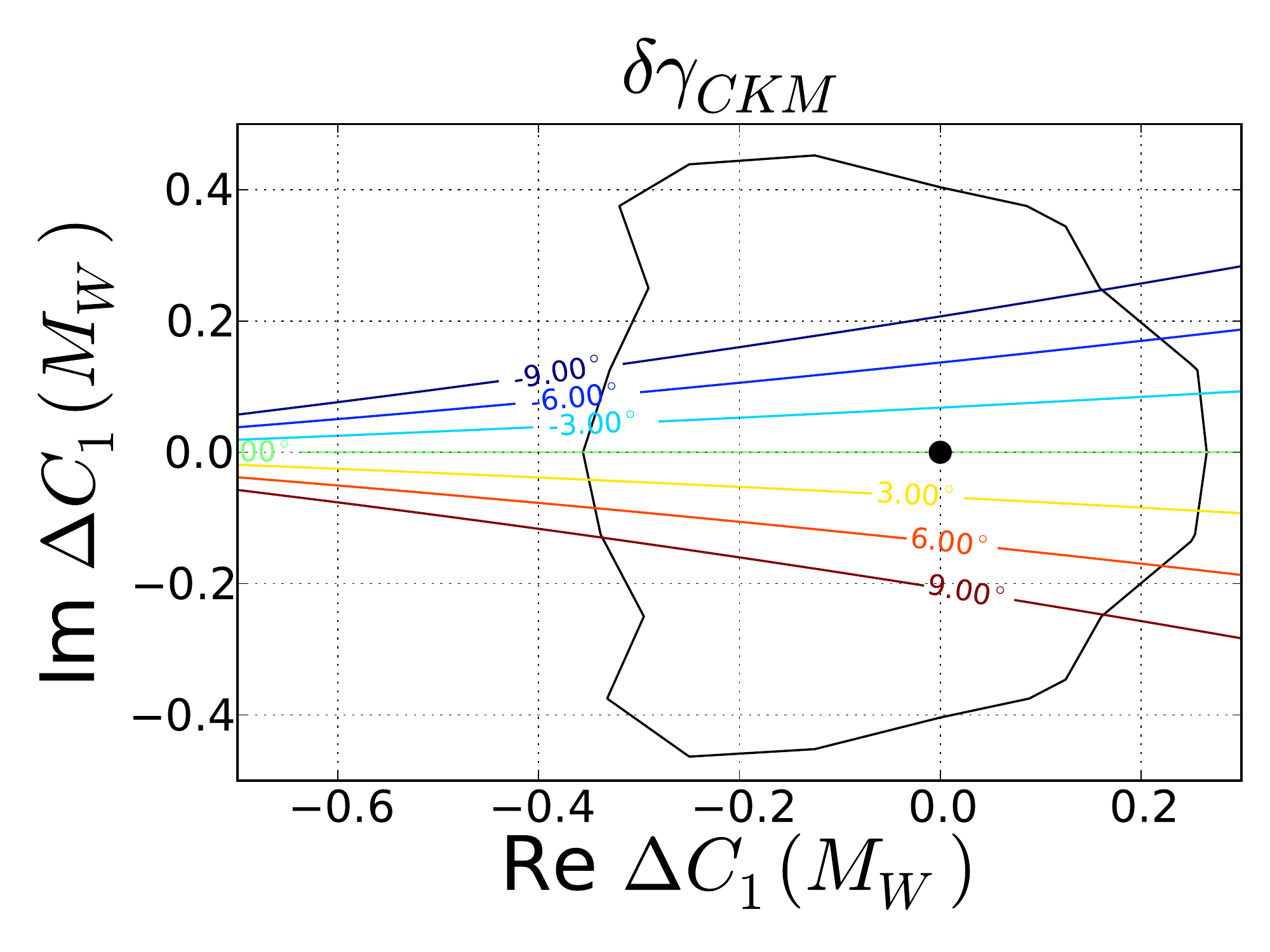}
\includegraphics[height=5.5cm]{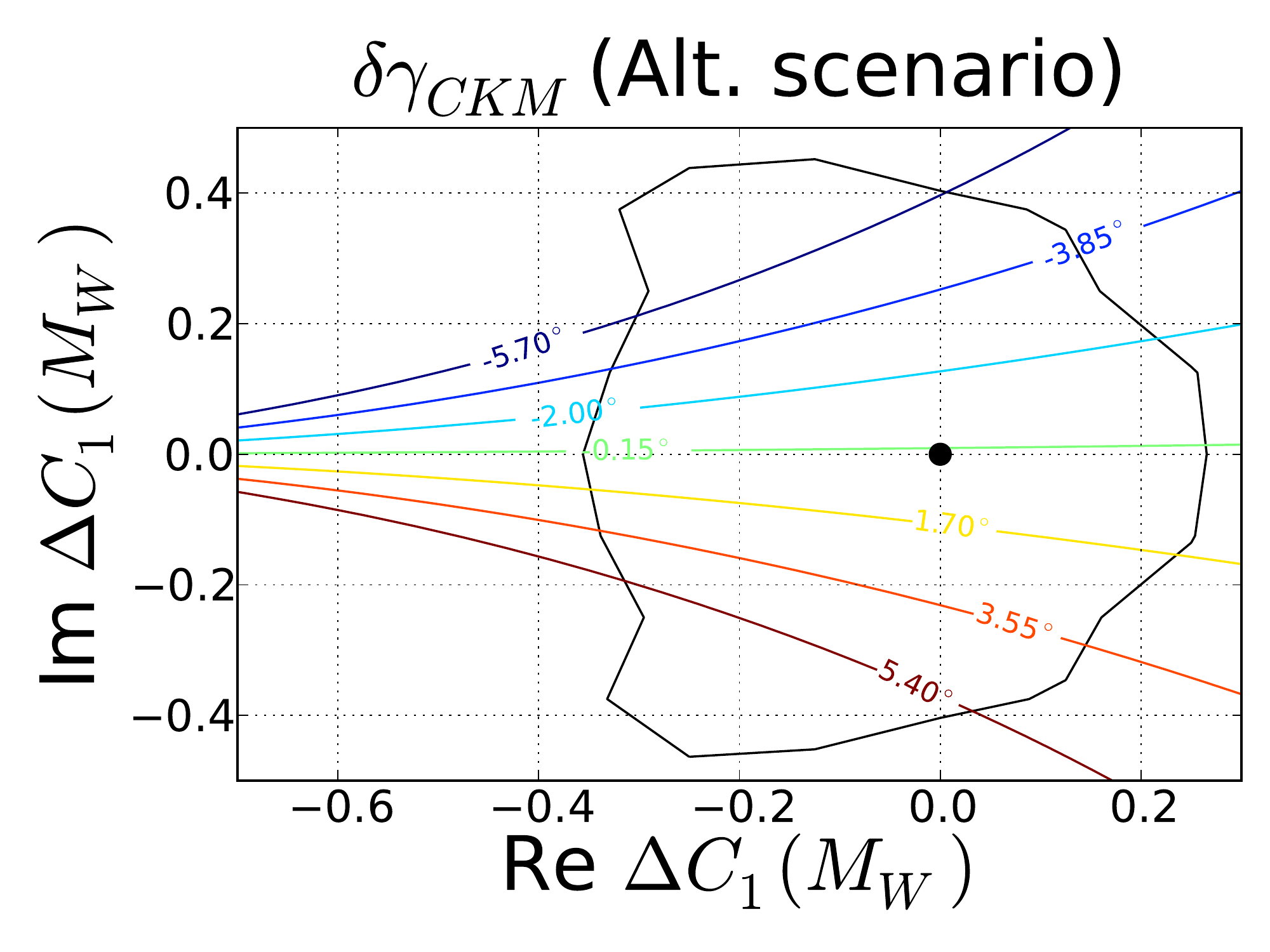}
	\caption{Possible deviations on the CKM phase $\gamma$ due to NP at tree level in $C_1(M_W)$ assuming $r_A=0.4$ (left) and $r_A=0.8$ 
(right). In both cases we have considered $r'_{A}=1$.
  The black dot corresponds to the SM result.}
\label{fig:CKM_gamma}
\end{figure}

As is well known \cite{Bigi:1981qs,Gronau:1990ra,Gronau:1991dp,Atwood:1996ci,Atwood:2000ck,Giri:2003ty} the CKM phase $\gamma$ can
be determined from the interference of the transition amplitudes associated
with the quark tree level decays $b\rightarrow c\bar{u}s$ and $b\rightarrow u\bar{c}s$ with negligible theory uncertainty within the SM
\cite{Brod:2013sga}\footnote{Due to the absence of penguins and the fact that the
relevant hadronic matrix elements  cancel, the extraction of CKM $\gamma$ is extremely clean. 
The irreducible theoretical uncertainty is due to higher-order electroweak corrections and has
been found to be negligible. For instance, when the modes
$B \to D K$ are used the correction effect is $|\delta \gamma/\gamma| <\mathcal{O}(10^{-7})$  ~\cite{Brod:2013sga}.
On the other hand, if CKM  $\gamma$ is obtained using 
$B \to D \pi$ decays instead, then $|\delta \gamma/\gamma| <\mathcal{O}(10^{-4})$  ~\cite{Brod:2014qwa}.}.
At the exclusive level, this can be done
with the decay channels $B^-\rightarrow D^0 K^-$ and $B^-\rightarrow \bar{D}^0K^-$.
The ratio of the two corresponding decay amplitudes
can be written as
\begin{equation}
\label{eq:gamma_theoretical}
r_B e^{i (\delta_B - \gamma)} = \frac{\mathcal{A} (B^- \to \bar{D}^0 K^-)}
                                     {\mathcal{A} (B^- \to       D^0 K^-)} \;,
\end{equation}
where the $r_B$ stands for the ratio of the modulus of the relevant amplitudes. The resulting phase has a strong component,
denoted as $\delta_B$, and a weak one, which is precisely CKM $\gamma$.
New effects in $C_1$ and $C_2$ can lead to huge shifts in $\gamma$. To
study this possibility we follow \cite{Brod:2014qwa} and assume
universal NP in $C_1$ and $C_2$. Thus 
$\Delta C^{s, uc}_1= \Delta C^{s, cu}_1$ and 
$\Delta C^{s, uc}_2= \Delta C^{s, cu}_2$. Then, the left side of Eq.~(\ref{eq:gamma_theoretical}) will be modified according 
to~\cite{Brod:2014bfa}
\begin{eqnarray}
r_B e^{i (\delta_B - \gamma)} &\to&
r_B e^{i (\delta_B - \gamma)} \cdot
\Biggl[
\frac{C_2 + \Delta C_2 + r_{A'} ( C_1 + \Delta C_1)}{C_2 + r_{A'} C_1}\nonumber\\
&&~~~~~~~~~~~~~~~\cdot
\frac{C_2 + r_A C_1}{C_2 + \Delta C_2 + r_A (C_1 +\Delta C_1)}
\Biggl] \; ,
\label{eq:exact}
\end{eqnarray}
where
\vspace{-0.2cm}
\begin{equation}\label{eq:rA}
\begin{split}
r_{A'} = \frac{\langle \bar{D}^0 K^-| Q_1^{\bar{u}cs} | B^- \rangle}
              {\langle \bar{D}^0 K^-| Q_2^{\bar{u}cs} | B^- \rangle} \; ,
\quad 
r_A = \frac{\langle       D^0 K^-| Q_1^{\bar{c}us} | B^- \rangle}
              {\langle       D^0 K^-| Q_2^{\bar{c}us} | B^- \rangle} \; .
\end{split}
\end{equation}
The ratios of matrix elements in Eq.~(\ref{eq:rA}) have not been determined from first principles, to provide an estimation we use naive 
factorization arguments and colour counting to obtain \cite{Brod:2014bfa},\cite{Brod:2014qwa}
\begin{eqnarray}
r_{A} = 0.4,&& r_A-r_A'=-0.6.
\label{eq:initial_values_matrix_elements}
\end{eqnarray}
Eq.(\ref{eq:exact}) gives a particularly strong dependence of the shift in $\gamma$ on the imaginary part of $C_1$; approximately we get~\cite{Brod:2014bfa}
\begin{equation}
  \delta \gamma = \left(r_A - r_{A'}\right) \frac{\rm Im  \left[\Delta C_1 \right] }{C_2} \, .
\label{eq:deltaCKMgamma}	
\end{equation}
We are now ready to update the study 
presented in \cite{Brod:2014bfa} on the effects of NP in $C_1$ and $C_2$ on the precision for the determination
of the CKM angle $\gamma$, our results are presented graphically in Fig.~\ref{fig:CKM_gamma}.
On the left hand side of Fig.~\ref{fig:CKM_gamma} we can see how for 
the
values of $r_A$ and $r'_A$ shown in Eq.~(\ref{eq:initial_values_matrix_elements}), the current uncertainties in our knowledge of $C_1$ seem to indicate an uncertainty in the extraction of
the CKM angle $\gamma$ of considerably more
than $10^{\circ}$. This much higher than the current experimental uncertainty of
around five degrees~\cite{Kenzie:2018oob,Amhis:2019ckw}. Interestingly direct measurements give typically larger values
than the ones obtained by CKM fits \cite{Charles:2004jd,Bona:2006ah}
or extracted from B-mixing \cite{King:2019rvk}. Even more interestingly, future measurements will dramatically improve the  precision of $\gamma$ to the one degree level \cite{Bediaga:2018lhg}
and our BSM approach would offer a possibility of explaining large deviations in the extraction of the CKM angle $\gamma$.
 We would like, however, to add some words of cautions: for a quantitative reliable relation between the deviations of
$C_1$ and the shifts in the CKM angle $\gamma$, the
non-perturbative parameter $r_A$ and $r_{A'}$ have to be known more precisely.
The values proposed in Eq.~(\ref{eq:initial_values_matrix_elements}) correspond to an educated ansatz.
We can explore the effects of modifying these values
on CKM-$\gamma$. For instance, consider an alternative scenario where $r_{A}$ is twice the value presented in 
Eq.~(\ref{eq:initial_values_matrix_elements}), while  $r'_A$ remains fixed. This is equivalent
to assigning an uncertainty of $100\%$ to $r_A$
and taking the upper limit. The results for this new scenario are presented on the right hand side of  Fig.~\ref{fig:CKM_gamma}, where
the shifts $\delta \gamma_{CKM}$ have been halved with respect to those found on the left hand side of the same figure, however the absolute numerical
values of about $\pm 5^{\circ}$, still represent huge effects on the CKM angle $\gamma$ itself.

Here clearly more theoretical work leading to a more precise understanding of  $r_A$ and $r_{A'}$ is highly desirable.

%
%
%
%
%
%
%
%
%
%
%
%
%
%
%
%
%
%
%
%
%
%
%
%
%
%
%
%
%
\section{Future prospects}
\label{sec:future}
In this section we will present projections for observables, that are particularly promising to
further shrink the allowed regions of NP contributions to non-leptonic tree-level decays.
We have already studied the impact of BSM effects in non-leptonic tree-level decays on the
observables
$\Delta \Gamma_d$ and the CKM angle $\gamma$ in detail.
More precise experimental data on $\Delta \Gamma_d$ will immediately lead to stronger bounds
on the $\Delta B = 1$
Wilson coefficients, it could also exclude the possibility of solving the D0 dimuon asymmetry
with an enlarged value of $\Delta \Gamma_d$.
Alternatively, if the measured values of $\Delta \Gamma_d$ will not be SM-like, we could get
an intriguing hint for BSM physics.
In order to make use of the extreme sensitivity of the CKM angle $\gamma$ on an imaginary part
of $C_1$ more theory work is required to
make this relation quantitatively reliable. If this is available, then already the current
experimental uncertainty on $\gamma$  will exclude a large part
of the allowed region on $\Delta C_1$ - or it will indicate the existence of NP effects.
Below we will show projections for improved experimental values on the lifetime ratio
$\tau_{B_s}/\tau_{B_d}$ and the semi-leptonic CP asymmetries, as well
as commenting on consequences of our BSM approach to the recently observed flavour anomalies.
\subsection{ $\tau_{B_s}/\tau_{B_d}$}
As already explained, the lifetime ratio $\tau_{B_s}/\tau_{B_d}$ can pose very strong constraints on the
Wilson coefficients $C_1$ and $C_2$, if we e.g.
assume that BSM effects are only acting in the $b \to c \bar{c} s$ channel.
\begin{figure}
\centering
\includegraphics[height=5.0cm]{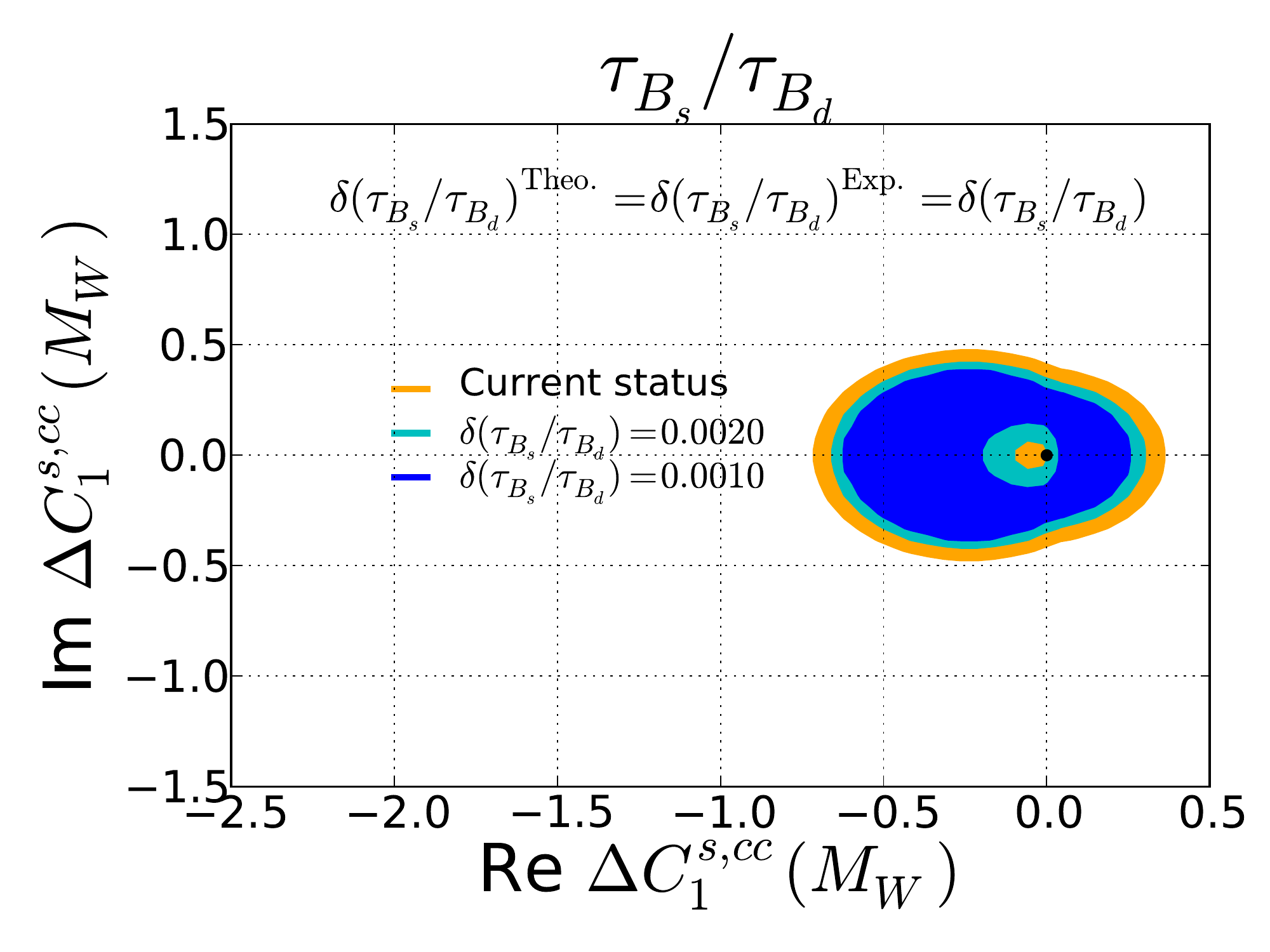}
\includegraphics[height=5.0cm]{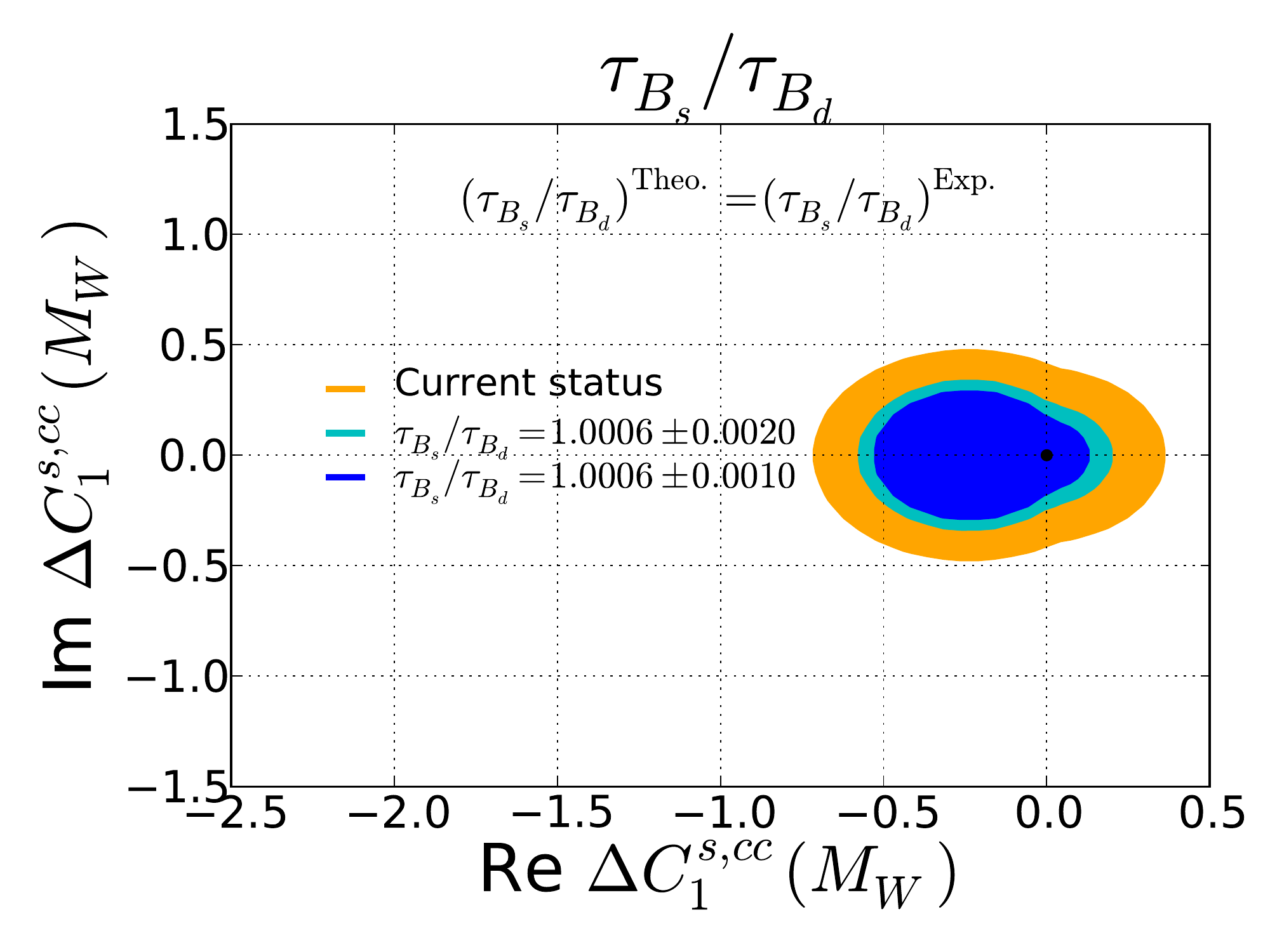}
\caption{Future scenarios concerning the behaviour of $\tau_{B_s}/\tau_{B_d}$. In the left panel the central experimental value of the lifetime
  ratio is assumed to remain unchanged in the future whereas the uncertainties will be reduced. In the right panel,
  the theoretical and experimental
	values for the lifetime ratio are supposed to become equal. The black dot corresponds to the SM result.}
\label{fig:Life_time_future}
\end{figure}
In Fig. \ref{fig:Life_time_future} we show future projections, assuming the errors will go down to 2 per mille or even one per mille.
On the l.h.s. of  Fig. \ref{fig:Life_time_future}
we assume that the current experiment value will stay - in this case a tension between the SM value and the experimental measurement will emerge.
On the r.h.s. of  Fig. \ref{fig:Life_time_future} we assume that the future experimental value perfectly agrees with the SM prediction. 

\subsection{Semi-leptonic CP asymmetries}
\begin{figure}
\centering
\includegraphics[height=5cm]{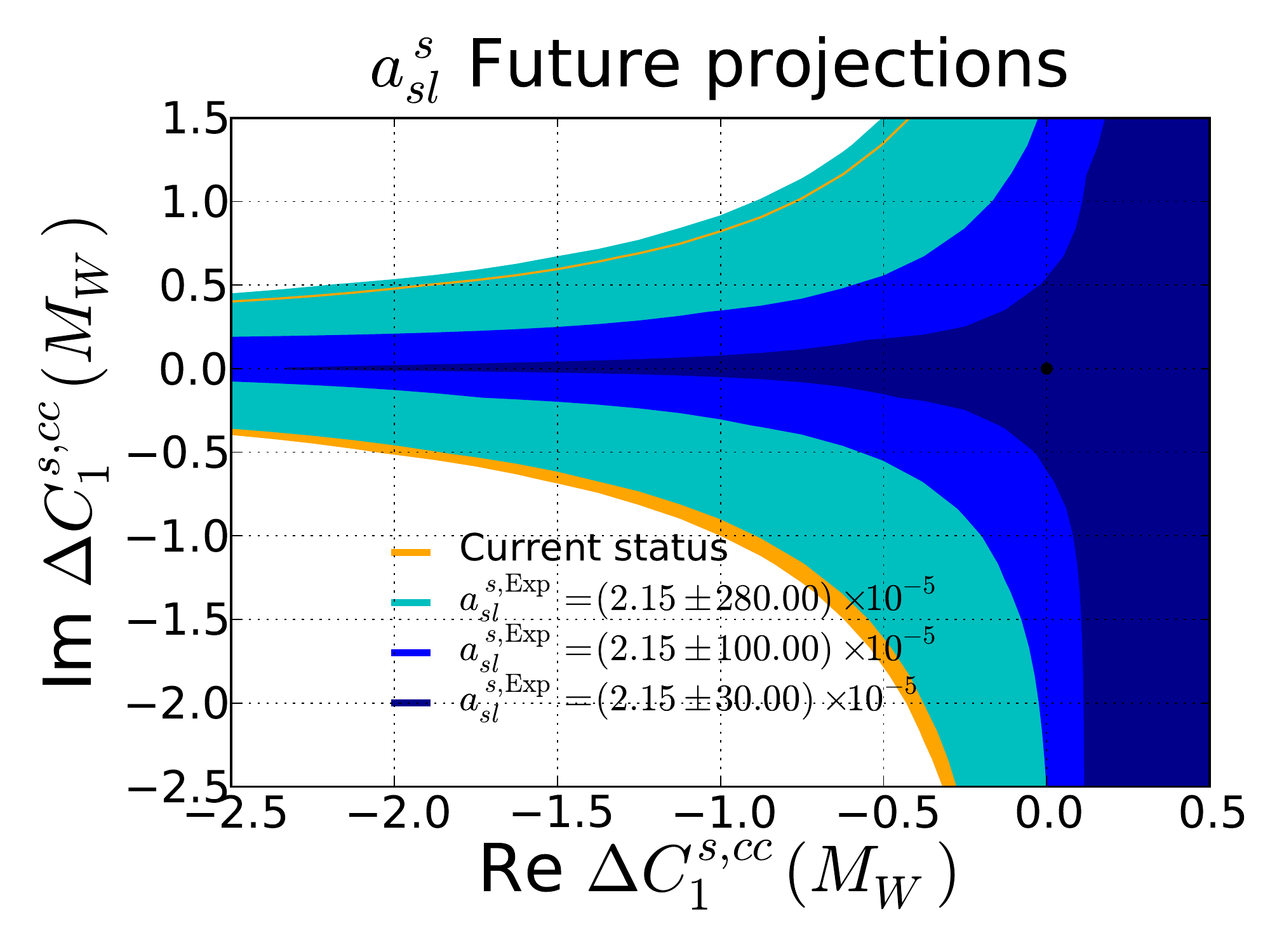}
\includegraphics[height=5cm]{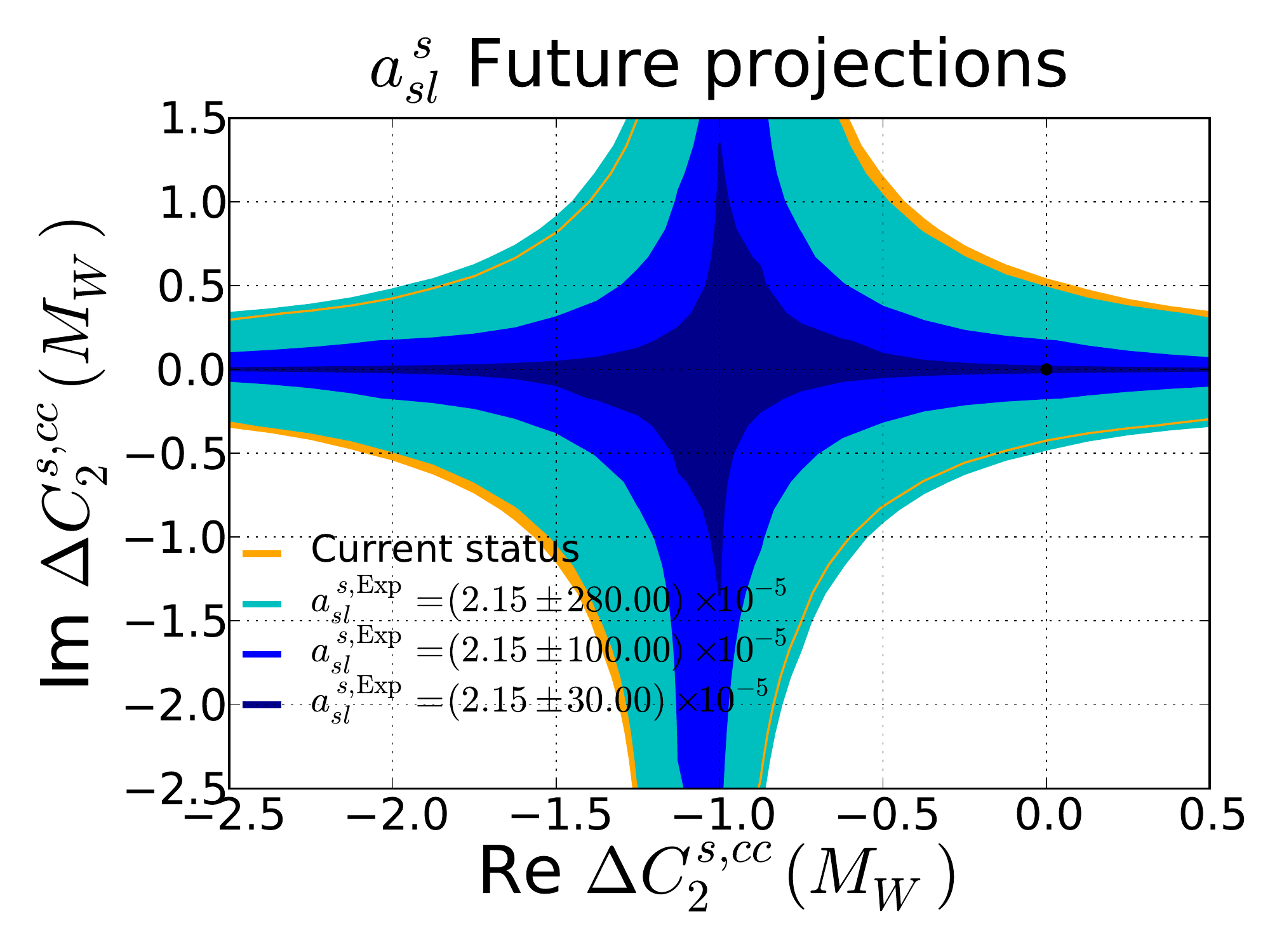}
\caption{Future scenarios for the precision in the observable $a_{sl}^s$ and resulting constraints on $\Delta C_1$ and $\Delta C_2$.
  The current uncertainty is expected to be reduced down to
1 per mille and later even to 0.3 per mille. }
\label{fig:asls_future}
\end{figure}

The experimental precision for the semi-leptonic CP asymmetries is still much larger than the tiny SM values for these
quantities. Nevertheless already at this stage $a_{sl}^q$ provide important bounds on possible BSM effects in the
Wilson coefficients.
The experimental precision in the semi-leptonic CP-asymmetries will rise considerable in the near future, see e.g.
Table 1 of \cite{Cerri:2018ypt} from where we take:
\begin{eqnarray}
  \delta \left( a_{sl}^s \right) & = & 1 \cdot 10^{-3} \hspace{1cm} \mbox{LHCb 2025}
  \\
   \delta \left( a_{sl}^s \right) & = & 3 \cdot 10^{-4} \hspace{1cm} \mbox{Upgrade II}
  \end{eqnarray}
We show the dramatic impact of these future projections on the BSM bounds on the Wilson coefficients in
Fig.~\ref{fig:asls_future}.
\subsection{Rare decays}
\begin{figure}
	\centering
	\includegraphics[height=5cm]{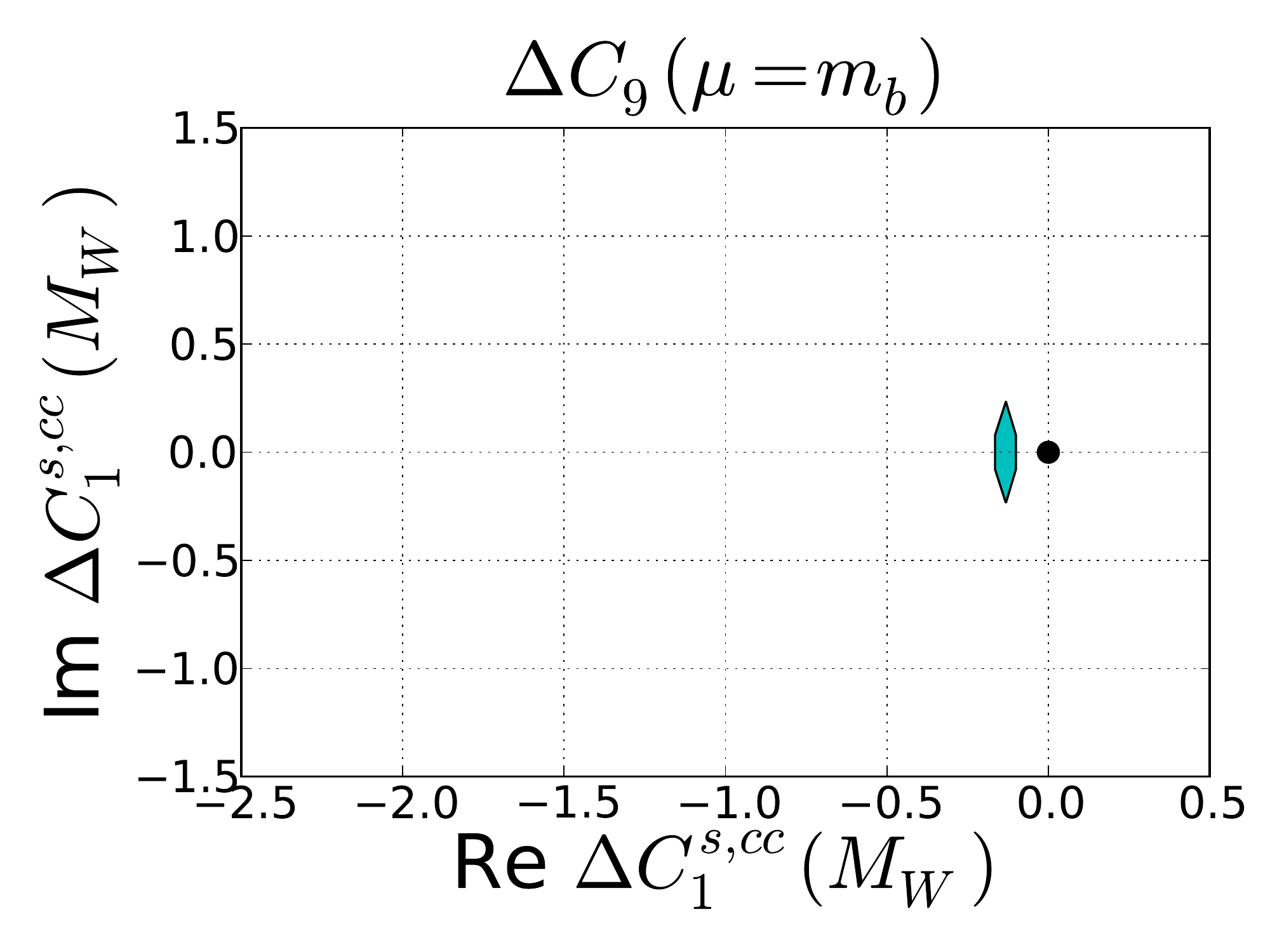}
    \includegraphics[height=5cm]{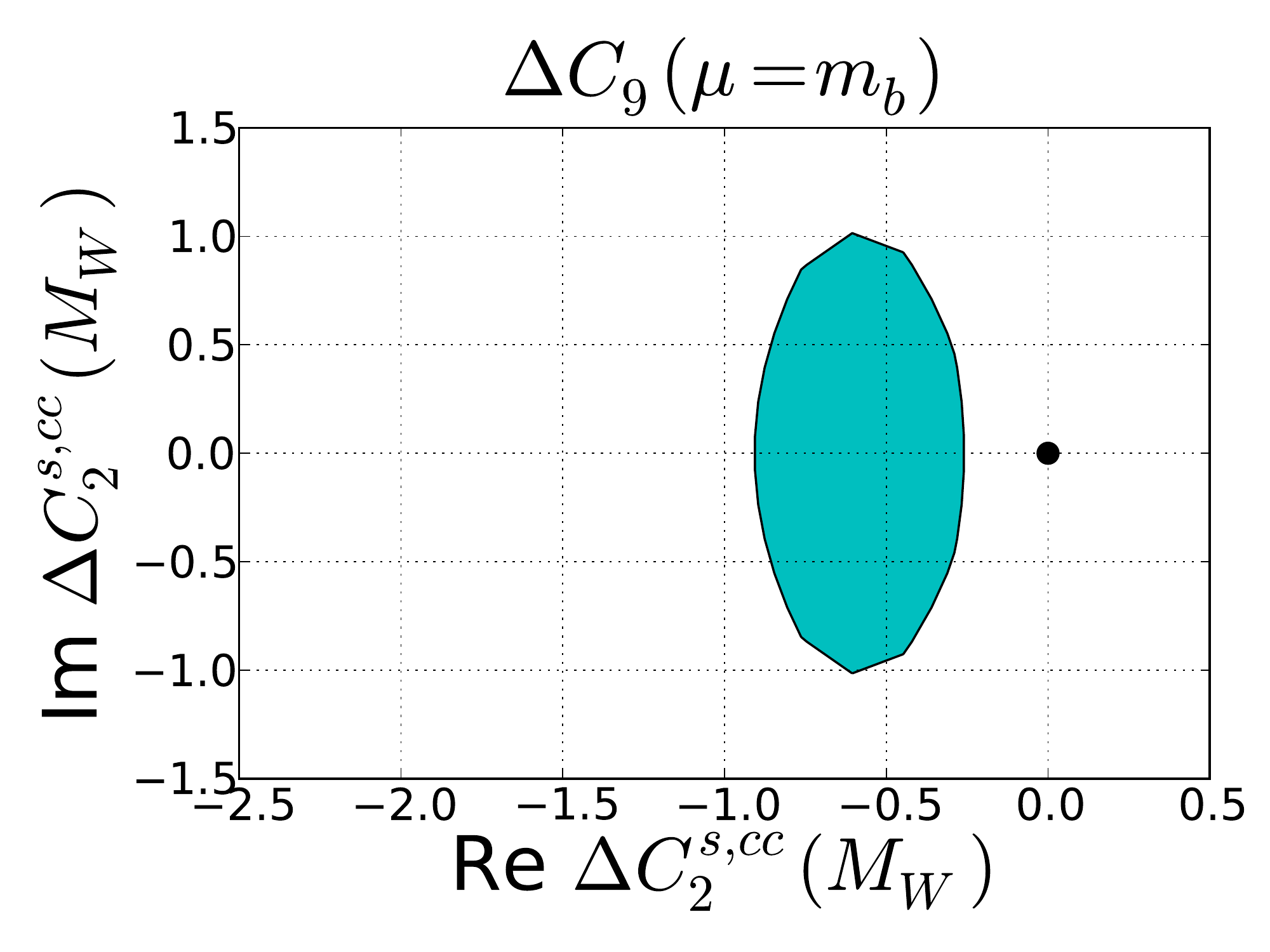}
	\caption{Regions for NP, at $90\%$ C. L., in the
		${\rm Re}~\Delta C^{s, cc}_1$ - ${\rm Im}~\Delta C^{s, cc}_1$ (left)
		and ${\rm Re}~\Delta C^{s, cc}_2$ - ${\rm Im}~\Delta C^{s, cc}_2$ (right)
		planes allowed by the B physics anomalies related
		with the decay $B\rightarrow K^{(*)}\mu^+\mu^-$. The black dot corresponds to the SM result.}
	\label{Fig:dC9}
\end{figure}

As discussed in \cite{Jager:2017gal,Jager:2019bgk} NP effects in the $b\rightarrow c\bar{c} s$ transitions can induce shifts
in the Wilson coefficient of the operator
\begin{eqnarray}
\hat{Q}_{9 V}=\frac{\alpha}{4\pi}(\bar{\hat{s}}_L \gamma_{\mu} \hat{b}_L)
(\bar{\hat{\ell}} \gamma^{\mu} \hat{\ell}),	
\end{eqnarray}
leading to
\begin{eqnarray}
\Delta C^{\rm eff}_9\Bigl |_{\mu=m_b} =\Bigl[8.48~\Delta C_1 + 1.96~ \Delta C_2 \Bigl]\Bigl |_{\mu=M_W}. 
\end{eqnarray}
This result offers an interesting link with the anomalous deviations in observables associated with the
decay $B\rightarrow K^{(*)}\mu^+\mu^-$, where model independent explanations with physics only in $C_9$
require $\Delta C^{\rm eff}_9\Bigl |_{\mu=m_b} =-\mathcal{O}(1)$. In order to account for NP phases
we use the results presented in \cite{Alok:2017jgr} where $\Delta C_9$ is allowed to take
complex values leading to the constraints shown in Fig.~\ref{Fig:dC9}. Here both $C_1$ and $C_2$ get a shift
towards negative values.
BSM in effects in non-leptonic tree-level can in principle explain the deviations seen in lepton-flavour
universal observables, like the branching ratios or $P_5'$; they can, however,  not explain the anomalous values of lepton flavour 
universality violating observables like $R_K$.
Future measurements will show, whether the bounds, obtained  in Fig.~\ref{Fig:dC9} should be included in our full fit.
%
%
%
%
%
%
%
%
%
%
%
%
%
%
%
%
%
%
%
%

%
%
%
%
%
%
\section{Conclusions and outlook}
\label{sec:conclusion}

In this work we have questioned the well accepted
assumption of having no NP in tree level decays, in particular we explored for possible
deviations with respect to the SM
values in the dimension six current-current operators $\hat{Q}_1$ (colour suppressed) and $\hat{Q}_2$ (colour allowed) associated
with the quark level transitions $b\rightarrow q \bar{q}' s$ and $b\rightarrow q \bar{q}' d$  ($q, q'=u,c$). We evaluated the size of the NP effects by modifying the
corresponding Wilson coefficients according to $C_1 \rightarrow  C_1 + \Delta C_1$, $C_2 \rightarrow  C_2 +\Delta C_2$,
for $\Delta C_{1,2}\in \mathcal{C}$; we found that sizeable deviations in $\Delta C_{1,2}$ are not ruled out by the recent experimental data.
\\
Our analysis was based on a $\chi^2$-fit where we included different B-physics observables involving the decay processes:
$\bar{B}^0_d\rightarrow D^{*}\pi$, $\bar{B}^0_d\rightarrow \pi\pi$, $\bar{B}^0_d\rightarrow \pi \rho$, $\bar{B}^0_d\rightarrow \rho\rho$,
$\bar{B}\rightarrow X_s \gamma$, $\bar{B}_s\rightarrow J/\psi \phi$ and  $\bar{B}\rightarrow X_d \gamma$. We also considered neutral B mixing observables:
the semi-leptonic asymmetries $a^s_{sl}$ and $a^d_{sl}$ as well as the decay width difference $\Delta \Gamma_s$ of $B^0_s$ oscillations and the lifetime
ratio of $B_s$ and $B_d$ mesons. Finally we also studied the CKM angles $\beta$, $\beta_s$ and $\gamma$.
\\
For the  amplitudes of the hadronic transitions $\bar{B}_d^0\rightarrow D^{*}\pi$, 
$\bar{B}^0_d\rightarrow \pi\pi$, $\bar{B}^0_d\rightarrow \pi\rho$ and $\bar{B}^0_d\rightarrow \rho\rho$ and $\bar{B}_s\rightarrow J/\psi \phi$  we used the 
formulas calculated within the QCD factorization framework. We have identified a high sensitivity on $\Delta C_{1,2}$ with respect to the power corrections
arising in the annihilation topologies and in some cases in those for the hard-spectator scattering as well. It is also important to mention
that the uncertainty in the parameter  $\lambda_B$ used to describe the inverse moment of the light cone distribution for the neutral B mesons
is of special importance in defining the size of $\Delta C_1$ and $\Delta C_2$.
For the mixing observables and the lifetime ratios we have benefited from the enormous progress achieved in the precision of the hadronic input parameters, 
thus we have also updated the corresponding SM predictions:
\begin{eqnarray}
  \Delta M_s = (18.77 \pm 0.86 ) \, \mbox{ps}^{-1},
  &&
  \Delta M_d = (0.543 \pm 0.029) \, \mbox{ps}^{-1},
  \nonumber
  \\
  \Delta \Gamma_s = (9.1 \pm 1.3 ) \cdot 10^{-2} \, \mbox{ps}^{-1},
  &&
  \Delta \Gamma_d = (2.6 \pm 0.4 ) \cdot 10^{-3} \, \mbox{ps}^{-1},
  \nonumber
  \\
  a_{sl}^s = (2.06 \pm 0.18) \cdot 10^{-5},
  &&
a_{sl}^d = (-4.73 \pm 0.42) \cdot 10^{-4}. 
\end{eqnarray}
We have made a channel by channel study by combining different
constraints for the decay chains $b\rightarrow u\bar{u}d$, $b\rightarrow c\bar{u}d$, $b\rightarrow c\bar{c}s$ and $b\rightarrow c\bar{c}d$;
we also performed a universal $\chi^2$-fit where we have included observables mediated by $b\rightarrow  q q's$ decays as well.  The universal $\chi^2$-fit provides
the strongest bounds on the NP deviations, we found that
\begin{eqnarray}
  |\hbox{Re} (\Delta C_1)|\leq \mathcal{O}(0.4),
  &&
  |\hbox{Re} (\Delta C_2)|\leq \mathcal{O}(0.1),
  \\
 |\hbox{Im} (\Delta C_1)|\leq \mathcal{O}(0.5),
 &&
 |\hbox{Im} (\Delta C_2)|\leq \mathcal{O}(0.04),
\end{eqnarray}
 whereas for the independent channel analyses the corresponding deviations can much larger.
\\
We have analysed the implications of having NP in tree level b quark transitions on the decay width difference of neutral $B^0_d$ mixing
$\Delta \Gamma_d$ - note, that the most recent experimental average is still consistent with zero. We found
that enhancements in $\Delta \Gamma_d$ with respect to its SM value of up to a factor of five
are consistent with the current experimental data. Such a huge enhancement could solve the tension between experiment and theory in the D0
measurement for the dimuon asymmetry. Thus we strongly encourage further
experimental efforts to measure $\Delta \Gamma_d$, see also \cite{Gershon:2010wx}.
\\
Next we evaluated the impact of our allowed NP regions
for $\Delta C_1$ and $\Delta C_2$ on the  determination of the CKM phase $\gamma$,
where the absence of penguins leads in principle to an
exceptional theoretical cleanness. We found that $\gamma$ is highly sensitive to the imaginary components of $\Delta C_1$ and $\Delta C_2$
and our BSM effects could lead to deviations in this quantity by up to $10^{\circ}$.
It has to be stressed, however, that for quantitative statements about the size of the shift 
 $\delta \gamma$ the ratios of the matrix
elements $\langle \bar{D}^0 K^-| Q_1^{\bar{u}cs} | B^- \rangle/
          \langle \bar{D}^0 K^-| Q_2^{\bar{u}cs} | B^- \rangle$
and      $\langle       D^0 K^-| Q_1^{\bar{c}us} | B^- \rangle/
          \langle       D^0 K^-| Q_2^{\bar{c}us} | B^- \rangle$
          have to be determined in future with more reliable methods. So far only naive estimates are available for these ratios.
          \\
          Finally we studied future projections for observables that will shrink the allowed region for NP effects - or identify a BSM region -
          in non-leptonic tree-level decays. Here $\tau(B_s) / \tau(B_d)$ and the semi-leptonic CP asymmetries seem to be very promising.
 %
%
%
%
%
%
%
%
%
%
%
%
%
%
%
%
%
%
%
%

\section*{Acknowledgements}
We would like to thank Martin Wiebusch for collaborating at early stages of the project. This work benefited from physics discussions
with Christoph Bobeth, Tobias Huber, Joachim Brod, Leonardo Vernazza, Michael Spannowsky, Marco Gersabeck, Thomas Rauh, Marcel Merk, Niels Tuning, 
Patrick Koppenburg and Laurent Dufour. We acknowledge Luiz Vale for his help with CKM-Fitter Live.  We thank to Bert Schellekens for his help and support in accessing the Nikhef computing cluster ``Stoomboot''
during the development of this project and to Patrick Koppenburg for supporting the access in the final stages of the project as well. 
GTX acknowledges support from the NWO program 156, ``Higgs as Probe and Portal'' and by the Deutsche
Forschungsgemeinschaft (DFG, German Research Foundation) under grant  396021762 - TRR 257.
The work of AL was supported by STFC via the IPPP grant.
%
%
%
%
%
%
%
%
%
%
%
%
%
%
%
%
%
%
%
%

\appendix

\section{Numerical Inputs}
\label{Sec:Inputs}
In this section we collect the numerical values of the input parameter used
in this work.
\begin{center}
\begin{table}
  \renewcommand{\arraystretch}{1.1}
\scriptsize{	
\begin{tabular}{|lccc|lccc|}
\hline
  Parameter
& Value
& Unit
& Ref.
& Parameter
& Value
& Unit
& Ref.
\\
\hline
\hline
\multicolumn{8}{|l|}{\centering{\textbf{Lepton masses, gauge boson masses and couplings}}}
\\
\hline
  $m_\mu$            & {$0.1056583745(24)$}                   & GeV        & \cite{Tanabashi:2018oca}
& $G_F$              & {$1.166 378 7(6)\cdot 10^{-5}$}        & GeV$^{-2}$  & \cite{Tanabashi:2018oca}
\\
  $m_\tau$           & ${1.77686(12)}$                        & GeV        & \cite{Tanabashi:2018oca}
& $\alpha_s(M_Z)$    & 
                      { $0.1181 \pm 0.0011$}                  &            & \cite{Tanabashi:2018oca}
\\
  $M_Z$              & $91.1876(21)$                                      & GeV        & \cite{Tanabashi:2018oca}
& $\alpha$           & $ { 7.297 352 5664(17) \cdot 10^{-3}}$ &            & \cite{Tanabashi:2018oca}
\\
  $M_W$              & $80.379(12)$                                       & GeV        & \cite{Tanabashi:2018oca}
& $\Lambda^{QCD}_{5}$ &   { $0.210 \pm 0.014$}                 & GeV        & \cite{Tanabashi:2018oca}
\\
                     &                                                   &             & 
& $\hbar $           &   $6.582119514(40) \cdot 10^{-25}    $             & GeV s       & \cite{Tanabashi:2018oca}
\\
\hline\hline
\multicolumn{8}{|l|}{\textbf{CKM}}
\\
\hline
$|V_{us}|$  & 
{ $0.224746^{+0.000253}_{-0.000058}$}
&       &  &   $\gamma$  
&
{ $(65.17^{+0.26}_{-3.05})^{\circ}$}
&        & \\
$|V_{cb}|$  &
{ $0.04243^{+0.00036}_{-0.00088}$}
&       &  &
$\sin(2\beta)_{dir.}$     &
{  $0.699 \pm 0.017 $}&       & \cite{Amhis:2016xyh} \\
$|V_{ub}/V_{cb}|$  & 
{  $0.08833\pm 0.00218$}
&      &  &
$\sin(2\beta)_{indir.}$    &
{ $0.732\pm 0.029$}& & \\
\hline
\hline
\multicolumn{8}{|l|}{\textbf{Quark masses}}
\\
\hline
$m_d$            & $0.00467^{+0.00048}_{-0.00017}$              & GeV & \cite{Tanabashi:2018oca}
& $m^{1S}_b$ & $4.65\pm 0.03$         & GeV & \cite{Tanabashi:2018oca}
\\
$m_s(2~\hbox{GeV})$ & {  $0.093^{+0.011}_{-0.005}$ }            & GeV & \cite{Tanabashi:2018oca}
&    &           & & 
\\
$\bar{m}_c (\bar{m}_c)$             &
{ $1.27\pm 0.02$}
& GeV & \cite{Tanabashi:2018oca}
&    &  &  &
\\

 $\bar{m}_c(\bar{m}_b)$       & { $0.96 \pm 0.02$} & GeV &
& $m^{pole}_t$      &  $173.1 \pm 0.9$     & GeV & \cite{Tanabashi:2018oca, ATLAS:2014wva}
\\
$\bar{m}_b(\bar{m}_b)$     &   
  { $4.214^{+0.042}_{-0.043}$}       & GeV & \cite{Rauh:2018vsv}
& $\bar{m_t}(\bar{m_t})$               &   $163.3\pm 0.9$     & GeV &
\\
$m^{pole}_b$               &   {  $4.61\pm 0.05$}         & GeV & 
& $m_t(m_W)$    &   $172.6\pm 1.0$     & GeV &
\\
\hline
\hline
\end{tabular}
\renewcommand{\arraystretch}{1.0}
\nonumber  \label{tab:numeric:input2}
}
\caption{Values of the input parameters used for our numerical evaluations.}
\end{table}
\end{center}

\begin{table}
\scriptsize{
\begin{tabular}{|lccc|lccc|}
\hline
  Parameter
& Value
& Unit
& Ref.
& Parameter
& Value
& Unit
& Ref.
\\
\hline
\hline
	\multicolumn{8}{|l|}{\centering\textbf{$B$- and light meson properties (cont.)}}
\\
\hline
$m_{B^+}$          & $5279.33(13)$      & MeV & \cite{Tanabashi:2018oca}
& $a^{\pi}_{1}$  &  $0.0$       &     & \cite{Ball:2005vx, Ball:2006wn}
\\
{  $m_{B_d}$}         &{  $5279.64(13)$   }    & {MeV}& \cite{Tanabashi:2018oca}
& $a^{\pi}_{2}$  &  $0.17\pm 0.10$       &     & \cite{Ball:2006wn}
\\
  $m_{B_s}$                 & { $5366.88(17)$}       & MeV & \cite{Tanabashi:2018oca}
&  $a^{\rho}_{1}$  &  $0.0$       &     & \cite{Ball:2007zt}
\\
  $m_{\pi^+}$                 & { $139.57061(24)$}        & MeV & \cite{Tanabashi:2018oca}
& $a^{\rho}_{1\perp}$  &  $0.0$       &     & \cite{Ball:2007zt} 
\\
  $m_{\pi^0}$   & { $134.9770(5)$}       & MeV & \cite{Tanabashi:2018oca}
& $a^{\rho}_{2}$  &  $0.1\pm 0.3$       &     & \cite{Beneke:2003zv} 
\\
{ $m_{\rho^+}$   }  & { $775.11 \pm 0.34$  }     & { MeV} & \cite{Tanabashi:2018oca}
& $a^{\rho}_{2\perp}$  &  $0.11\pm 0.05$       &     & \cite{Ball:2007zt, Ball:2007rt}
\\
$m_{\omega}$                         &  $782.65 \pm 0.12$                & MeV & \cite{Tanabashi:2018oca}
& $a^{\phi}_{1 ||}$& $0$            &   &  \cite{Ball:1998fj}
\\
{$m_{D^{*+}}$}                      & { $2010.26 \pm 0.05$}                & MeV & \cite{Tanabashi:2018oca}
&  $a^{\phi}_{2 ||}$& $0 \pm 0.1$      &   &  \cite{Ball:1998fj}
\\ 
 $m_{K^0}$     & $497.611 \pm 0.013$       & MeV & \cite{Tanabashi:2018oca}
& $B^s_3/B^s_1$ &  $1.006 \pm 0.066$  &  & 
\\
$f_{B_{u,d}}$       & $190.0 \pm 1.3$ & MeV & \cite{Aoki:2019cca}
& $B^s_{R_0}/B^s_1$ &  $0.377 \pm 0.154$  &  &
\\
$f_{B_{d}}^2 B_1^d$       & $(0.0305\pm0.0011)$  & GeV$^2$ & 
& $B^s_{R_1}/B^s_1$  & $1.193\pm 0.052$  & & 
\\
$f_{B_{s}}^2 B_1^s$       & $(0.0452\pm0.0014)$  & GeV$^2$ & 
&$B^s_{R_2}/B^s_1$  & $0.318\pm 0.118$  & &
\\
  $f_{\pi}$           &   { $130.2 \pm 0.8$}       & MeV & \cite{Aoki:2019cca}
& $B^s_{R_3}/B^s_1$ &  $0.389 \pm 0.130$  &  &
\\
{ $f_{\rho}$ }        & {  $216 \pm 3$ }              & MeV & \cite{Bell:2009fm, Ball:2006eu}
&$B^s_{\tilde{R}_1}/B^s_1$  & $1.130 \pm 0.047$ & &
\\
{ $f^{\perp}_{\rho}(1~\hbox{GeV})$  }       & {  $165 \pm 9$}              & MeV & \cite{Ball:2006eu, Ball:2007zt}
 &  &      &      & 
\\
{ $f_{\omega}$  }       & { $195 \pm 3$ }              & MeV & \cite{Ball:2004rg}
& $B^d_3/B^d_1$ &  $0.928 \pm 0.072$  &  & 
\\
{$f_{D^{*}}$}  & {$223.5 \pm 8.4$}        & { MeV }  & \cite{Lubicz:2017asp}
&$B^d_{R_0}/B^d_1$ &  $0.383 \pm 0.156$  &  &
\\
{$F^{B\rightarrow \pi}_+(0)$  }  & {$0.261 \pm 0.023$ }       &     & \cite{Gonzalez-Solis:2018ooo, Bharucha:2012wy}
&$B^d_{R_1}/B^d_1$  & $1.190\pm 0.060$  & & 
\\
{  $A_0^{B\rightarrow \rho}(0)$ } & {  $0.36 \pm 0.04$ }        &     & \cite{Straub:2015ica}
&$B^d_{R_2}/B^d_1$  & $0.323\pm 0.120$  & &
\\
{ $A_0^{B \rightarrow D^{*} }$ } & { $0.66 \pm 0.02$}        &     & \cite{Bernlochner:2017jka, Sakaki:2013bfa}
&$B^d_{R_3}/B^d_1$ &  $0.395 \pm 0.132$  &  &
\\
$A^{B\rightarrow \phi}_0(m^2_{J/\psi}) $  &  $0.68\pm0.07$         &   & \cite{Straub:2015ica}
&$B^d_{\tilde{R}_1}/B^d_1$ &  $1.190 \pm 0.060$  &  &
\\
$A^{B\rightarrow \phi}_1(m^2_{J/\psi}) $  &  $0.37\pm0.04$         &   & \cite{Straub:2015ica}
& &   &  &
\\
$A^{B\rightarrow \phi}_2(m^2_{J/\psi}) $  &  $0.40\pm0.14$         &   & \cite{Straub:2015ica}
& $\tau(B^0_s)$                       & { $1.509\pm 0.004$ }       & ps & { \cite{Amhis:2016xyh}}
\\
$V^{B\rightarrow \phi}_2(m^2_{J/\psi}) $  &  $0.70\pm0.06$         &   & \cite{Straub:2015ica}
& $\tau(B^0_d)$   & { $1.520 \pm 0.004$}  & ps & { \cite{Amhis:2016xyh}}
\\
$\Lambda_h$  &  $500$           & MeV  & \cite{Beneke:2003zv}
&{ $\Gamma_{\omega}$       }  &{ $8.49 \pm 0.08$}           & MeV  & \cite{Tanabashi:2018oca}
\\
$\lambda_B$         & $400 \pm 150$           & MeV & \cite{Bell:2009fm}
&{ $\Gamma_{\rho}$       }  &{  $149.5 \pm 1.3$}           & MeV  & \cite{Tanabashi:2018oca}
\\
\hline
\hline
\end{tabular}
}
\caption{Values of the input parameters used for our numerical evaluations (cont.).}
\end{table}
{
Using the PDG value for the strong coupling
\begin{equation}
\alpha_s(M_Z)  =  0.1181 \pm 0.0011
\end{equation}
we derive with  $M_Z = 91.1876 \pm 0.0021$ GeV at NLO-QCD
\begin{equation}
\Lambda_{QCD}^{(5)} = 228 \pm 14 \; \mbox{MeV}\; ,
\end{equation}
while PDG gives
\begin{equation}
\Lambda_{QCD}^{(5)} = 210 \pm 14 \; \mbox{MeV}\; ,
\end{equation}
using 4-loop running, 3-loop matching. We decided to use the latter value, the effects on $\alpha_s(m_b)$ are very small.
\\
For quark masses we use the PDG values in the MSbar definition, except for the b-quark, where we use a more conservative determination.
The PDG value reads for comparison
\begin{eqnarray}
m_b(m_b) & = & 4.18^{+0.04}_{-0.03} \; \mbox{GeV}
\end{eqnarray}
The PDG value for $m_c(m_c)$ correspond to  $m_c(m_b)= 0.947514$, which 
will be used \footnote{Actually $\bar{z} := m_c^2(m_b)/m_b^2(m_b) = 0.0505571$
is used.} for the analysis of the mixing quantities $\Delta \Gamma_q$ 
and $a_{sl}^q$.
\\
For the top quark pole mass we use the result obtained from cross-section
measurements given in \cite{Tanabashi:2018oca}
\begin{eqnarray}
m_t^{\rm Pole} & = &  173.1 \pm 0.9 \; \mbox{GeV}
\label{eq:mtpole}
\end{eqnarray}
which is an average including measurements from D0, ATLAS and CMS.
\\
Entering  Eq.~\ref{eq:mtpole} in the version 3 of the software {\tt RunDec} \cite{Herren:2017osy} we obtain
\begin{eqnarray}
\bar{m}_t (\bar{m}_t) & = & 163.3 \pm 0.9~\mbox{GeV}, 
\end{eqnarray}
and
\begin{eqnarray}
m_t (M_W) & = & 172.6 \pm 1.0~\mbox{GeV}.
\end{eqnarray}
}
We use the averages of the $B$ mixing bag parameters obtained in  \cite{DiLuzio:2019jyq}
based on the HQET sum rule calculations in \cite{King:2019lal,Kirk:2017juj,Grozin:2016uqy,Grozin:2008nu} and
the  corresponding lattice studies in \cite{Carrasco:2013zta,Aoki:2014nga,Bazavov:2016nty,Boyle:2018knm,Dowdall:2019bea}:
\begin{equation}
 \begin{array}{ll}
  B_1^s(\mu_b) = 0.849\pm0.023\,, &\hspace{1cm}B_1^d(\mu_b) = 0.835\pm0.028\,,\nonumber\\
  B_2^s(\mu_b) = 0.835\pm0.032\,, &\hspace{1cm}B_2^d(\mu_b) = 0.791\pm0.034\,,\nonumber\\
  B_3^s(\mu_b) = 0.854\pm0.051\,, &\hspace{1cm}B_3^d(\mu_b) = 0.775\pm0.054\,,\nonumber\\
  B_4^s(\mu_b) = 1.031\pm0.035\,, &\hspace{1cm}B_4^d(\mu_b) = 1.063\pm0.041\,,\nonumber\\
  B_5^s(\mu_b) = 0.959\pm0.031\,, &\hspace{1cm}B_5^d(\mu_b) = 0.994\pm0.037\,,
 \end{array}
 \label{eq:AberageBagParameters}
\end{equation}
at the scale $\mu_b=\bar{m}_b(\bar{m}_b)$.
For the first time we do not have to rely on vacuum insertion approximation for the dimension seven operators,
instead we can now use the values obtained in \cite{Davies:2019gnp,Dowdall:2019bea}
\begin{eqnarray}
  B^q_{R_0}        & = & 0.32  \pm 0.13  \, ,
  \nonumber
  \\
  B^q_{R_1}        & = & 1.031 \pm 0.035 \, ,
  \nonumber
  \\
  B^q_{\tilde{R}_1} & = & 0.959 \pm 0.031 \, ,
  \nonumber
  \\
  B^q_{R_2}        & = & 0.27 \pm 0.10 \, ,
  \nonumber
  \\
  B^q_{R_3}        & = & 0.33 \pm 0.11 \, .
\end{eqnarray}
Note that our notation for the dimension seven Bag parameter $B^q_{R_2}$ and $B^q_{R_3}$ corresponds to the primed bag parameter
of \cite{Davies:2019gnp}.
For the remaining two operators we are using equations of motion  \cite{Beneke:1996gn}
\begin{eqnarray}
  B^q_{\tilde{R_2}} & = & -B^q_{R_2}  
  \nonumber
  \\
  B^q_{\tilde{R_3}} & = & \frac75 B^q_{R_3}  - \frac25 B^q_{R_2}  \, .
\end{eqnarray}
For the determination of the uncertainties of the ratios of Bag parameter, we first symmetrized the errors of the individual
bag parameter.
Based on the updated value for the bag parameter $B^q_1$ given above and the lattice average ($N_f = 2+1+1$) for $f_{B_q}$ presented
in \cite{Aoki:2019cca} - based on \cite{Christ:2014uea,Bussone:2016iua,Hughes:2017spc,Bazavov:2017lyh}
\begin{eqnarray}
  f_{B_s}&=& (230.3 \pm 1.3)~\hbox{MeV},
  \nonumber
  \\
  f_{B_d}&=&(190.0\pm 1.3)\hbox{MeV},
\end{eqnarray}
we obtain after symmetrizing the uncertainties
\begin{eqnarray}
  f_{B_s}^2  B^s_1&=&(0.0452 \pm 0.0014)~\hbox{GeV}^2 ,
  \nonumber
  \\
  f_{B_d}^2  B^d_1&=&(0.0305 \pm 0.0011)~\hbox{GeV}^2.
\end{eqnarray}
Additionally, for the determination of the contributions of the double insertion of the  $\Delta B=1$ effective Hamiltonians to $M^d_{12}$
we require the following Bag parameters at the scale $\mu_c=1.5~\hbox{GeV}$ (see \cite{Kirk:2017juj})
\begin{eqnarray}
B^d_1(1.5~\hbox{GeV})=0.910^{+0.023}_{-0.031},\quad\quad B^d_2(1.5~\hbox{GeV})=0.923^{+0.029}_{-0.035}.
\end{eqnarray}
To calculate the CKM-elements in Eq.~(\ref{eq:CKMElements}) we require the renormalization group invariant bag parameter $\hat{B}^s_1$ which in the
$\overline{\text{MS}}$-NDR scheme relates with $B^s_1$, via (see e.g. \cite{Buchalla:1995vs})

\begin{eqnarray}
  \hat{B}^s_1&=&\alpha_s(\mu)^{-\gamma_0/(2\beta_0)}\Bigl[1 + \frac{\alpha_s(\mu)}{4\pi}\Bigl(\frac{\beta_1\gamma_0-\beta_0\gamma_1}{2\beta^2_0}\Bigl)\Bigl]B^s_1
  \\
  &=&\alpha_s(\mu)^{-\frac{6}{23}}\Bigl[1 + \frac{\alpha_s(\mu)}{4\pi} \frac{5165}{3174}  \Bigl]B^s_1 = 1.52734 \, B^s_1,
\end{eqnarray}
where we have used
\begin{eqnarray}
C_F & = & \frac{N_c^2 - 1}{2 N_c},
\\
\beta_0 & = & \frac{11 N_c - 2 n_f}{3}, \hspace{0.35cm}
\beta_1 = \frac{34}{3} N_c^2 - \frac{10}{3} N_c n_f - 2 C_F n_f,
\\
\gamma_0& = & 6 \frac{N_c-1}{N_c} ,\hspace{1.cm}
\gamma_1= \frac{N_c-1}{2 N_c} \left(-21 +\frac{57}{N_c} -\frac{19}{3} N_c +\frac{4}{3} n_f \right)  . 
\end{eqnarray}
Finally we take the lifetime bag parameter from the recent HQET sum rule evaluation in \cite{Kirk:2017juj} - here no corresponding up to date lattice
evaluation exists
\begin{eqnarray}
B_1(\mu=m_b)=1.028^{+0.064}_{-0.056},&&
B_2(\mu=m_b)=0.988^{+0.087}_{-0.079},\nonumber\\
~\epsilon_1(\mu=m_b)=-0.107^{+0.028}_{-0.029},&&
~\epsilon_2(\mu=m_b)=-0.033^{+0.021}_{-0.021}.
\end{eqnarray}
Using CKMfitter-Live \cite{Charles:2004jd} online, we perform a fit to the CKM elements $|V_{us}|$, $|V_{ub}|$,  $|V_{cb}|$ 
and the CKM
angle $\gamma$ excluding in all the cases the direct determination of the CKM angle $\gamma$ itself. Our inputs coincide mostly with the CKMfitter-Summer 2018 analysis,
however in order to be consistent with our main study we modify the following entries $\bar{m}_t(\bar{m}_t)$, $\bar{m}_c(\bar{m}_c)$, $\hat{B}^s_1$ and the ratios
\begin{eqnarray}
	\frac{\hat{B}^s_1}{\hat{B}^d_1}=0.987\pm 0.008 \hbox{\cite{King:2019lal}},&\quad&\frac{f_{B_s}}{f_{B_d}}=1.212\pm 0.011.
\end{eqnarray}
Our results are
\begin{eqnarray}
\label{eq:CKMElements}
|V_{us}|=0.224746^{+0.000253}_{-0.000058}, && |V_{ub}|=0.003741^{+0.000082}_{-0.000061}\nonumber \\
|V_{cb}|=0.04243^{+0.00036}_{-0.00088}, && \gamma=(65.17^{+0.26}_{-3.05})^{\circ},
\end{eqnarray}
from which we obtain 
\begin{eqnarray}
\frac{|V_{ub}|}{|V_{cb}|}=0.08833\pm 0.00218.
\end{eqnarray}
The full set of CKM matrix elements is then calculated under the assumption of the  unitarity
of the $3\times 3$ CKM matrix.

\section{QCD-Factorization formulas}
\label{Sec:QCDFact}

\subsection{Generic parameters}

\begin{eqnarray}
f^{\perp}_{V}(\mu)=f^{\perp}_{V}(\mu_0)\Bigl(\frac{\alpha_s(\mu)}{\alpha_s(\mu_0)}\Bigl)^{\frac{C_F}{\beta_0}},
&& r^{\pi}_{\chi}(\mu)=\frac{2m^2_{\pi}}{m_b(\mu)2m_q(\mu)},\nonumber\\
r^{\rho}_{\chi}(\mu)=\frac{2m_{\rho}}{m_b(\mu)}\frac{f^{\perp}_{\rho}(\mu)}{f_{\rho}},
&&r^{D^*}_{\chi}(\mu)=\frac{2m_{D^*}}{m_b(\mu)}\frac{f^{\perp}_{D^*}(\mu)}{f_{D^*}},\nonumber\\
r^{K}_{\chi}(\mu)=\frac{2m^2_K}{m_b(\mu)\Bigl(m_q(\mu) +  m_s(\mu)\Bigl)},
&&A_{\pi\pi}=i\frac{G_F}{\sqrt{2}}m^2_BF^{B\rightarrow \pi}_{0}(0)f_{\pi},\nonumber\\
A_{\pi\rho}=-i\frac{G_F}{\sqrt{2}}m^2_{B}F^{B\rightarrow \pi}_{0}(0)f_{\rho},
&&A_{\rho\pi}=-i\frac{G_{F}}{\sqrt{2}}m^2_BA_{0}^{B\rightarrow \rho}(0)f_{\pi},\nonumber\\
A_{\rho\rho}=i\frac{G_F}{\sqrt{2}}m^2_BA^{B\rightarrow \rho}_{0}(0)f_{\rho},
&&B_{\pi\pi}=i\frac{G_F}{\sqrt{2}}f_Bf_{\pi}f_{\pi},\nonumber\\
B_{\pi\rho}=B_{\rho\pi}=-i\frac{G_F}{\sqrt{2}}f_Bf_{\pi}f_{\rho},
&&B_{\rho\rho}=i\frac{G_F}{\sqrt{2}}f_{B}f_{\rho}f_{\rho}\nonumber,\\
\tilde{\alpha}_4^{p,\pi\pi/\pi\rho}=\alpha_4^{p,\pi\pi/\pi\rho} + r^{\pi/\rho}_{\chi}\alpha^{p, \pi\pi/\pi\rho}_{6},
&&\tilde{\alpha}_4^{p,\rho\pi}=\alpha_4^{p,\rho\pi} - r^{\pi}_{\chi}\alpha^{p, \rho\pi}_{6},\nonumber\\
\tilde{\alpha}_{4,EW}^{\pi\pi/\pi\rho}=\alpha^{p, \pi\pi /\pi\rho}_{10} + r^{\pi/\rho}_{\chi}\alpha^{p, \pi\pi/\pi\rho}_{8},
&&\tilde{\alpha}_{4,EW}^{\rho\pi}=\alpha^{p, \rho\pi}_{10}-r^{\pi}_{\chi}\alpha^{p, \rho\pi}_{8}.
\label{eq:GenPar}
\end{eqnarray}

Following \cite{Beneke:2001ev, Beneke:2003zv} we take

\begin{eqnarray}
m_q(\mu)=\frac{m^2_{\pi}}{(2 m^2_K - m^2_{\pi})}  m_s(\mu),
\end{eqnarray}

which leads to the condition  $r^{\pi}_{\chi}(\mu)=r^{K}_{\chi}(\mu)$.

\subsubsection{Vertices for the $B\rightarrow \pi\pi, \rho\pi, \pi\rho, \rho \rho$ decays}
\begin{align}
V^{\pi}_{1, 2, 4, 10}&=12 \hbox{ln} \frac{m_b}{\mu}-18 +\Bigl[-\frac{1}{2} - 3 i \pi + \Bigl(\frac{11}{2} - 3i\pi \Bigl)a^\pi_1
          -\frac{21}{20}a^\pi_2\Bigl],\nonumber\\
V^{\pi}_{6,8}&=-6,\nonumber\\
V^{\rho}_{1,2,3,9}&= V^{\rho}=  12 \hbox{ln} \frac{m_b}{\mu}-18 +\Bigl[-\frac{1}{2} - 3 i \pi + \Bigl(\frac{11}{2} - 3i\pi \Bigl)a^{\rho}_1
          -\frac{21}{20}a^{\rho}_2\Bigl],\nonumber\\
V^{\rho}_{4} &= 
\begin{cases}
    V^{\rho} &\text{ for } \bar{B}^0\rightarrow \pi^{+}\rho^{-}, \\
    \\
    V^{\rho}-\frac{C_{5}}{C_{3}}r^{\rho}_{\chi}V^{\rho}_{\perp} &\text{ for } B\rightarrow \rho\rho, \\
\end{cases}\nonumber\\
V^{\rho}_{\perp}&= 9 - 6 i\pi +\Bigl(\frac{19}{6} - i\pi\Bigl) a^{\rho}_{2,\perp}, \nonumber\\
V^{\rho}_{7}&=-12 \hbox{ln} \frac{m_b}{\mu}+6 -\Bigl[-\frac{1}{2} - 3 i \pi - \Bigl(\frac{11}{2} - 3 i \pi\Bigl) a^{\rho}_1 - \frac{21}{20} a^{\rho}_{2} \Bigl], \nonumber\\
V^{\rho}_{6,8}&=9-6i\pi + \Bigl(\frac{19}{6}-i\pi\Bigl)a^{\rho}_{2,\perp},\nonumber\\
V^{\rho}_{10} &= 
\begin{cases}
    V^{\rho} &\text{ for } \bar{B}^0\rightarrow \pi^{+}\rho^{-}, \\
    \\
    V^{\rho}-\frac{C_7}{C_9}r^{\rho}_{\chi}V^{\rho}_{\perp} &\text{ for } B\rightarrow \rho\rho. \\
\end{cases}\nonumber\\
\label{eq:Vertices}
\end{align}

\subsubsection{Vertices for the $B\rightarrow J/\psi \phi$ decay.}
\label{eq:vertexJPsiPhi}

\begin{eqnarray}
V^i_{J/\psi \phi}&=
\begin{cases}
-18 -12 \ln\frac{\mu}{m_b}+ f^h_{I}& \text{ for $i=1,3,9$}\\
\\
-6 -12 \ln\frac{\mu}{m_b}+ f^h_{I}& \text{ for $i=5, 7$}\\
\end{cases}\nonumber\\
\end{eqnarray}

\begin{eqnarray}
f^h_I&=&
\begin{cases}
	f_I + g_I\cdot (1-\tilde{z})\frac{A^{BK*}_0}{A^{BK*}_3}& \text{ for $h=0$}\\
\\
f_I& \text{ for $h=\pm$}\\
\end{cases}\nonumber\\
\end{eqnarray}

\begin{eqnarray}
f_{I}&=&\int^{1}_{0}d\xi \Phi_{||}^{J/\psi}(\xi)\Bigl\{\frac{2 \tilde{z} \xi}{1-\tilde{z}(1-\xi)}+ (3-2\xi)\frac{\hbox{ln}\xi}{1-\xi} \nonumber\\
&& + \Bigl( -\frac{3}{1-\tilde{z}\xi} + \frac{1}{1-\tilde{z}(1-\xi)} - \frac{2 \tilde{z} \xi}{[1-\tilde{z}(1-\xi)]^2}\Bigl)\times
\tilde{z}\xi \ln[\tilde{z} \xi] \nonumber\\
&&+ \Bigl(3(1-\tilde{z}) + 2 \tilde{z} \xi + \frac{2 \tilde{z}^2 \xi^2}{1-\tilde{z}(1-\xi)}\Bigl)\times \frac{\ln(1-\tilde{z})-i \pi }{1-\tilde{z}(1-\xi)}\Bigl\}\nonumber\\
&&+\int^{1}_{0}d\xi \Phi_{\perp}^{J/\psi}(\xi)\Bigl\{-4r\frac{\ln\xi}{1-\xi}+\frac{4\tilde{z} r\ln[\tilde{z}\xi]}{1-\tilde{z}(1-\xi)}\nonumber\\
&&-4\tilde{z}r\frac{\hbox{ln}(1-\tilde{z})-i\pi}{1-\tilde{z}(1-\xi)}\Bigl\}
\label{eq:VIh}
\end{eqnarray}

\begin{eqnarray}
g_I&=&\int^1_0 d\xi \Phi^{J/\Psi}_{||}(\xi)\Bigl\{
	\frac{-4\xi}{(1-\tilde{z})(1-\xi)}\ln\xi + \frac{\tilde{z}\xi}{(1-\tilde{z}(1-\xi))^2}\ln(1-\tilde{z})\nonumber\\
	&&+\Biggl(\frac{1}{(1-\tilde{z}\xi)^2} - \frac{1}{(1-\tilde{z}(1-\xi))^2}
	+ \frac{2(1+\tilde{z}-2\tilde{z}\xi)}{(1-\tilde{z})(1-\tilde{z}\xi)^2}\Biggl)\times\tilde{z}\xi
	\ln[\tilde{z}\xi]\nonumber\\
	&&-i\pi\frac{\tilde{z}\xi}{(1-\tilde{z}(1-\xi))^2}\Bigl\}
        +\int^1_0 d\xi \Phi^{J/\Psi}_{\perp}(\xi)\Bigl\{\frac{4r}{(1-\tilde{z})(1-\xi)}\ln\xi\nonumber\\
	&&-\frac{4r\tilde{z}}{(1-\tilde{z})(1-\tilde{z}\xi)}\ln[\tilde{z}\xi]\Bigl\}	
\label{eq:VIh2}
\end{eqnarray}

for 

\begin{eqnarray}
\tilde{z}= \frac{m^2_{J/\Psi}}{m^2_B}, && r=2\cdot \Bigl(\frac{m_c}{m_{J/\Psi}}\Bigl)^2.
\end{eqnarray}

\subsubsection{Penguin functions}

To simplify the following equations we have denoted $M=\pi, \rho$ when the corresponding expressions apply to both $\pi$ and $\rho$ mesons. In addition we have used

\begin{eqnarray}
s_p=\Bigl(\frac{m_p}{m_b}\Bigl)^2,
\end{eqnarray}

for $p=u,c$, although in practice we consider $s_u=0$. Notice that in the following equations the symbol ``hat'' does not denote an operator and is used 
to distinguish the different kind of functions under consideration.

\begin{align}
P^{p,M}_{1, 2, 3}&=P^{M}_{1, 2, 3} =0,\nonumber\\
P^{p,\pi}_{4}&=\frac{C_F \alpha_s}{4\pi N_c}\Bigl\{C_2\Bigl[\frac{4}{3}\hbox{ln}\frac{m_b}{\mu}+\frac{2}{3} - G_{\pi}(s_p) \Bigl] + C_{3}\Bigl[\frac{8}{3}\hbox{ln}\frac{m_b}{\mu} + \frac{4}{3} - G_{\pi}(0) -  
G_{\pi}(1) \Bigl] \nonumber\\
&+ \Bigl(C_4 + C_6 \Bigl)\Bigl[\frac{4 n_f}{3}\hbox{ln}\frac{m_b}{\mu} - (n_f -2) G_{\pi}(0) -G_{\pi}(s_c) - G_{\pi}(1) \Bigl] \nonumber\\
&- 6C^{eff}_{8g}\Bigl( 1 +\alpha^{\pi}_{1} +\alpha^{\pi}_2  \Bigl)\Bigl\},\nonumber\\
P^{p,M}_{6}&= \frac{C_F \alpha_s}{4\pi N_c}\Bigl\{ C_2\Bigl[\frac{4}{3}\hbox{ln}\frac{m_b}{\mu}+\frac{2}{3} - \hat{G}_{M}(s_p) \Bigl] + C_{3}\Bigl[\frac{8}{3}\hbox{ln}\frac{m_b}{\mu} + \frac{4}{3} - \hat{G}_{M}(0) -  
\hat{G}_{M}(1) \Bigl] \nonumber\\
&+ \Bigl(C_4 + C_6 \Bigl)\Bigl[\frac{4 n_f}{3}\hbox{ln}\frac{m_b}{\mu} - (n_f - 2) \hat{G}_{M}(0) -\hat{G}_{M}(s_c) - \hat{G}_{M}(1) \Bigl] 
- 2C^{eff}_{8g}\Bigl\},\nonumber\\
P^{p,\pi}_{8}&= \frac{\alpha}{9\pi N_c}\Bigl\{\Bigl(N_c C_1 + C_2\Bigl) \Bigl[\frac{4}{3}\hbox{ln}\frac{m_b}{\mu} + \frac{2}{3}- 
\hat{G}_{\pi}(s_p) \Bigl] - 3C^{eff}_{7} \Bigl\},\nonumber\\
P^{p,M}_{10}&= \frac{\alpha}{9\pi N_c}\Bigl\{\Bigl(N_c C_1 + C_2\Bigl)\Bigl[\frac{4}{3}\hbox{ln}\frac{m_b}{\mu} + \frac{2}{3}- 
G_{M}(s_p)\Bigl] - 9C^{eff}_{7} \Bigl(1 + \alpha^{M}_1 + \alpha^{M}_2   \Bigl)\Bigl\},\nonumber\\
P^{p, \rho}_{4} &= 
\begin{cases}
    P'^{p,\rho}_{4}  &\text{ for } \bar{B}^0\rightarrow \pi^{+}\rho^{-}, \\
    \\
    P'^{p,\rho}_{4} - r^{\rho}_{\chi}P''^{p,\rho}_{4} &\text{ for } B\rightarrow \rho\rho, \\
\end{cases}\nonumber\\
P'^{p,\rho}_{4}&=\frac{C_F \alpha_s}{4\pi N_c}\Bigl\{C_2\Bigl[\frac{4}{3}\hbox{ln}\frac{m_b}{\mu}+\frac{2}{3} - G_{\rho}(s_p) \Bigl] + C_{3}\Bigl[\frac{8}{3}\hbox{ln}\frac{m_b}{\mu} + \frac{4}{3} - G_{\rho}(0) -  
G_{\rho}(1) \Bigl]\nonumber\\
&+ \Bigl(C_4 + C_6 \Bigl)\Bigl[\frac{4 n_f}{3}\hbox{ln}\frac{m_b}{\mu} - (n_f -2) G_{\rho}(0) -G_{\rho}(s_c) - G_{\rho}(1) \Bigl] \nonumber\\
&- 6C^{eff}_{8g}\Bigl( 1 +\alpha^{\rho}_{1} +\alpha^{\rho}_2  \Bigl)\Bigl\},\nonumber\\
P''^{p,\rho}_{4}&=-\Bigl[C_2 \hat{G}_{\rho}(s_p) + C_{3}\Bigl(\hat{G}_{\rho}(0)+\hat{G}_{\rho}(1)\Bigl) \nonumber\\
&+ \Bigl(C_4 + C_6\Bigl)\Bigl(3\hat{G}_{\rho}(0) + \hat{G}_{\rho}(s_p) + \hat{G}_{\rho}(1)\Bigl) \Bigl],\nonumber\\
P_{7,9}^{u,\rho}&=\frac{\alpha}{9\pi}\Bigl\{\Bigl(N_c C_1 + C_2\Bigl) \Bigl[\frac{ 4}{ 3}\frac{m_b}{\mu} - \frac{10}{9} +\frac{4\pi^2}{3}\sum\limits_{r=\rho,\omega}\frac{f^2_r }{m^2_{\rho} - m^2_{r} + i m_r \Gamma_r }\nonumber\\
& -\frac{2\pi}{3}\frac{m^2_{\rho}}{ t_c} i + \frac{2}{ 3}\hbox{ln}\frac{m^2_{\rho}}{m^2_{b}} + \frac{2}{3}\frac{t_c - m^2_{\rho} }{t_c}\hbox{ln}\frac{t_c -m_{\rho}^2}{m^2_{\rho}}\Bigl] - 3C^{eff}_{7, \gamma}\Bigl\},\nonumber
\end{align}

\begin{align}
P_{7,9}^{c,\rho}&= \frac{\alpha}{9\pi}\Bigl\{ \Bigl(N_c C_1 + C_2\Bigl)\Bigl[\frac{4}{3}\hbox{ln}\frac{m_b}{\mu} +\frac{2}{3}  + \frac{4}{3}\hbox{ln}\frac{m_c}{m_b}\Bigl]-3C^{eff}_{\gamma}\Bigl\},\nonumber\\
P^{p,\rho}_{8}&=-\frac{\alpha}{9\pi N_c}\Bigl( N_c C_1  + C_2\Bigl) \hat{G}_{\rho}(s_p), \nonumber\\
P^{p,\rho}_{10}&=\frac{\alpha}{9\pi N_c }\Bigl( P'^{p, \rho}_{10} + r^{\rho}_{\chi}P''^{p,\rho}_{10}  \Bigl),\nonumber\\
P'^{p, \rho}_{10}&=\Bigl(N_c C_1 + C_2 \Bigl)\Bigl[\frac{4}{3}\hbox{ln}\frac{m_b}{\mu} + \frac{2}{3} -G_{\rho}(s_p)\Bigl]-9C^{eff}_{7,\gamma}\Bigl( 1 +\alpha^{\rho}_{1}
+ \alpha^{\rho}_{2}\Bigl),\nonumber\\
P''^{p,\rho }_{10}&=\Bigl( N_c C_1 + C_2\Bigl)\hat{G}_{\rho}(s_p).
\label{eq:Penguins}
\end{align}

For the calculation of $P_{7,9}^{u,\rho}$ above the symbol $t_c$ denotes

\begin{eqnarray}
t_c=4\pi^2 (f^2_{\rho} +  f^2_{\omega}).
\end{eqnarray}

Extra functions required for the evaluation of the penguin contributions 

\begin{eqnarray}
G_{M}(s_c)&=&\frac{5}{3}-\frac{2}{3}\hbox{ln}(s_c) +\frac{\alpha^{M}_1}{2}+\frac{\alpha_{2}^{M}}{5} + \frac{4}{3}\Bigl( 8 +9\alpha^{M}_{1}+9\alpha^{M}_{2}\Bigl)s_c\nonumber\\
    &&+2\Bigl(8 + 63\alpha^{M}_1 + 214 \alpha^{M}_{2} \Bigl)s_c^2 - 24 \Bigl( 9\alpha^{M}_{1}+80\alpha^{M}_{2}\Bigl)s_c^3 \nonumber\\
&&+ 2880\alpha^{M}_{2}s_c^4-\frac{2}{3}\sqrt{1-4s_c}\Bigl(2\hbox{arctanh}\sqrt{1-4s_c} -i \pi   \Bigl)    \Bigl[1+2s_c  \nonumber\\
&&+ 6\Bigl(4 + 27 \alpha^{M}_{1} + 78 \alpha^{M}_{2}\Bigl)s_c^2- 36 \Bigl( 9 \alpha_1^{M}+ 70 \alpha^{M}_{2}\Bigl)s^3_{c} + 4320 \alpha_2^{M}s_c^4\Bigl] \nonumber\\
    &&   + 12 s^2_c \Bigl(2\hbox{arctanh}\sqrt{1-4s_c} -i \pi   \Bigl)^2   \Bigl[1+3\alpha^{M}_{1}+6\alpha_2^{M}-\frac{4}{3}\Bigl(1+9\alpha^{\rho}_{1} \nonumber\\
&&+ 36\alpha^{M}_2\Bigl)s_c + 18 \Bigl(\alpha_1^{M} + 10 \alpha^{M}_{2}\Bigl)s^2_c -240 \alpha^{M}_2s_c^3\Bigl],\nonumber\\
G_{M}(0)&=&\frac{5}{3}+\frac{2 i \pi}{3}+\frac{\alpha^{M}_{1}}{2}+\frac{\alpha^{M}_2}{5},\nonumber\\
G_{M}(1)&=&\frac{85}{3}- 6\sqrt{3}\pi+\frac{4\pi^2}{9}-\Bigl( \frac{155}{2}-36\sqrt{3}\pi +12\pi^2\Bigl)\alpha^{M}_1+\Bigl(\frac{7001}{5} \nonumber\\
&& -504\sqrt{3}\pi + 136\pi^2\Bigl)\alpha^{M}_2,\nonumber\\
\hat{G}^{p}_{\pi}(s_c)&=& \frac{16}{9} \Bigl( 1 - 3 s_c\Bigl) - \frac{2}{3}\Bigl[\hbox{ln}(s_c) + \Bigl(1-4s_c\Bigl)^{3/2}\Bigl(2\hbox{arctan}\sqrt{1-4s_c}-i\pi \Bigl)\Bigl],\nonumber\\
\end{eqnarray}

\begin{eqnarray}
\hat{G}^{p}_{\pi}(0)&=& \frac{16}{9} + \frac{2\pi i}{3}, \nonumber\\
\hat{G}^{p}_{\pi}(1)&=& \frac{2\pi}{\sqrt{3}}-\frac{32}{9},\nonumber\\
\hat{G}_{\rho}(s_c) &=& 1 + \frac{\alpha^{\rho}_{1,\perp}}{3} + \frac{\alpha^{\rho}_{2,\perp}}{6} - 4 s_c \Bigl( 9 + 12 \alpha^{\rho}_{1,\perp} + 
14 \alpha^{\rho}_{2,\perp} \Bigl)\nonumber -6 s^2_c \Bigl(8 \alpha^{\rho}_{1,\perp}  \nonumber\\
&&+ 35 \alpha^{\rho}_{2,\perp} \Bigl) + 360 s^3_c \alpha^{\rho}_{2,\perp} + 12 s_c \sqrt{1-4s_c}\Bigl(1 +\Bigl[1 + 4 s_c\Bigl] \alpha^{\rho}_{1,\perp}\nonumber\\
&&+ \Bigl[1+ 15 s_c  - 30 s^2_c \Bigl] \alpha^{\rho}_{2,\perp} \Bigl) \Bigl(2\hbox{arctanh}\sqrt{1-4s_c} -i \pi \Bigl)\nonumber\\
&&- 12 s^2_c \Bigl(1 + \Bigl[3-4 s_c\Bigl]\alpha^{\rho}_{1,\perp}+ 2 \Bigl[3-10 s_c + 15 s^2_c \Bigl] \alpha^{\rho}_{2,\perp} \Bigl)  \nonumber\\
&&\times\Bigl(2\hbox{arctanh}\sqrt{1-4s_c} -i \pi \Bigl)^2, \nonumber\\
\hat{G}_{\rho}(0)&=&1 + \frac{1}{3}\alpha^{\rho}_{1,\perp} + \frac{1}{6}\alpha^{\rho}_{2,\perp},\nonumber\\
\hat{G}_{\rho}(1)&=&-35 + 4\sqrt{3}\pi + \frac{4\pi^2}{3} +\Bigl(-\frac{287}{3} + 20\sqrt{3}\pi - \frac{4\pi^2}{3}\Bigl)\alpha^{\rho}_{1,\perp}\nonumber\\
&&+\Bigl(\frac{565}{6} - 56\sqrt{3}\pi + \frac{64\pi^2}{3} \Bigl)\alpha^{\rho}_{2,\perp}.
\end{eqnarray}

\subsubsection{Hard Scattering functions for the $B\rightarrow \pi\pi, \rho\pi, \pi\rho, \rho \rho$ decays.}
\begin{eqnarray}
H_{1, 2, 4, 10}^{\pi\pi}(\mu)&=&\frac{B_{\pi\pi}}{A_{\pi\pi}}\frac{m_B}{\lambda_B}\Bigl( 9 \Bigl[1+ a^{\pi}_1 + a^{\pi}_2\Bigl]^2 + 3r^{\pi}_{\chi}(\mu)\Bigl[1-a^{\pi}_1+a^{\pi}_{2}\Bigl]X_H \Bigl),\nonumber\\
H_{6,8}^{\pi\pi}(\mu)&=&0,\nonumber\\
H_{2, 4, 10}^{\pi\rho}(\mu)&=& \frac{B_{\pi\rho}}{A_{\pi\rho}}\frac{m_B}{\lambda_B}\Bigl( 9\Bigl[1+a^{\pi}_1 + a^{\pi}_2\Bigl]\Bigl[1+a^{\rho}_1 + a^{\rho}_2\Bigl] + 3 r^{\pi}_{\chi}(\mu) \Bigl[1 - a^{\rho}_1\nonumber\\
&& + a^{\rho}_2\Bigl]X_H\Bigl),\nonumber\\
H_{6,8}^{\pi\rho}(\mu)&=&0,\nonumber\\
H_{2, 4, 10}^{\rho\pi}&=& \frac{B_{\rho\pi}}{A_{\rho\pi} }\frac{m_B}{\lambda_B}\Bigl( 9\Bigl[1+a^{\pi}_1 + a^{\pi}_2\Bigl]\Bigl[1+a^{\rho}_1 + a^{\rho}_2\Bigl] 
+ 3 r^{\rho}_{\chi}(\mu) \Bigl[1 - a^{\pi}_1 \nonumber\\
&&+ a^{\pi}_2\Bigl]\Bigl[3(1 + a^{\rho}_{1,\bot} + a^{\rho}_{2,\bot})X_H -(6 +9 a^{\rho}_{1,\bot} + 11 a^{\rho}_{2,\bot})\Bigl]\Bigl),\label{eq:Hi_rhopi}\nonumber
\end{eqnarray}
\begin{eqnarray}
H_{6,8}^{\rho\pi}(\mu)&=&0,\nonumber\\
H_{1,2,4,9,10}^{\rho\rho}(\mu)&=&\frac{B_{\rho\rho}}{A_{\rho\rho}}\Bigl[\frac{m_{B_d}}{\lambda_B}\Bigl]\Bigl[9\Bigl(1+a_1^{\rho}+a^{\rho}_2\Bigl)^2 +  9r^{\rho}_{\chi}(\mu)\Bigl(1-a^{\rho}_{1}+a_{2}^{\rho}\Bigl)\nonumber\\
&&\times\Bigl(X_{H}-2\Bigl)\Bigl],\nonumber\\
H_{7}^{\rho\rho}(\mu)&=&-\frac{B_{\rho\rho}}{A_{\rho\rho}}\Bigl[\frac{m_{B_d}}{\lambda_B}\Bigl]\Bigl[9\Bigl(1+a_1^{\rho}+a^{\rho}_2\Bigl)\Bigl(1-a_1^{\rho}+a^{\rho}_2\Bigl)  +  9r^{\rho}_{\chi}(\mu)\nonumber\\
&&\times \Bigl(1+a^{\rho}_{1} +a_{2}^{\rho}\Bigl)\Bigl(X_{H}-2\Bigl)\Bigl].
\label{eq:HardScattering}
\end{eqnarray}

\subsubsection{Hard scattering function for the $B\rightarrow J/\psi \phi$}

For the amplitudes of the decay  $B\rightarrow J/\psi \phi$, the spectator interaction functions depend on the polarization of the final states,
for $h=0,\pm$ we have

\begin{eqnarray}
\label{eq:HJPsiphi}	
H^{J/\psi \phi, 0}_{1,3,9}&=&\frac{f_B f_{J/\psi} f_{\phi}}{\tilde{h}^0}
\int^1_0d\xi
	\frac{\Phi^B_1(\xi)}{\xi}\int^1_0 d\tilde{\xi}\frac{\Phi^{J/\Psi}(\tilde{\xi})}{\tilde{\xi}}\int^1_0 d\bar{\eta}\frac{\Phi^{\phi}(\bar{\eta})}{\bar{\eta}},\nonumber\\	
H^{J/\psi \phi, \pm}_{1,3,9}&=&
	\frac{2f_B f_{J/\Psi} f_{\phi} m_{J/\Psi}m_{\phi}}{m^2_B \tilde{h}^{\pm}(1-\tilde{z}) }\int^1_0d\xi
	\frac{\Phi^B_1(\xi)}{\xi}\int^1_0 d\tilde{\xi}\frac{\Phi^{J/\Psi}(\tilde{\xi})}{\tilde{\xi}}\cdot\nonumber\\
	&&\int^1_0 d\bar{\eta}\Biggl[\frac{\Phi^{\phi, v}_{\perp}(\bar{\eta})}{\bar{\eta}} \pm \frac{\Phi^{\phi, a}_{\perp}(\bar{\eta})}{4\bar{\eta}^2}\Biggl],\nonumber\\	
	H^{J/\psi \phi, h}_{5,7}&=&-H^{J/\psi \phi, h}_{1,3,9}.	
\end{eqnarray}

The helicity functions in the denominators of Eqs.~(\ref{eq:HJPsiphi}) are

\begin{eqnarray}
	\tilde{h}^0&=&\frac{f_{J/\psi}}{2m_{\phi}}\Biggl[\Bigl(m^2_{B} - m^2_{J/\Psi} -m^2_{\phi}\Bigl)\Bigl(m_{B} + m_{\phi}\Bigl)
	A^{B\rightarrow \phi}_{1}(m^2_{J/\psi}) - \frac{4 m^2_{B} p^2_c}{m_B + m_{\phi}} A^{B\rightarrow \phi}_{2}(m^2_{J/\psi})   \Biggl],\nonumber\\
	\tilde{h}^{\pm}&=&m_{J/\psi}f_{J/\psi}\Biggl[ \Bigl(m_B + m_{\phi} \Bigl)A^{B\rightarrow \phi}_{1} (m^2_{J/\psi}) 
	\pm  \frac{2m_B p_c}{m_B + m_{\phi}} V^{B\rightarrow \phi}(m^2_{J/\psi}) \Biggl],
\end{eqnarray}

with

\begin{eqnarray}
p_c=\frac{\sqrt{\Bigl(m^2_{\phi}-m^2_{J/\psi}\Bigl)^2 + m^2_B\Bigl(m^2_B-2\Bigl[m_{J/\psi} + m^2_{\phi}\Bigl]\Bigl)}}{2m_B}.	
\end{eqnarray}

The form factors $A^{B\rightarrow \phi}_{1,2}(m^2_{J/\psi})$ and $V^{B\rightarrow \phi}(m^2_{J/\psi})$ used for the evaluation 
of the functions $\tilde{h}^0$ and $\tilde{h}^{\pm}$  were calculated based on \cite{Bharucha:2012wy}, the corresponding numerical values 
can be found in Appendix \ref{Sec:Inputs}.\\

The twist-3 distribution amplitudes of the $\phi$ meson in Eqs.~(\ref{eq:HJPsiphi}) have been denoted by $\Phi^{\phi, a}_{\perp}(x)$ and $\Phi^{\phi, v}_{\perp}(x)$, 
they are given explicitly by

\begin{eqnarray}
	\Phi^{\phi, a}_{\perp}(x)&=&6x(1-x)\Biggl[1 + a^{||}_1 \Bigl[2x-1\Bigl] +\Biggl\{\frac{1}{4}a^{||}_2 +
	\frac{5}{3} \zeta_3 \Bigl(1-\frac{3}{16}\omega^{A,\phi}_3\nonumber\\
	&&+ \frac{9}{16}\omega^{V,\phi}_3 \Bigl)\Biggl\}\Biggl(5\Bigl[2x-1\Bigl]^2 - 1\Biggl) + 6\delta_+\Biggl\{3x(1-x)\nonumber\\
	&&+ (1-x)\ln(1-x)
	+ x\ln x\Biggl\} + 6\delta_{-}\Biggl\{(1-x)\ln(1-x) -x\ln x \Biggl\}\Biggl],\nonumber\\
	\Phi^{\phi, v}_{\perp}(x)&=&\frac{3}{4}\Biggl\{1+\Bigl[2x-1\Bigl]^2  \Biggl\}    + \frac{3}{2}a^{||}_{1}\Bigl[2x-1\Bigl]^3
	+ \Biggl\{\frac{3}{7}a^{||}_{2} + 5\zeta_3\Biggl\}\Biggl\{3 \Bigl[2x-1\Bigl]^2 - 1\Biggl\}\nonumber\\
	&& + \Biggl\{\frac{9}{112}a^{||}_{2} + \frac{15}{64}\zeta_3\Biggl[3\omega^V_3 - \omega^A_3\Biggl]\Biggl\}
	    \Biggl\{3 - 30 \Bigl[2x-1\Bigl]^2 + 35 \Bigl[2x-1\Bigl]^4\Biggl\} \nonumber\\
	&& + \frac{3}{2}\delta_{+}\Biggl\{2 + \ln x + \ln[1-x]\Biggl\} +\frac{3}{2}\delta_{-}\Biggl\{ 2 \Bigl[2x-1\Bigl] +\ln(1-x)
	-\ln x \Biggl\}.\nonumber\\
	\label{eq:gphi}
\end{eqnarray}

For the rest of the LCD amplitudes of the vector mesons $J/\psi$ and $\phi$ in Eqs.~(\ref{eq:VIh}), (\ref{eq:VIh2}) and (\ref{eq:HJPsiphi}) we use the leading term in 
the Gegenbauer expansion

\begin{eqnarray}
\Phi^V(\xi)&=&6\xi(1-\xi).
\end{eqnarray}

For different hadronic parameters required for the numerical evaluation of Eq.~(\ref{eq:gphi}) we use \cite{Ball:1998fj}

\begin{eqnarray}
	\zeta_3=0.023,\quad \quad\omega^{A}_{3}=0,\quad\quad\omega^V_3=3.7,\quad \quad \delta_{+}=0.41,\quad \quad \delta_{-}=0. 
\end{eqnarray}

The divergences encountered when integrating the twist-3 distribution amplitudes in Eqs.~(\ref{eq:HJPsiphi}) are parameterized following 
the model in Eq.~(\ref{eq:XH}).

\subsection{Annihilation coefficients}

\begin{eqnarray}
\beta_{i}^{p, M_1 M_2 }&=&\frac{B_{M_1M_2}}{A_{M_1M_2}}b^{p, M_1 M_2}_{i}\nonumber\\
b^{M_1M_2}_{1}&=&\frac{C_F}{N^2_c}C_{1}A^{i, M_1 M_2}_1 \nonumber\\
b^{M_1M_2}_{2}&=&\frac{C_F}{N^2_c}C_{2}A^{i, M_1 M_2}_1 \nonumber\\
b^{p, M_1 M_2}_{3}&=&\frac{C_F}{N^2_c}\Bigl[C_3 A^{i,M_1 M_2}_1 + C_5\Bigl(A^{i, M_1 M_2}_3+A^{f, M_1 M_2}_3\Bigl)+N_cC_6A^{f, M_1 M_2}_3\Bigl]\nonumber\\
b^{p, M_1 M_2}_{4}&=&\frac{C_F}{N^2_c}\Bigl[C_4A^{i, M_1 M_2}_1+C_6A^{i, M_1 M_2}_2\Bigl]\nonumber\\
b^{p, M_1 M_2}_{3,EW}&=&\frac{C_F}{N^2_c}\Bigl[C_9A^{i, M_1 M_2}_1+C_7\Bigl(A^{i, M_1 M_2}_3+A^{f, M_1 M_2}_3\Bigl)+N_c C_8 A^{f, M_1 M_2}_3\Bigl]\nonumber\\
b^{p, M_1 M_2}_{4,EW}&=&\frac{C_F}{N^2_c}\Bigl[C_{10}A^{i, M_1 M_2}_1+C_8A^{i, M_1 M_2}_2\Bigl]
\label{eq:annihilation}
\end{eqnarray}

\subsection{Annihilation kernels}

\begin{eqnarray}
A^{i,\pi\pi}_{1}&\approx& A^{i, \pi\pi}_{2}\approx 2\pi\alpha_s(\mu_h)\Bigl[9\Bigl(X_A-4+\frac{\pi^2}{3}\Bigl) + r^{\pi}_{\chi}r^{\pi}_{\chi}X^2_{A}\Bigl]\nonumber\\
 A^{i, \pi\rho}_{1}&=&A^{i, \rho\pi}_{1}\approx6\pi\alpha_s\Bigl[3\Bigl(X_A-4+\frac{\pi^2}{3}\Bigl)+
                            r^{\rho}_{\chi}r^{\pi}_{\chi}\Bigl(X_A^2-X_A\Bigl)\Bigl]\nonumber\\
A^{i, \pi\rho}_{2}&=&A^{i,\rho\pi}_{2}\approx-A^{i, \pi\rho}_{1} \nonumber\\
A^{i,\pi\pi}_{3}&\approx& 0\nonumber\\ 
A^{i,\pi\rho}_{3}&=& A^{i,\rho\pi}_{3}\approx 6\pi\alpha_s\Bigl[-3r^{\rho}_{\chi}\Bigl(X^2_A-2X_A-\frac{\pi^2}{3}+4\Bigl)+
r^{\pi}_{\chi}\Bigl(X^2_A-2X_A+\frac{\pi^2}{3}\Bigl)\Bigl]\nonumber\\
A^{f,\pi\rho}_{1}&=&A^{f,\pi\rho}_{2}= A^{f,\rho\pi}_{1}=A^{f,\rho\pi}_{2}= 0\nonumber\\
A^{f,\pi\pi}_{3}&\approx& 12\pi\alpha_sr^{\pi}_{\chi}\Bigl(2X^2_A-X_A\Bigl)\nonumber\\
A^{f,\pi\rho}_{3}&\approx&-6\pi\alpha_s\Bigl[3r^{\pi}_{\chi}\Bigl(2X_A-1\Bigl)\Bigl(X_A-2\Bigl)+r^{\rho}_{\chi}\Bigl(2X^2_A-X_A\Bigl)\Bigl]\nonumber\\
A^{f,\rho\pi}_{3}&=&-A^{f,\pi\rho}_{3}\approx 6\pi\alpha_s\Bigl[3r^{\rho}_{\chi}\Bigl(2X_A-1\Bigl)\Bigl(2 - X_A\Bigl) -r^{\pi}_{\chi}\Bigl(2X^2_A-X_A\Bigl)\Bigl]\nonumber\\
A^{i,\rho\rho}_{1}&=& A^{i,\rho\rho}_{2} \approx 18\pi\alpha_s\Bigl[\Bigl(X_A -4 + \frac{\pi^2}{3} \Bigl) + (r^{\rho}_{\chi})^2(X_A-2)^2 \Bigl]\nonumber\\
A^{i,\rho\rho}_{3}&=&0  \nonumber\\
A^{f,\rho \rho}_{3}&\approx& -36\pi \alpha_s r^{\rho}_{\chi}\Bigl(2X^2_A -5 X_A + 2  \Bigl)\nonumber\\
\label{eq:annihilation2}
\end{eqnarray}

\bibliographystyle{utphys}
\bibliography{paper_jhep}

\end{document}